\documentclass[11pt,openright,oneside,letterpaper,onecolumn]{report}  
\pdfoutput=1

\newcommand{\thesistitle}{Topics in vacuum decay}
\newcommand{\thesisauthor}{Ali Masoumi}
\newcommand{\thesisyear}{2013}

\usepackage{amsmath}
\usepackage{amssymb}
\usepackage{mathtools}
\usepackage{fancyhdr}
\usepackage{afterpage}
\usepackage{lineno}

\usepackage{slashed}
\usepackage{mathrsfs}
\usepackage{graphicx} 
\usepackage{multirow}
\usepackage{braket}
\usepackage{subfig}

\usepackage[numbers,sort&compress]{natbib}
\newcommand*\patchAmsMathEnvironmentForLineno[1]{%
  \expandafter\let\csname old#1\expandafter\endcsname\csname #1\endcsname
  \expandafter\let\csname oldend#1\expandafter\endcsname\csname end#1\endcsname
  \renewenvironment{#1}%
     {\linenomath\csname old#1\endcsname}%
     {\csname oldend#1\endcsname\endlinenomath}}%
\newcommand*\patchBothAmsMathEnvironmentsForLineno[1]{%
  \patchAmsMathEnvironmentForLineno{#1}%
  \patchAmsMathEnvironmentForLineno{#1*}}%
\AtBeginDocument{%
\patchBothAmsMathEnvironmentsForLineno{equation}%
\patchBothAmsMathEnvironmentsForLineno{align}%
\patchBothAmsMathEnvironmentsForLineno{flalign}%
\patchBothAmsMathEnvironmentsForLineno{alignat}%
\patchBothAmsMathEnvironmentsForLineno{gather}%
\patchBothAmsMathEnvironmentsForLineno{multline}%
}
\usepackage[pdftitle={\thesistitle},pdfauthor={\thesisauthor},pdfpagemode={UseOutlines},letterpaper,bookmarks,bookmarksopen=true,pdfstartview={FitH},bookmarksnumbered=true]{hyperref}

\newcommand{\singlespace}{\renewcommand{\baselinestretch}{1.15} \small \normalsize}

\newcommand{\doublespace}{\renewcommand{\baselinestretch}{1.5} \small \normalsize}
\newcommand{\normalspace}{\doublespace}
\newcommand{\thesistitlepage}{
    \normalspace
    \thispagestyle{empty}
    \begin{center}
        \textbf{\LARGE \thesistitle} \\[1cm]
        \textbf{\LARGE \thesisauthor} \\[8cm]
        Submitted in partial fulfillment of the \\
        requirements for the degree \\
        of Doctor of Philosophy \\
        in the Graduate School of Arts and Sciences \\[4cm]
        \textbf{\Large COLUMBIA UNIVERSITY} \\[5mm]
        \thesisyear
    \end{center}
    \clearpage
}

\newcommand{\thesiscopyrightpage}{
    \thispagestyle{empty}
    \strut \vfill
    \begin{center}
      \copyright \thesisyear \\
      \thesisauthor \\
      All Rights Reserved
    \end{center}
    \cleardoublepage
}

\begin{document}

\pagestyle{empty}

\thesistitlepage
\thesiscopyrightpage

    \thispagestyle{empty}
    \begin{center}
    \textbf{\LARGE ABSTRACT} \\[1cm]
     \textbf{\LARGE \thesistitle} \\[1cm]
     \textbf{\LARGE \thesisauthor} \\[1cm]
    \end{center}
    If a theory has more than one classically stable vacuum, quantum tunneling and thermal jumps make the transition between the vacua possible. The transition happens through a first order phase transition started by nucleation of a bubble of the new vacuum. The outward pressure of the truer vacuum makes the bubble expand and consequently eat away more of the old phase.  In the presence of gravity this phenomenon gets more complicated and meanwhile more interesting. It can potentially have important cosmological consequences.  Some aspects of this decay are studied in this thesis. Solutions with different symmetry than the generically used $O(4)$ symmetry are studied and their actions calculated.  Vacuum decay in a spatial vector field is studied and novel features like kinky domain walls are presented. The question of stability of vacua in a landscape of potentials is studied and the possible instability in large dimension of fields is shown. Finally a compactification of the Einstein-Maxwell theory is studied which can be a good lab to understand the decay rates in compactification models of arbitrary dimensions. 
    \cleardoublepage


\pagenumbering{roman}
\pagestyle{plain}

\setlength{\footskip}{0.5in}

\setcounter{tocdepth}{3}
\renewcommand{\contentsname}{Table of Contents}
\tableofcontents
\cleardoublepage

\listoffigures
\cleardoublepage
\listoftables 
\cleardoublepage


%
~\\[1in] 
\textbf{\Huge Acknowledgments}\\

First I express my deepest appreciation and thanks to my advisor Professor Erick J. Weinberg for his advice, friendship, wisdom, patience and generosity. Without his help, this thesis would have been impossible. He is much more than an advisor for me,  a role model I tried to learn from and become similar to one day. His extraordinary patience with me during the sluggish progress in the beginning of my PhD program helped me to not give up and try harder. His sensitivity to the problems his students are facing with is beyond what I can express  in this short acknowledgement. I am very grateful for the enjoyable discussions we had on physics, history, politics, economy and linguistics in long afternoons which he patiently shared his vast knowledge with me. It has been a great honor    to learn from  such a  great  scientist.

I thank Professors Gabadadze and Greene  for giving me the opportunity to collaborate with and learn from and Professor Vilenkin for the opportunity to continue what I enjoy the most, physics.

I am thankful to Professors Aleiner, Ardalan, Arfaei, BahmanAbadi,   Beloborodov, Blaer, Christ, Hedayati,  Hui, Kabat, Karimipour, Ponton, Mawhinney, Nicolis,  Shahshahani, Tuts, Vesaghi \ldots whom I learned physics from and enjoyed long discussions with.

I am grateful to Professors Blaer and Dodd for their help in arranging the TA responsibilities in a very pleasant schedule. I  thank Lalla Grimes and Randy Torres for their invaluable help.
I   thank Kimyeong Lee and Piljin Yi for their hospitality at KIAS.

I want to thank my academic brothers Adam Brown, Xiao Xiao and HakJoon Lee for their friendship and support. I am thankful to my collaborators I-Sheng Yang, Xiao Xiao, Adam Brown, Alex Dahlen, David Kagan  and Dhagash Mehta for the enjoyable collaboration and sharing their experience and knowledge. I want to thank my friends at Columbia physics department  Aaron Angerami, Andrej Ficnar,  Kurt Hinterbichler and Bart Horn for the nice and friendly environment. I also want to thank my friends in Iran especially Mahmoud Safari, Salman Beiki, Ali Sharifi, Omid Kokabee,  Mohammad Asadollahi and Amir Zabet for the nice time I had in college. Of course there are many others I gained from and owe to which  there is not enough space to thank them here. I am grateful to all of them.

Last but not the least  I want to thank my family. My wife Samaneh whose love and support kept me going and her extraordinary patience helped me to cope with the dark times I faced  in graduate school. She kept me hopeful for a brighter coming future. My parents Hossein and Monir who sacrificed everything they had for the sake of the education of their children. Their unconditional love kept the very difficult childhood in a  war time  not only bearable, but also happy and pleasant. I want to thank my brothers Ehsan, Iman and Mehdi for the unforgettable memories of an exciting  childhood.

\cleardoublepage


\thispagestyle{plain}
\strut \vfill 
\centerline{\LARGE 
To Samaneh and my parents}
\vfill \strut 
\cleardoublepage

\noindent``The desire for knowledge, I could not forego \newline
Few secrets remained that I did not know\newline
For seventy-two years, I thought night and day \newline
Until I came to Know, I had nothing to show."\\
Omar Khayyam
\pagestyle{headings}
\pagenumbering{arabic}

%
%
\setlength{\textheight}{8.5in}
\setlength{\footskip}{0in}

\fancypagestyle{plain} {%
\fancyhf{}
\fancyhead[LE,RO]{\thepage}
\fancyhead[RE,LO]{\itshape \leftmark}
\renewcommand{\headrulewidth}{0pt}
}
\pagestyle{plain}

\chapter{Introduction}
\label{section:introduction}

The vacuum has to be the most boring  place on Earth (in the Universe), completely empty with no interesting events. 
This was the  notion of  people living in the pre-historic era of the 19th century, before they had learned to understand  quantum mechanics.

\begin{figure}[htbp] 
   \centering
   \includegraphics[width=2.5in]{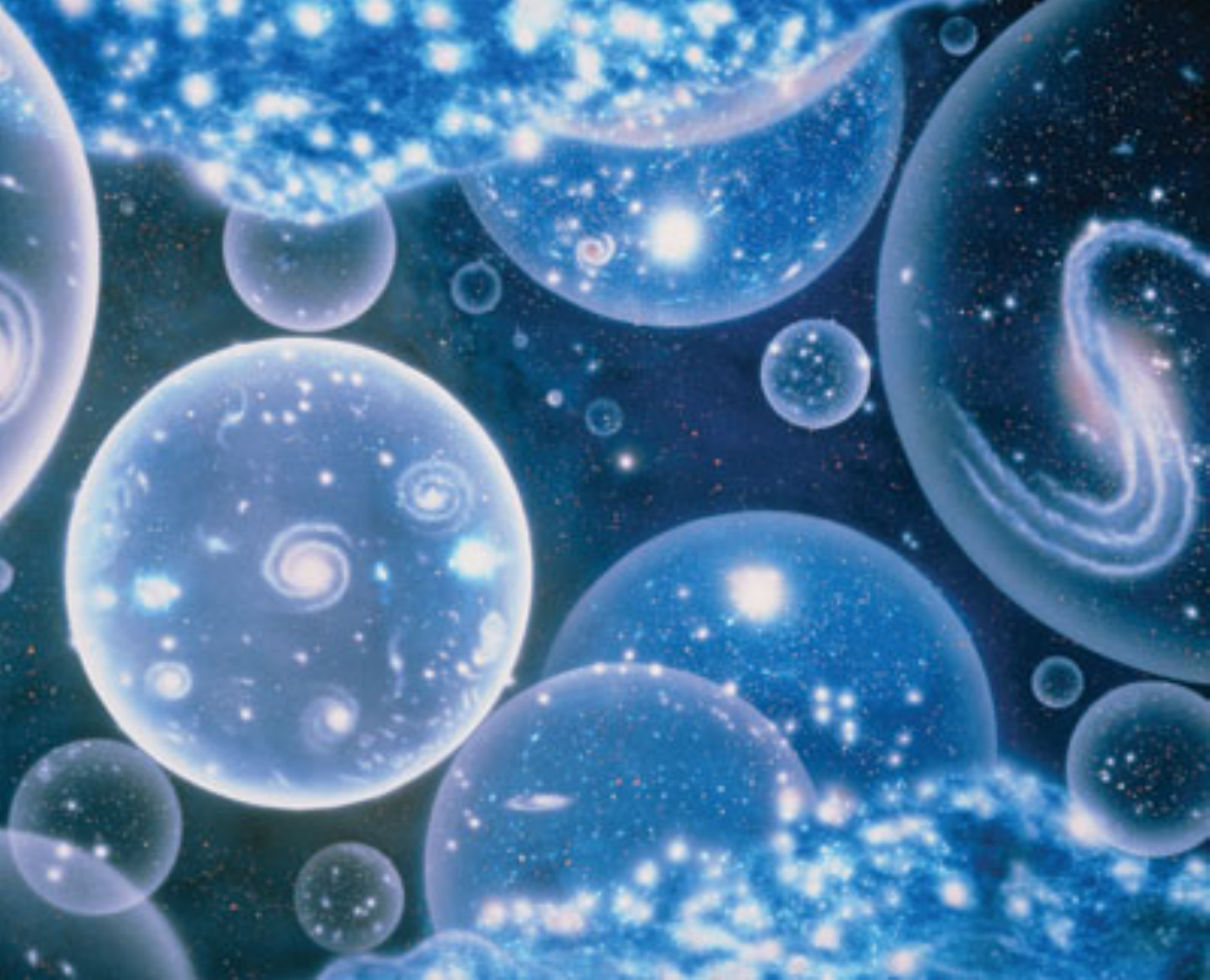} 
   \caption{An artistic depiction of the multiverse. There is a whole universe inside each bubble.  }
   \label{Intro-MultiversePic}
\end{figure}

Over the last quarter  of  century we started discovering   how fascinating  and interesting this place is.  In this previously dull environment, all of a sudden a bubble is born. It grows fast, in fact very fast, almost at the speed of light, and before we get alarmed it engulfs us. This bubble gives birth to new bubbles and we end up with a universe full of  universes, a multiverse (See Fig.\ref{Intro-MultiversePic}). \noindent In this thesis I will explain some topics on this ``road to the multiverse"\footnote{Family Guy, season 8, episode 1.}. 

\begin{figure}[htbp] 
   \centering
   \includegraphics[width=2in]{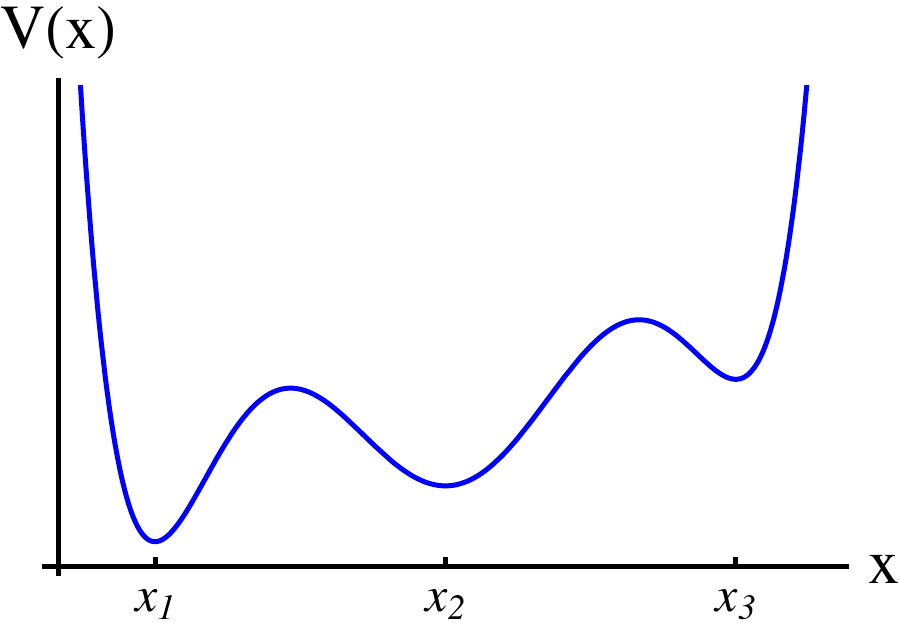} 
   \includegraphics[width=2in]{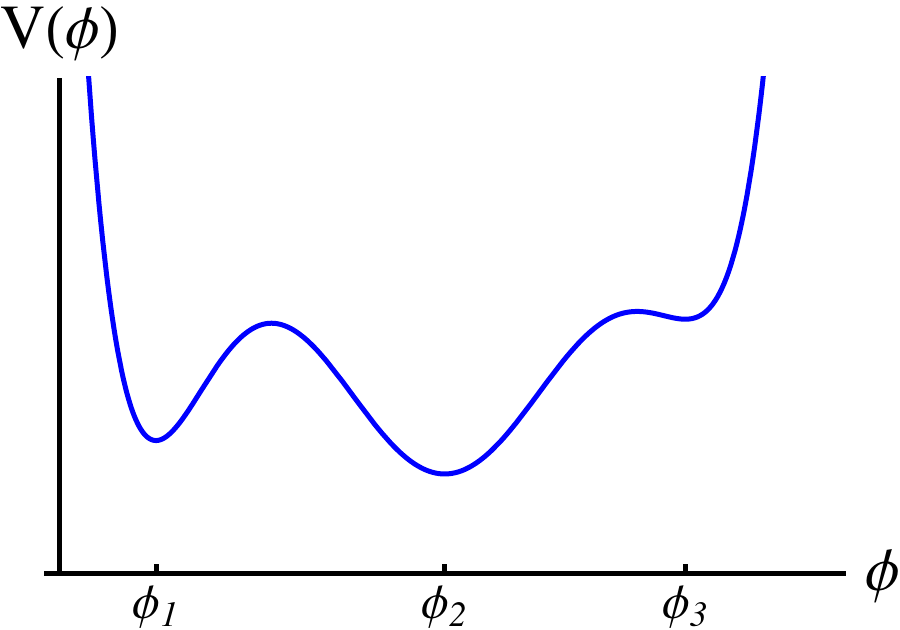} 
   \caption{Left, the potential  for a particle in one dimensions. By vacuum here we simply mean $x_1, x_2$ and $x_3$. Right, the field potential for a single scalar field. Here by vacuum we mean a spatially homogeneous configuration where the field is resting at $\phi_1$, $\phi_2$ or $\phi_3$. }
   \label{Intro1-pic-vacua}
\end{figure}

Let's start from the simple case of a particle moving in one dimension under the influence of a potential $V(x)$ shown in Fig.\ref{Intro1-pic-vacua}. If the particle is at rest at any of the minima of the potential,  classical mechanics predicts that it will remain there forever.

Similarly the vacua of a field $\phi(\vec x)$ are the local minima of its potential $V(\phi)$. If  $V(\phi)$, the field potential, takes its  local minima at $\phi_1$, $\phi_2$ and $\phi_3$ (see Fig.\ref{Intro1-pic-vacua}), the minimum energy configurations are the states where the field is spatially homogenous and  takes values $\phi_1$, $\phi_2$ or $\phi_3$. If the field starts at rest in any of these vacua, according to classical mechanics it remains there forever. However the situation is different when we include quantum mechanics. Quantum tunneling allows the particle or field to tunnel through the potential barriers and move out of a metastable minimum.
The same is true if  we put these closed systems in thermal contact with a heat bath. They can absorb enough energy to overcome the barrier and move towards a more stable minimum. In fact they will also have the opportunity to go from a more stable to a less stable minimum, but at a lower rate.
For a single particle, there is a nonzero probability for these events to happen. For a field in an infinite universe, the situation is trickier. The chance that  the  field tunnels through or jumps over the barrier everywhere at once is zero. From the WKB perspective the tunneling exponent is infinite and  thermal jumps would need an infinite energy, so according to Boltzmann the transition rate ought to  be zero. What happens in reality for quantum tunneling is that a bubble of the new phase is formed and expands. This is the familiar prescription for the first order phase transitions we see in the everyday life. When we boil water, it does not evaporate all at once. Instead small bubbles form near nucleation centers  and then expand until the whole liquid phase disappears. The same thing will happen in flat space vacuum decay and after a long enough time no trace of the old vacuum is left.
When we add gravity to the picture, the situation gets more obscure and meanwhile more interesting. Because of existence of  a horizon there is a nonzero Hawking temperature which makes  thermal tunneling feasible. The old vacuum loses some of its volume to the decay and meanwhile it gains more volume because of the cosmic expansion.  Depending on the rate of the  bubble nucleation  and the cosmic expansion, we may have different scenarios\cite{Guth:1982pn}. If the tunneling rate is very fast, the transition will complete in a finite time and the whole space converts into the new truer vacuum. On the other hand, if the expansion rate is much faster than the decay rate, more and more of the true vacuum bubbles nucleate and chip away parts of the false vacuum. These bubbles grow and there may be new bubbles created inside each of them. Meanwhile the false vacuum itself grows much faster and the overall picture is a universe filled with the false vacuum and with a large number of isolated true vacuum bubbles in the false vacuum. The transition never completes. Here there are a variety of interesting problems about ``what fraction" of the space is in the new or old vacuum. Since the space is infinite, defining a measure on this set  is an immediate question necessary before giving any probabilistic or anthropic description of nature. We do not know the answer to this question yet\cite{Winitzki:2006rn, Guth:2007ng,Vilenkin:2006xv,Linde:2007fr}.

In this thesis I will cover several  topics regarding vacuum decay. 

In  Chapter \ref{section:intro} I will review the basics of tunneling and describe the formalism needed for the following chapters. I start by describing the meaning of the decay in a simple quantum mechanical system and then generalize this concept to a field theory. There are two approaches available on the market to deal with the problem of vacuum decay. One is based on the WKB approximation. The other  uses a Euclidean path integral approach. I will briefly review both and then  describe  thermal tunneling.

The tunneling rate is determined by the Euclidean action of the solution that carries the decay. The higher the action of this solution, the less probable the decay described by it. We  know that generically the highly symmetric solutions have lower action. It was shown that  in flat spacetime the solution with lowest action and therefore highest rate has an $O(4)$ symmetry  \cite{Coleman:1977th} . However there is no such proof for  field theories in curved spacetime. The notion that the higher the symmetry, the lower the action seems to be wrong in some cases. For example the solution with the highest possible spacetime symmetry, the Hawking-Moss solution  which has an $O(5)$ symmetry, has a higher action than the $O(4)$-symmetric solutions in most cases. In  Chapter \ref{O(3)} I will describe  solutions with the next highest symmetry, solutions periodic in Euclidean time and spherically symmetric in spatial directions. These solutions have an $O(3) \times O(2)$ symmetry. I will obtain such solutions numerically and analytically. I show that for a wide range of potentials these solutions have higher action than the $O(4)$ symmetric solutions and therefore  are subdominant. 

The usual scenario for vacuum decay is  bubble nucleation in a  scalar field $\phi$ whose potential has more than one local minimum. In   Chapter \ref{VectorBubble} I study bubble nucleation in a spatial vector field theory. This vector field has  different speeds of sound in the longitudinal and transverse directions. I show that in this case the bubble does not have a spherical symmetry. This work is motivated by the phenomenon of A-B phase transition in the liquid $^3$He. I show that if the ratio of the longitudinal to transverse speeds of sound goes beyond  a threshold, the critical bubble develops a kink and the flat domain wall breaks into zigzag segments.

In the last few years anthropic principles  have become   very popular as a paradigm for explaining  the cosmological constant problem. The general belief is that it is possible to have an enormous number of vacua  in compactifications of string theory. It is assumed that there are  huge number of vacua, perhaps ``$10^{500}$" (500  is a  number quoted  frequently, it could be several thousands as well).  If there are  so many vacua, we should not be very surprised to see a cosmological constant which is 120 orders of magnitudes smaller than the Planck mass. Having $10^{500}$ uniformly distributed vacua, we will get something like $10^{380}$ vacua with cosmological constants as small as the observed value. But it is not proven  that all of these vacua are stable. Recently there has been work that shows that many  of these ``vacua" are indeed saddle points rather than   local minima. Of course even if ``most" of these vacua were unstable, let's say 99.99\% of them, still we would end up with $10^{496}$ stable ones. However we need to be worried about the lifetime of these vacua. Our Universe is an old one with  a long lifetime. On the other hand, using string theories with many moduli means that the effective potential depends on many variables. In a potential with many variables, there are many different directions for tunneling. Some of these directions are crossing  low barrier heights. This makes it possible for the metastable vacua to tunnel easily. I present   the analysis of the lifetime of the vacua with large number of fields in Chapter \ref{Tumbling}. Because performing the exact calculations in the string theory context  is a nightmare, we used a simple model of a  theory of $N$ scalar fields. We looked at the behavior of their potential near the vacua. Our conclusion is that the vacua become  highly unstable with a  large number of fields. In fact we show that the lifetime drops like an exponential of a power of $N$. Therefore if our analysis applies to the landscape of string theory, we cannot find enough diversity of vacua to describe the smallness of the cosmological constant in a natural way.

Our analysis in  Chapter \ref{Tumbling} is based on a model of $N$ scalar fields with a random potential. However we need to be very conservative in applying  it to the landscape of the string theory. There are many points that need to be clarified and taken into account before such an application. Among them the non-canonical kinetic terms, the possibility of brane nucleation and keeping higher order terms can be listed. In Chapter \ref{EMLandscape} I  present a simple model of compactification of the Einstein-Maxwell model which presents many of  the features we expect from the landscape of string theory. I describe the landscape of the vacua of this theory which includes $N$ moduli  which are the radii of the extra-dimensional spaces. I present some of the nice results we obtained about this landscape.

Finally I conclude in   Chapter \ref{section:conclusion}.

\clearpage

\chapter{Vacuum decay, background}
\label{section:intro}
\section{Vacuum decay in one-dimensional quantum mechanics}
\begin{figure}[h!]  
   \centering
   \includegraphics[width=2.8in]{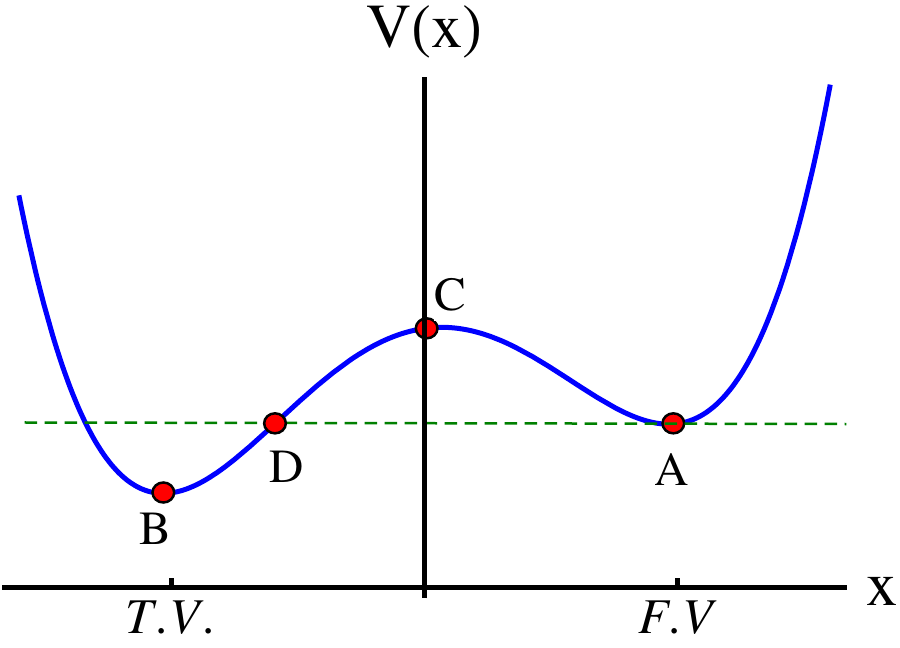} 
   \caption{An example of a potential with a metastable vacuum at A and a stable vacuum at B.}
   \label{Intro:OneDQuantumMechanics}
\end{figure}

Although the minima of a potential are considered to be stable states in classical mechanics, the  story changes drastically when we deal with quantum or thermal physics. In quantum mechanics, the phenomenon of barrier penetration allows escaping from a classically forbidden region. In thermal systems, excitation through absorption of energy from a heat bath makes this jump possible.  Figure  \ref{Intro:OneDQuantumMechanics} shows a simple potential for a particle moving in one dimension. If the particle is originally at the right minimum, it can penetrate through the barrier and go to the left side and emerge at D. The exact treatment of such a system is utterly difficult and we have to use the WKB  approximations to solve it.  The tunneling probability is given by
\begin{equation}
	\Gamma \approx e^{- 2\int _D^ A {dx\over \hbar} \sqrt{2m (V(x) - V(A))}}~.
\end{equation}
The outcome of the tunneling is  a particle at rest emerging at $D$.  Similarly, thermal fluctuations may render the right minimum unstable by boosting the particle to C with a probability 
\begin{equation}
	P \approx  e^{-{ V(C)- V(A) \over k_B T}}~.
\end{equation}
The outcome will be a particle emerging at rest at $C$ that may then  classically roll down to the more stable minimum $B$.

\section{Imaginary energy states and decay}
\begin{figure}[htbp] 
   \centering
   \includegraphics[width=2.5in]{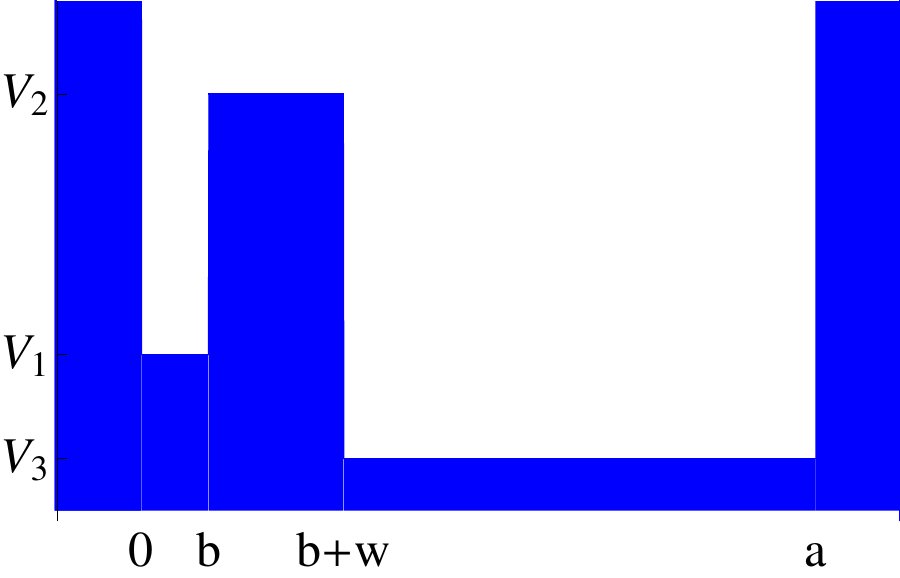} 
   \caption{An infinite well sectioned into two pieces, left a false vacuum of width $b$ and potential $V_1$ separated by a wall of width $w$ and potential $V_2$ from the rest of the well.}
   \label{Intro:InifiniteWell}
\end{figure}

The time evolution of a particle with an imaginary energy is not unitary. The probability of its survival decays exponentially over time. It is unstable. This is similar to the situation we have in hand in the decay of a false vacuum. Let's start with a simpler system shown in Fig. \ref{Intro:InifiniteWell}. If $V_2$ was infinite, the left side would admit its own ground state with a wave function

\begin{equation} \label{Intro:ApproxGroundState}
	\Psi_{\rm left}  = \sqrt{2 \over b} \sin {\pi x \over b}~.
\end{equation} 

\begin{figure}[htbp] 
   \centering
   \includegraphics[width=2.5in]{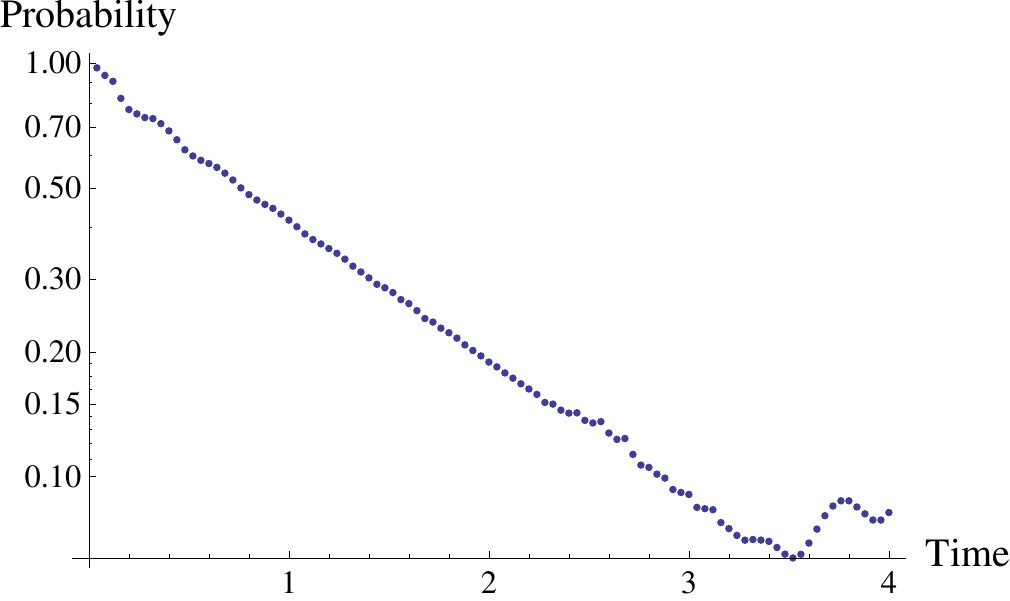} 
   \includegraphics[width=2.5in]{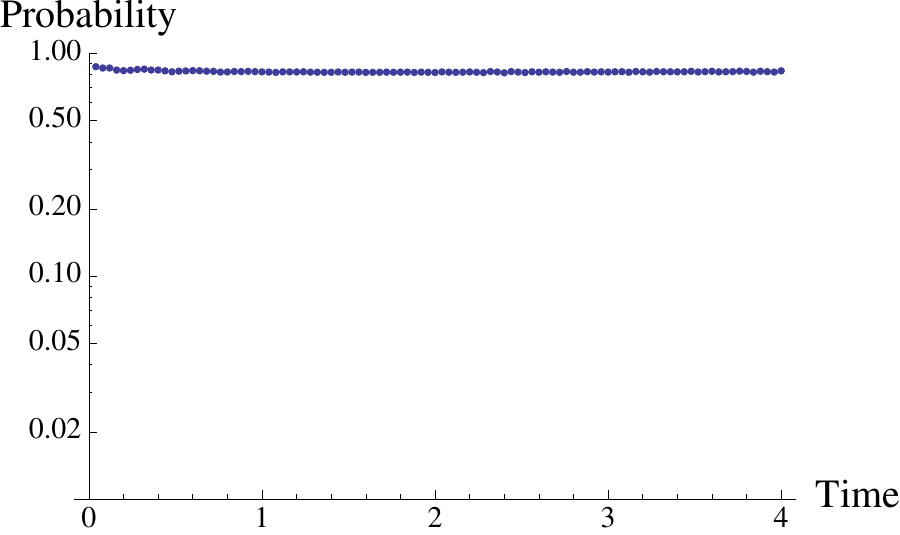} 
   \caption{Chance of finding the particle at the left of the barrier as a function of time. Left is for the case $\upsilon_1=4, \upsilon_2=40, \upsilon_3=0$, Right for the case  $\upsilon_1=-8, \upsilon_2=40 \,{\rm and } \, \upsilon_3=0$. Time is measured in units of $2 m b^2 \hbar^{-1}$ and the vertical axis has  logarithmic scale. In both cases, $a=20$, $w=0.2$ and $b=1$~.}
   \label{Intro:SurvivalChance}
\end{figure}

However the finiteness of the barrier makes this state an approximately stationary one.  Let's choose the values of the potential  in  units of $\frac{\hbar^2}{2 m b^2}$
\begin{equation}
	V_i= \upsilon_i \frac{\hbar^2}{2 m b^2}~.
\end{equation}
We can use  the Schrodinger equation to evolve the approximate ground state [Eq. \eqref{Intro:ApproxGroundState}] in time and  calculate the chance of finding the particle on the left side of the wall, i.e $x\le b$. These probabilities are shown  in Fig. \ref{Intro:SurvivalChance}, for the cases where $V_1>V_3$ (left) and $V_1<V_3$ (right).  As  is seen, the former is unstable and the latter is  quite stable. \footnote{To be more accurate, we need to make a comparison between the ground state energies on the left and right sides of the barrier. This may or may not be the same as the comparison of the potentials.} In addition, for the left case we see that the chance of survival of the particle drops exponentially with time (at least for a short period of time), which is reminiscent of the decay of unstable particles  with imaginary energies. Of course, this exponential decay cannot continue forever in this simple system because after a while  the particle starts oscillating between the left and right sides of the well. We can see this in Fig. \ref{Intro:SurvivalChance} where the linear behavior on the log plot gets obscured after some time has passed. To understand this better, let's repeat the same simulation for  fixed potential parameters and different values of  $a$, the location of the right part of the well.

\begin{figure}[htbp] 
   \centering
   \includegraphics[width=4.5in]{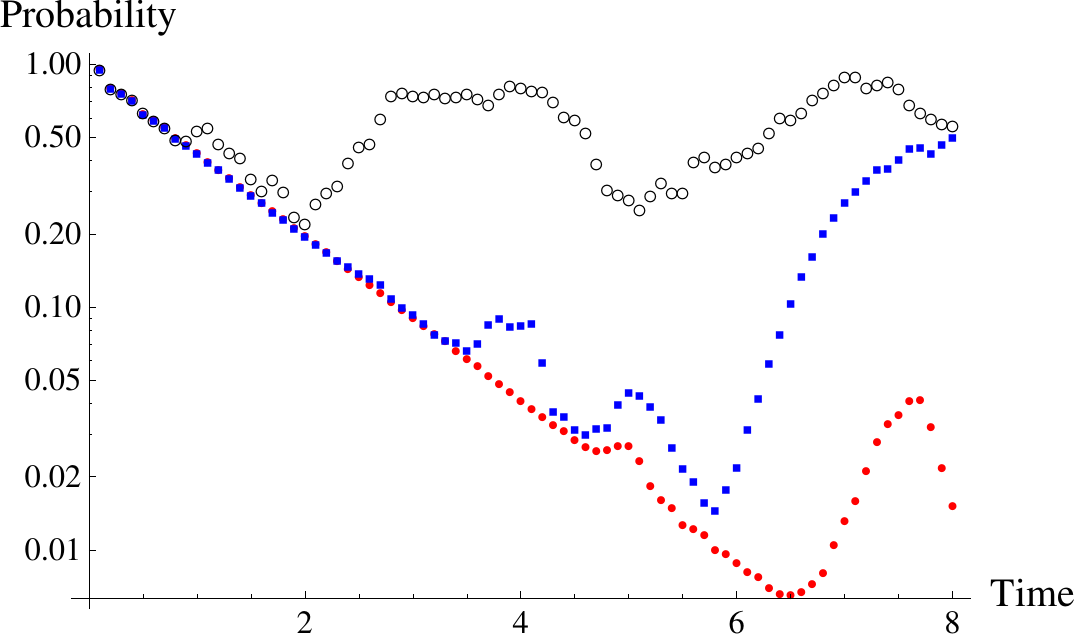} 
  
   \caption{Probability of finding the particle on the left side of the barrier as a function of time. The potential is described by $\upsilon_1=4, \upsilon_2=40, \upsilon_3=0$, $b=1$ and $w=0.2$. Time is measured in units of $2 m b^2 \hbar^{-1}$ and the vertical axis has a logarithmic scale. The red filled circles, blue squares and black empty circles correspond  to $a=40, 20$ and $5$.}
   \label{Intro:SurvivalChance2}
\end{figure}

\begin{figure}[htbp] 
   \centering
   \includegraphics[width=2in]{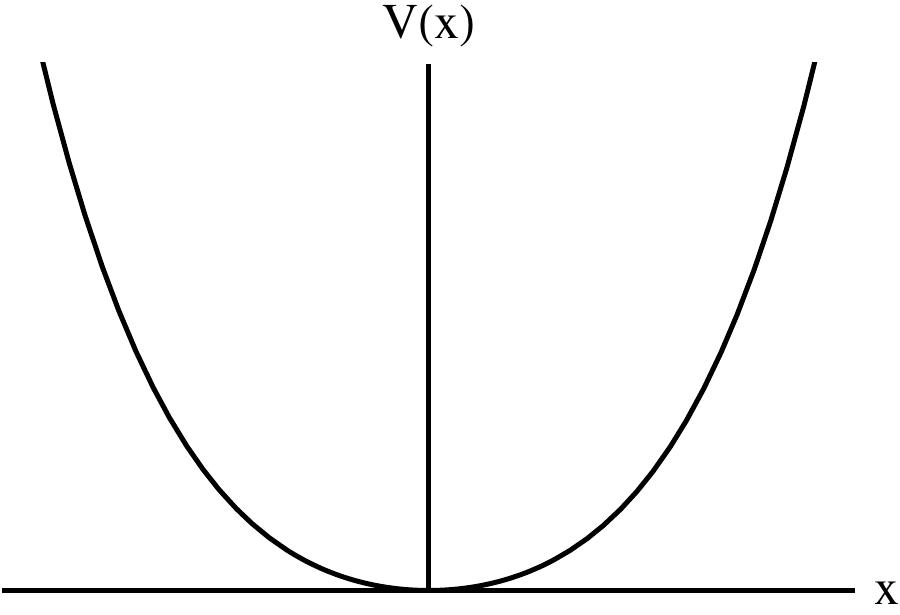} 
   \includegraphics[width=2in]{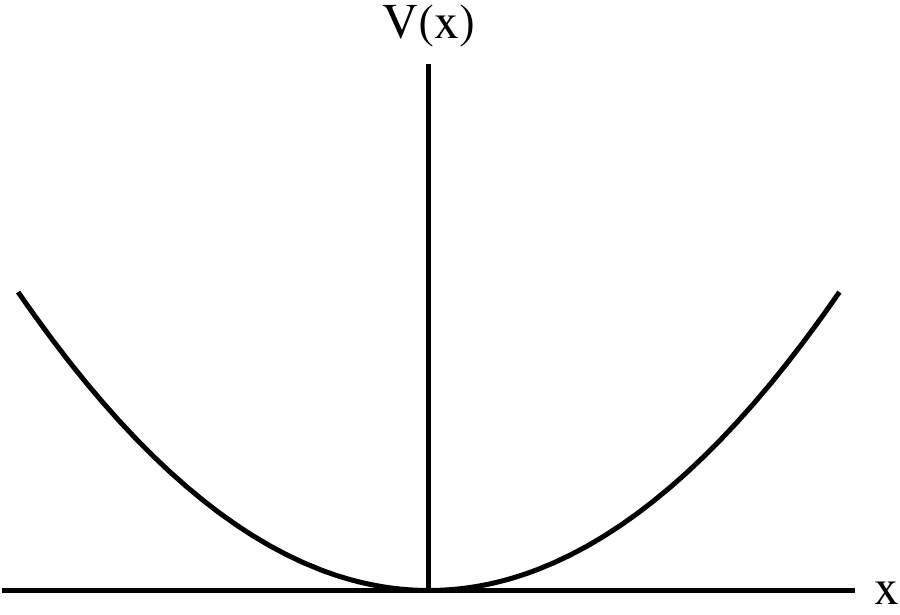} 
   \includegraphics[width=2in]{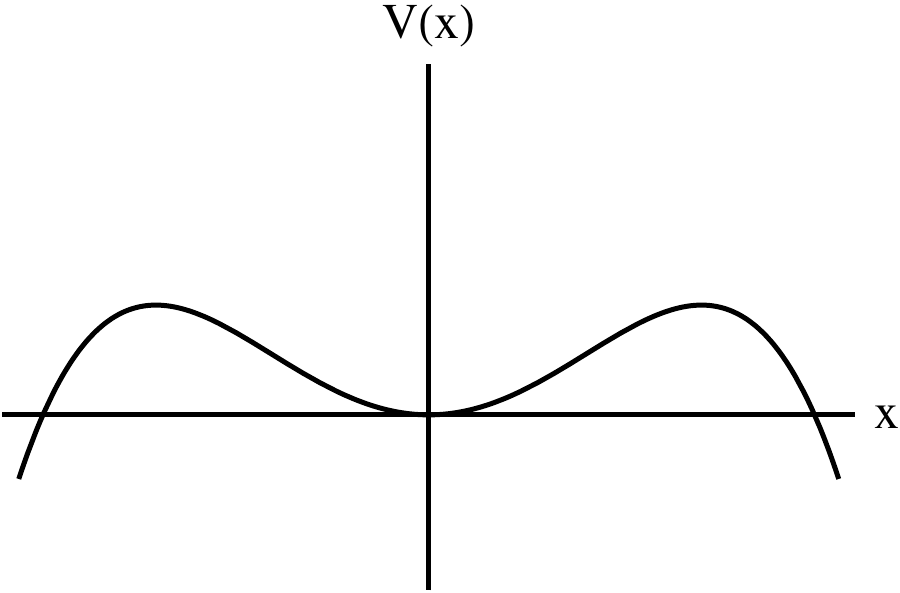} 
   \caption{The anharmonic oscillator potentials defined in Eq. \eqref{Intro:AnharmonicOscillator} for different values of $\lambda$. Left, middle and right graphs correspond to $\lambda=1,0,-1$~. }
   \label{Intro:DifferentLambda}
\end{figure}

The survival chance for different values of $a$  is shown in Fig. \ref{Intro:SurvivalChance2}. Interestingly enough, the exponentially decaying part does not depend on $a$ and its slope is the same for all of them. But the bigger  $a$ is,  the longer the time that the exponential decay approximation is correct. Therefore the imaginary part of the energy only depends on the barrier and the state on the left side and  is independent of whatever happens on the right side of the wall. But making more space available on the right of the barrier results in a higher density of states which in turn increases the  time  it takes for the particle to return to its initial state. For the case that we will be considering  the most, quantum field theory, the number of states available is infinite and therefore the decay process will be a real decay with no return to the original state. This means that we are dealing with states with an imaginary part of energy. But wait! The Hamiltonian in the form $ p^2/2m + V$ is a Hermitian operator and its spectrum can only admit real energy states. How can it accommodate an imaginary energy? To understand this, let's look at the quantum anharmonic oscillator. The Hamiltonian for this system is 
\begin{equation} \label{Intro:AnharmonicOscillator}
	H = -\frac{d^2}{dx^2} + \frac{1}{4} x^2 + \frac{1}{4} \lambda x^4~,
\end{equation}
with the associated  boundary conditions 
\begin{equation} \label{Intro:AnharmonicOscillatorBCs}
	\lim_{|x|\rightarrow \infty} \Psi(x) = 0~.
\end{equation}
 For simplicity, we set $m = \frac{1}{2}, \omega = 1, \hbar=1$~. This potential for different values of $\lambda$ is shown in Fig. \ref{Intro:DifferentLambda}~. If there are unstable states for local but not global minima, we should expect to get states with  imaginary energies  for negative $\lambda$.  
 
\begin{figure}[htbp] 
   \centering
   \includegraphics[width=3in]{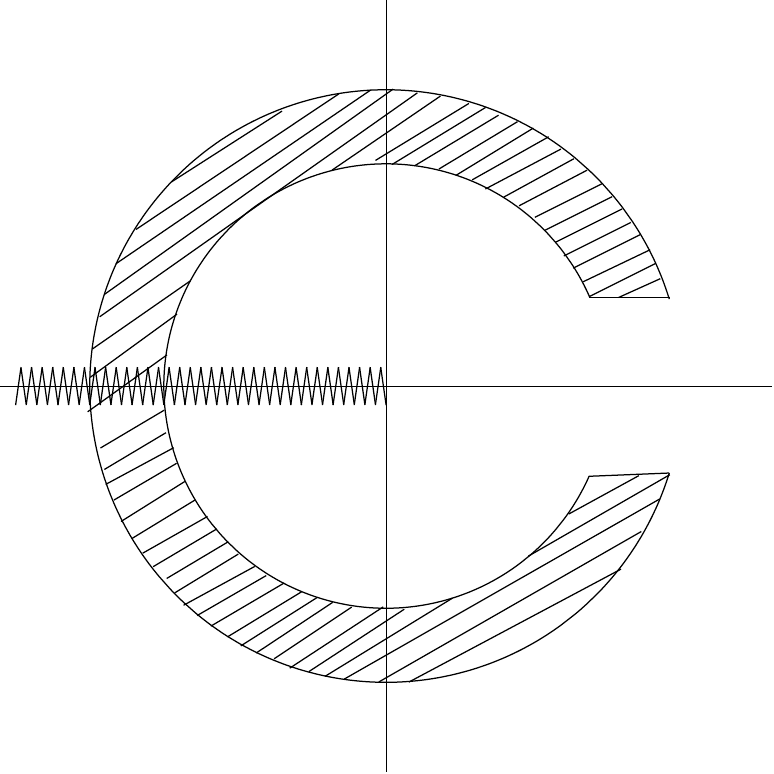} 
   \caption{Analyticity domain of $E(\lambda)$ in the complex $\lambda$ plane. The inner product is not defined for all  vectors and it has a nontrivial analytic structure. Therefore these eigenvalues  evade the usual proof for the reality of the eigenvalues of the Hermitian operators}
   \label{Intro:AnalyticityDomain}
\end{figure}

Let's define   $E(\lambda) $ to be the energy eigenvalues of the Hamiltonian in Eq. \eqref{Intro:AnharmonicOscillator} constrained to the boundary conditions of Eq. \eqref{Intro:AnharmonicOscillatorBCs}. Using Symanzik transformations, we can show that $|E(\lambda)|$  grows like $|\lambda|^{1/3}$ for very large $\lambda$. Loeffel and Martin \cite{Loeffel:1970zz} showed that $E(\lambda)$ is analytic in the complex $\lambda$ plane except for the cut and  the dashed region in Fig. \ref{Intro:AnalyticityDomain}.
 Loeffel, Martin, Simon and Wightman  \cite{Loeffel:1970fe} showed that the $K$'th  energy level can be written as an asymptotic series 
\begin{equation}
	E^K(\lambda) = K + \frac{1}{2} + \sum_{n=1}^\infty A_n^K \lambda^n~.
\end{equation}
It is apparent from Fig. \ref{Intro:AnalyticityDomain} that moving from positive to negative $\lambda$, which  corresponds  to moving  from a stable to an unstable potential requires landing on  a branch cut. We should expect that something nontrivial happens here. This makes the physical Hamiltonian for negative $\lambda$ the limit of a non-Hermitian operator and the branch cuts and discontinuities make  the transition to the real potential non-smooth. This causes the energy levels to be  complex numbers.  Bender and Wu in a series of papers \cite{Bender:1969si,Bender:1971gu,Bender:1973rz} showed that $E(\lambda)$ acquires an imaginary energy part  and  were able to calculate this imaginary energy for  small negative $\lambda= - \epsilon$ to be
\begin{equation} \label{IntroAnharmonicImaginaryEnergy}
	{\rm Im} E^K(\lambda) = \frac{4^{K+\frac{1}{2}}}{K! \sqrt{2 \pi}} e^{- \frac{1}{3 \epsilon}} \epsilon^{-K - \frac{1}{2}} ~.
\end{equation}
This number is very tiny  for small $\epsilon$,  which corresponds to a tall wall. For a general one-dimensional potential the exponential part of the energy in Eq. \eqref{IntroAnharmonicImaginaryEnergy} can also  be calculated using the standard WKB methods for  finding  the transmission coefficients.

\section{Vacuum decay in multi-dimensional quantum mechanics}
\begin{figure}[htbp] 
   \centering
   \includegraphics[width=4in]{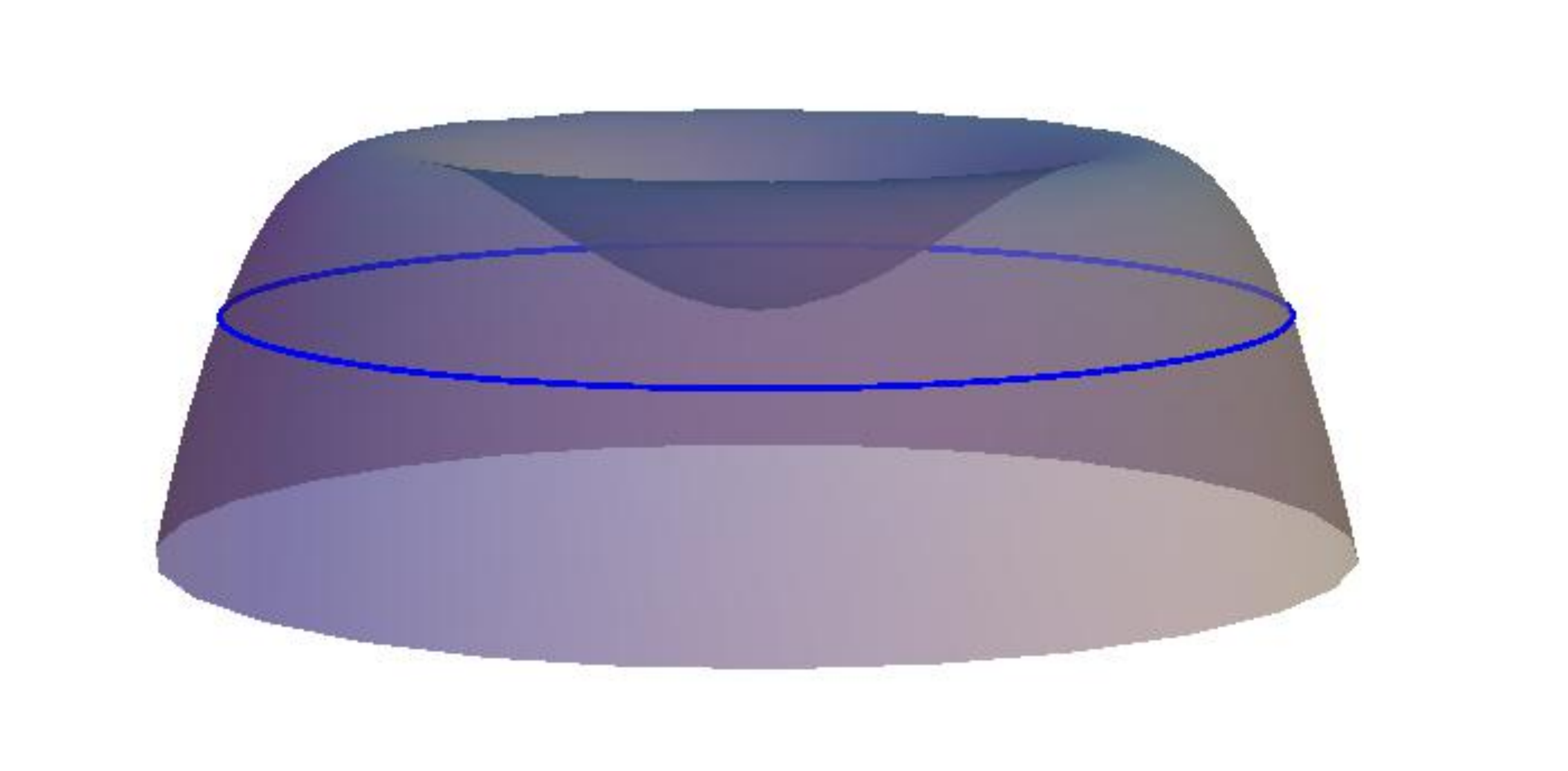} 
   \caption{Possible exit points in a multi-dimensional potential. }
   \label{Intro:ExitPoints}
\end{figure}

Vacuum decay in many dimensions has slightly different qualitative features.  Let's look at the  two-dimensional potential in Fig. \ref{Intro:ExitPoints}. When the particle tunnels out of the central minimum, it can go to any of the points on the (blue) line which have the same energy as the central minimum and respect the conservation of energy. After the tunneling, the particle can emerge with zero kinetic energy at any of these points. In addition, any path that goes from the minimum to the exit point is an acceptable path. These paths do not contribute equally and they are exponentially suppressed. When we add them, we only need to take into account the ones which have the least damping (resistance).

In a series  of papers, Banks, Bender and Wu (BBW) \cite{Banks:1973ps,Banks:1974ij} generalized the one-dimensional WKB methods to N-dimensional systems of coupled harmonic oscillators. Their idea is   to break  down the tunneling problem to a one-dimensional  WKB approximation  along the  classical path to get the exponential  part of the tunneling rate. The prefactor is calculated using a small tube that encompasses this one-dimensional path. The amplitude for a particle to take a path is proportional to $e^{i S \over \hbar}$, where $S$ is the  action along the path. This exponent has an  imaginary part for the case of   tunneling and therefore the probability is exponentially small. Therefore to get the largest contribution we can minimize the action along the path. Using the Maupertuis principle, the problem of minimizing the action becomes equivalent to the  variational problem 

\begin{equation} \label{Intro:Maupertuis}
	\delta \int ds (V-E)^{1/2} =0~.
\end{equation}  
Here $s$ is the path length along the trajectory. The corresponding variational  equations are
\begin{equation}
	2 (V-E) \frac{d^2 x_i}{d s^2} + \frac{d x_i}{d s} \sum_j \frac{dx_j}{ds } \frac{\partial V}{\partial x_j} = \frac{\partial V}{ \partial x_i}~.
\end{equation}
If we had chosen  time as a parameter, the sign of the potential gradient on the right side would be the opposite of what we would expect. In ordinary cases, we  expect  $\ddot{x}_i = - \partial_iV$. This means that the potential should be treated as an upside down potential or, equivalently,  we should solve  the problem in a Euclidean spacetime.  We should have seen this if we were more careful in the context of barrier penetration. There the momentum became purely imaginary and to get a normal momentum, we should have made a transformation $t \rightarrow i \tau$. This is again changing to Euclidean spacetime. Solving these equations, we get a set of MPEPs, the most probable escape paths,   which carry most of the decay. Then we  look at the variations of the  wave function in narrow "tubes" around these trajectories. Here we should distinguish between the two levels of approximation involved. The first is treating the amplitude of the wave function as  a constant and only looking at the damping phase of it along the trajectory. This is similar to geometrical optics and zeroth order WKB and it gives the exponential damping. Then we look at the variation of the amplitude around this trajectory which is the first order WKB   that gives the non-exponential part of the decay rate.  Using this method, BBW could calculate the imaginary part of the  energy of coupled anharmonic oscillators with straight MPEPs could give  a recipe for a general MPEP. However the calculation is very difficult and generalization to many dimensions is a challenge.

\begin{figure}[htbp] 
   \centering
   \includegraphics[width=4in]{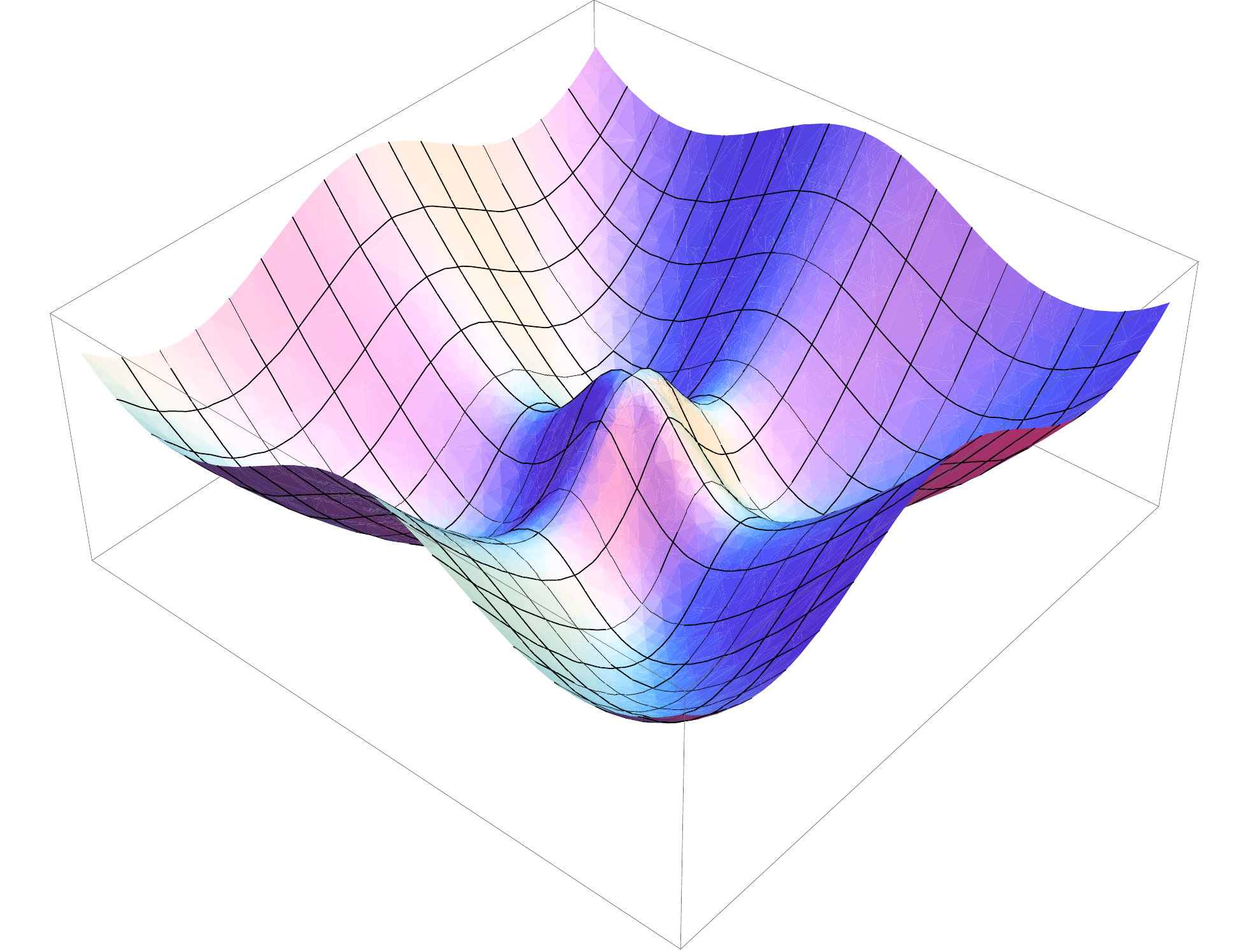} 
   \caption{The dominant configuration  for thermal tunneling in a 2D potential is jumping to a saddle point with one negative eigenvalue of the Hessian matrix. }
   \label{Intro:MultiDThermal}
\end{figure}

As mentioned earlier there is another type of decay,  thermal tunneling, which is possible in the existence of a heat bath.  In thermal tunneling, the path of least resistance will be a path which has the lowest barrier height. As  is clear from Fig. \ref{Intro:MultiDThermal}, the point which has the lowest barrier height is a stationary point where the potential  is increasing in all directions except for the one that leads to the neighboring minimum. The particle jumps to this saddle point and emerges there. It may then classically roll down to  the other minimum. 

\section{Vacuum decay in field theory}
Vacuum decay in field theory was studied by Langer \cite{Langer:1967ax} in the context of statistical physics  of saturation of droplets and also by Kobzarev, Okun and  Voloshin  \cite{Kobzarev:1974cp}. However the modern accepted view appeared  in two seminal works of  Coleman \cite{Coleman:1977py} and Callan and Coleman \cite{Callan:1977pt}. In this section, we follow the approach of the  latter  two works. The decay rate is given by $\Gamma = A e^{- B /\hbar}$, where $B$ is shown to  be  the Euclidean action of the configuration that leads to  tunneling. In first part of this section, a WKB generalization to field theory is presented and the exponential part of the tunneling is calculated. In the second part, we use a path integral approach  to calculate the prefactor of the tunneling rate.

\subsection{WKB approach}
\begin{figure}[htbp] 
   \centering
   \includegraphics[width=2.5in]{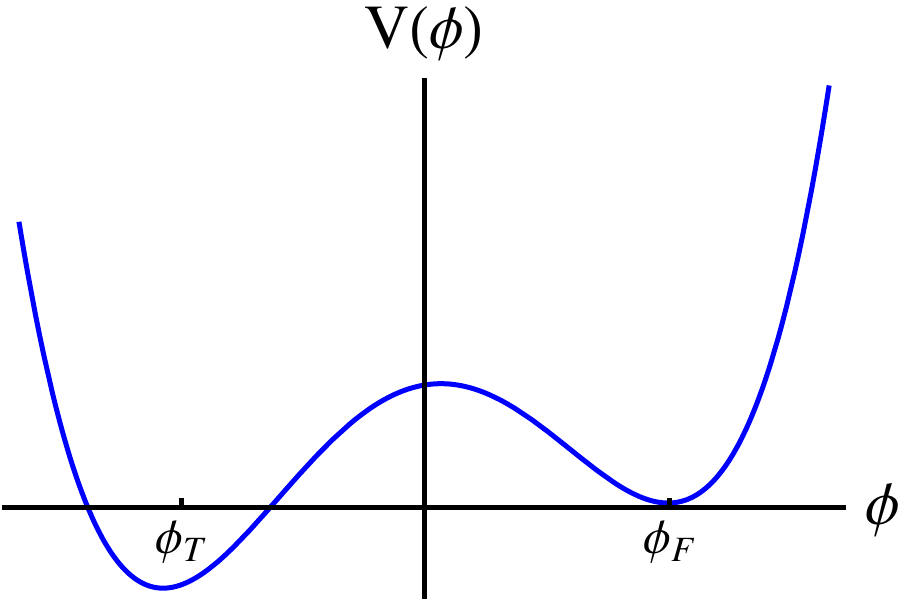} 
   \caption{True and false vacua in a scalar field potential. For convenience, we set $V(\phi_F) = 0~.$}
   \label{Intro:TrueFalseField}
\end{figure}

Let's start from a scalar field theory with a Lagrangian 
\begin{equation} \label{Intro:Lagrangian}
	{\mathcal L} = \frac{1}{2 } \partial_\mu \phi \partial^\mu \phi -V(\phi)~,
\end{equation}
where $V$ is shown in Fig. \ref{Intro:TrueFalseField} and  we have shifted the potential so that the false vacuum has a zero potential. We want to find a configuration in which the field rests in its false vacuum at a long (Euclidean) time ago and a bubble emerges at rest at time $\tau = 0$.  These correspond to 
\begin{eqnarray} \label{Intro:BoundaryConditions}
	\lim _{\tau \rightarrow -\infty} \phi(\tau, \vec{x}) &=& \phi_F ~, \\
	\frac{\partial \phi}{ \partial \tau } (0, \vec{x}) &=& 0~.
\end{eqnarray}

\begin{figure}[htbp] 
   \centering
   \includegraphics[width=2.5in]{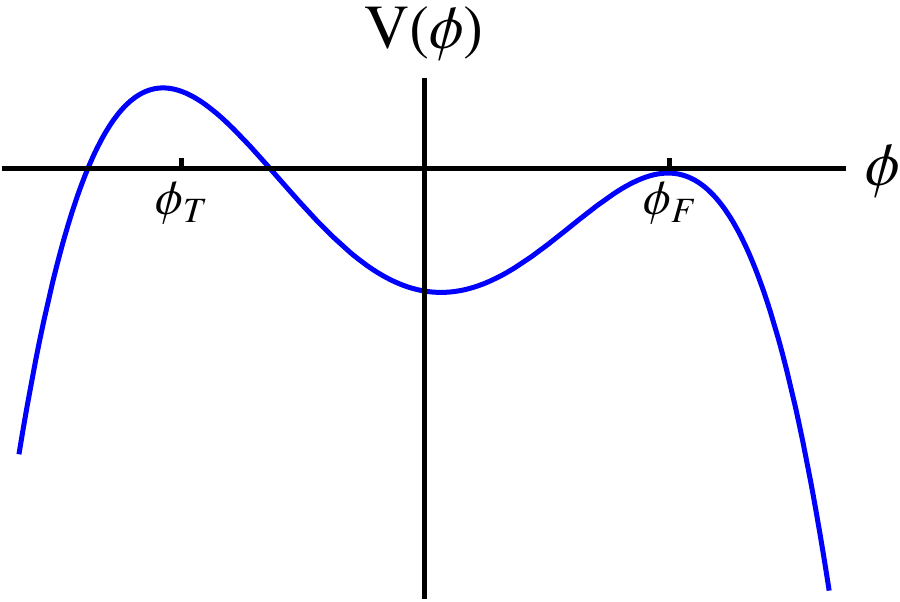} 
   \caption{The motion in Euclidean time is equivalent to motion in an upside down potential~.}
   \label{Intro:UpsideDown}
\end{figure}

\begin{figure}[htbp] 
   \centering
   \includegraphics[width=1.5in]{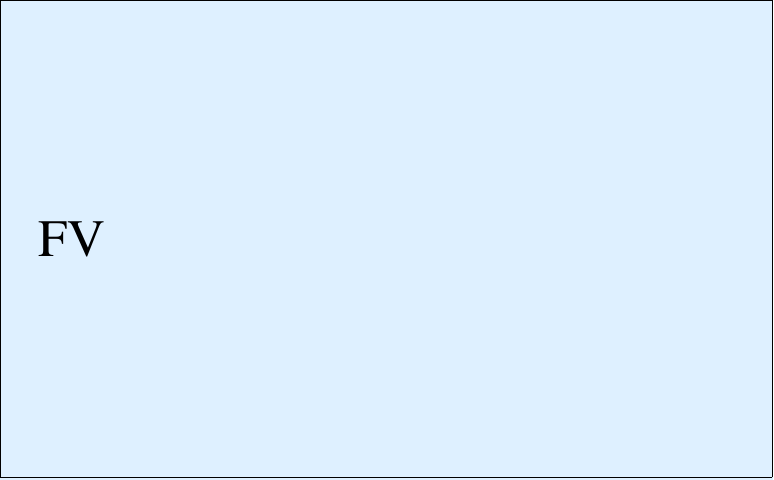} 
   \includegraphics[width=1.5in]{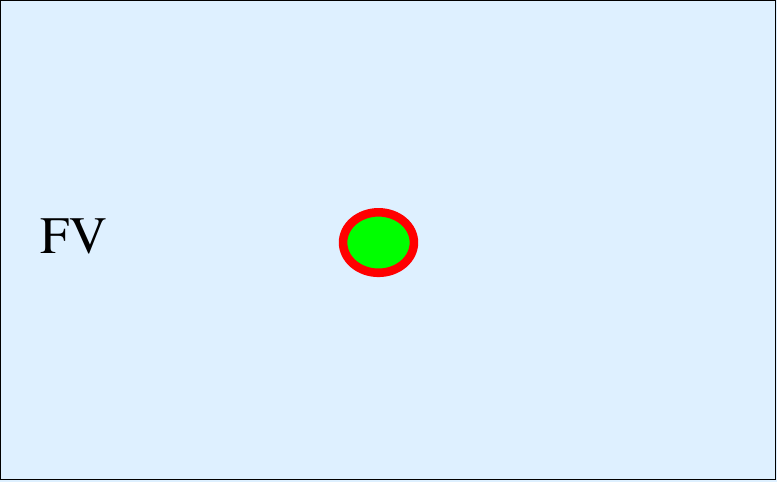}
   \includegraphics[width=1.5in]{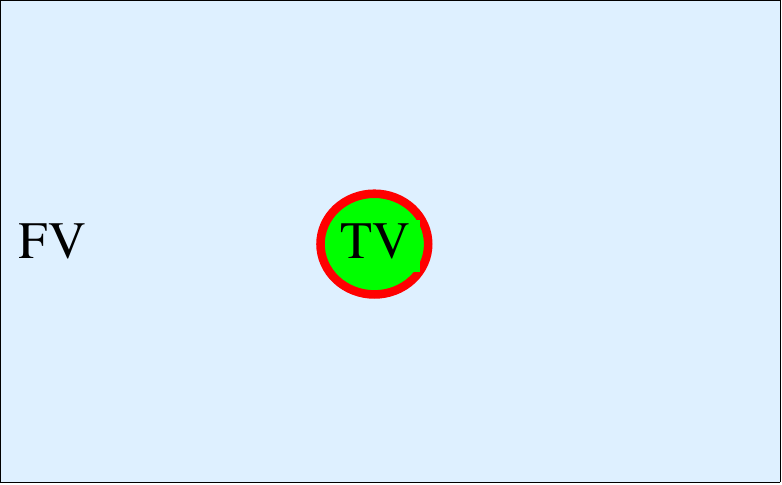}
   \includegraphics[width=1.5in]{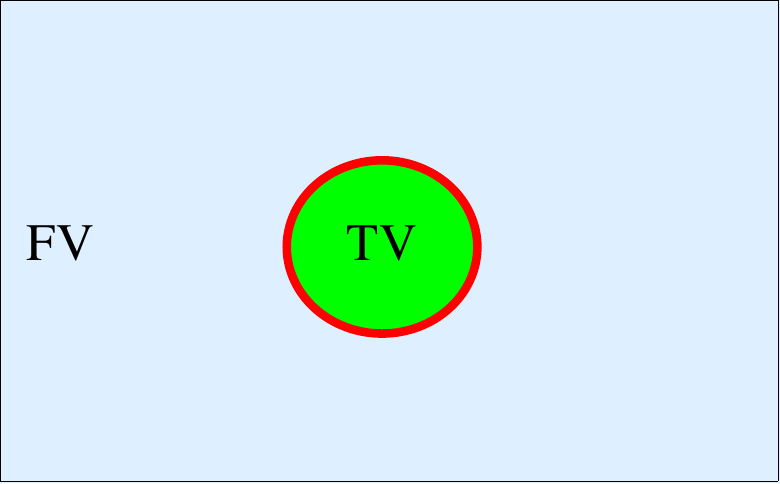}
   \includegraphics[width=1.5in]{Seq3.pdf}
   \includegraphics[width=1.5in]{Seq2.pdf}
   \includegraphics[width=1.5in]{Seq1.pdf}
   \caption{Bubble nucleation as seen in Euclidean time from top left to the bottom right. In the beginning the whole space is filled with the false vacuum (shaded light blue region) and then a small bubble of true vacuum (green region inside the circles) with a wall (red region which is the boundary of the circle ) appears and grows. The top right configuration shows the configuration at the moment of the bubble nucleation. For large Euclidean times, it bounces back to  a space filled with the false vacuum.}
   \label{Intro:Sequence}
\end{figure}

\begin{figure}[htbp] 
   \centering
   \includegraphics[width=2.5in]{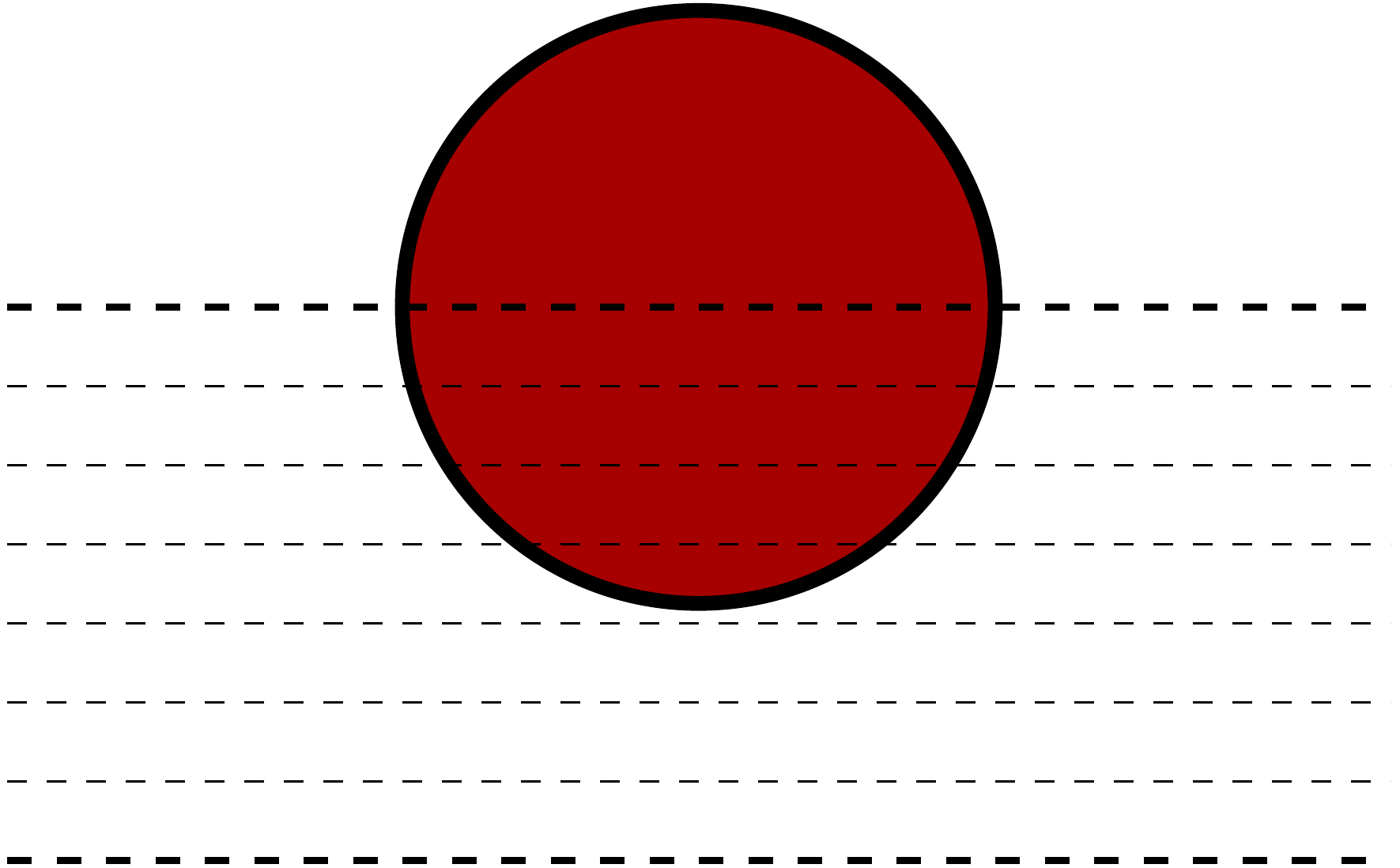} 
   \caption{Another view of the bounce. In this picture the Euclidean time progresses upwards, the false vacuum is pictured as a white background and the true vacuum is in red. Any horizontal line cuts the bounce at a spatial hypersurface. In the far past (the lower thick dashed line), the field is almost at its false vacuum, and the cut through the center of the bounce (upper thick dashed line) shows the field configuration right after the nucleation of the bubble.}
   \label{Intro-Slicing1}
\end{figure}

To get the dominant tunneling path, we need to find the stationary points of  the Euclidean action  
\begin{equation}
	S_E = \int d\tau d^3 x \left[ \frac{1}{2} \left( \frac{\partial \phi}{\partial \tau}\right)^2 + \frac{1}{2} \left( \frac{\partial \phi}{\partial x_i}\right)^2 +V(\phi) \right]~.
\end{equation}
Therefore we have to solve the Euclidean equations of  motion
\begin{equation} \label{Intro:EuclideanEOMS}
	\left( \frac{\partial^2}{\partial \tau^2 } + \nabla^2\right) \phi = \frac{\partial V}{\partial \phi}~.
\end{equation}
Because this equation is invariant under time reversal, the first boundary condition in Eq. \eqref{Intro:BoundaryConditions} can be consistently written as  
\begin{eqnarray} \label{Intro:BoundaryConditions2}
	\lim _{\tau \rightarrow \pm\infty} \phi(\tau, \vec{x}) &=& \phi_F ~.
\end{eqnarray}
This means that we look for field configurations which start from the false vacuum, evolve  to the field configuration for tunneling at $\tau=0$ and then bounce back to the false vacuum. These are called bounce solutions. To get a finite Euclidean action and therefore  a nonzero tunneling rate, we need to make sure that at any moment of time. The field rests at its false vacuum for large spatial distances
\begin{equation}
	\lim_{|\vec{x}| \rightarrow \infty} \phi(\tau, \vec{x}) = \phi_F~.
\end{equation}
The center of the bubble can appear anywhere  in  space and, by shifting the time origin, we can arrange the tunneling to occur at any desired  Euclidean time. This means that the nucleation nucleation probability  obtained for a specific bubble is in fact the nucleation probability  per unit  time per unit  volume and   the total nucleation rate takes a  factor of spacetime volume. Coleman proved that the configuration with the least Euclidean action has an $O(4)$ spherical symmetry. Let's assume that the field is only a function of the four-dimensional distance $\rho = \left( \tau^2 + |\vec{x}|^2\right)^{1/2}$. Equation \eqref{Intro:EuclideanEOMS} simplifies to 
\begin{equation} \label{Intro:SphericalSymmetricEOMS}
	\frac{d^2 \phi}{d \rho^2} + \frac{3}{\rho} \frac{d \phi }{d \rho} = \frac{d V}{d \phi}~.
\end{equation}
This equation resembles the motion of a classical  particle moving in an upside down potential shown in Fig. \ref{Intro:UpsideDown}.
Boundary conditions in Eq. \eqref{Intro:BoundaryConditions2} correspond to a motion starting at time $\rho=0$ at somewhere close to the true vacuum $\phi_T$ with zero speed ($\frac{d \phi}{d \rho}=0$) and landing on the false vacuum $\phi_F$ at $\rho=\infty$. This motion is affected by  the presence of a time-dependent  friction force proportional to velocity ($\frac{3}{\rho} \frac{d \phi} {d \rho}$). If $\phi(0)$ is very far from $\phi_T$, it will not have enough energy  to reach $\phi_F$. If it starts too close to $\phi_T$, it will stay there for a long time. This suppresses the  friction term so due to the difference between the potentials and lack of enough friction, it will pass  $\phi_F$ in a finite time. Therefore there must be some point in between for which the particle does not overshoot or undershoot and  reaches $\phi_F$ in an infinite time. This proves that  the Euclidean equations of motion always admit a solution. Except for special cases, Eq. \eqref{Intro:SphericalSymmetricEOMS} does not have a closed form solution.  However if the potential difference between the true and false vacua $\epsilon = V(\phi_F) - V(\phi_T)$ is small, we can solve this equation using  a thin-wall approximation. If $\epsilon$ is small, in order for the field to make it to $\phi_F$, the effect of friction should be very tiny. Therefore the field should remain close to $\phi_T$ until $\rho$ becomes equal to some  large radius $R$ and therefore friction gets suppressed. This is only possible if it starts very close to $\phi_T$. Then for   $\rho>R$ we can neglect the friction term and Eq. \eqref{Intro:SphericalSymmetricEOMS} simplifies to
\begin{equation}
		\frac{d^2 \phi}{d \rho^2} = \frac{d V}{d \phi}~.
\end{equation}
This equation has a solution of the form
\begin{equation}
	\rho = \int_0^\phi \frac{d\phi}{ \sqrt2\sqrt{  V(\phi) - V(\phi_T)}}~.
\end{equation}
The Euclidean action for the spherically symmetric solution  is
\begin{equation} \label{Intro-FirstEuAction}
	S_E = 2\pi^2 \int_0^\infty d\rho \, \rho^3 \left[ \frac{1}{2} \left( \frac{d \phi}{d \rho}\right)^2  + V(\phi) \right]~.
\end{equation}
The decay exponent $B$ is the difference between the Euclidean action of the bounce  and the pure false vacuum solution. We can break this integral into three parts. The first part is  where the field stays very close to the true vacuum ($\rho<R$) and we can neglect the derivative terms. After subtraction of the pure false vacuum action we get
\begin{equation}
	B_1= S_{1E} - S_{1F} = 2\pi^2 \int_0^R d\rho \; \rho^3 \left[ V(\phi_T) - V(\phi_F)\right] = -\frac{1}{2} \pi^2 R^4 \epsilon~.
\end{equation}
In a short transition region, which is the wall region, we can treat $\rho$ to be constant and we get

\begin{align}
	&B_2= S_{2E} - S_{2F} = 2\pi^2 \int_0^R d\rho \; \rho^3 \left[  \frac{1}{2} \left(\frac{d \phi}{d \rho}\right)^2  + V(\phi)  - V(\phi_F)\right] \nonumber \\
	&\qquad \approx 2\pi^2 R^3  \int_0^R d\rho \left[  \frac{1}{2} \left(\frac{d \phi}{d \rho}\right)^2  + V(\phi)  - V(\phi_F)\right] = \pi^2 R^3 \sigma ~.
\end{align}
Here $\sigma$ is a number that depends on the explicit form of the potential and  represents a surface tension.  For $R<\rho$, where the field get very close to its false vacuum value (in fact exponentially close for any potential), there is no difference between the false vacuum action and the Euclidean action. Therefore the net contribution to the decay exponent is only due to the wall and the interior region of the bubble and is given by
\begin{equation}
	B= -\frac{1}{2} \pi^2 R^4 \epsilon +2 \pi^2 R^3 \sigma~.
\end{equation}
We are left with one free parameter $R$ which determines  for how long the field stays close to its true vacuum value. We can vary this number to get the largest possible decay rate. Doing so we get
\begin{equation}
	0 = \frac{\partial B }{\partial R } =- 2\pi^2 R^3 \epsilon + 6 \pi^2 R^2 \sigma ~.
\end{equation}
This leads to 
\begin{eqnarray}
	R &=& \frac{3 \sigma}{\epsilon}~, \\
	B &=& \frac{27 \pi^2 \sigma^4}{2 \epsilon^3}~.
\end{eqnarray}
It is important to notice that the transition happens in the configuration space of the field and not through a straight path through the barrier. The term potential barrier may be misleading since the spatial gradient term is effectively a potential if we think about the field as points on a lattice. We will get back to this point in more detail in Sec. \ref{Intro-sec-HM}. The family of field configuration that leads to tunneling is shown in Fig. \ref{Intro:Sequence} where a bubble of true vacuum  pops out of the false vacuum and reaches to a maximum size at $\tau=0$ and again disappears when $\tau$ gets very large .  Stacking all of these spatial slices together, gives the picture in Fig. \ref{Intro-Slicing1}.

\subsection{Growth of the bubble.}
\label{intro-sub-Growth}
Until now we have described the formation of a bubble through  quantum tunneling. Inside a space filled with  a false vacuum, a bubble of true vacuum pops out and the field configuration is given by 
\begin{eqnarray}
	&\phi(\vec x, t=0) = \phi(\vec x, \tau=0)~, \cr
	&\frac{\partial}{\partial t}\phi(\vec x, t=0) =0~.
\end{eqnarray}
If we assume that the size of the bubble is not too small, we can describe the subsequent evolution of the field by its classical (Lorentzian) equations of motion
\begin{equation}\label{Intro-LorentzianEOMS}
	\left( - \frac{\partial^2}{ \partial t^2} + \nabla^2 \right) \phi = {\partial \over \partial \phi} V(\phi)~.
\end{equation}

Comparison between  Eq. \eqref{Intro:EuclideanEOMS} and  Eq. \eqref{Intro-LorentzianEOMS} shows that the solutions of  the Lorentzian equations of motion are the analytic continuations of their Euclidean counterparts and can be obtained by replacing $\tau\rightarrow i t$ in the Euclidean solutions. Therefore 
\begin{equation}\label{Intro-LorentzExp1}
	\phi(\vec x,t) = \phi(\vec x, i \tau) = \phi(\rho) = \phi(|\vec x|^2-t^2)~.
\end{equation}

As seen from Eq. \eqref{Intro-LorentzExp1}, the $O(4)$ symmetry of the Euclidean solution translates into the $O(3,1)$ symmetry for the consequent expansion of the bubble. Again we can get intuition about the solution by going to the thin-wall limit. In this approximation, the location of the wall is at 
\begin{equation}\label{WallLocation}
	|\vec{x}|^2 - t^2 = R^2~.
\end{equation}
 The bubble radius $R$ is  determined by the potential as described in the previous subsection. It should be of the same order  as the energy scales of the scalar field and therefore a  relatively short length compared to macroscopic lengths. This means that immediately after the nucleation of the bubble, the wall moves almost at the speed of light and starts eating away more and more of the false vacuum. The wall's Lorentz factor  $\gamma = (1-v^2)^{-1/2}$ from Eq. \eqref{WallLocation} is
 \begin{equation}
	\gamma= {x \over R}~.
 \end{equation}
During the conversion of the false vacuum into the true vacuum, some energy is released. This  energy is spent on accelerating the wall. It is easy to show that in the thin-wall limit, all the energy gained from the transition is exactly converted into the kinetic energy of the wall. This means that when the bubble passes a point, there is no ripple or radiation left behind it and only the true vacuum region at rest is created. 

\subsection{Path integral approach}
\label{Intro-sec-PathIntegral}
Coleman and Callan \cite{Callan:1977pt} used a  path integral approach in Euclidean spacetime to calculate the tunneling exponent and prefactor for the decay of metastable vacua. In this section we follow their calculation.  For simplicity let's start again from a one-dimensional quantum mechanics in imaginary time  and then generalize it to field theories. The amplitude for a particle to move from $x_i$ at time $-T/2$ to $x_f$ at time $T/2$ is
\begin{equation}\label{Intro-Path1}
	\langle x_f  | e^{-H T/\hbar} | x_i\rangle = N \int [dx] e^{-S_E/\hbar}~,
\end{equation}
where in the path integral side  the sum is over all the paths that satisfy the boundary conditions $x(-T/2) = x_i$ and $x(T/2)= x_f$ and $N$ is a normalization factor. The left hand side of this equation has a simple expansion in terms of  the energy eigenstates of the Hamiltonian:
\begin{equation}
	\langle x_f  | e^{-H T/\hbar} | x_i\rangle = \sum_n e^{-E_n T/\hbar } \langle x_f | n \rangle \langle  n | x_i \rangle~.
\end{equation}
In the limit where $T\rightarrow \infty$, the ground state contribution dominates  and
\begin{equation}
	\langle x_f  | e^{-H T/\hbar} | x_i\rangle = e^{-E_0 T/\hbar } \langle x_f | 0 \rangle \langle  0 | x_i \rangle~.
\end{equation}
This gives a simple expression for the ground state energy $E_0$. In this section we  evaluate the ground state energy $E_0$ by calculating the right side of Eq. \eqref{Intro-Path1} using  standard path integral methods in the semiclassical approximation. We will see that indeed it has an imaginary part if the energy eigenstate is localized around a metastable vacuum. This imaginary part of the energy gives the decay (nucleation) rate. 

\begin{figure}[htbp] 
   \centering
   \includegraphics[width=2.in]{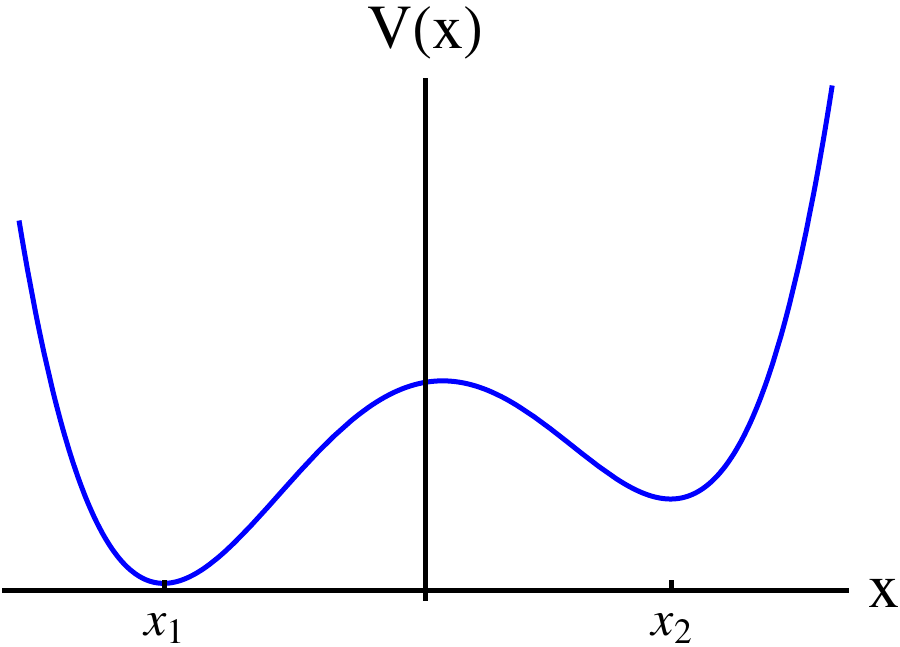} \hspace{40pt}
   \includegraphics[width=2.in]{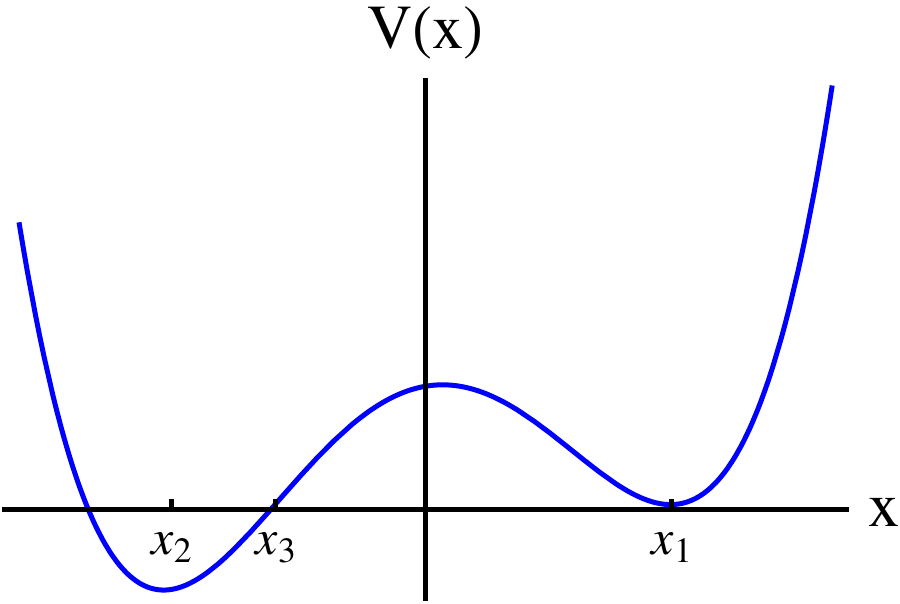} \\ \vspace{5pt} 
   \includegraphics[width=2.in]{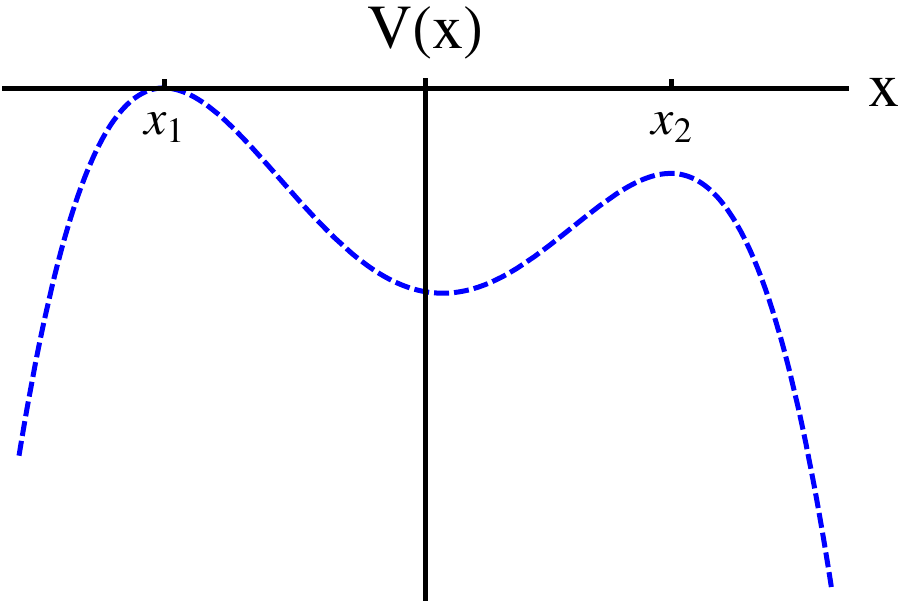} \hspace{40pt}
   \includegraphics[width=2.in]{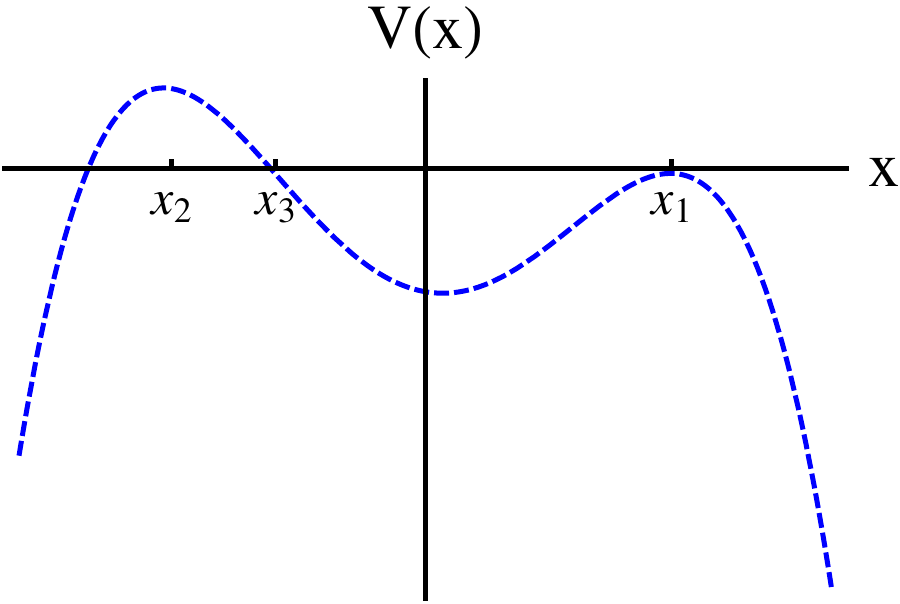} 
   \caption{The Euclidean equations of motion have drastically different behaviors for paths starting from the true and false vacua. On the top, the two potentials with minima at $x_1$ and at the bottom their upside-down versions which are more relevant in Euclidean spacetimes. In the left potential, $x_1$ is located at a true  vacuum and there is no solution for the Euclidean equations of motion except for $x(t)=x_1$.  But for the left, in addition to this constant solution, there is another solution that starts from $x_1$, moves to $x_3$  and then returns to $x_1$ after an infinite time. This solution is called a bounce. }
   \label{Intro-Twocases}
\end{figure}

\begin{figure}[htbp] 
   \centering
   \includegraphics[width=2.in]{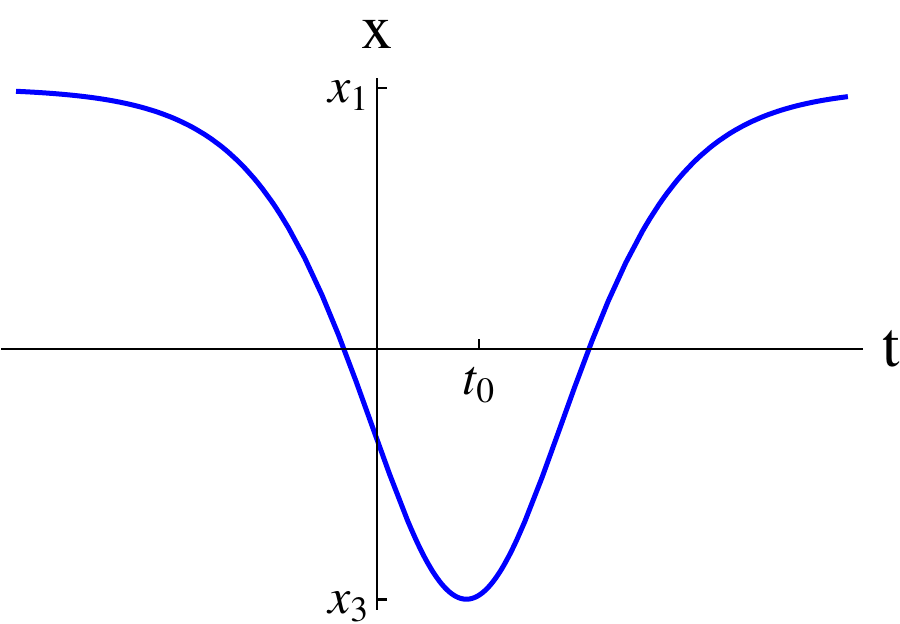} \hspace{40pt}
   \includegraphics[width=2.in]{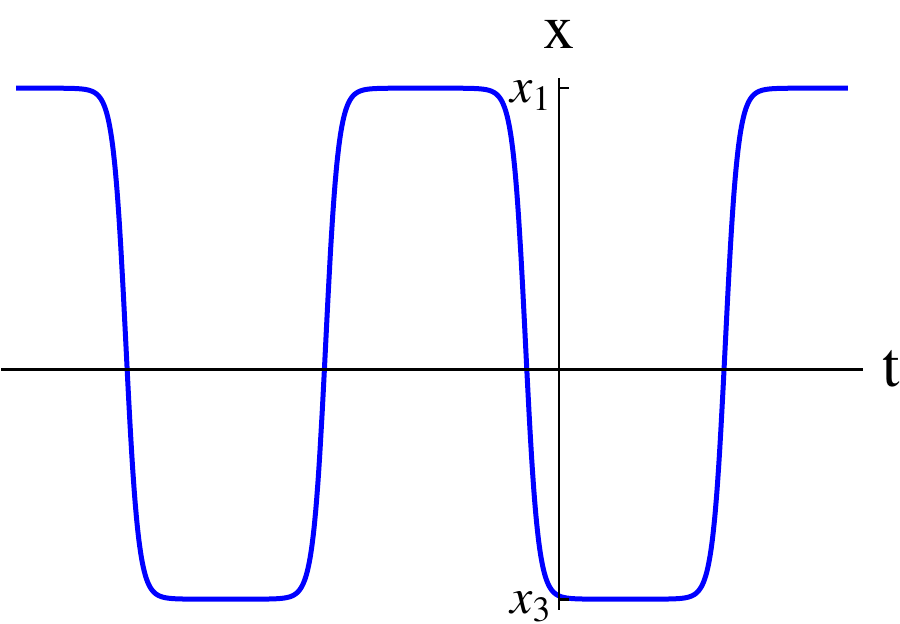} 
   \caption{Left is a single bounce centered at $t_0$ and right a multi-bounce configuration. Both of these are approximate stationary points of the action and will contribute to the path integral in the semiclassical level.(The left bounce would be an exact stationary point if the bounce was  centered at $t=0$). }
   \label{Intro-BouncePic1}
\end{figure}

\begin{figure}[htbp] 
   \centering
   \includegraphics[width=2in]{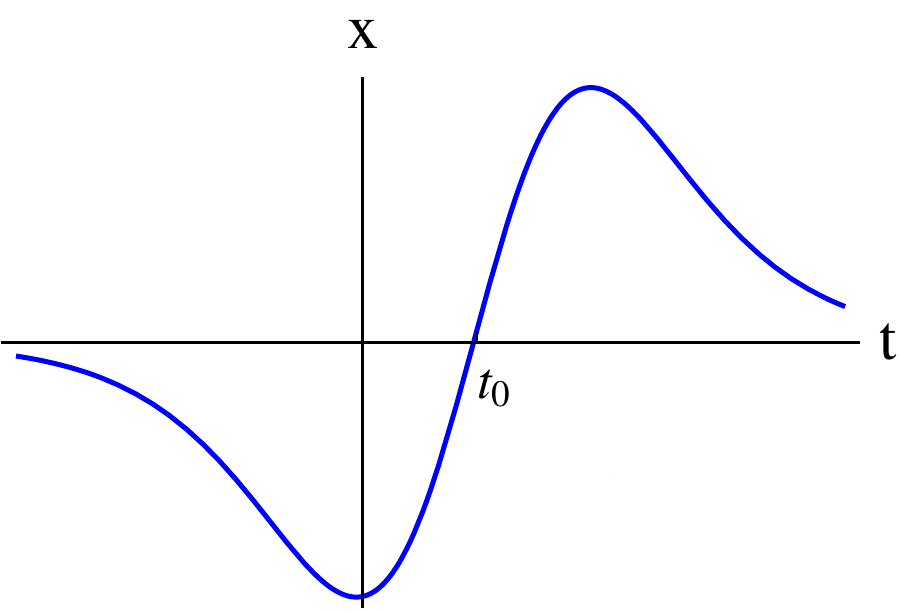} 
   \caption{The zero eigenfunction of the second variation of the action. This eigenfunction has a node so  there must be another lower (negative) eigenvalue in the spectrum of these variations.}
   \label{Intro-ZeroModeGraph} 
\end{figure}

\begin{figure}[htbp] 
   \centering
   \includegraphics[width=2in]{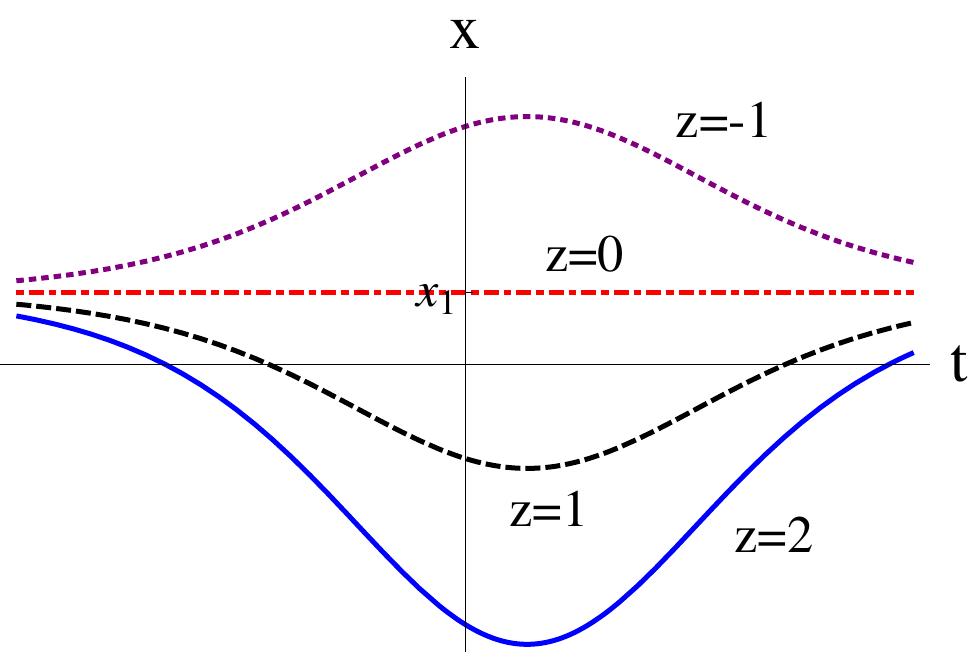} 
   \includegraphics[width=2in]{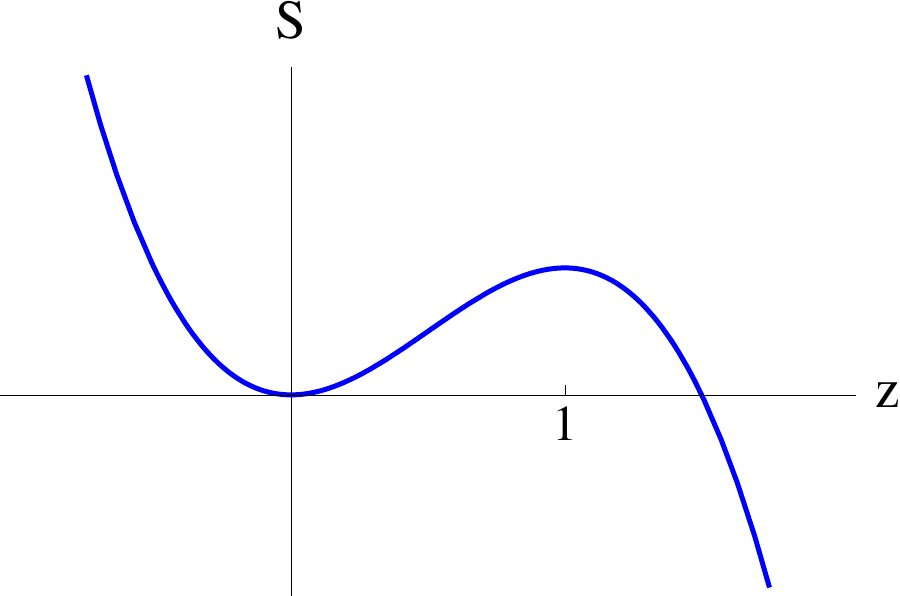} 
   \caption{ Left, a class of paths parametrized by $z$ and right, their Euclidean actions as a function of $z$. The $z=0$ path corresponds to the constant path $x(t) = x_1$ and  has zero action. The  $z=1$ path is the bounce. We chose the paths in a way that the tangent to the $z=1$ path is the negative mode and therefore we are sampling the divergent region of the path integral. }
   \label{Intro-Graph-Action1}
\end{figure}

\begin{figure}[htbp] 
   \centering
   \includegraphics[width=2in]{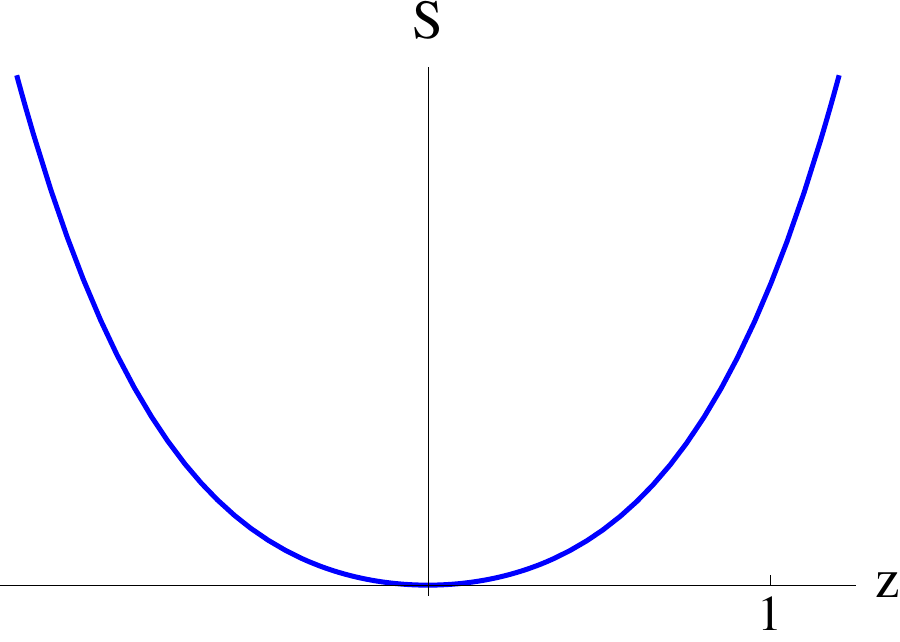} 
   \caption{ Euclidean action as a function of $z$ for the potential shown on the left graph in Fig. \ref{Intro-Twocases}. This function is positive and the path integral is well-defined. To get the correct result for the potential on the right side of Fig. \ref{Intro-Twocases}, we analytically continue it to the potential on the left.  }
   \label{Intro-Graph-Action2}
\end{figure}

\begin{figure}[htbp] 
   \centering
   \includegraphics[width=2in]{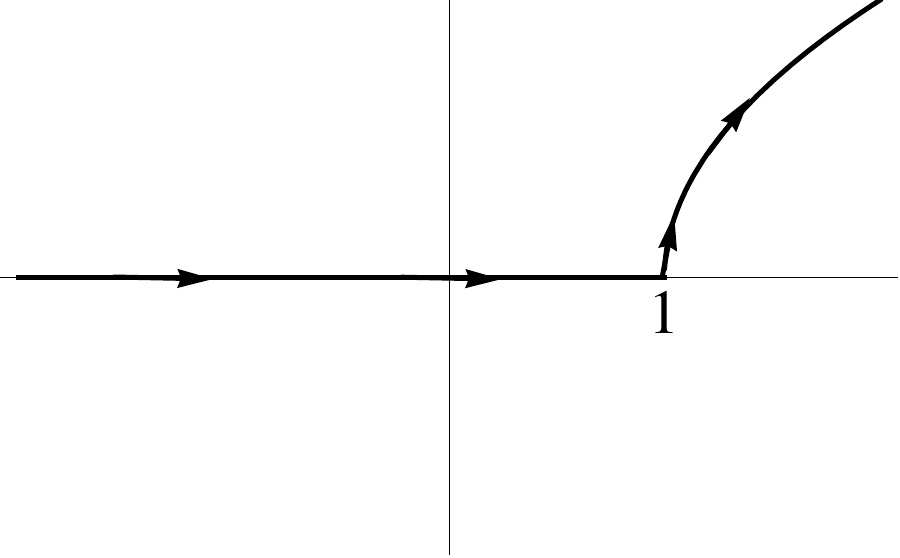} 
   \caption{The contour of integration chosen for the path integral along $z$ which continues from $-\infty$ on the real axis and extends to the saddle point at $z=1$ where the bounce is located. Then it distorts to the upper half plane where the Euclidean action is again positive. }
   \label{Intro-NewContour}
\end{figure}

In the limit $\hbar \rightarrow 0$, the main contribution to the path integral comes from regions close to paths $\bar{x}(t)$, the stationary points of the Euclidean action  where
\begin{equation} \label{Intro-Variation1}
	\frac{\delta S_E}{\delta \bar{x}} = -{d^2 \bar{x} \over d t^2 } + V'(\bar x) = 0~.
\end{equation}
This is  the equation of motion of a classical particle moving in an upside down potential $-V$. We use  the method of steepest descent to evaluate the path integral in a neighborhood of the classical path. But before proceeding further, let's look at two different  cases  where  $x_i = x_f=x_{\rm min}$, one for a global minimum and one for a local minimum. These two cases are shown in Fig. \ref{Intro-Twocases}. If $x_1$ is a true (global) minimum, the only solution to Eq. \eqref{Intro-Variation1} is a constant solution in time. But if it is a local minimum and not a global one, there is another solution which starts from $x_1$ at time $-\infty$  and then moves to the left and after an infinite time returns to $x_1$. For obvious reasons this solution is called a bounce and is shown in Fig. \ref{Intro-BouncePic1}. 
At the semiclassical level, we can evaluate the path integral by looking at paths close to $x_1$. This can be done easily by expanding these paths in terms of a complete set of orthonormal functions $x_n(t)$ which satisfy the boundary conditions $x_n(\pm T/2) = 0$:
\begin{eqnarray} \label{Intro-EigenExpansion1}
	x(t) = \bar x(t) + \sum_n c_n x_n(t) ~, \cr 
	\int_{-T/2}^{T/2} dt \, x_n(t)\, x_m(t) = \delta_{mn}~.
\end{eqnarray} 
The integration over all paths now simplifies to an integration over the coefficients $c_n$. We define the measure to be
\begin{equation}
	[dx] = \prod_n (2 \pi \hbar)^{-1/2} dc_n~.
\end{equation}
We choose $x_n(t)$ to be the eigenstates of the second variation of the Euclidean action
\begin{equation} \label{Intro-Eigen1}
	{d^2 x_n \over d t^2 } + V''(\bar x) x_n = \lambda_n x_n~.
\end{equation}
Now the path integral simplifies to Gaussian integrals over $c_n$'s
\begin{eqnarray}
	\int [dx] e^{-S_E/\hbar} &=& e^{-S_E(\bar x)/\hbar} \prod_n \lambda_n^{-1/2} \left[ 1 + \mathcal O(\hbar)\right] \cr
	 &=& e^{-S_E(\bar x)/\hbar} \left[ {\rm det} \left( -\partial_t^2 + V''(\bar x)\right)\right]^{-1/2}  \left[ 1 + \mathcal O(\hbar)\right] ~.
\end{eqnarray}
We made an important assumption here that the $\lambda_n$'s are all positive. Otherwise the integrals would diverge. For the case of a true vacuum, shown in the left part of Fig. \ref{Intro-Twocases}, the only contribution comes from the path $\bar x=0$ and therefore $V''(\bar x)= V''(x_1) = \omega^2$ is a constant number. A simple calculation (for example \cite{coleman1988aspects}) leads to
\begin{equation}
	N  \left[ {\rm det} \left( -\partial_t^2 + V''(\bar x)\right)\right]^{-1/2}  = \left( {\omega \over \pi \hbar} \right)^{1/2} e^{-\omega T/2}.
\end{equation}
The energy of the ground state has shifted by $\hbar \omega/2$, the familiar zero energy of a harmonic oscillator. The situation is different for the false vacuum depicted in the right panel of Fig. \ref{Intro-Twocases}. Here we have to take into account the contribution from the other stationary point of the action, the bounce.\footnote{ The bounce solution is  only an exact stationary point of the action in the limit  $T\rightarrow \infty$ and we should justify using this approximate stationary point in the sum over all paths. However there is a more accurate treatment of this issue  in page 275 of \cite{coleman1988aspects} by comparing the approximate stationary points  with having stationary points at infinity.} For the bounce solution, we have  $\bar x(\pm T/2) = x_1$ and at some time $t_0$  we have $x(t_0)= x_3$ and $\dot x(t_0)=0$. This is the turning point. First we calculate the action for a single bounce. In Euclidean time, the energy is a constant of motion
\begin{equation}
	E = \frac12 \left({d \bar x \over d t}\right)^2 - V(\bar x)~.
\end{equation}
We shifted  the potential so that $V(x_1)=0$. This makes $E$ vanish for the bounce. Let's call the Euclidean action of this bounce  $B$. 
\begin{equation}
	B = \int_{-\infty}^\infty dt  \left[ \frac12 \left({d \bar x \over dt}\right)^2 + V(\bar x) \right] = \int_{-\infty}^\infty dt \; \left({d \bar x \over dt}\right)^2= \int_{x_3}^{x_1} dx \sqrt{2 V(x)}~.
\end{equation}
We also need to sum over multi-bounce configurations.  If a multi-bounce configuration is composed of $n$ separate bounces, its action is $n B$ and the ${\rm det} [-\partial_t^2 + V''(\bar x) ]$ is the product of the determinant of $n$  single bounces which are separated by large time intervals and the determinant is 
\begin{equation} \label{Intro-KDef}
	\left( \omega \over \pi \hbar \right)^{1/2} e^{-\omega T /2 } K^n~.
\end{equation}
Here $K$ is a  factor chosen so that this expression  is correct for a single bounce. 
For  multi-bounce solutions, the turning points can take different values and we have to sum over the location of all of them. This sum contributes a factor 
\begin{equation}
	\int_{-T/2}^{T/2} dt_1 \int_{-T/2}^{t_1} dt_2 \ldots \int_{-T/2}^{t_{n-1}} dt_n = \frac1{n!} T^n~.
\end{equation}
Now we can sum over all  the contributions from bounces:
\begin{equation}
	\sum_{n=0}^\infty \left( \omega \over \pi \hbar \right)^{1/2} e^{-\omega T/2} {(K e^{-B/\hbar} T )^n \over n! } =   \left( \omega \over \pi \hbar \right)^{1/2} \exp{\left( - \ \frac{\omega T} 2  + K e^{-B/\hbar} T \right)}~.
\end{equation}
Therefore the energy of the ground state becomes 
\begin{equation}
E_0 = \left( {\hbar \omega \over 2} - \hbar K e^{-B/\hbar} \right) \left[ 1 + {\mathcal O}(\hbar)\right]~.
\end{equation}
Please notice that the second term in the parentheses is much smaller than $\hbar^2$, so in general it does not make sense to keep it here. However, as we will see shortly, $K$ is imaginary and this term is in fact the first nonzero contribution to the imaginary part of the energy. Before calculating $K$, we should clarify some points about the eigenvalues of Eq. \eqref{Intro-Eigen1}. In order to perform the Gaussian integrals in the path integral, we needed to assume that all $\lambda_n$'s were positive numbers. But because of the time invariance of the equations of motion, there is one zero eigenvalue, corresponding to 
\begin{equation}
	x_1= B^{-1/2} {d \bar x \over dt}~.
\end{equation}
The factor $B$ is introduced here to satisfy the convention in Eq. \eqref{Intro-EigenExpansion1}.  Integrating over this zero eigenvalue  yields a factor proportional to $T$. Fortunately, we already have taken care of it when we integrated over the location of the turning point $t_0$. The eigenfunction $x_1$ is shown in Fig. \ref{Intro-ZeroModeGraph}. Because the zero eigenfunction has a node, there must be a lower eigenvalue which is negative. This is worrisome, because now the path integral diverges. This should not be very surprising, because from the beginning we were trying to calculate the energy of a state localized near the false vacuum. We know that such a state  should not be stable and therefore it should not be part of the spectrum of the Hamiltonian. The correct way to treat this energy is by analytic continuation.
To make the computation as simple as possible, we restrict ourselves to a subspace of paths  parametrized by a real parameter $z$. These paths are shown in Fig. \ref{Intro-Graph-Action1}. The  $z=0$ path  is the constant path that stays at the false vacuum and therefore  has zero action. The $z=1$ path is the bounce and we chose the paths in such a way that the tangent to it   is the negative mode. For large $z$, where the path remains in the negative potential region   near $x_2$ for a long time, the action can get arbitrarily large and negative.
If we evaluate the path integral over $z$, which is 
\begin{equation}
	J = (2 \pi \hbar)^{-1}\int dz e^{- S(z)/\hbar}~,
\end{equation}
we will end up with a badly divergent integral for $z \gg 1$. To remedy this divergence we use an analytic continuation of the potential. If instead of the potential on the right side of Fig. \ref{Intro-Twocases}, we had used the potential on the left side, the action as a function of $z$ would look like Fig. \ref{Intro-Graph-Action2}, where the integral is convergent. Now let's devise an analytic continuation of the potential on the left side of Fig. \ref{Intro-Twocases} which maps it to the potential on the right side. We choose this continuation in such a way that along the real $z$ axis, we recover the right potential and on the upper complex $z$ plane we continue to the potential on the left. To take the $z$ integral, we can follow a contour which is shown in Fig. \ref{Intro-NewContour}. This contour  extends from $-\infty$ along the positive $z$ axis to the saddle point at $z=1$ and distorts to the upper half plane. The main contribution comes from the region near $z=1$ and we can use the steepest descent method to evaluate the path integral:
\begin{equation}
	{\rm Im} J = (2 \pi \hbar)^{-1} {\rm Im} \int_1^{1 + i \infty}  e^{{-1/2\hbar }[2S(1)+ S''(1) (z-1)^2]} = {1\over 2\sqrt{|S''(1)|}} e^{-S(1)/\hbar} ~.
\end{equation}
Because  the integration took a path only half-way  around the saddle point, a factor of $1/2$ appeared here. 
By generalizing this  analysis  to the path integral over the whole function space, we get
\begin{equation}
	{\rm Im} \left( N \int [dx] e^{-S/\hbar}\right)_{\text{ one bounce} } = \frac12 N e^{-B/\hbar} \left( \frac{B} {2 \pi \hbar}\right)^{1/2} T  \left|{\rm det'}\left[ -\partial_t^2 + V''(\bar x)\right] \right|^{-1/2}~,	
\end{equation}
where $\rm det'$ means evaluating the determinant ignoring the zero mode, which we have already taken care of by integrating over the turning points. By comparing this with Eq. \eqref{Intro-KDef}, we  find that the imaginary part of $K$ is
\begin{equation}
	{\rm Im } K = \frac12 \left( {B \over 2 \pi \hbar} \right)^{1/2} \left|{ {\rm det }' \left[ -\partial_t^2 + V''(\bar x)\right]\over {\rm det} \left[ - \partial_t^2 + \omega^2 \right]} \right|^{-1/2}~.
\end{equation}
The decay rate per unit of time is 
\begin{equation}\label{Intro-ParticleDecayRate1}
	\Gamma = - 2{\rm Im} {E_0 \over \hbar} =  \left( {B \over 2 \pi \hbar} \right)^{1/2} e^{-B/\hbar}\left|{ {\rm det }' \left[ -\partial_t^2 + V''(\bar x)\right]\over {\rm det} \left[ - \partial_t^2 + \omega^2 \right]} \right|^{-1/2}~.
\end{equation}
The generalization to field theory is very straightforward. We  need to notice that there are four zero modes, corresponding to four translations in spacetime. This causes four factors of   $ \left( {B \over 2 \pi \hbar} \right)^{1/2}$ in Eq. \eqref{Intro-ParticleDecayRate1}. The decay (nucleation) rate in this case is  per unit  volume and we get 
\begin{equation}
	\frac\Gamma V = {B^2 \over 4 \pi^2 \hbar^2}  e^{-B/\hbar}\left|{ {\rm det }' \left[ -\Box + V''(\phi(t,\vec x))\right]\over {\rm det} \left[ - \Box + V''(\phi_{\rm fv}) \right]} \right|^{-1/2}
\end{equation}
where $\phi(t, \vec x)$ is the bounce solution. 
In the case of a scalar field theory in four-dimensional spacetime with $O(4)$ symmetric solution, Coleman proved that there is a single negative mode. It was crucial for our argument that there is a single negative mode. If there are an even number of negative modes, there will be no imaginary part contribution to the ground state energy and it cannot predict a decay. Also a higher number of negative modes for a bounce is usually an indication that there is a nearby bounce with lower Euclidean action which is going to be the dominant path. Therefore the bounce with more than one negative mode will be a subdominant decay mode. 

\section{Tunneling at finite temperature}
\label{Intro-sec-HM}
In this section we consider the effect of a nonzero temperature on  tunneling. We assume that the system is in equilibrium with a thermal bath at a temperature $T$. One of the interesting cases that we will encounter in the future is the nonzero temperature due to the existence of a horizon. 

\begin{figure}[htbp] 
   \centering
   \includegraphics[width=3in]{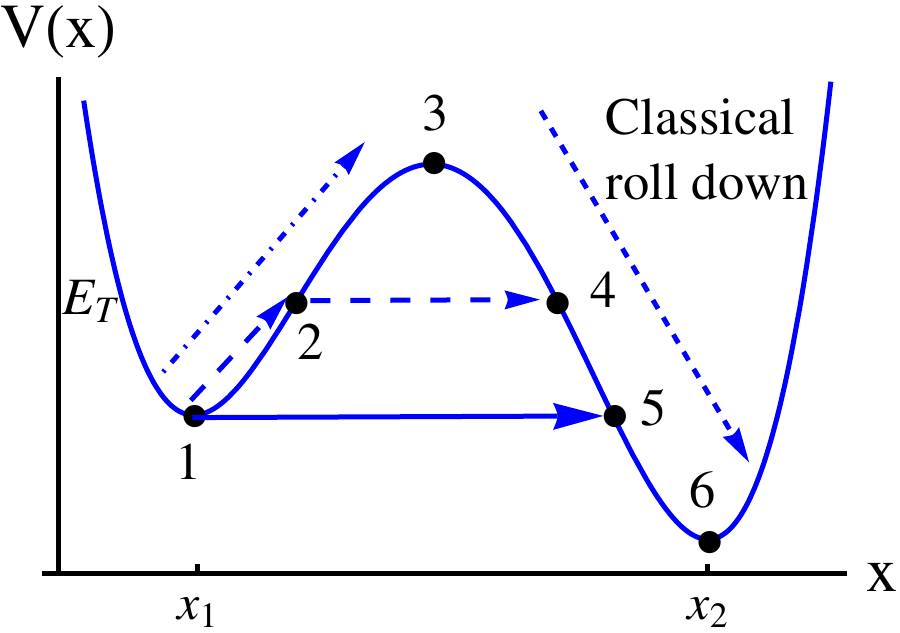} 
   \caption{At a nonzero temperature we can distinguish three different paths for tunneling. The horizontal arrow shows a purely quantum mechanical tunneling. The dashed arrows show a thermally assisted tunneling in which the particle first jumps thermally from 1 to 2 and then quantum mechanically tunnels to 4. The dotted dashed arrows show a purely thermal tunneling in which the particle jumps to a stationary point with one negative eigenvalue. In all cases, after tunneling to the right of the barrier, the particle evolves classically towards the lower minimum.}
   \label{Intro-pic-Thermal1}
\end{figure}

\begin{figure}[htbp] 
   \centering
   \includegraphics[width=3.8in]{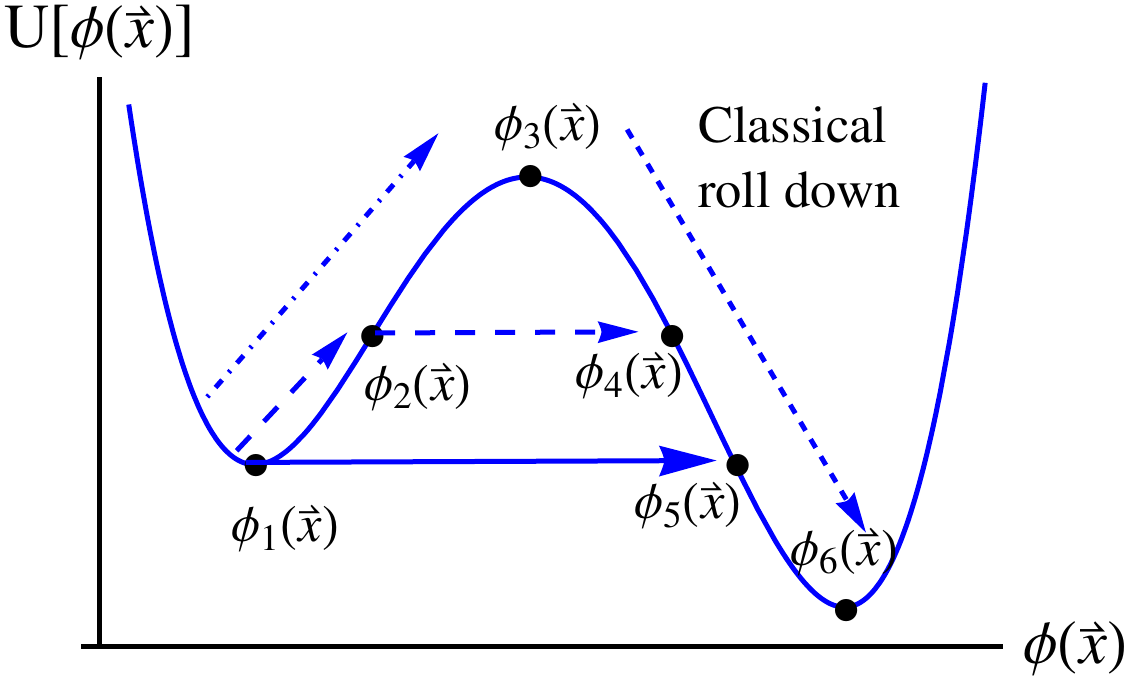} 
   \caption{The same as Fig. \ref{Intro-pic-Thermal1}, but for a field theory. This time the horizontal axis represents the (infinite-dimensional) field configuration space. Points $\phi_1(\vec x)$ and $\phi_6(\vec x)$ show the two homogenous configurations where the field is spatially uniform and rests in its vacua. $\phi_3(\vec x)$ is a configuration which has only one negative eigenvalue and it is in fact a bubble of true vacuum inside the false vacuum. Its negative eigenvalue is along the expansion/contraction directions for the bubble. $\phi_2(\vec x)$  shows the bounce configuration at $\tau =0$ and $\tau=\beta$  in Fig. \ref{Intro-Slicing3} and $\phi_4(\vec x)$ is the configuration after the nucleation for a thermally assisted bubble.}
   \label{Intro-pic-Thermal2}
\end{figure}

To treat the problem properly we have to use the finite temperature effective potential $V_{\rm eff}(\phi,T)$. To simplify  the notation we drop  $T$ in the text, but have it implicitly in mind. Also we need to take into account the new ways for tunneling  at nonzero temperature \cite{Affleck:1980ac,Linde:1981zj}. Again our starting point is a one-dimensional quantum mechanical system and then we will look at the generalization to field theory. There are two new possibilities for tunneling. We can call them thermal tunneling and thermally assisted tunneling. These are shown in Fig.   \ref{Intro-pic-Thermal1}. The horizontal arrow shows the purely quantum tunneling that we covered in the previous sections. The dashed arrows show a thermally assisted tunneling in which the particle jumps thermally to point 2 which has energy $E_T$ and then quantum mechanically tunnels to point 4. The dot-dashed arrow shows a purely thermal jump from point 1 to  3, which is a stationary point with one negative eigenvalue. 
Purely thermal tunneling is the easiest one to explain. The particle jumps from $x_1$ to $x_3$ and then rolls classically down to $x_6$. In more than one dimension, the point $x_3$ is replaced by a stationary point of the potential which has a single negative mode. This point  is the lowest point on a ridge that surrounds the false vacuum. The rate for this tunneling is given by
\begin{equation}
	\Gamma_{\rm tunn} \sim e^{- \beta (V(x_3)- V(x_1))}~.
\end{equation}
To calculate the tunneling rate, we do not need a path in configuration space. We only need an escape point. The easiest way to generalize this concept to a field theory is using the energy functional 
\begin{equation}
	E[\phi(\vec x)] = \int d^3 x \left[  \frac12\left({\partial \phi\over \partial t}\right)^2 + \frac12 (\nabla \phi)^2 + V(\phi)\right]~.
\end{equation}
The potential energy part is 
\begin{equation}
	U[\phi(\vec x)] = \int d^3 x \left[  \frac12 (\nabla \phi)^2 + V(\phi)\right]~,
\end{equation}
which is different from $V(\phi)$. This is depicted in Fig. \ref{Intro-pic-Thermal2}.

\begin{figure}[htbp] 
   \centering
    \includegraphics[width=1.5in]{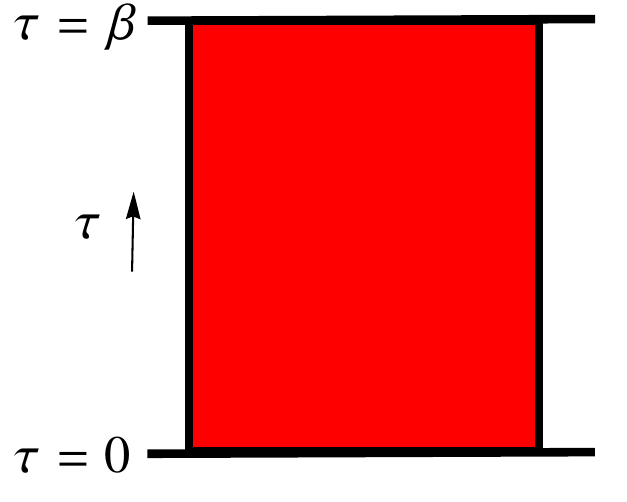}  
   \caption{Bounces representing thermal tunneling are cylinders in four dimension with a period of $\beta$. Each slice through this bounce gives a three-dimensional ball which is a bubble of true vacuum separated from the false vacuum by  a wall (the solid vertical  edges of the rectangle).}
   \label{Intro-Slicing2}
\end{figure}

As mentioned earlier, the tunneling happens  through an infinite-dimensional configuration space with a potential $U[\phi(\vec x)]$ not $V(\phi)$. The correct tunneling configuration is  a stationary point of $U[\phi(\vec x)]$ in the configuration space with a single negative mode. This stationary point is a  static solution of the full equations of motion. A solution $\phi(\vec x)$ to these equations is not spatially homogenous and in fact is a three-dimensional bubble of true vacuum which is separated from the false vacuum by a wall. This solution is time independent and  it has an $O(3) \times O(2 )$ symmetry where the  $O(2)$ symmetry comes from the periodicity in the time direction. The negative mode corresponds to expansion/contraction of the bubble. The energy of the critical bubble in the thin-wall limit  is given as the sum of the contributions from the surface tension of the wall  and the energy density inside the bubble. 
\begin{equation}
	E = 4 \pi R^2 \sigma - {4 \pi\over 3} R^3 \epsilon~,
\end{equation}
\begin{figure}[htbp] 
   \centering
   \includegraphics[width=2in]{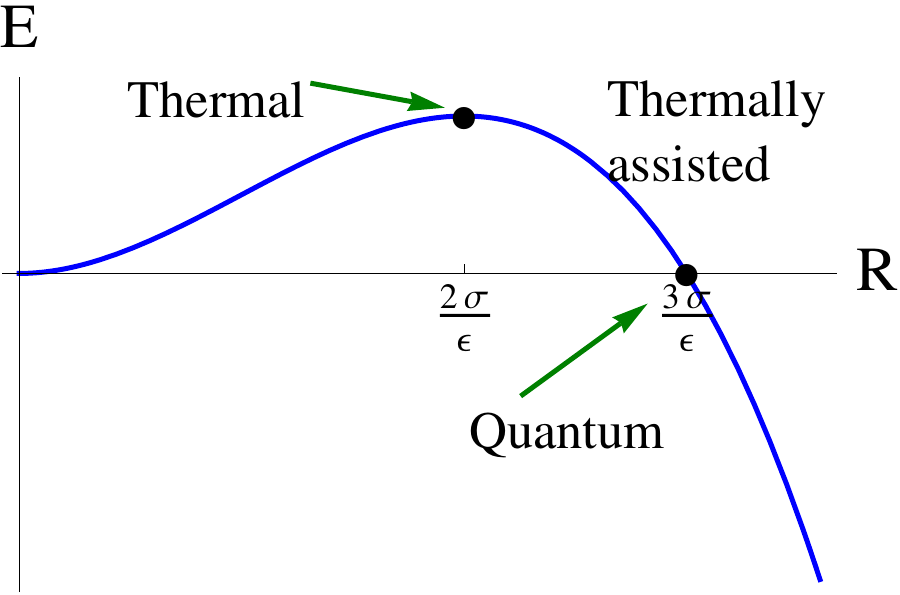} 
   \caption{A quantum bubble appears with zero energy, as expected from the conservation laws. A thermal bubble appears with energy absorbed from the heat bath and in fact its energy is a maximum with respect to the radius. A  thermally assisted tunneling is between these two extremes. }
   \label{Intro-ThermalQuantumEnergy}
\end{figure}
where $\epsilon= U(\phi_{\rm fv}) - U(\phi_{\rm tv})$. We can find the energy of the critical bubble by maximizing this expression with respect to $R$. This situation is shown in Fig. \ref{Intro-ThermalQuantumEnergy}. The bubble emerges with nonzero energy that it absorbs from the heat bath. It is clear from the picture that the radius of the three-dimensional thermal bubble is smaller than the radius of the four-dimensional bubble for quantum tunneling. We can think about this static solution as a cylinder in four dimensions where each cross section  corresponds to the critical bubble. The shape of this bounce solution is given in Fig. \ref{Intro-Slicing2}.

\begin{figure}[htbp] 
   \centering
   \includegraphics[width=1.5in]{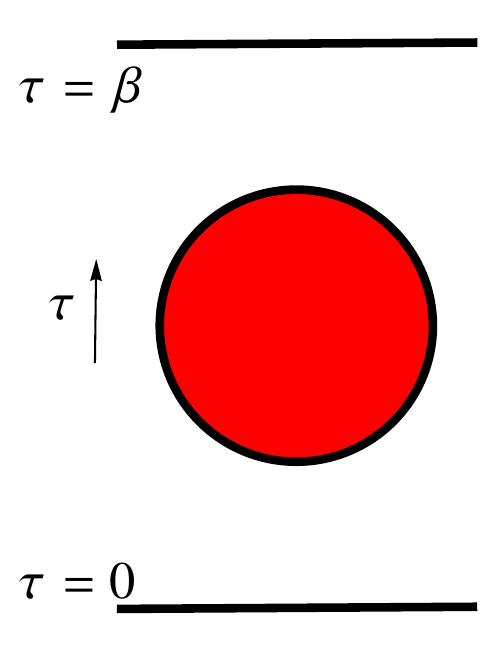}
    \includegraphics[width=1.5in]{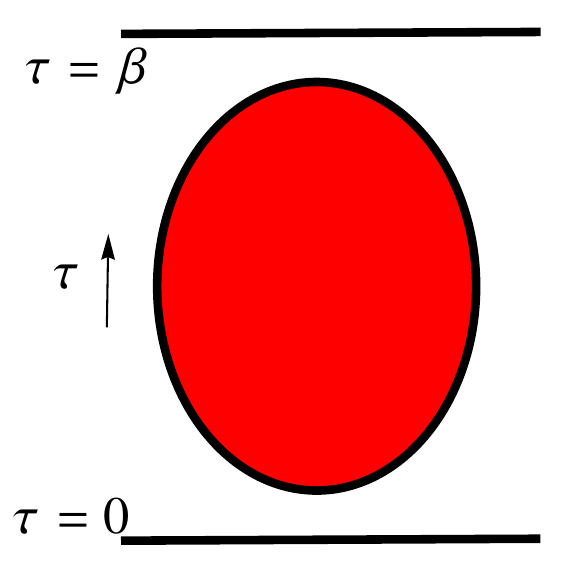} 
   \caption{The left picture shows a low-temperature thermally assisted tunneling and the right a  high-temperature thermally assisted tunneling. The Euclidean time $\tau$  progresses upwards. The field configuration  at $\tau = 0$ and $\tau=\beta$ is given by $\phi_2(\vec x)$ and a spatial slice through the center gives $\phi_4(\vec x)$. (See Fig. \ref{Intro-pic-Thermal2}). In the limit $\tau \rightarrow \infty$, this pictures reduces to the graph in Fig. \ref{Intro-Slicing1}, as expected. }
   \label{Intro-Slicing3}
\end{figure}

Now we get back to the thermally  assisted tunneling for a single particle. The probability for thermal transition from 1 to 2 is given by the Boltzmann factor
\begin{equation}
	{\rm P}_{1\rightarrow2} = e^{-\beta (E_{\rm T}- E_{\rm fv})}~.
\end{equation}
$B(E_{\rm T})$, the Euclidean action for the transition from 2 to 3,  is given by 
\begin{equation}
	B(E_{\rm T}) = 2 \int_{x_2}^{x_4} dx \sqrt{V(x) - E_{\rm T}}~.
\end{equation}
Therefore  the total rate of the transition is an integral over $E_{\rm T}$
\begin{equation}\label{ThermalAssisted-EquationForTunneling}
	\Gamma_{\rm tunn} \sim \int_{E_{\rm fv}}^{E_{\rm T}} d E_{\rm T} \,e^{-\beta (E_{\rm T}- E_{\rm fv})  -B(E_{\rm T})} ~.
\end{equation}
The main contribution comes from the region close to the $E_{\rm T^*}$ that maximizes the integrand (minimizes the exponent). Minimizing the exponent with respect to $E_{\rm T}$ leads to 
\begin{eqnarray}
	\beta &=& -2 {d \over dE_{\rm T}} \int_{x_2}^{x_4} dx \sqrt{V(x) - E_{\rm T}}  = 2 \int_{x_2}^{x_4} {  dx \over \sqrt{V(x) - E_{\rm T}}} \cr \cr
	&=& 2 \int_{x_2}^{x_4} {dx \over \sqrt{\left({d\tau \over dx}\right)^2}} = 2 \int_{\tau_2}^ {\tau_4} d\tau = 2(\tau_4 - \tau_2)~.
\end{eqnarray}
This is the Euclidean  time for the motion from point 2 to point 4. But the full bounce takes twice this Euclidean time. Therefore, along a bounce  the change in Euclidean time should be the same as $\beta$. This means that the bounce solution must be periodic in Euclidean time with a period of $\beta$. The same argument goes through for the case of quantum mechanics with $n$ degrees of freedom. Similarly the solutions for field theory should have a period of $\beta$. This is what we expect from a finite temperature field theory. Keeping in mind that the integral for the bounce action is over a Euclidean time $\beta$, we can interpret $\beta(E_{\rm T^*}  - E_{\rm fv})$ in Eq. \eqref{ThermalAssisted-EquationForTunneling} as part of the Euclidean action and we get the familiar result for the tunneling rate
\begin{equation}
	\Gamma _{\rm tunn} \sim e^{- \left[S_{\rm bounce} - S_{\rm fv} \right]}~.
\end{equation}
The shapes of the thermally assisted bounces are different from their quantum counterparts shown in Fig. \ref{Intro-Slicing1}. They do not have  the same symmetry as the $O(4)$-symmetric quantum bubbles. In addition, the time direction is periodic. However, at very low temperatures the two bounces converge to each other. The analog of Fig. \ref{Intro-Slicing1} is shown in Fig. \ref{Intro-Slicing3}.

At low temperature the thermal bounces acquire multiple negative modes and they do not  correspond to tunneling processes.

\section{Vacuum decay in curved space time}
\label{Intro-sec-CurvedSpace}

Vacuum decay in curved spacetime is a much more formidable problem. Despite all the progress made over the last few decades, this phenomenon is not completely understood. On top of the usual difficulties associated with the large number of new metric variables and the complexity of the Einstein equations, the lack of a well-defined energy in curved spacetime is the main source of the difficulty. The problems  of the number of the negative modes, the  symmetry of the solution with lowest action and also the interpretation of the  bounce solutions are not very clear. On the other hand, this feature also makes a rich new variety of phenomena possible which are not allowed in flat spacetime. For example, not only can the false vacuum  decay to a truer (lower) one, it is also possible to tunnel from a true vacuum to a false one\cite{Lee:1987qc}. The existence of a finite volume horizon makes it possible for the whole field to jump to a saddle point of the potential $V(\phi)$ (not $U[\phi(\vec x)]$) and roll down\cite{Hawking:1981fz}.  This solution is called a Hawking-Moss bounce.

Initially the field is at its false vacuum at every point. The vacuum energy in this state is $V(\phi_{\rm fv})$ and it is constant in spacetime. Therefore the energy-momentum tensor and the curvature of the spacetime are constant. Depending on the sign of $V(\phi_{\rm fv})$, the corresponding space  is called a de Sitter, Minkowski or Anti-de Sitter space. These spaces and their basic properties are briefly described in Appendix \ref{Appendix-DeSitter}.

The first step towards a calculation of tunneling rates in curved spacetime was taken by Coleman and De Luccia (CDL)  \cite{Coleman:1980aw}. They  used the analogy with the previous flat space results and borrowed the formalism developed there. The only change they made was to add the Einstein-Hilbert term to  the action in Eq. \eqref{Intro-FirstEuAction}. The action of a scalar field in curved spacetime is then given by
\begin{equation}\label{Intro-CurvedLagrangian1}
	S = \int d^4x \, \sqrt{|g|} \left[-{1 \over 16 \pi G} {\mathcal R} + \frac12 \partial_\mu \phi \partial^\mu \phi + V(\phi)\right]~.
\end{equation}
In flat space, the tunneling is dominated by solutions with $O(4)$ symmetry. Although unproven, CDL conjectured that the dominating curved spacetime solutions  have the same symmetry. This means that the metric takes the very simple form 
\begin{equation}
	ds^2 = d \xi^2 + \rho(\xi)^2 d \Omega_3^2~,
\end{equation}
where $\xi$ is a radial parameter and $d\Omega_3^2$ is the metric of a unit three-sphere. The scalar curvature of this metric is
\begin{equation}
	{\mathcal R}= {6 \over \rho^2} (1 - \rho \rho'' - \rho'^2)~,
\end{equation}
where  the primes denote  differentiation with respect to $\xi$.  The scalar field  also  only depends  on $\xi$ so we can write it as $\phi(\xi)$. The variational equations for $\phi$ and $\rho$ are
\begin{eqnarray}
	&&\phi''+ {3 \rho' \over \rho} \phi' = \frac{d V}{d \phi} ~, \label{Intro-Curved-PhiEq1}\\  \cr
	&&\rho'^2= 1 + {8 \pi G \over 3} \rho^2 \left( \frac12 \phi'^2 - V\right)\label{Intro-Curved-RhoEq1}~,
\end{eqnarray}
We still have the  freedom to shift $\xi$ by a constant  and we choose it so that a zero of $\rho$ occurs at $\xi=0$. If $\rho$ only has one zero, then $\xi$ ranges from 0 to infinity. In this case the boundary conditions are
\begin{eqnarray}
	\phi'(0) &=& 0 ~,\cr
	\rho(0) &=& 0 ~,\cr
	\phi(\infty) &=& \phi_{\rm fv} ~.
\end{eqnarray}
The first condition is to avoid a singularity at $\xi=0$ in Eq. \eqref{Intro-Curved-PhiEq1}, the second one is the way we chose the origin of $\xi$ and the third condition is to ensure the finiteness of the Euclidean action. These boundary conditions completely determine both fields. The solution obtained this way has the topology of $\mathbb R^4$. If $\rho$ has a second zero, let's say at $\xi_{\rm max}$, the third boundary condition is replaced by $\phi'(\xi_{\rm max})=0$ to avoid  a singularity in Eq. \ref{Intro-Curved-PhiEq1}. This space has the topology of a four-sphere. 

Again, in general it is not possible to solve these equations in a closed form, but fortunately in the weak gravity limit, they reproduce their flat space counterparts. 

\noindent Let's look at the main qualitative differences between the flat space and curved space solutions.

For the case of tunneling from a false to a true vacuum in flat space, the argument using undershoot and overshoot solutions ensures the existence of a solution. But in curved spacetime there is no guarantee for the existence of these solutions. One famous example is the case of a potential  $V(\phi)$ which gets too flat on top. In this case the CDL bounces disappears\cite{Jensen:1988zx,Hackworth:2004xb,Batra:2006rz}.

\begin{figure}[htbp] 
   \centering
   \includegraphics[width=2in]{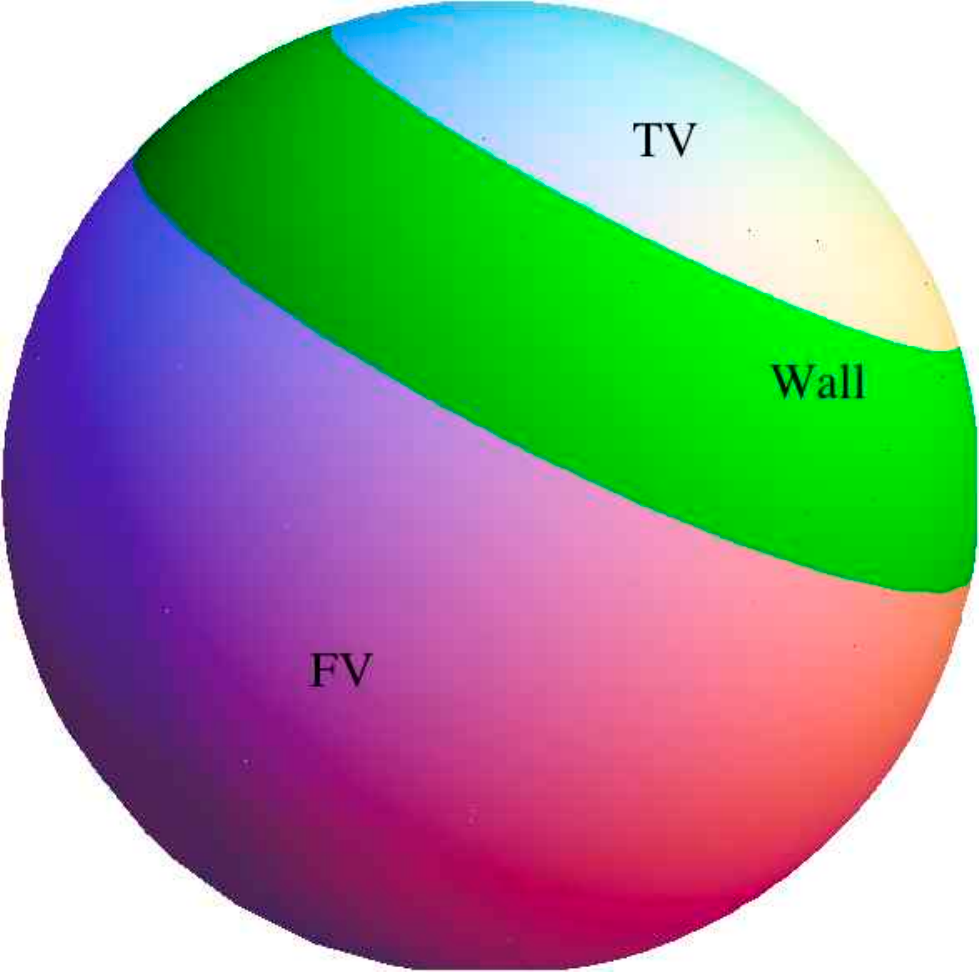} \hspace{20pt}
   \includegraphics[width=2in]{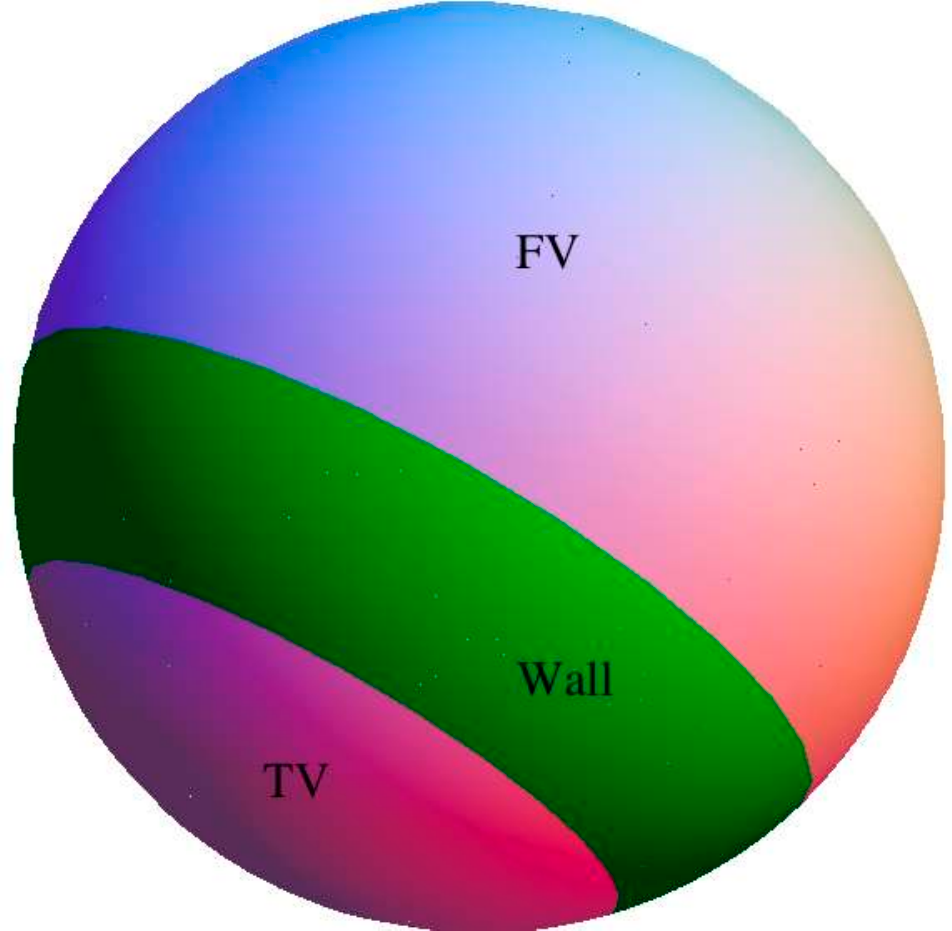} 
   \caption{Left is  CDL instanton that describes tunneling from a false vacuum to a true one. Right is the same instanton, but the poles interchanged by a rotation. It describes the CDL instanton for the decay of the true vacuum.}
   \label{Intro-TFFT}
\end{figure}

When the CDL bounce has a four-sphere topology, the instanton that describes the transition from a true vacuum to a false vacuum has a region close to the true vacuum in its north pole (assuming the center of the bubble is at the north pole). It approaches the false vacuum at the south pole. We can rotate this instanton and get another instanton that describes the tunneling from false to true vacuum\cite{Lee:1987qc}. These are shown in Fig. \ref{Intro-TFFT}. Because the instanton action is the same, the ratio of the tunneling rates is given by
\begin{equation}
	\frac{\Gamma(\text{True} \rightarrow \text{False})}{\Gamma(\text{False} \rightarrow \text{True})} = \frac{e^{- (S_{\rm bounce}- S_{\rm fv})}}{e^{- (S_{\rm bounce}- S_{\rm tv})}}= e^{S_{\rm fv}- S_{\rm tv}}=e^{ -\frac\pi G \left(\Lambda_{\rm tv}^2 - \Lambda_{\rm fv}^2\right)}~.
\end{equation}
Again in the limit where $\Lambda_{\rm tv} - \Lambda_{\rm fv} \ll \Lambda_{\rm tv}$ and $ \Lambda_{\rm fv}$, it simplifies to the familiar thermal expression. It should not be surprising that in the curved spacetime, the true vacuum can also decay. In contrast to the flat space where a bubble of false vacuum has  to collapse because of the surface tension and the pressure from the outside, the bubble of false vacuum in curved spacetime  can be  saved from collapsing by the cosmological expansion. Another way to think about it is the existence of the thermal bath which provides the energy needed for creation of the bubble of higher energy (false vacuum bubble). But this upward decay is only allowed for a de Sitter to de Sitter transition.  Bubbles of Anti-de Sitter or Minkowski have an infinite $S_{\rm fv}$ and therefore cannot up-tunnel. 

\begin{figure}[htbp] 
   \centering
   \includegraphics[width=2in]{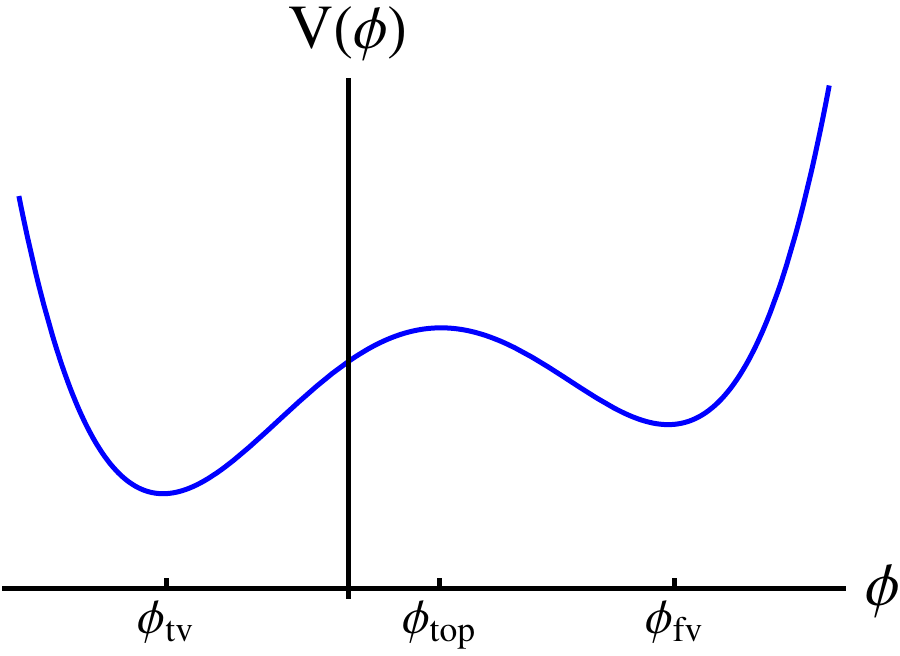} 
   \caption{One possible mode of decay is the fluctuation of the field in every point to $\phi_{\rm top}$. It has a finite action and nonzero probability in de Sitter space.}
   \label{Intro-pic-HM}
\end{figure}

Another difference between   flat spacetime  and curved spacetime tunneling is the existence of a new type of solutions which are called  Hawking-Moss bounces. Because of the infinite volume of the flat space, the field cannot fluctuate at once in every point to reach the top of the barrier. The Euclidean action of such a solution is infinite and the associated decay rate is exactly zero. However in curved spacetimes with positive energy density, namely the de Sitter spaces, a horizon volume is finite and this transition is possible. The field fluctuates to the top of the barrier shown in Fig. \ref{Intro-pic-HM} and then classically rolls down to the true vacuum. The Euclidean action of this solution is 
\begin{equation}
	S_E= -{A_{\rm Horizon} \over 4 G} = -{\pi  \over G }\Lambda_{\rm top}^2~,
\end{equation}
where $\Lambda_{\rm top}$ is the horizon radius of a de Sitter space with energy $V_{\rm top}$. It is the inverse of the Hubble parameter of the space and is given by
\begin{equation}
	\Lambda_{\rm top}= H_{\rm top}^{-1} = \sqrt{3 \over 8 \pi G V_{\rm top}}~.
\end{equation}

After subtracting the Euclidean action of the false vacuum, we get the tunneling exponent $B$ 
\begin{equation}
	B = \frac\pi G \left(\Lambda_{\rm top}^2 - \Lambda_{\rm fv}^2\right)~.
\end{equation}
In the limit that $\Lambda_{\rm top} - \Lambda_{\rm fv} \ll \Lambda_{\rm top}$ and $ \Lambda_{\rm fv}$, this expression takes a very simple form
\begin{equation}
	B \approx e^{- \left[V_{\rm top} - V_{\rm fv}\right] \times {\text{Horizon Volume}\over T_{\rm fv}}} = e^{- {\Delta U \over T_{\rm fv}}}~,
\end{equation}
where $\Delta U$ is the difference of the energy in a horizon volume of $\phi_{\rm top}$ and $\phi_{\rm fv}$. $T_{\rm fv}$ is the temperature of the de Sitter space filled with the false vacuum and this suggests that the Hawking-Moss has a thermal nature. 

Since the calculations are much easier in the thin-wall limit, we study them here to get a better understanding of the CDL solutions. So again we assume that the bounce has three regions. Region one  extends from the center at $\xi=0$ to a value $\bar \xi$ where the wall is located. In this region the field is exactly at its true vacuum. There is a narrow transition region at $\xi = \bar \xi$ and outside of this region the bounce  is filled with the false vacuum. Let's first use Eq. \eqref{Intro-Curved-RhoEq1} in Eq. \eqref{Intro-CurvedLagrangian1} and integrate by part. This leads to
\begin{equation}
	S_E = 4 \pi^2 \int d\xi \left( \rho^3 V - {3 \rho \over 8 \pi G} \right)~.
\end{equation}
In the thin-wall limit,  the action of the bounce inside the wall is the action of a de Sitter or Anti-de Sitter space and the action of the wall comes  from the surface tension. The general expressions are not very simple. But in the simple case where the initial space is a low-energy de Sitter space and the transition is to a Minkowski vacuum, the tunneling exponent is 
\begin{equation}
	B = {B_{\rm flat} \over \left(1 + \left( \rho(\bar \xi) \over 2 \Lambda_{\rm fv}\right)^2\right)}~.
\end{equation}
The tunneling exponent for decay from a Minkowski space to a low-curvature Anti-de Sitter space is given by
\begin{equation}
	B = {B_{\rm flat} \over \left(1 - \left( \rho(\bar \xi) \over 2 \Lambda_{\rm tv}\right)^2\right)}~.
\end{equation}
These show that the tunneling exponents for the decay from the de Sitter space are smaller than the flat space case. Therefore gravity enhances the decays from de Sitter space, as we might   expect from the new  possibility of  thermally assisted tunneling. On the other hand it decreases the decay rates from  Minkowski or Anti-de Sitter spaces.

\begin{figure}[htbp] 
   \centering
   \includegraphics[width=2.5in]{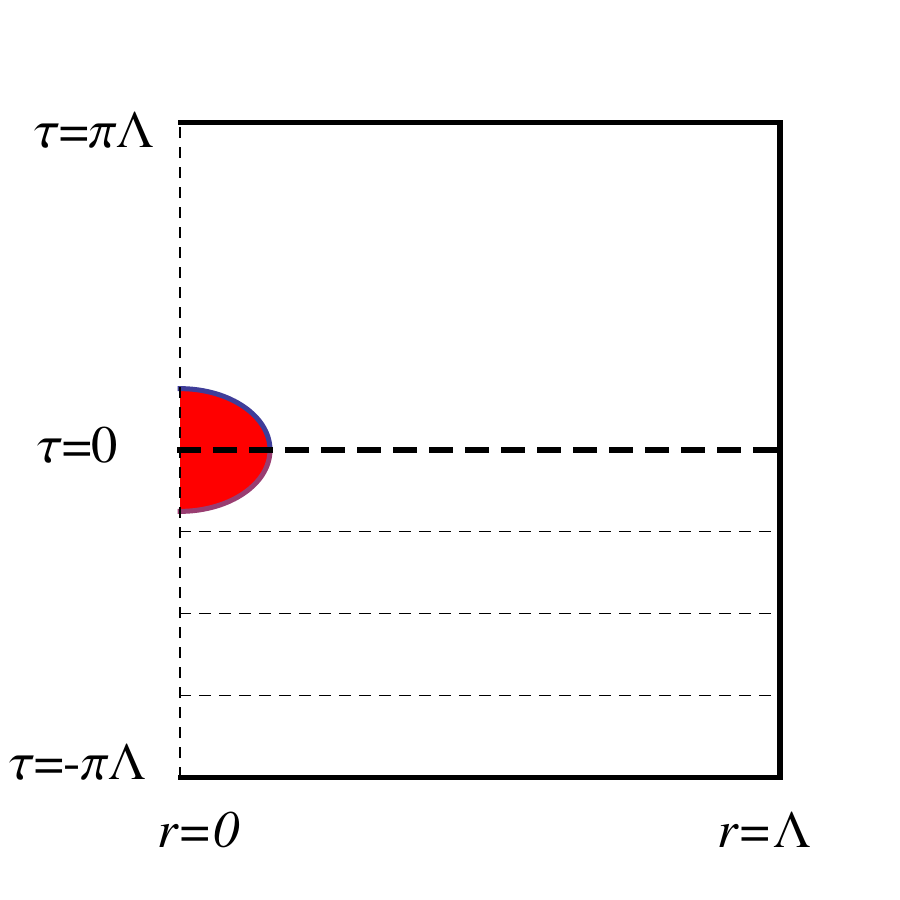} \hspace{20pt}
   \caption{A CDL bounce as it is seen in the static patch. The horizontal dashed lines are the constant $\tau$ hypersurfaces, the shaded region the true vacuum and $r$ ranges from 0 at the center of the bubbles to $\Lambda$ at the horizon.}
   \label{Intro-Slicing4}
\end{figure}

Until now all the formalism developed for curved spacetime was by analogy with the flat space case. But a more accurate treatment of the solutions is necessary for the interpretation of the bounces. For large bounces the interpretation is not very clear. In the flat space case shown in Fig. \ref{Intro-Slicing1}, at very large negative Euclidean time we get the configuration before the tunneling and at $\tau=0$ the configuration afterwards. But in curved spacetime with  large bubbles, there is no such time which shows the configuration before the tunneling.  Brown and Weinberg \cite{Brown:2007sd} clarified the interpretation for the case of  a fixed background approximation\footnote{Here we follow the approach in \cite{WeinbergBook}}. In this approximation the variation of the potential in the horizon volume is much smaller than the potential itself. Therefore we can assume that the metric is the one of a de Sitter space.  This metric in the (Euclidean) static patch   is
\begin{eqnarray} \label{Intro-StaticPatchMetric}
	ds^2 &=&  P(r)^2 d \tau^2 + h_{ij}dx^i dx^j \cr
	&=&  \left(1 -{r^2 \over \Lambda^2} \right) d\tau^2 + \left(1 -{r^2 \over \Lambda^2} \right) ^{-1} dr^2 + r^2 (d\theta^2 + \sin^2 \theta d\phi^2)~.
\end{eqnarray}
In the static patch, $r$ starts from 0 and ends at the horizon radius $\Lambda$ and the Euclidean time $\tau$ has a period of $2\pi\Lambda$. In the fixed background approximation, the constant $\tau$ surfaces are three-dimensional balls bounded by  two-spheres at the horizon. In this static background, we can define an energy functional with respect to the time-like Killing vector $\partial_\tau$
\begin{equation}
	E = \int d^3 x \sqrt{{\rm det} h} \left[ {1 \over 2 P(r)} \left( {\partial \phi \over \partial \tau}\right)^2+ \frac12 P(r) h^{ij} \partial_i \phi \partial_j \phi + P(r) V(\phi)\right]~,
\end{equation}
where $P(r)$ and $h_{ij} $ are defined in Eq. \eqref{Intro-StaticPatchMetric}. Now we can  treat the tunneling problem as a flat space case with a nonzero temperature (the time coordinate has a finite period) and there is no ambiguity. The bounce is shown in Fig. \ref{Intro-Slicing4}. 

\begin{figure}[htbp] 
   \centering
   \includegraphics[width=2.5in]{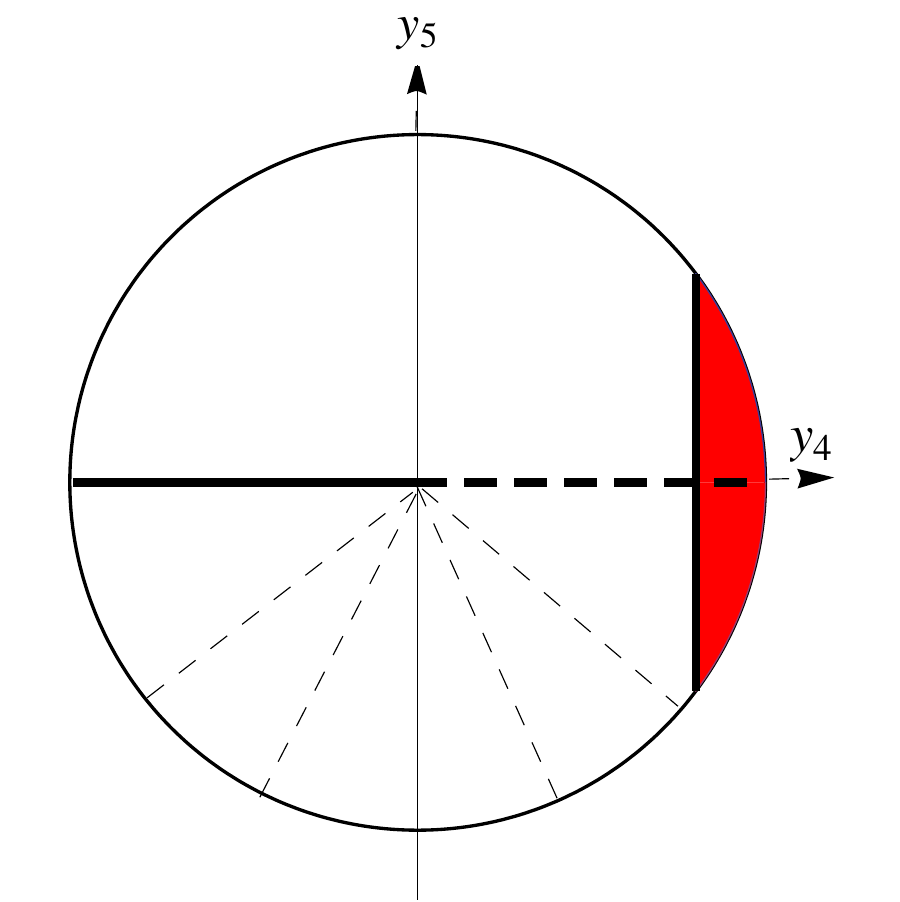} \hspace{20pt}
   \includegraphics[width=2.5in]{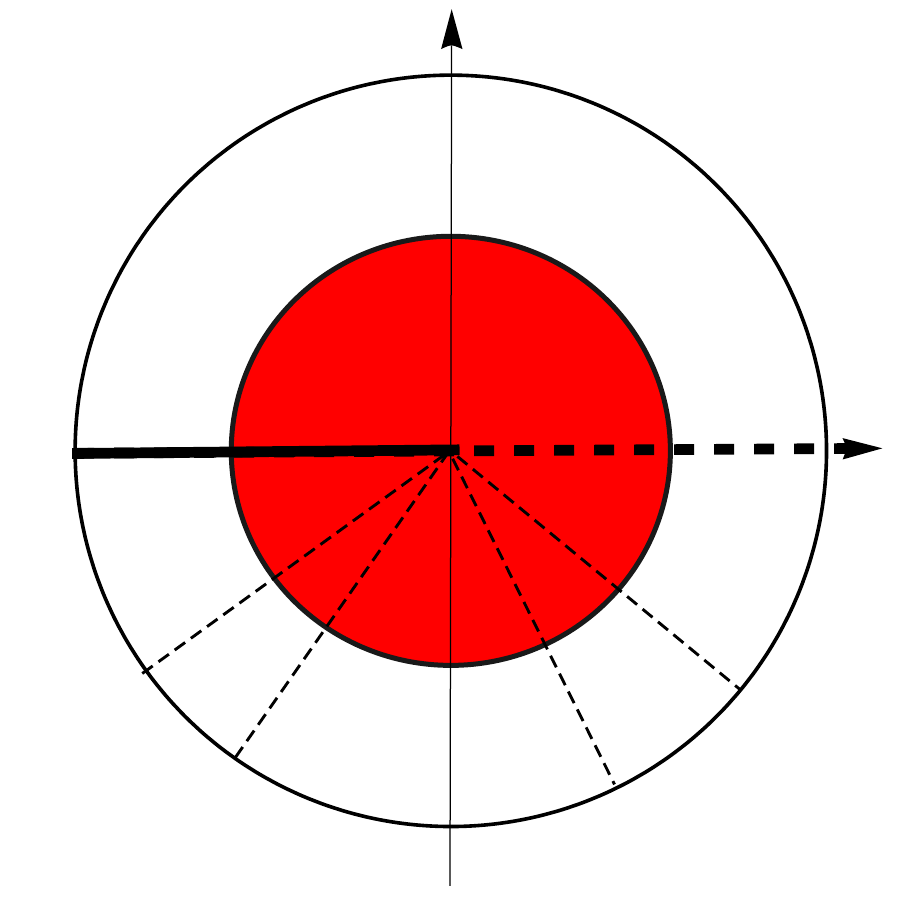} 
   \caption{The bounce of Fig. \ref{Intro-Slicing4} is shown in the $y_4\text{-}y_5$ plane on the left. The lines $\tau = -\pi \Lambda$ and $\tau = \pi \Lambda$  are identified and they show the radial directions. The configuration after tunneling is given by half of the hyperplane that goes through the center of the bounce. The other half shows the configuration  before the nucleation. On the right the  same bounce is shown after a rotation. This picture shows a $\tau$-independent bounce which resembles a purely thermal tunneling.}
   \label{Intro-Slicing5}
\end{figure}

These bounces have the topology of a four-sphere, as expected from the Euclidean version of the de Sitter space. Let's use a coordinate system which shows the four-sphere structure of the bounce more clearly 
\begin{eqnarray}
	y_1 &=& r \sin \theta \cos \phi ~,\cr
	y_2 &=& r \sin \theta \sin \phi ~,\cr
	y_3 &=& r \cos \theta ~,\cr 
	y_4 &=& \sqrt{\Lambda^2 - r^2 } \cos \left(\tau / \Lambda\right) ~,\cr
	y_5 &=& \sqrt{\Lambda^2 - r^2 } \sin \left(\tau / \Lambda\right)~.
\end{eqnarray}
Here $\tau = -\pi \Lambda$ and $\tau = \pi \Lambda$ are identified. If we look at the bounce of Fig. \ref{Intro-Slicing4} in the $y_4\text{-}y_5$ plane we see the left picture of Fig. \ref{Intro-Slicing5}. Rotating it we get the right picture of Fig. \ref{Intro-Slicing5}. This one shows a $\tau$-independent bounce which resembles the purely thermal bounces for the flat space. This means that the CDL bounce can be viewed as a thermal bounce. This should not be very surprising, since the thermal nature of the de Sitter space is a consequence of  quantum mechanics. 

%
%

\clearpage

\chapter{Bounces with ${\rm O}(3) \times {\rm  O}(2)$ symmetry} 
\label{O(3)}
We studied tunneling in curved spacetime in Sec.\ref{Intro-sec-CurvedSpace}  by starting  from the argument for the  flat spacetime  case  \cite{Coleman:1977py} and generalizing  it by analogy to curved spacetime  \cite{Coleman:1980aw}. In this  picture, tunneling is carried by bounce solutions which are the solutions of the Euclidean equations of motion of gravity coupled to matter fields. In flat space and zero temperature decay of a single scalar field, the solutions with the minimum  Euclidean action, and therefore the highest rate,  have an $O(4)$ symmetry \cite{Coleman:1977th}. There is no such  statement for  decays in curved spacetime or decays in theories with more than one scalar field. However it is generally believed that $O(4)$ symmetric solutions are also dominant in curved spacetime or multi-scalar field theories. In this chapter, we study solutions which have the next highest symmetry, $O(3) \times O(2)$, where $O(3)$ is the rotational invariance in  spatial coordinates and $O(2)$ shows that the solutions are independent of  the periodic Euclidean time $\tau$. 

This chapter is based on \cite{Masoumi:2012yy}. We provide more evidence for the dominance of $O(4)$ symmetric solutions in curved spacetime and also shed more light on the reinterpretation of the Coleman-De Luccia bounces  in de Sitter space introduced in  \cite{Brown:2007sd} where these bounces were understood as tunneling in a finite horizon volume and finite temperature. This interpretation was described in length in Sec. \ref{Intro-sec-CurvedSpace}.

Static self-gravitating rotationally symmetric bounces have been previously studied in various contexts, such as  the false vacuum decay in the presence of a black hole (for example \cite{Farhi:1989yr,Farhi:1986ty,Maeda:1981gw,Samuel:1991dy,Berezin:1991qz}  and the references  in these papers ). 
Our interest in these bounces was first raised by the work of Garriga and Megevand  \cite{Garriga:2004nm}. Motivated by the analogy with the thermal production of bubbles in flat spacetime, they studied bounces with $O(3) \times O(2)$ symmetry and interpreted them as the nucleation of a pair of bubbles. Their discussion is in the context of brane nucleation \cite{Brown:1987dd,Brown:1988kg}, in which the bubble walls are two-branes that separate the true vacuum at the center of a horizon volume from the false vacuum that extends to the horizon. This  calculation can be applied to the case of a field theory in the thin-wall approximation. In Sec.\ref{O(3)-sec-Garriga} we discuss the thin-wall solution and review the results of Garriga-Megevand.  In Sec.\ref{O(3)-sec-Background}, we set up the formalism that we use for studying the decay of a single scalar field in the context of the new symmetry. Our  analysis is  not restricted  to the thin-wall limit. The limiting cases are studied in  Section \ref{O(3)-sec-LimitingCases} and we show that when the bubble gets large, the thin-wall approximation eventually breaks down. We then summarize in Sec. \ref{O(3)-sec-Summary} and conclude that the $O(4)$ symmetric solutions are the dominant configuration  of tunneling in all the cases we studied.

\section{Review of the thin-wall results for the case of brane nucleation}
\label{O(3)-sec-Garriga}
\begin{figure}[htbp] 
   \centering
   \includegraphics[width=2.8in]{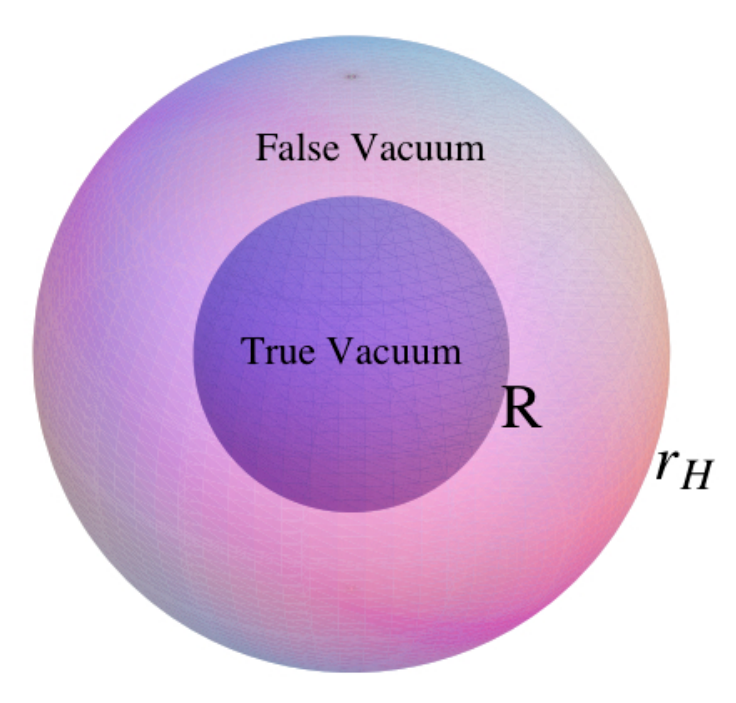} 
   \caption{A schematic view of the thin-wall solution. The wall is located at the radius $R$ and inside is the true vacuum. The horizon is located $r_H$ and between the wall and the horizon is the false vacuum.  The bounce solution is constant in the periodic Euclidean time $\tau$ and apparently rotationally invariant. }
   \label{O(3)-fig-thinwall}
\end{figure}

\begin{figure}[htbp] 
   \centering
   \includegraphics[width=3in]{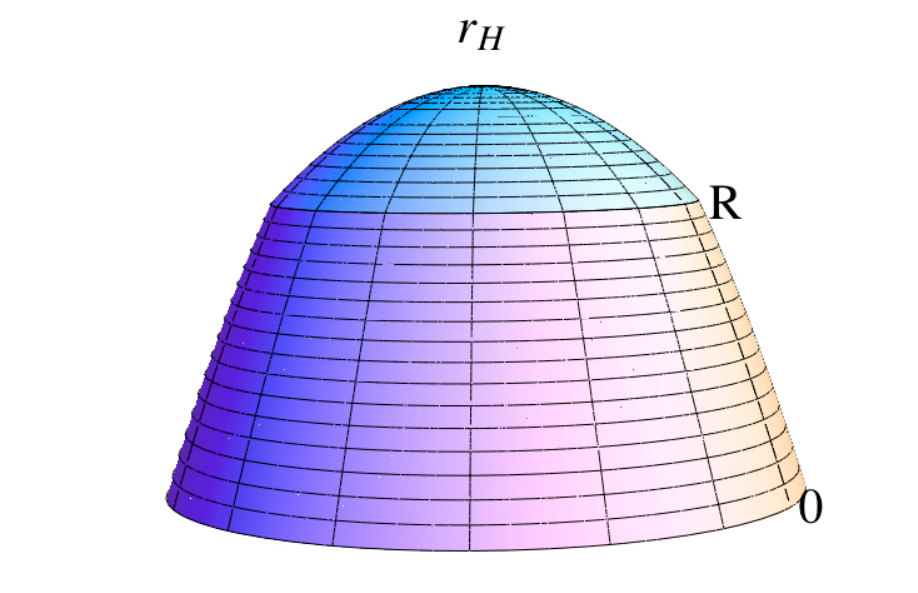} 
   \includegraphics[width=3.3in]{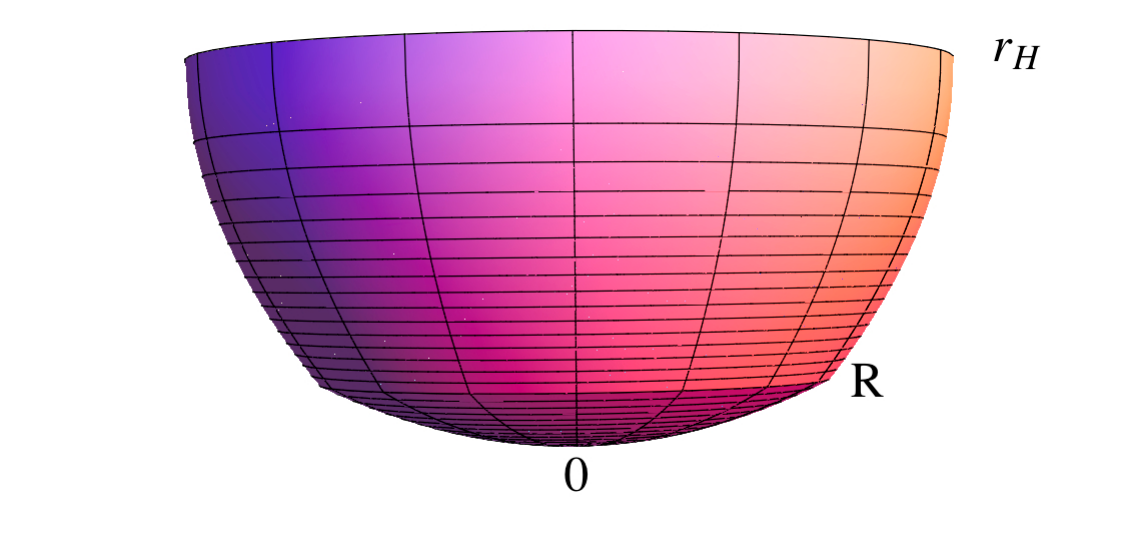} 
   \caption{The induced geometry of the static instanton in two different planes. Left is the $r-\tau$ plane where it is described by a periodic angular coordinate $\tau$ and the coordinate along the vertical axis is $r$.  Horizon is at the tip of this object  $r_H$ and the wall is at the location that there is a discontinuity in the first derivative due to the energy density of the wall.  Right, the same picture, but  in $r-\phi$ plane. Again $\phi$ is the periodic coordinate and $r$ is along the vertical axis. To avoid  a conical singularity, the period of $\tau$ must be a very specific number.  }
   \label{O(3)-fig-ThinwallGeometry}
\end{figure}

The thin-wall solutions we are considering here are static solutions with rotational symmetry\footnote{This section and the figures in it are following \cite{Garriga:2004nm}.}. We can think about them as a three-ball in spatial coordinates.  There is a wall at a radius $R$ and the horizon is located at a radius $r_H>R$. This geometry is shown in Fig.\ref{O(3)-fig-thinwall}. 
Inside the wall, the metric is that of a de Sitter (dS) space. Because the mass inside the wall is not zero, the metric on the outer part is a Schwarzschild-de Sitter (SdS) metric. Therefore the metrics inside  and outside of the wall  (in static patch coordinates\footnote{These coordinates are described in Appendix \ref{Appendix-DeSitter} ~.}) are
\begin{eqnarray} 	\label{O(3)-eq-MetricsInOut}
	ds_i^2 &=& C^2 f_i(r) d\tau^2 + f^{-1}_i(r) dr^2 + r^2 d \Omega^2 \nonumber ~,\\
	ds_o^2 &=& f_o(r) d\tau^2 + f^{-1}_o(r) dr^2 + r^2 d \Omega^2 ~,
\end{eqnarray}
where
\begin{eqnarray}
	f_i(r) &=&1 - H_i^2 r^2 \nonumber ~,\\
	f_o(r) &=&1 - H_o^2 r^2 - {2 G M \over r} ~,
\end{eqnarray}
Here $d\Omega^2 = d\theta^2 + \sin^2  \theta  \,d\phi^2$ is the element of area on a unit two-sphere, $H_i$ and $H_o$ the Hubble parameters  inside and outside of the wall, and $M$ the mass parameter in the SdS metric which is related to the total mass of the inside bubble plus the mass of the wall. $C$ is a parameter to make the metrics smooth at the tangential directions at  the wall. The projections of this geometry in the $r-\tau$ and $r-\phi$ planes  are shown in Fig.\ref{O(3)-fig-ThinwallGeometry}.

The two metrics are glued smoothly at the location of the wall using the Israel junction conditions\cite{Israel:1966rt}. Knowing $\sigma$, the surface tension of the wall, and $H_{\rm tv}$ and $H_{\rm fv}$, the Hubble parameters of the true and false vacua  is sufficient to determine the mass and  radius of the region inside the wall. In terms of the extrinsic curvature $K_{ab} $, the junction conditions can be written as
\begin{equation}
	\left[ K_{ab}\right] = - 4 \pi G \sigma \gamma_{ab}~,
	\label{O(3)-eq-JunctionConditions1}
\end{equation}
where $\left[ K_{ab}\right]$  is the difference between the extrinsic curvatures on both sides of the wall and $\gamma_{ab}$ is the world-sheet metric. The extrinsic curvature is
\begin{equation}
	K_{ab} = {1 \over 2} f^{1/2} \partial_r g_{ab}~.
\end{equation}
We can rewrite Eq.~\eqref{O(3)-eq-JunctionConditions1} in terms of metric components in Eq.~\eqref{O(3)-eq-MetricsInOut} and obtain
\begin{equation}
	g_o - g_i = - 4 \pi G \sigma~, \qquad \qquad  g'_o-g'_i=0~,
\end{equation}
where 
\begin{equation}
	g_o(r) = \frac{f_o^{1/2}(r)}{r} ~, \qquad \qquad g_i(r) = \frac{f_i^{1/2}(r)}{r} ~,
\end{equation}
and all the derivatives are with respect to $r$. These equations completely determine $M$ and $R$ in terms of $\sigma$, $H_o$ and $H_i$:
\begin{eqnarray}
	R^{-2} &=& x^2 + H_i^2 ~, \nonumber \\
	M &=& {4 \pi \sigma R \over 3 x}~,
\end{eqnarray}
where $x$ is defined by
\begin{equation}
	x = {\epsilon \over 4 \sigma } + {3 \sigma \over 16 M_P^2 } + \left[ \left( {\epsilon \over 4 \sigma} + {3 \sigma \over 16 M_P^2}  \right)^2 + {H_i^2 \over 2} \right]^{1/2}~.
\end{equation}
In this equation, $\epsilon$ is the difference between the true and false vacuum energies. These equations imply that the tension cannot get too large, which is what we would naively expect. To compensate for the increase of the tension, we have to increase the radius of the bubble. But because of the existence of  a horizon, there is not enough room for the radius to get too large and therefore there should be a maximum allowed tension $\sigma_N$ which is shown to be
\begin{equation}
	\sigma_N= 2 M_P ^2  \sqrt{3 H_o^2 - H_i^2}~.
\end{equation}
When $\sigma$ approaches $\sigma_N$, the outside metric starts developing a double root at the horizon and beyond  $\sigma_N$, there is no solution. We will see in Sec.\ref{O(3)-sub-NoNariai} that this picture gets modified in the context of a scalar field theory and that  before reaching this limit the thin-wall approximation breaks down. 

The periodicity of $\tau$ is a measure of the temperature felt by the system. It can be determined by demanding that there be no conical singularity at the horizon. Near the horizon, we can expand the metric in powers of distance from the horizon  
\begin{equation}
	f_o(r) = B^2 (r_H - r)~,
\end{equation}
where $B^2$ is positive to make the metric positive inside the horizon. It is straightforward to calculate it:
\begin{equation}
	B^2 = 3 H_o^2 r_H - {1 \over r_H}~.
\end{equation}
Using the standard change of variables 
\begin{equation}
	\rho = {2  \over B } \sqrt{r_H-r}~,  \qquad \gamma= {B^2 \over 2} t~,
\end{equation}
we can rewrite the metric near horizon as
\begin{equation}
	ds^2=\rho^2 d \gamma^2 + d\rho^2 + r_H^2 d\Omega^2~.
\end{equation}
To avoid a conical singularity, $\gamma$ must have a period of $2\pi$ and therefore the period of the time coordinate is given by
\begin{equation}
	\beta  = {4 \pi r_H \over 3 H_o^2 r_H^2 -1} = {2 \pi r_H^2 \over r_H- 3 G M}~.
\end{equation}
In this approximation, the action of the bounce solution is
\begin{equation}
	S = - {\pi r_H^2 \over G}~.
\end{equation}
We will show in Sec. \ref{O(3)-sec-Background} that this result is in fact exact. Numerical calculations in \cite{Garriga:2004nm} show that the action  for the $O(3)\times O(2)$ in the thin-wall approximation is higher than their $O(4)$ symmetric counterparts in thin-wall limit.

\section{Inclusion of a scalar field}
\label{O(3)-sec-Background}
In this section,  we set up the formalism for the rest of this chapter. Our goal is to  study the decay of a scalar field $\phi$ with a potential $V(\phi)$ in de Sitter space. The initial Hubble parameter is set by  $V_{\rm fv}$ (or $V_{\rm tv}$ for decays from a true to a false  vacuum) and the bounce solution has an $O(3)\times O(2)$ symmetry. Later in Sec. \ref{O(3)-sec-Numerical} we will  scan over possible $V$'s. The scalar field and the metric are independent of the Euclidean time $\tau$ and only depend on the spherical coordinate radius $r$. The most general $\tau$-independent metric compatible with spherical symmetry can be written as 
\begin{equation} \label{O(3)-MetricAnsatz}
	ds^2 = B(r)^2 d \tau^2 + A(r) dr^2 + r^2 d\Omega^2~.
\end{equation}
We still have the  freedom to rescale  $\tau$. This will modify $B(r)$ by a rescaling. The normalization for $B$ is explained in the next  paragraph. Each spatial slice is a three-ball and the horizon radius $r_H$ is located at the zero of $g^{rr}= 1/A$. 

\begin{figure}[htbp] 
   \centering
   \includegraphics[width=4in]{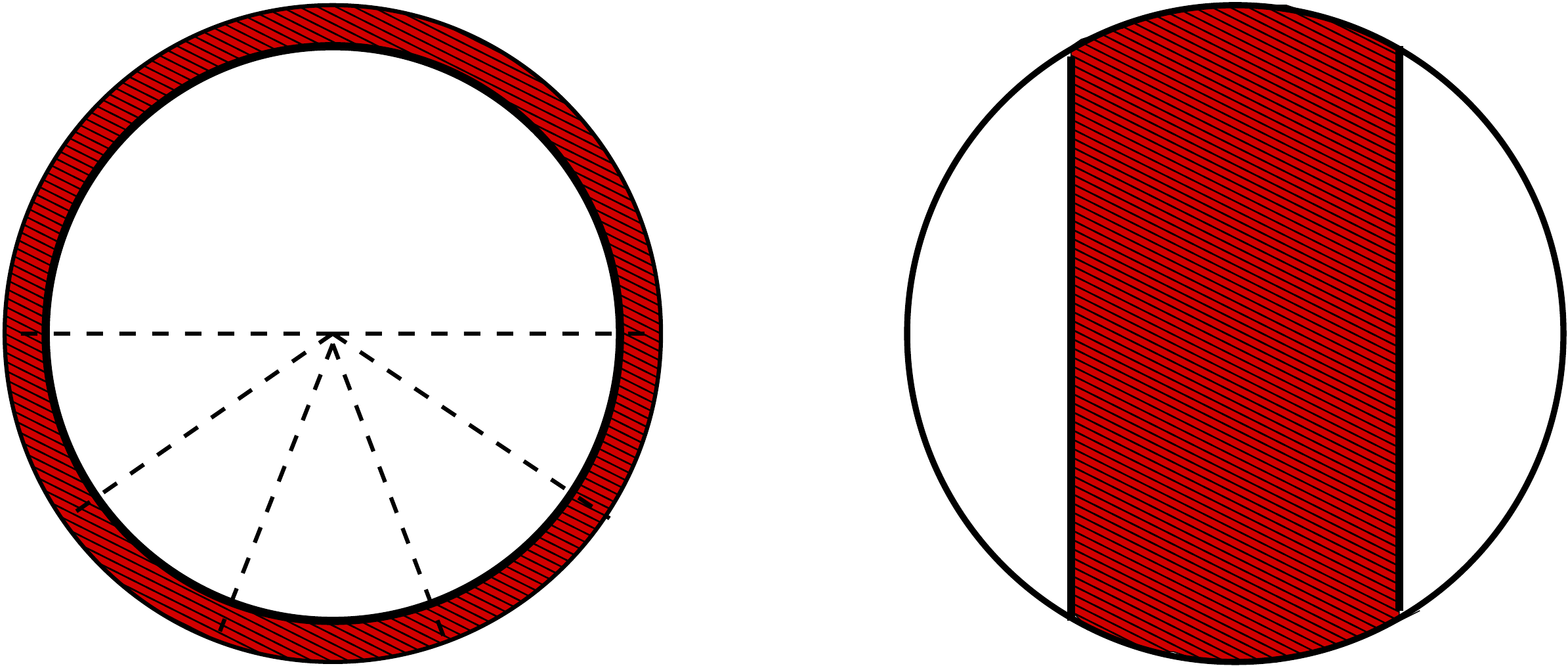} 
   \caption{Two views of an $O(3) \times O(2)$-symmetric bounce. These are from the same viewpoints as those of the CDL bounce in  \ref{Intro-Slicing5}. (The views on the right would be the same if viewed from along any axis in the $x_4{\textrm -}x_5$ plane.) On the left-hand view the radial dashed lines correspond to the similar lines in Fig. \ref{Intro-Slicing5}. The bounce is independent of the Euclidean time. The three- dimensional  configuration corresponding to any of these lines contains a single true vacuum critical bubble in the center of a false vacuum horizon volume.}
   \label{O3O2picture}
\end{figure}

In order to get a non-singular solution at the origin, we need $A(0)=1$ to avoid  a conical singularity. We choose the normalization of $B$ by demanding that $A(r_H) B(r_H) = 1$~. This configuration corresponds to thermal nucleation of a bubble at the center of the horizon volume. The bounce solution is drawn in Fig. \ref{O3O2picture} on a four-sphere. The analysis of \cite{Brown:2007sd} predicts the production of a single bubble, while the conventional interpretation is the production of  a pair of bubbles. The former seems to be a more natural interpretation. 
\subsection{Periodicity of Euclidean time}
\label{O(3)-sub-periodicity}
In thermal tunneling in flat space, the temperature determines the period of the Euclidean time and can be chosen arbitrarily. In curved spacetime, however, we expect the temperature and hence $\beta$, the period of the Euclidean time to be determined by the metric\footnote{$\beta$ only enters in our calculations as an overall multiplicative factor in the action and does not otherwise affect the solution. It would be a straightforward matter to extend our results to an arbitrary temperature, but the physical origin of the temperature would be less compelling.}. Therefore we assume that $\beta$ is determined by the surface gravity at the horizon of the Lorentzian counterpart of our Euclidean metric:
\begin{equation} \label{O(3)-SurfaceGravity}
	\beta = {1 \over T} = 4\pi \left[ B' \left( {1\over A}\right)'\right]_{r_H}^{-1/2}
\end{equation}
where primes denote differentiation with respect to $r$ and the subscript indicates that all the quantities are to be evaluated at the horizon. Our normalization condition that $A(r_H) B(r_H)=1$, the fact that at the horizon $A^{-1}=0$, and  Eq.~\eqref{O(3)-SurfaceGravity} lead to 
\begin{equation}
	B' = {d \over dr } \left[ (AB) \left( {1 \over A}\right)\right] = (AB)' \left( { 1 \over A} \right) + (AB) \left( {1 \over A}\right)' = \left( {1 \over A}\right)' ~.
\end{equation} 
 Therefore $\beta$ simplifies to
 \begin{equation}
	\beta = - 4 \pi \left[ B'(r_H)\right]^{-1}~,
 \end{equation}
with the minus sign coming from the fact that $B'(r_H)$ is negative. If the space were  pure de Sitter, $\beta$ would be $2 \pi r_H$. It should remain the same order of magnitude for the bounce solution. 

There is another way to deduce Eq.~\eqref{O(3)-SurfaceGravity}. At the horizon  $1/ A = 0$ and $A B =1$. This means that $B(r_H)=0.$ Therefore both $\left( {1 \over A}\right)$ and $B$ vanish at the horizon. Assuming that they do not develop  double roots, to the first order of Taylor expansion
\begin{eqnarray}
	B(r) &=& - B' (r_H - r)~, \nonumber \\
	{1 \over A(r)} &=& C^2 (r_H -r )~,
\end{eqnarray}
where $C^2 = - \left( {1 \over A}\right)'$. In terms of the new variable  $y = \sqrt{4(r_H-r) \over C^2}$, the metric near the horizon is
\begin{equation}
	ds^2 = dy^2 + {B'  C^2 \over 4} y^2 d\tau^2 + r^2 d \Omega^2~.
\end{equation}
Avoiding a conical singularity in the $y\textrm{-}\tau$ plane leads to 
\begin{equation}
	\beta = 4\pi \left( B' C^2\right)^{-1/2}=4\pi \left[ B' \left(\frac{1}{A}\right)'\right]^{-1/2}~,
\end{equation}
which is consistent with Eq.~\eqref{O(3)-SurfaceGravity}~.

\subsection{Equations of motion and boundary conditions}
\label{O(3)-sub-EOMS}

The Euclidean action for  a scalar field coupled to gravity is\footnote{Please notice that the $g$ is a positive quantity in Euclidean spacetimes.}
\begin{equation} \label{O(3)-eq-GeneralAction}
	S = \int d^4x \sqrt{g} \left( -{1 \over 16 \pi G} {\mathcal R} + {\mathcal L}_{\rm matter} \right)~,
\end{equation}
where $\mathcal R$ is the scalar curvature and ${\mathcal L}_{\rm matter}$ is
\begin{equation}
	{\mathcal L}_{\rm matter} = \frac{1}{2 } \partial_\mu \phi \partial^\mu \phi + V(\phi)~.
\end{equation}
The nontrivial components of the Einstein tensor and the energy-momentum tensor for the metric in Eq.~\eqref{O(3)-MetricAnsatz} and the action in Eq.~\eqref{O(3)-eq-GeneralAction} are 
\begin{eqnarray}\label{O(3)-eq-O(3)1}
	G_{\tau\tau} &=& -\frac{B}{r^2 A^2} (- A + A^2 + r A') \nonumber ~,\\ 
       G_{rr} &=& \frac{ B - A B + r B'}{r^2 B} \nonumber ~, \\
       T_{\tau\tau} &=& - B \left( \frac{\phi'^2}{2A} + V(\phi) \right) ~, \nonumber \\
       T{rr} &=& A    \left(\frac{\phi'^2}{2A} - V(\phi)\right)~, \nonumber \\
       {\mathcal R} &=& { 4 A^2 B^2 + r B A'( 4 B + r B' ) + A \left(- 4 B^2 + r^2 B'^2 -2 r B (2 B' + r B'')\right)  \over  2 r^2 A^2 B^2}  ~.
 \end{eqnarray}
The gravitational and matter contributions to the action inside a horizon volume are ($\kappa = 8 \pi G$)
\begin{eqnarray}\label{O(3)-eq-O(3)2}
	 S_{\rm gravity} &=&- \int_0^{r_H} \pi dr {\left[ 4 A^2 B^2 + r B A' (4 B + r B') -4 A B^2 + r^2 A B'^2 - 2 r A B ( 2 B' + r B'')\right]\over \kappa (A   B)^\frac{3}{2}}~, \nonumber \\
       S_{\rm matter} &=& \int_0^{r_H} dr  4 \pi r^2 \sqrt{A B} \left(    \frac{\phi'^2}{2A} + V(\phi) \right)~. 
\end{eqnarray}
The equations of motion arising from varying the action in Eq.~\eqref{O(3)-eq-O(3)2} with respect to $\phi$, $A$ and $B$ are\footnote{These equations can also be obtained by plugging the ansatz in Eq.~\eqref{O(3)-MetricAnsatz} into the full field equations. }
\begin{eqnarray}
	0&=&  {d \over dr } \left[ r\left({1\over A} - 1\right)\right] + \kappa r^2 \left[ {1 \over 2} {(\phi')^2 \over A} + V(\phi)\right]    \label{O(3)-EOMSA} ~,    \\  
       0&=& {(A B)' \over AB} - \kappa r (\phi')^2 ~,  \label{O(3)-EOMSB} \\ 
     	0&=&\frac{d}{dr}\left(\frac{r^2 \sqrt{A B} \phi'}{A}\right)-r^2 \sqrt{A B} \frac{d  V}{d \phi}  \label{O(3)-EOMSC}~.
\end{eqnarray}
We can integrate the second of these equations and get
\begin{equation}
	A(r) B(r) = A(r_H) B(r_H) \exp{\left(- \kappa \int_r^{r_H} dr' \,r'\, (\phi')^2 \right)}~.
\end{equation}
We chose $A(r_H) B(r_H)=1$ in \ref{O(3)-sub-periodicity}. Therefore 
\begin{equation} \label{O(3)-eq-EliminatingB}
	A(r) B(r) = \exp{\left(- \kappa \int_r^{r_H} dr' \,r'\, (\phi')^2 \right)}~.
\end{equation}
We can eliminate $B$ from these equations so the only remaining fields to determine are $A(r)$  and $\phi(r)$~. Before proceeding further, it is  convenient to define new variables $f(r)$ and ${\mathcal M}(r)$ by
\begin{eqnarray}
	f(r) = {1 \over A(r)} = 1 - {2 G {\mathcal M}(r) \over r} ~.
\end{eqnarray}
We can rewrite  Eq.~\eqref{O(3)-EOMSA}  in terms of $f$ as
\begin{equation} 
	0=  {d \over dr } \left[ r\left(f - 1\right)\right] + \kappa r^2 \left[ {1\over 2} {(\phi')^2f } + V(\phi)\right] \label{O(3)-EOMSD}~.
\end{equation}
This can also be written in the familiar form
\begin{equation}\label{O(3)-EOMSE}
	{\mathcal M}' = 4 \pi r^2 \left[ {1 \over 2} (\phi')^2 f + V(\phi)\right]~.
\end{equation}
Substituting Eq.~\eqref{O(3)-EOMSB} in Eq.~\eqref{O(3)-EOMSC} gives
\begin{equation} \label{O(3)-EOMSF}
	0 = f \left[ \phi'' + {2 \over r} \phi' + 4 \pi G r (\phi')^3\right]+ f' \phi' - {dV \over d \phi}~.
\end{equation}
This equation together with Eq.~\eqref{O(3)-EOMSD} gives two equations involving only $f$ and $\phi$. They are second order in $\phi$ and first order in $f$ and need three boundary conditions. To avoid a conical singularity at the origin and guarantee the smoothness of $\phi$, we would need
\begin{eqnarray} \label{O(3)-eq-BCA}
	f(0) &=& 1 \nonumber~,\\
	\phi'(0)&=&0~.
\end{eqnarray}
The first boundary condition prevents a conical singularity at the origin (since $d\Omega^2$ is the standard two-sphere metric) and the second one is a consequence of Eq.~\eqref{O(3)-EOMSF}. To completely determine the fields, we need one more boundary condition. At the horizon, $f=0$ and if we assume that $\phi''$ and $\phi'$ are finite, it imposes
\begin{equation} \label{O(3)-eq-BCA2}
	\left.\left[f' \phi' - {d V \over d \phi}\right] \right|_{r=r_H}=0
\end{equation}
In fact this condition holds even for singular $\phi'$ if $\phi'$ grows slower than $(r_H-r)^{-1/2} $ as we approach the horizon. Such a divergence does not imply a divergent action density and  is merely a coordinate singularity. In fact, choosing the proper distance from the horizon as a new coordinate makes $\phi'$ finite. 

There is another subtlety that we have to address here. With the periodicity introduced in Eq.~\eqref{O(3)-SurfaceGravity}, the four-sphere obtained by identifying all of the boundary  two-spheres is a smooth manifold, just as in the CDL case. However, as in CDL, we do not require that $\phi$ to be smooth on this two-sphere. Smoothness of $\phi$ on the four-sphere requires $\phi'(r_H)=0$. Equation ~\eqref{O(3)-EOMSF} implies that $dV / d \phi=0$ at the horizon which in turn means that the field must be precisely at a vacuum value (or else exactly at the top of the barrier, as in the Hawking-Moss solution). This is only possible in the (unattainable) limit in which the thin-wall approximation is exact. Having $\phi'(r_H)\neq0$ causes a discontinuity in slope along a path passing through the two-sphere. However, this discontinuity does not cause any cost in action and therefore it is fine for $\phi'$ to be non-smooth. Knowing $\phi$ and $f$ and using Eq.~\eqref{O(3)-eq-EliminatingB} determines $B$. 
\subsection{Bounce action}
\label{O(3)-sub-Action}
One of the  interesting features of these $O(3)\times O(2)$-symmetric bounces is the remarkable coincidence that the on-shell Lagrangian density is a total derivative and therefore the action only depends on the boundary values of the fields. Starting from Eq.~\eqref{O(3)-eq-O(3)2} and integrating by parts, we can obtain 
\begin{equation}
	S = {\beta \over 4 G}\left[ r^2 {B' \over \sqrt{AB}}\right]_{r=0}^{r=r_H} + 4 \pi \beta \int_0^{r_H} dr \sqrt{AB} \left\{ {d \over \kappa dr} \left[ r\left({1 \over A} -1\right)\right] + r^2 \left[ {1\over 2} {(\phi')^2 \over A} + V(\phi)\right]\right\}~.
\end{equation}
Upon using Eq.~\eqref{O(3)-EOMSA}, the terms in the curly brackets vanish. Now using Eq.~\eqref{O(3)-SurfaceGravity}, we can simplify the first term 
\begin{equation}\label{O(3)-SimplifiedAction}
	S_{\rm bounce} = - {\pi \over G} r_H^2~.
\end{equation}
The action only depends on the horizon area. The tunneling exponent $\mathcal B$ is the difference between the bounce action and the false vacuum action. 
\begin{equation} \label{O(3)-ExactTunnelingExponent}
	{\mathcal B} = S_{\rm bounce} - S_{\rm fv} = {\pi \over G } \left( \Lambda_{\rm fv}^2 - r_H^2\right)~,
\end{equation}
where 
\begin{equation}
	\Lambda_{\rm fv} = H_{\rm fv}^{-1} = \sqrt{3 \over 8 \pi G V_{\rm fv}}~.
\end{equation}
It agrees with the thin-wall results of \cite{Garriga:2004nm} and is expected on more general grounds. As shown in \cite{Banados:1993qp,Hawking:1995fd}, the action of a static Euclidean solution with periodicity $\beta$ is $S_E = \beta E - {\mathcal S}$, where $E$ is the total energy and ${\mathcal S}$ is the entropy. For a solution without a boundary the total energy vanishes and the action depends on the entropy, which in turn depends on the horizon area.
\subsection{Comments on the  thin-wall approximation}
\label{O(3)-sec-thinwall2}
The thin-wall approximation is a  useful tool  for gaining an intuitive understanding of  tunneling processes. In this section we describe  the conditions for validity of this approximation in $O(3)\times O(2)$-symmetric solutions. In this approximation, the bounce is approximated by a region of pure true vacuum inside the wall, a transition region which is the wall and the  outside region which is in  the pure false vacuum. To use this approximation, we require two essential conditions. First, the thickness of the wall (transition region) must be much smaller than the radius of the wall so we can approximate the wall as being locally planar. The second condition concerns  the position dependence of the surface tension $\sigma$ of the wall. Although in most of the previous tunneling calculations, $\sigma$ was considered to be position-independent, it can depend on the position. To make the thin-wall approximation valid, the fractional variation of $\sigma$ must be small through the wall. 

In flat spacetime, the thickness of the wall is determined by the shape of the potential barrier separating true and false vacua. The wall radius is determined by the ratio of  $\sigma$ to  $\epsilon$, the difference between the false and true vacuum energy densities. Reducing $\epsilon$ and meanwhile keeping the shape of the potential almost unchanged  increases the wall radius and therefore it is always possible to obtain a thin-wall by making $\epsilon$ small enough. This is the case for both $O(4)$-symmetric zero-temperature and for high-temperature solutions that describe a critical bubble in three dimensions. 

The situation is different in the presence of gravity because it introduces a new length scale, the horizon radius. Therefore the radius of the wall gets an upper limit equal to the horizon length and the true vacuum region cannot get arbitrarily large. To even have a possibility of a thin-wall approximation, the shape of the potential must be chosen in such a way that the natural width of the wall is small compared to the horizon radius. If this criteria is satisfied, then by making $\epsilon$ small, we can achieve a thin-wall solution. In Sec.\ref{O(3)-sub-NoNariai} we argue that this is not always the case for our $O(3)\times O(2)$ solutions.

\subsection{No Nariai  limit in the thin-wall approximation}
\label{O(3)-sub-NoNariai}
In the context of brane nucleation it was claimed in \cite{Garriga:2004nm}  that as the surface tension $\sigma$ approaches the critical surface tension $\sigma_N$, the wall approaches the horizon and can get arbitrarily close to it. Meanwhile, the metric develops a double-root at the horizon. This is the  Nariai limit\cite{Nariai1951,Ginsparg:1982rs}. Here we show that  it is not possible in a field theoretical set up to achieve a thin-wall solution which approaches the Nariai solution.  If there is such a solution, near the horizon the metric can be approximated as $f(r) = B (R_H - r)^2$, where $B$ is a positive constant.  Because of the thin-wall approximation, the field near the horizon should be very close to the false vacuum. Expanding the potential near the false vacuum ($P$ and $Q$ are  determined by $V$),
 \begin{align}
        V(\phi) = V_{\rm fv} + \frac{1}{2} P (\phi- \phi_{\rm fv})^2 + \frac{1}{3} Q (\phi-\phi_{\rm fv})^3 + \ldots~.
 \end{align}
It is more convenient  to write  Eq.~\eqref{O(3)-EOMSD} and Eq.~\eqref{O(3)-EOMSF} in terms of $y = R_H- r$ and $\Phi= \phi - \phi_{\rm fv}$. In the remainder of this subsection the primes denote differentiation with respect to $y$ we obtain
\begin{align}
        &1 + R_H f' - y f' - f = 8 \pi G (R_H-y)^2 \left(f \frac{\Phi'^2}{2} + V\right) ~, \\
        &\Phi'' f + \Phi' f' - \frac{2}{R_H-y} \Phi' f - 4 \pi G (R_H-y) \Phi'^3 f = \frac{\partial V}{\partial \Phi} ~.
\end{align}
The boundary conditions are
\begin{align}
	& \Phi' f' \left.= \frac{\partial V}{\partial \Phi}\right|_{y=0}~,  \\
	&\Phi'(R_H) = 0~, \\  
	& f(R_H) = 1~.
\end{align}
Notice that $y=R_H$ is the same as $r=0$.
At the horizon, the equations of motion simplify to
\begin{align}
         &1 = 8 \pi G R_H^2 ( f \frac{\Phi'^2}{2} + V_H)~, \\
        &BR_H^2 = -1 + 4 \pi G R_H^3 \left( f \frac{\phi'^2}{2} + V\right)'~, \\
        &\frac{\partial U}{\partial \Phi} = f' \Phi' = 0~.
 \end{align}
Here we get a second order and a first order differential equation and two unknowns, $B$ and $R_H$, and five boundary conditions which make it possible to solve the equations near the horizon. Expanding $\Phi$ in terms of $y$   to the first nonzero order,
we get $ \Phi(y) = A \sqrt{y}$ where
\begin{equation} 
	A = \sqrt{\frac{2}{\pi G R}\left( \frac{3}{4} - \frac{P}{B}\right)} ~.
\end{equation}
Plugging back this solution for $\Phi$ into the metric equation gives 
\begin{align}
	&B R_H^2 = -1 + 4 \pi G R_H^3 \left( f {(\phi')^2 \over 2} + U\right)' = -1 + 4 \pi G R_H^3 A^2 \left( \frac{B}{8} + \frac{P}{2}\right) ~.
\end{align}
This leads to 
\begin{align}
	\frac{B}{4} \left(1 - \frac{4 P}{B}\right)^2 + \frac{1}{R_H^2}=0~,
\end{align}
where  $B$ is a positive number. This shows that the metric cannot develop a double root at the horizon. Because the left side is always a nonzero positive number, adding a very small linear term to the metric does not make the situation better. Therefore not only can the metric not develop a double root in the thin-wall limit, but also it cannot get arbitrarily close to having a double root.

\section{Limiting cases}
\label{O(3)-sec-LimitingCases}
We could not find  closed form solutions of  the field equations Eq.~\eqref{O(3)-EOMSA}-Eq.~\eqref{O(3)-EOMSC} and had to use numerical methods to understand the behavior of bounce. However there are limiting cases in which  we could get more analytic insight. We present them in this section.
\subsection{Weak gravity limit}
\label{O(3)-sub-WeakG}
If the gravitational constant $G$ is small, we should expect to recover the flat space results with small corrections due to the nonzero temperature of the horizon. If $G$ is sufficiently small, $\phi$ varies from a value near the true vacuum to one which is exponentially close to $\phi_{\rm fv}$ in an interval $0 \le r \le \tilde{r}$, where $\tilde r$ is much smaller than the horizon radius $\Lambda_{\rm fv}$. In region $\tilde r \le r \le \Lambda_{\rm fv} $, the field almost assumes the false vacuum value and the space is well approximated by a SdS metric. In region $0 \le r \le \tilde{r}$, the field  does not feel the curvature of the spacetime and Eq.~\eqref{O(3)-EOMSE} should be well approximated by the flat spacetime equations. Outside this region, we can integrate Eq.~\eqref{O(3)-EOMSE} to get
\begin{equation}
	{\mathcal M} = E_{\rm flat} + {1 \over 2 G} {r^3 \over \Lambda_{\rm fv}^2},
\end{equation} 
where $E_{\flat}$ is the energy of the critical bubble in flat spacetime. The horizon appears at the place where $f$ vanishes or, equivalently, $2 G {\mathcal M}(r) = r$. To first nonzero order in $G$, we get
\begin{equation}
	r_H = \Lambda_{\rm fv} - G E_{\rm flat}~.
\end{equation}
Since the bounce action only depends on the horizon radius, the tunneling exponent Eq.~\eqref{O(3)-ExactTunnelingExponent} is given by
\begin{equation}
	{\mathcal B}= 2 \pi \Lambda_{\rm fv} E_{\rm flat} = {E_{\rm flat} \over T_{\rm dS}}~,
\end{equation}
where from, Appendix\ref{Appendix-DeSitter}, $T_{\rm fv}= 1/2\pi \Lambda_{\rm fv}$. This is the expected Boltzmann exponent that we would obtain for nucleation of a critical bubble in flat spacetime at a temperature $T= T_{\rm dS}$. However, because $T \ll 1/ \tilde r$, the high temperature bounce has multiple negative modes and must be discarded. 

\subsection{Strong gravity and oscillating bounces}
\label{O(3)-sub-Oscillating}
\begin{figure}[htbp] 
   \centering
   \includegraphics[width=2.5in]{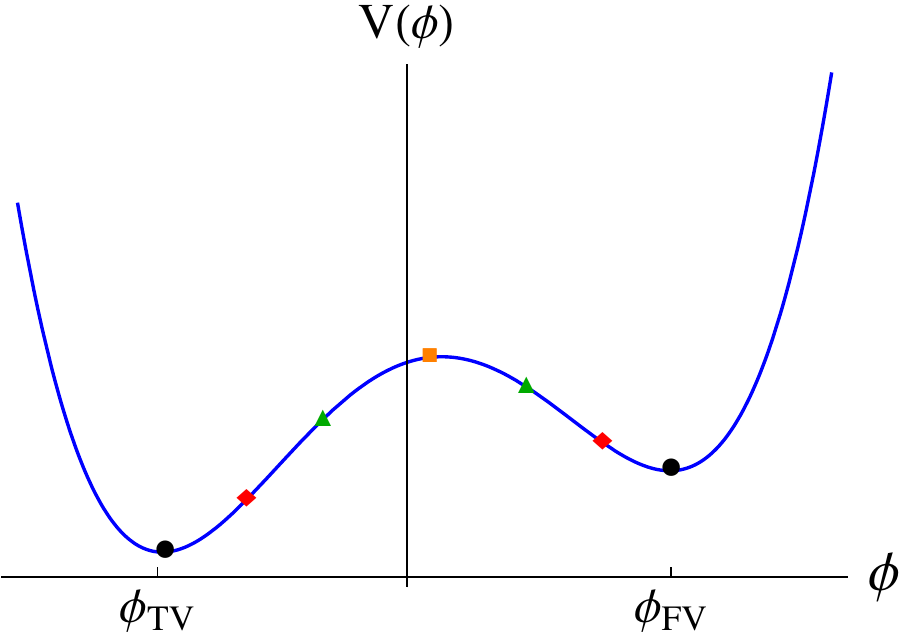} 
   \caption{Increasing $G$ leaves less and less room for the wall. As a result $\phi$ varies less  over the horizon volume. In this graph the markers on the left of the barrier show $\phi(0)$ and  the markers on the right of the barrier $\phi(r_H)$  for different $G$. Black circles, red diamonds, green triangles and the orange square show $\phi(0)$ and $\phi(r_H)$ in order of increasing $G$. For small $G$ (black circles), the field starts very close to the true vacuum at the center and ends very close to the false vacuum at the horizon. Increasing $G$ makes the change in $\phi$ smaller. Eventually, for a critical value of $G$ which is calculated in this section, the two points merge(orange square). This corresponds to a Hawking-Moss solution.}
   \label{O(3)-Fig-IncreasingGravityFieldSpace}
\end{figure}

Increasing the gravitational constant shrinks the horizon radius $\Lambda_{\rm fv}$ and  leaves less and less space for the wall separating the two vacua. As $\Lambda_{\rm fv}$ becomes comparable to or smaller than the natural width of the wall, $\phi$ is restricted to an increasingly narrow range of values around the top of the barrier. Eventually $\phi(0)$ and $\phi(r_H)$ merge and the field becomes spatially homogeneous. This is the usual Hawking-Moss bounce whose $O(5)$ symmetry contains both $O(4)$ and $O(3)\times O(2)$ as subgroups. This is illustrated in Fig.\ref{O(3)-Fig-IncreasingGravityFieldSpace}. This is very similar to the  behavior of $O(4)$-symmetric bounces described in \cite{Jensen:1988zx,Hackworth:2004xb,Batra:2006rz}.

When the solution approaches  the Hawking-Moss bounce, $\delta \phi = \phi(r) - \phi_{\rm top}$ becomes small and this enables us to use an expansion in powers of $\delta \phi$ in Eq.~\eqref{O(3)-EOMSA}-Eq.~\eqref{O(3)-EOMSC}. Because the variation of the field is small, the background metric remains very close to a de Sitter metric  with $\Lambda = \Lambda_{\phi_{\rm top}}$. 

The analysis presented here is along the lines of  \cite{Hackworth:2004xb}. To zeroth order in $\delta \phi$, the solution to field equations are
\begin{eqnarray}
	\phi(r)&=& \phi_{\rm top}  \nonumber  ~, \\
	f(r) &=& 1 - H^2 r^2~, \quad \text{where} \quad H^2 =\frac{8 \pi G }{3} U_{\rm top} ~.
\end{eqnarray}

Expanding $\phi$ and $f$ as
\begin{eqnarray} \label{O(3)-eq-Expansion}
	\phi(r) &=& \phi_{\rm top} + \delta \phi(r) \nonumber ~, \\
	f(r) &=& f_0(r)+ \delta f(r) = 1 - H^2 r^2 + \delta f(r) ~, \nonumber\\ 
  	V(\phi) &=& V_{\rm top} + H^2 \left( -\frac{\beta}{2} \phi^2 + \frac{b}{3} \phi^3 + \frac{\lambda}{4} \phi^4 \right) ~.
\end{eqnarray}
The boundary conditions for $\delta f$ and $\delta \phi$ are 
\begin{equation} \label{O(3)-BCSA}
	\delta \phi' (0) =0~, \quad \delta f(0) = \delta f( \Lambda_{\rm top}) = 0~, \quad \delta \phi'(\Lambda_{\rm top}) \left[- 2 H_{\rm top} + \delta f'(\Lambda_{\rm top})\right]= \left.\frac{d V}{d\phi}\right|_{\Lambda_{\rm top}}~.
\end{equation}
Using Eq.~\eqref{O(3)-eq-Expansion} in Eq.~\eqref{O(3)-EOMSD} and Eq.~\eqref{O(3)-EOMSF} leads to
\begin{align} 
    {d \over r^2 dr}\left[ r^2 \delta \phi' ( f_0 + \delta f ) \right]+ 4 \pi G r (\delta\phi')^3 (f_0 + \delta f)   = H^2 ( - \beta \delta\phi + b (\delta\phi)^2 + \lambda (\delta \phi)^3)~, \label{O(3)-eq-FirstOrderA} \\
     {d \over dr } \left[ r\left(1-f_0- \delta f \right)\right] = \kappa r^2 \left[ {1\over 2} {(\delta \phi')^2f } +V_{\rm top} + H^2 \left( -\frac{\beta}{2} (\delta \phi)^2 + \frac{b}{3} (\delta \phi)^3 + \frac{\lambda}{4} (\delta \phi)^4 \right)\right]\label{O(3)-eq-FirstOrderB}.
 \end{align}
Equation Eq.~\eqref{O(3)-eq-FirstOrderB} implies that the first correction in $\delta f$ is of order $\delta \phi^2$. This simplifies the first equation immensely:
\begin{equation}\label{O(3)-eq-ExpansionB}
 \delta\phi'' (1 -H^2 r^2) + \delta\phi' ( - 4 H^2 r + \frac{2}{r} ) + 4 \pi G r (\delta\phi')^3 (1 - H^2 r^2) =H^2 ( - \beta \delta\phi + b (\delta\phi)^2 + \lambda (\delta\phi)^3) ~.
\end{equation}

\begin{figure}[htbp] 
   \centering
   \includegraphics[width=2.8in]{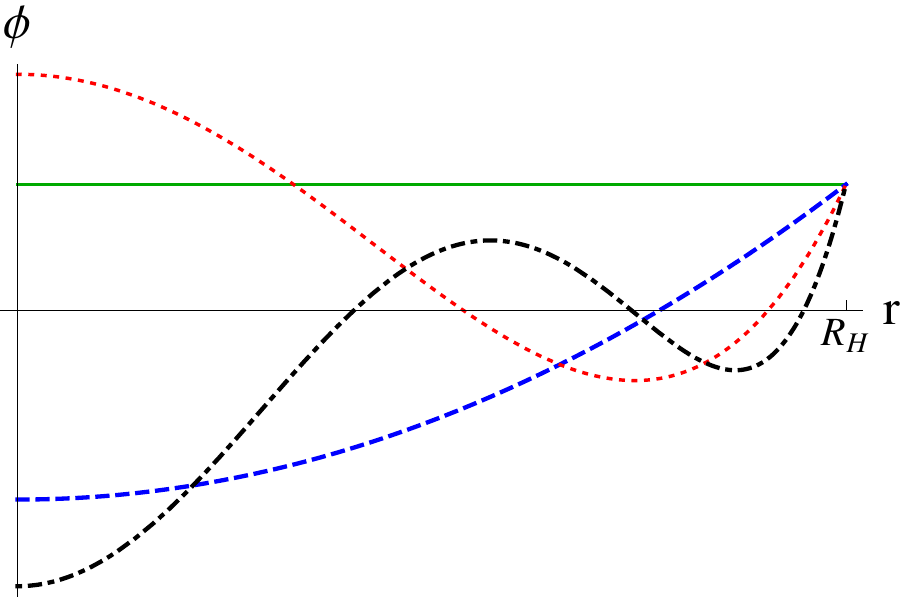} 
   \caption{Oscillating bounces for different values of $N$. Green solid line, blue dashed, red dotted and black dot-dashed lines represent solutions for $N=1,3,5$ and 7. Only the blue line represents a tunneling instanton.}
   \label{O(3)-OscillatingBounces}
\end{figure}

Defining $ y = r H_{\rm top}$, which brings the horizon to $y=1$,  and keeping only the first nonzero order in  $\delta\phi$ (the primes for the rest of this section denote differentiation with respect to $y$) leads to
\begin{equation}
 \delta \phi'' (1-y^2) - 4 y \delta \phi' + \frac{2}{y} \delta\phi' + \beta \delta\phi = 0~.
\end{equation}
To bring this equation into a more familiar form we change variables to 
\begin{equation}  \label{O(3)-eq-ChangeofVariable1}
	g(y)= y \delta\phi(y) ~.
\end{equation}
This leads to
\begin{equation} \label{O(3)-EOMSI}
 (1-y^2) g'' - 2 y g'  + ( \beta + 2 ) g = 0~,
\end{equation}
which is the Legendre equation of order $ -\frac{1}{2} + \frac{1}{2} \sqrt{ 4 \beta + 9}$. The boundary conditions Eq.~\eqref{O(3)-BCSA} in terms of $y$ are
\begin{equation} \label{O(3)-BCSB}
	\delta\phi'(0) = 0~,  \qquad  \left. \phi' f' \right|_{y=1} = \left. \frac{\partial U}{\partial \phi} \right|_{y=1}~. 
\end{equation}
After some simple manipulation using  Legendre's equation, the boundary condition at $y=1$ becomes
\begin{equation}
    (1 - y^2)  g'' = 0~,
\end{equation}
As long as  $g''$ is not more divergent than $1/(1-y)$, this is satisfied automatically. The most general solution to Eq.~\eqref{O(3)-EOMSI} is a linear combination of $P_n(y)$ and $Q_n(y)$.
Since all the $Q_n$'s are logarithmically divergent at $y=1$, they can't satisfy this equation. Therefore the solution must be proportional to $P_n(y)$. 
Let's look at the expansion of   $g$  around zero:
\begin{equation}
	 g(y) = a_0 + a_1 y + a_2 y^2 + {\cal O}(y^3)~.
\end{equation}
The boundary condition at $y=0$ in terms of $g$ and $y$ is
\begin{equation} 
	\left.\left(\frac{g}{y} \right)' \right|_{y=0}= 0~.
\end{equation}
To satisfy this boundary condition  we need $a_0 = a_2 =
0$. The Taylor expansion of $P_n(Q)$ is 

\begin{equation} 
P_n(y) = \frac{\sqrt{\pi}} {\Gamma(\frac{1-n}{2})\Gamma(\frac{2+n}{2})}+ \frac{n(n+1) \sqrt{\pi} }{2 \Gamma(\frac{2-n}{2}) \Gamma(\frac{3+n}{2})} y+ \frac{ n(n+2)(n^2-1)
\sqrt{\pi}}{8 \Gamma( \frac{3-n}{2}) \Gamma(\frac{4+n}{2})}y^2 +O(y^3) 
\end{equation}
The only possible way for $a_0$ and $a_2$ to vanish  is to choose  $n$ to be an odd integer. This  makes the Gamma function infinite. This restricts $\beta$ defined in Eq.~\eqref{O(3)-eq-Expansion} to $\beta  = n(n+1)-2$  where $n$ is an odd integer\footnote{This is very similar to the results obtained for $O(4)$-symmetric solutions in \cite{Hackworth:2004xb}. There $\beta$ had to be equal to $N(N+3)$ for integer $N$'s.}. Therefore, if we drop all the non-linear terms in Eq.~\eqref{O(3)-eq-ExpansionB}, the solutions  exist only for very special values of $\beta$. If we add the nonlinearities, it relaxes this condition. Since the Legendre polynomials form an orthogonal basis, we can expand any solution of the full nonlinear equation in this basis. We can rewrite Eq.~\eqref{O(3)-eq-ExpansionB} in terms of $g$
\begin{equation} \label{O(3)-eq-NonLinearEqForgA}
(1-y^2) g'' - 2 y g' + (\beta + 2) g - b g^2 - \lambda g^3 = 0~.
\end{equation}
Expanding the solution in terms of $P_M(Q)$ ($M$ is an odd integer):
\begin{equation} \label{O(3)-ExpandingInLegendreA}
	g(y) = \frac{1}{\sqrt{|\lambda| }} \sum_M A_M P_M(y)~,
\end{equation}
and plugging back into Eq.~\eqref{O(3)-eq-NonLinearEqForgA} leads to  
\begin{equation}
	\sum_M P_M(y) \left( \left(\beta +2 - M(M+1)\right)A_M- \frac{b}{\sqrt{|\lambda|}} \sum_{I,J} p_{IJM} A_I A_J -sgn({\lambda}) \sum_{I,J,K} q_{IJKM} A_I A_J A_K  \right)=0~,
\end{equation}
 where the $p$'s and $q$'s are defined by 
\begin{eqnarray}
    &p_{IJM} = \frac{(2 M +1)}{2} \int_{-1}^1 dy P_I(y) P_J(y) P_M(y) ~, \\
    &q_{IJKM} =\frac{(2 M +1)}{2}\int_{-1}^1 dy P_I(y) P_J(y) P_K(y) P_M(y)~.
\end{eqnarray}
Since the nonlinear terms are small, we expect the solution to the field equations to be close to the solutions for the linearized equation. Let's assume that the solution is very close to one of the $P_N(y)$ modes where $ \Delta  = \beta + 2 - N(N+1)$ is small (N is an odd integer). Therefore all the expansion coefficients $A_M(M \neq  N)$ in Eq.~\eqref{O(3)-ExpandingInLegendreA} are very small compared to $A_N$. Solving for $A_M$ to the lowest nonzero order leads to
\begin{eqnarray}
    &A_M ( \beta + 2 - M(M+1) ) - \frac{1}{\sqrt{|\lambda|} } b A_N^2 p_{NNM} = 0~,  \qquad M\neq N ~, \\
    & A_N \Delta - \frac{b}{\sqrt{|\lambda|} } \big( A_N^2 p_{NNN}+ 2 A_N A_M p_{NMN} \big) - sgn (\lambda) A_N^3q_{NNNN} = 0~.
\end{eqnarray}
From the first equation, we can solve for $A_M$. Plugging back into the second equation we have:
\begin{equation}
	c A_N^2 + \frac{b}{|\lambda|}p_{NNN} A_N  - \Delta + {\cal O}(A_N^3) = 0~,
\end{equation}
where 
\begin{equation} 
	c = sgn ( \lambda ) q_{NNNN} + \frac{2 b^2}{|\lambda|} \sum_M \frac{p_{NNM} \, \,p_{NMN}}{\beta + 2 - M(M+1) }~.
\end{equation}	
Since $N$ is an odd integer, $P_N(y)$ is an odd function and therefore $p_{NNN}$ vanishes. Solving for $A_N$ we have:
\begin{equation} 
	A_N = \sqrt{ \frac{ \Delta}{c}}~.
\end{equation}
Let's assume  that $ b=0$. The $N=1$ solution corresponds to $\Delta=\beta$, which is positive.This makes $A_1$ a real and acceptable coefficient.  In this case  $g(y) = P_1(y)=y$  which corresponds to a field configuration which is constant because  $\phi(y) = \frac{g(y)}{y} $. This solution is not  acceptable because it is inconsistent with the change of variable in  Eq.\eqref{O(3)-eq-ChangeofVariable1}.  So the first useful solution  corresponds  to the case $N=3$, which which makes $\Delta = \beta - 10$. Because $\Delta$ is under the square root, it cannot be negative and this sets a minimum value for $\beta$,
\begin{equation}\label{O(3)-CriticalBeta}
	\beta= \frac{V''(0)}{H^2} > 10~.
\end{equation}
These are the solutions that start from one side of the  barrier and end on the other side. They are acceptable tunneling configurations. $A_3$ gets smaller when $\beta$ approaches 10 from above and the bounce solution gets closer and closer to the top of the barrier and  merges with Hawking-Moss solution at $\beta=10$. For larger values of $N$ we will have similar situations. For example the $N=5$ solution exists only for $\beta \ge 28$. This solution crosses the barrier twice. It has more than one negative mode and therefore not an acceptable tunneling configuration. In general, solutions for $N=3+2k$ cross the barrier $k$ times and merge with Hawking-Moss when $\beta=2(k+1)(2k+5)$.None of them are acceptable as tunneling configurations. Some of these solutions are shown in Fig. \ref{O(3)-OscillatingBounces}.

\subsection{Large vacuum energy}
\label{O(3)-sub-LargeVacuum}
Another limiting case that we can study is to keep $G$ fixed and raise the vacuum energies by adding a large constant to $V(\phi)$. On a pure vacuum solution, increasing the vacuum energy has the same effect on the metric as increasing $G$ and  shrinks the horizon radius. This leaves less room for the wall and therefore the field spends more time near the top of the barrier and the  analysis of the previous section applies. If the vacuum energy gets too large, the $O(3) \times O(2)$-symmetric solution merges with Hawking-Moss at the critical value of $\beta$ found in Eq.~\eqref{O(3)-CriticalBeta}.

\section{Numerical results}
\label{O(3)-sec-Numerical}
Because it was not possible to solve the field equations in a closed form, we tried to understand the qualitative behavior of these bounces by choosing a quartic potential. If we look at the expansion of $V(\phi)$ near $\phi=0$, we need to at least keep terms to quartic order to capture the tunneling features. We considered quartic potentials with two minima. By shifting $\phi$, we set up the two vacua to be equally separated around the field space origin, with $\phi_{\rm tv} = v$ and $\phi_{\rm fv}= -v$. Any such potential can be written in the form 
\begin{equation} \label{O(3)-Defining}
	V(\phi)= \lambda \left( C_0 v^4 - k v^3 \phi - {1 \over 2} v^2 \phi^2 + \frac k3 v \phi^3 + \frac14 \phi^4\right)~,
\end{equation}
with $0<k<1$. The top of the barrier separating the two vacua is located at $\phi= - k v$. The theory is therefore characterized by four dimensionless quantities: $\lambda$, $C_0$, $k$ and 
\begin{equation}
	h = 8 \pi G v^2~.
\end{equation}
The dependence on $\lambda$ is very simple and we can scale it out. If $\phi(x)= g(x)$ is a bounce solution for a given value $\lambda$, then  $\phi(x) = g (\gamma x)$ will be a solution with $\lambda$ replaced by $\gamma^2 \lambda$.  Therefore the strong coupling limit (large $\lambda$) is related to the weak coupling limit by a rescaling of distances. 
The action $S$ of the bounce\footnote{This is explained in details in chapter 12 of \cite{WeinbergBook}.}is replaced by $\gamma^{-2} \lambda S$ . Therefore the classical action has a nontrivial dependence on only the  three remaining dimensionless parameters.We found it more useful to work with a different set of alternative parameters,
\begin{eqnarray}
	\epsilon&=& {1 \over \upsilon^4} \left(V_{\rm fv} - V_{\rm tv} \right) = {4 \over 3} \lambda k~,\cr\cr
	\alpha &=& {1 \over \upsilon^4} \left(V_{\rm top}- V_{\rm fv} \right) = {\lambda \over 12 }(3-k) (1+k)^3~, \cr\cr
	U_0 &=& {1 \over \upsilon^4}V_{\rm tv} = \lambda\left(- \frac12 - \frac{2k}{3}+ C_0 \right)~.
\end{eqnarray}
These parameters characterize the splitting between the energies of the true and false vacua, and between the top of the barrier and the false vacuum and the energy of the true vacuum, all in units of $\upsilon^4$. In the subsequent sections we study the dependence of the bounce solution on  each of these parameters. We change  one while keeping the others  fixed and solve  the bounce equations \eqref{O(3)-EOMSA}-\eqref{O(3)-EOMSC} numerically. Such a variation will correspond to a more complicated path in the $\lambda-C_0-k$ plane and  may extend to regions where $\lambda$ is large and therefore the coupling is strong. However, we are not concerned with this strong coupling since any large coupling solution can be mapped onto a weak coupling solution by the rescaling introduced in the beginning of this section. 

\subsection{Varying $G$}
\label{O(3)-sec-DifferentG}
 
\begin{figure}
\centering
\begin{tabular}{cc}
\includegraphics[height=2.0in]{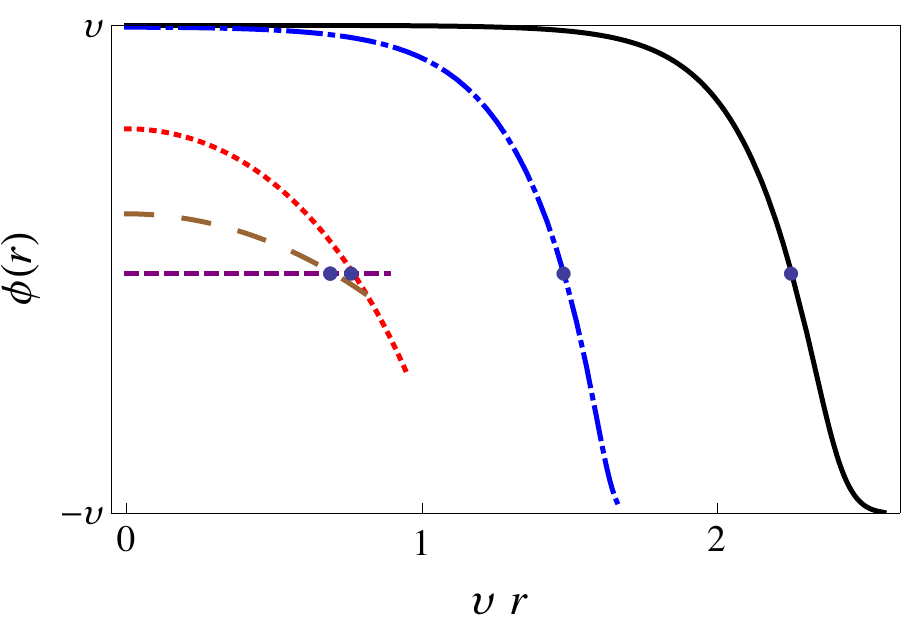}
\hskip .5in
\includegraphics[height=2.0in]{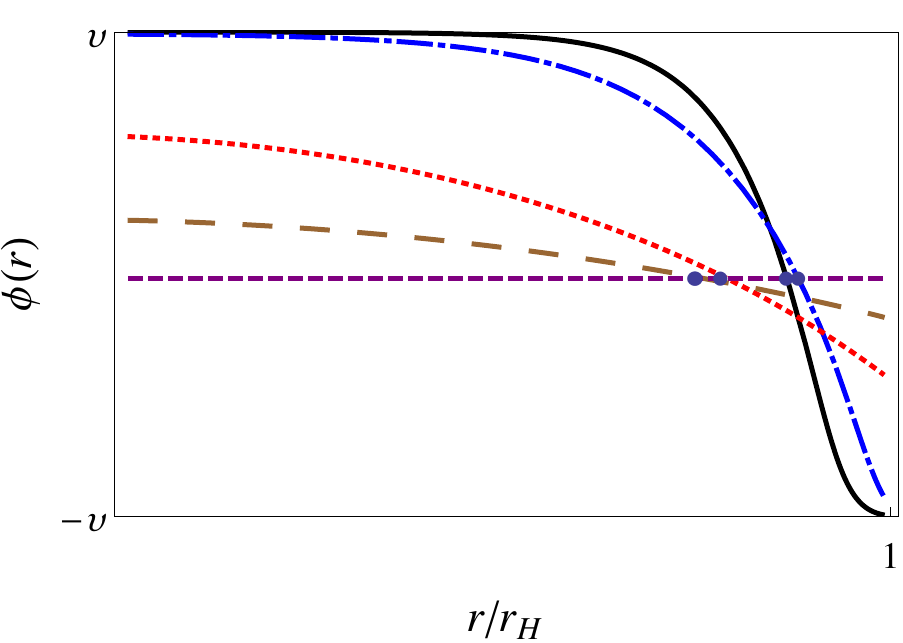}
\\(a) \hskip 3.0in (b) \\
\includegraphics[height=2.0in]{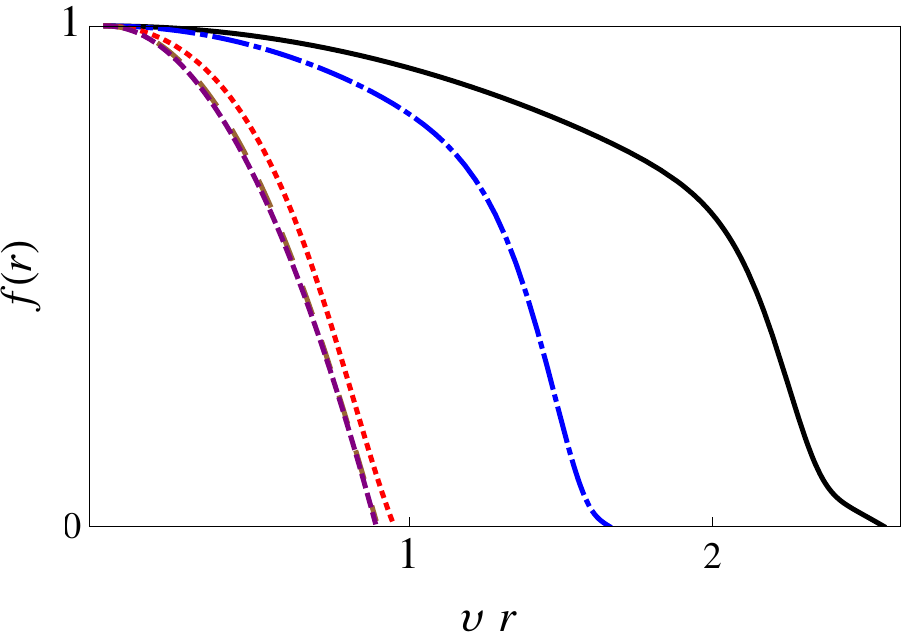}
\\  (c)
\end{tabular}
\caption{Evolution of the bounce as the gravitational constant is varied. The first two panels show $\phi$ as a function of (a) $r$ and (b) $r/r_H$.  In both cases the black dot indicates the value at the top of the barrier. The third panel shows $f=1/A$ as a function of $r$.  Reading from left to right,  the short-dashed purple, long-dashed brown, dotted red, dot-dashed blue, and  solid black lines correspond to $8\pi G v^2$ equal to 0.711, 0.704, 0.628, 0.251, and 0.126.  In all cases $U_0=2$, $\alpha=3$, and $\epsilon=0.3$.}
\label{varyG}
\end{figure}

 As mentioned in Sec. \ref{O(3)-OscillatingBounces}, increasing the gravitational constant causes the solution to approach the Hawking-Moss. In Fig.\ref{varyG}, the numerical solutions for different values of $G$ are shown. In these graphs $U_0$, $\epsilon$ and $\alpha$ are held fixed. Increasing $G$ causes the field to get close to the Hawking-Moss (purple dashed line) and eventually the $O(3)\times O(2)$ merges with the Hawking-Moss at a value of $\beta = 9.998$ which is in a very good agreement with what we found in Eq.~\eqref{O(3)-CriticalBeta}.  In this limit $\phi$ becomes spatially homogenous and $f$ corresponds to a pure de Sitter space with $\Lambda= \Lambda_{\rm top}$.

\subsection{Varying $U_0$}
\label{O(3)-sec-DifferentU0}
\begin{figure}
\centering
\begin{tabular}{cc}
\includegraphics[height=2.0in]{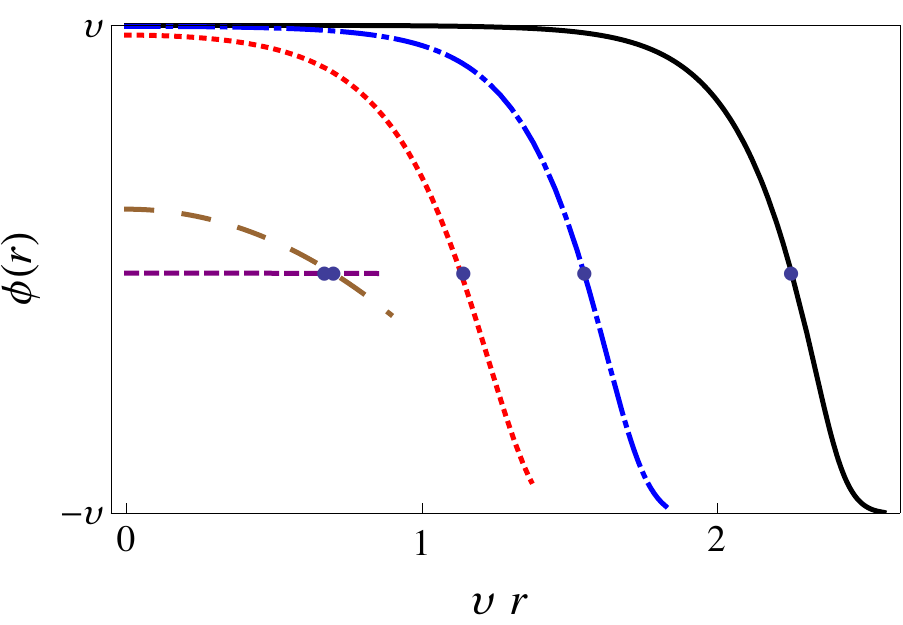}
\hskip .5in
\includegraphics[height=2.0in]{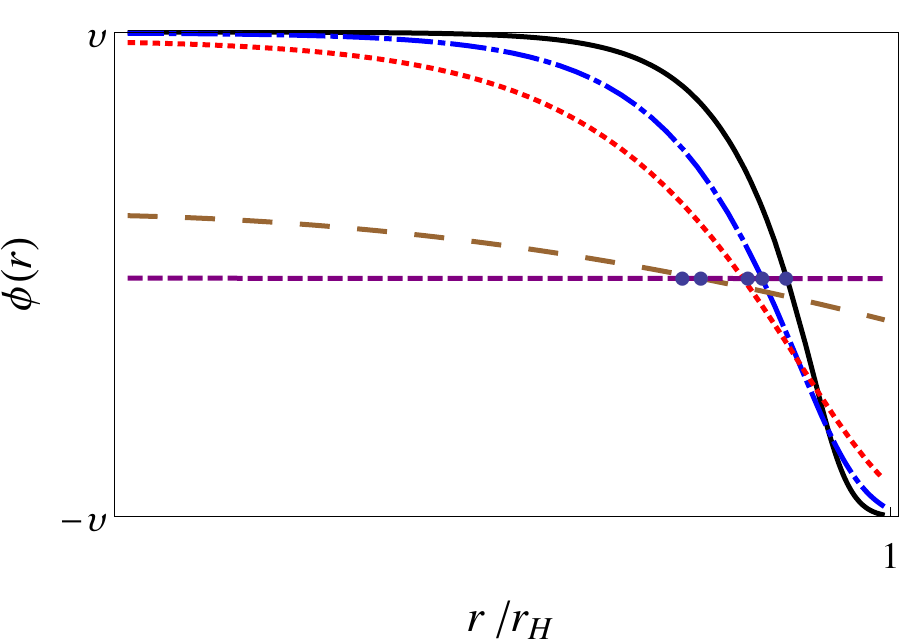}
\\(a) \hskip 3.0in (b) \\
\includegraphics[height=2.0in]{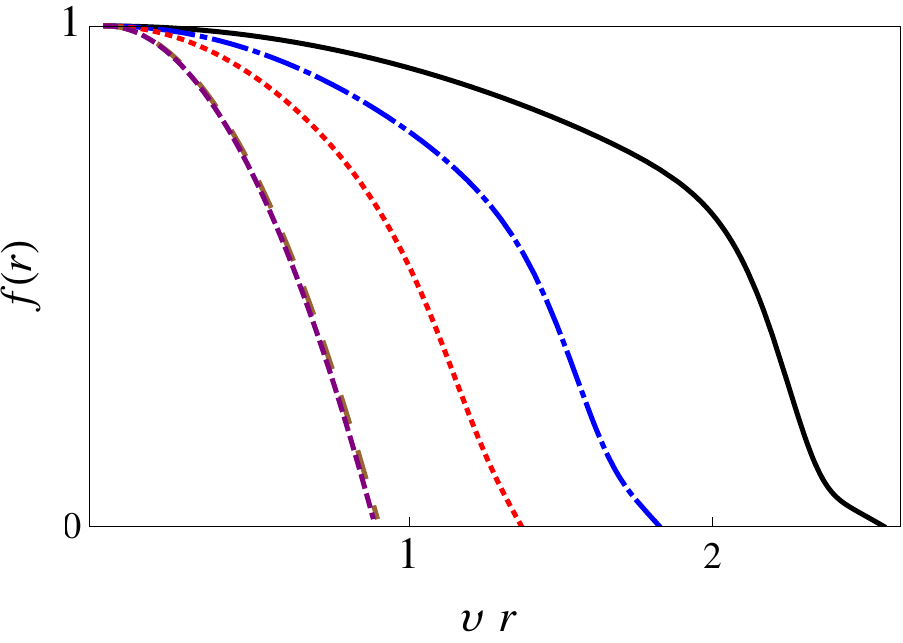}
\\  (c)
\end{tabular}
\caption{Evolution of the bounce as the true vacuum energy $U_0$ is  varied. The first two panels show $\phi$ as a function of (a) $r$ and (b) $r/r_H$.  In  both cases the black dot indicates the value at the top of the barrier. The third panel shows $f=1/A$ as a function of $r$.  Reading from left to right,  the short-dashed purple, long-dashed brown, dotted red, dot-dashed blue, and  solid black lines correspond to $U_0$ equal to 2, 5, 10, 2,  and 26.8  In all cases $8 \pi G \upsilon^2 =0.126$, $\alpha=3$, and $\epsilon=0.3$.}
\label{varyU0}
\end{figure}

We also argued in Sec.\ref{O(3)-sub-LargeVacuum} that the effect of increasing $V(\phi)$ by a constant density is qualitatively very similar to that of increasing $G$. Our numerical solutions show that this statement is correct and the solution merges with Hawking-Moss at a value given by Eq.~\eqref{O(3)-CriticalBeta}. These numerical solutions are shown in Fig.\ref{varyU0}.

\subsection{Varying $\epsilon$}
\label{O(3)-sec-DifferentEpsilon}
\begin{figure}
\centering
\begin{tabular}{cc}
\includegraphics[height=2.0in]{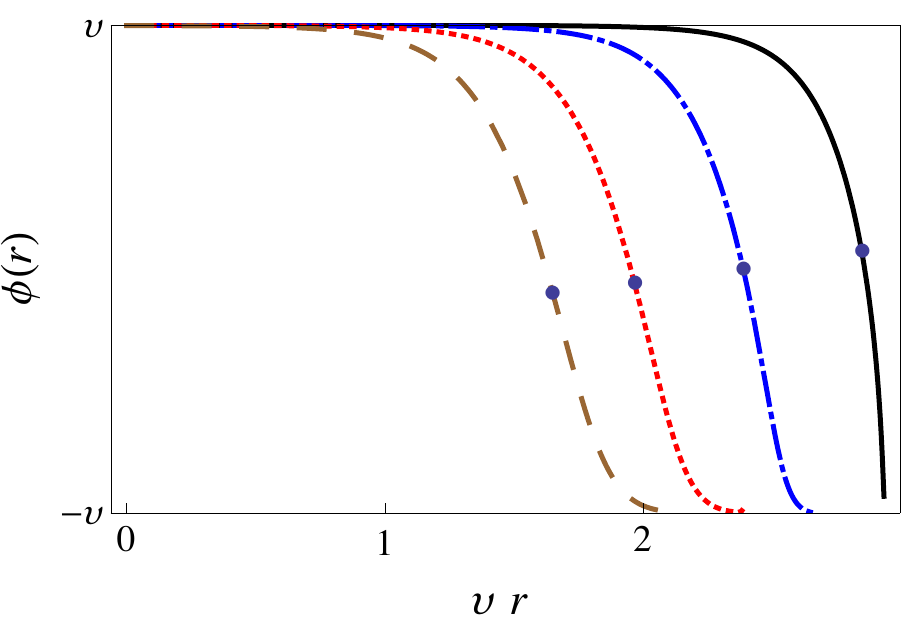}
\hskip .5in
\includegraphics[height=2.0in]{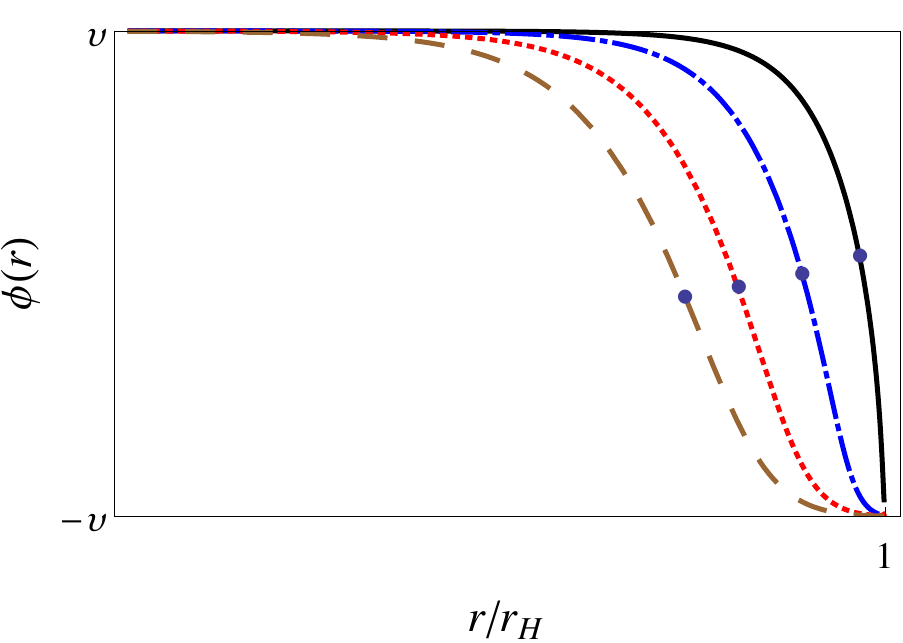}
\\   (a) \hskip 3.0in (b) \\
\includegraphics[height=2.0in]{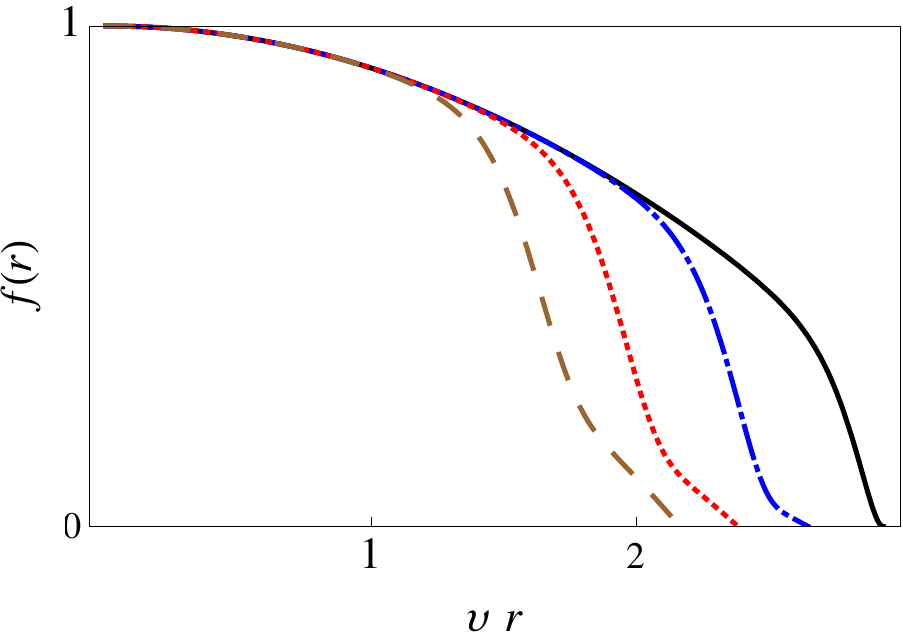}
\\   (c)
\end{tabular}
\caption{Behavior of the bounce as $\epsilon$ is varied.  Again,  panels (a) and (b) show $\phi$ as a function of $r$ and of $r/r_H$, with the black dot indicating the value at the top of the barrier. Panel (c) shows the metric function $f=1/A$.  Reading from left to right, the dashed brown, dotted red, dot-dashed blue, and solid black lines correspond to $\epsilon$ equal to 2, 1, 0, and $-1$. For all cases $U_0=2$, $\alpha=3$, and $8\pi G v^2 = 0.126$.}
\label{varyEps}
\end{figure}

In flat space, changing $\epsilon$, the dimensionless difference between the false and true vacuum energy densities, changes the radius of bubble. For smaller $\epsilon$ the bubble must get bigger to compensate for the surface tension of the bubble wall. In Fig. \ref{varyEps} we showed the field profile and metric for different values of $\epsilon$. We also included graphs for $\epsilon=0$, which corresponds to two degenerate minima, and also graphs for negative $\epsilon$'s which show tunneling from a true vacuum to a false vacuum. Neither  of these cases are possible in flat spacetime at zero temperature, where the bubble radius approaches  infinity as $\epsilon$ goes to zero. However, the situation is different in the presence of gravity. If the true and false vacua are both de Sitter, it is possible to tunnel upward \cite{Lee:1987qc}. What saves the bubble from collapsing due to both surface tension and vacuum energy differences in up tunneling is the Hubble flow of the de Sitter space provided that the initial bubble size is large enough. 

There are qualitative differences between the CDL ($O(4)$ symmetric) and $O(3)\times O(2)$-symmetric cases. In the CDL case,  tunneling from true to false vacuum proceeds  by nucleation of bubble of true vacuum (which we can assume to be centered about the ``north pole'' of the four-sphere  by using the de Sitter symmetry group\footnote{The geometry of de Sitter space and the terms north and south poles are explained  in Appendix. \ref{Appendix-DeSitter}.}) and the false vacuum region around the south pole. To describe the tunneling from false to true vacuum, we only need to to interchange  the south and north pole labels. The situation is different for  $O(3)\times O(2)$-symmetric bounces. The spacelike slices are three-balls with  the bubble of the new phase  at the center of a horizon volume and the old vacuum at the outer edge, extending to the horizon. There is no longer a symmetry between the two regions and therefore the bounce solutions are different, as can be seen in Fig. \ref{varyEps}.

\subsection{Varying $\alpha$}
\label{O(3)-sec-DifferentAlpha}

\begin{figure}
\centering
\begin{tabular}{cc}
\includegraphics[height=2.0in]{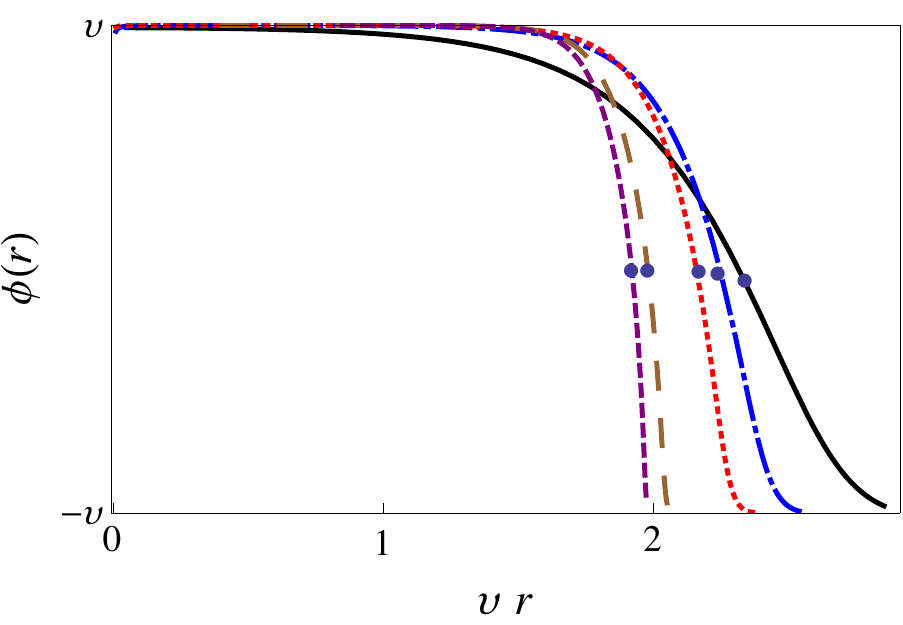}
\hskip .5in
\includegraphics[height=2.0in]{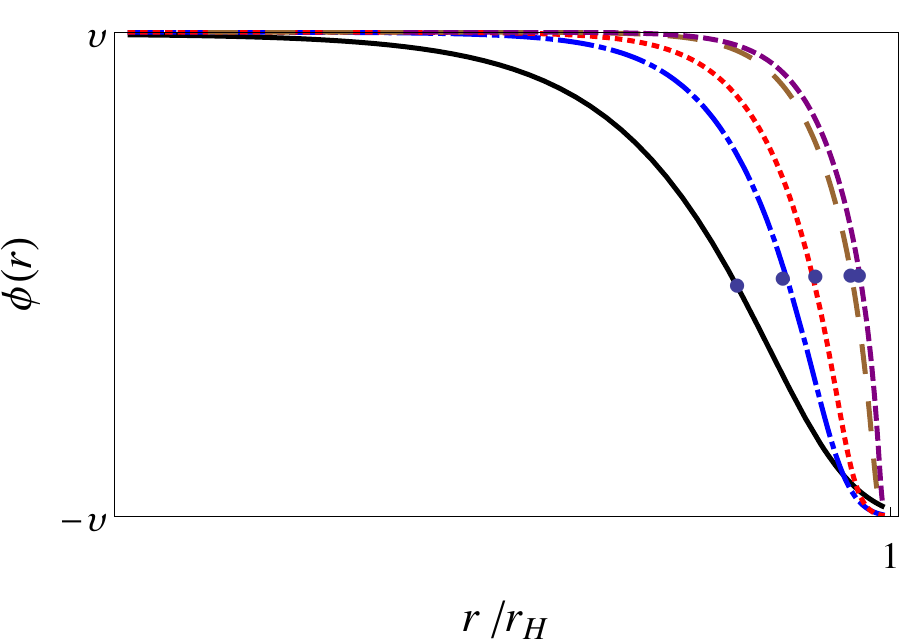}
\\   (a) \hskip 3.0in (b) \\
\includegraphics[height=2.0in]{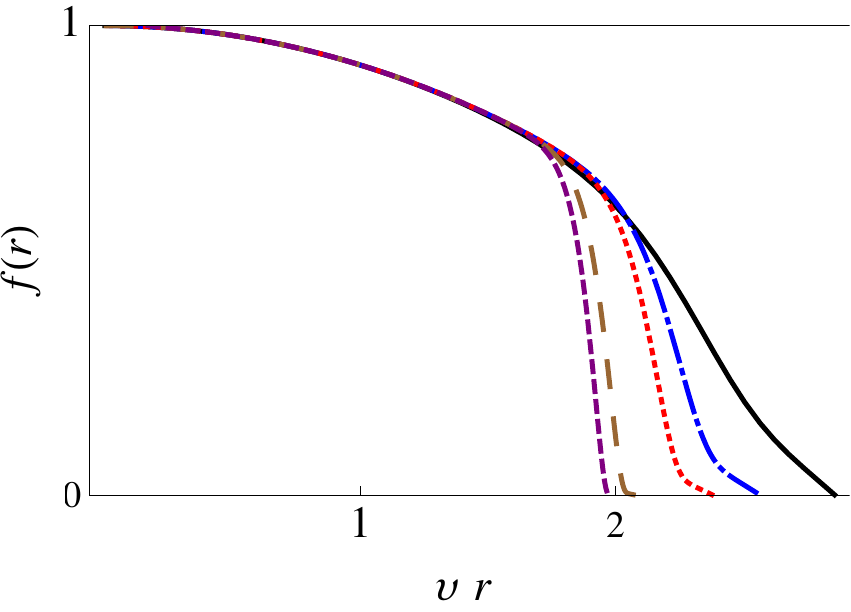}
\\   (c)
\end{tabular}
\caption{Variation of the bounce with barrier height for small $\alpha$ with the short-dashed purple, long-dashed brown, dotted red, dot-dashed blue, and solid black lines corresponding to $\alpha$ equal to 1, 3, 5, 12, and 15.  For all of these $U_0=2$,  $\epsilon=0.3$, and $8\pi G v^2 = 0.126$.  As in the previous figures, the black dot indicates the value of the field at the top of the barrier.}
\label{SmallAlpha}
\end{figure}

\begin{figure}
\centering
\begin{tabular}{cc}
\includegraphics[height=2.0in]{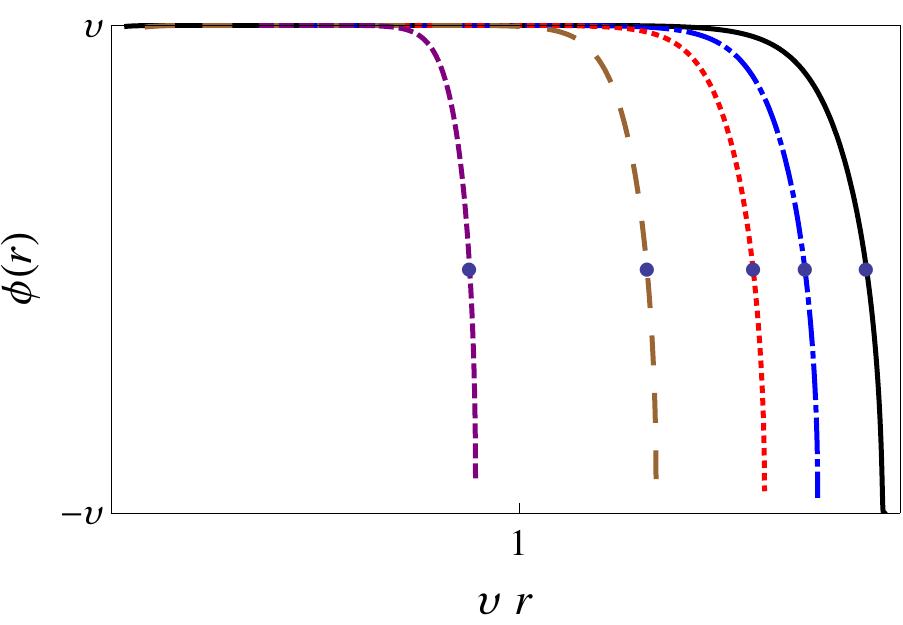}
\hskip .5in
\includegraphics[height=2.0in]{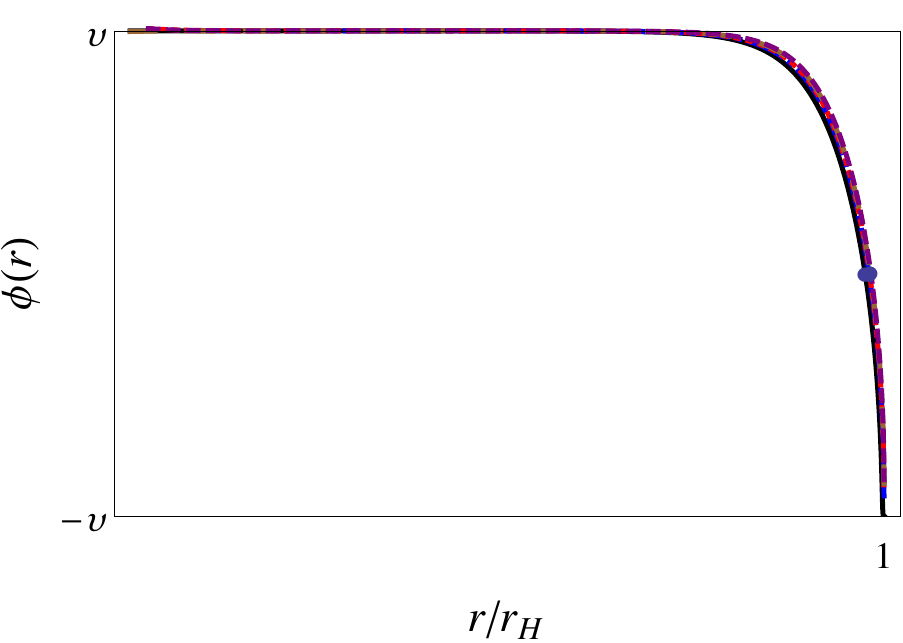}
\\   (a) \hskip 3.0in (b) \\
\includegraphics[height=2.0in]{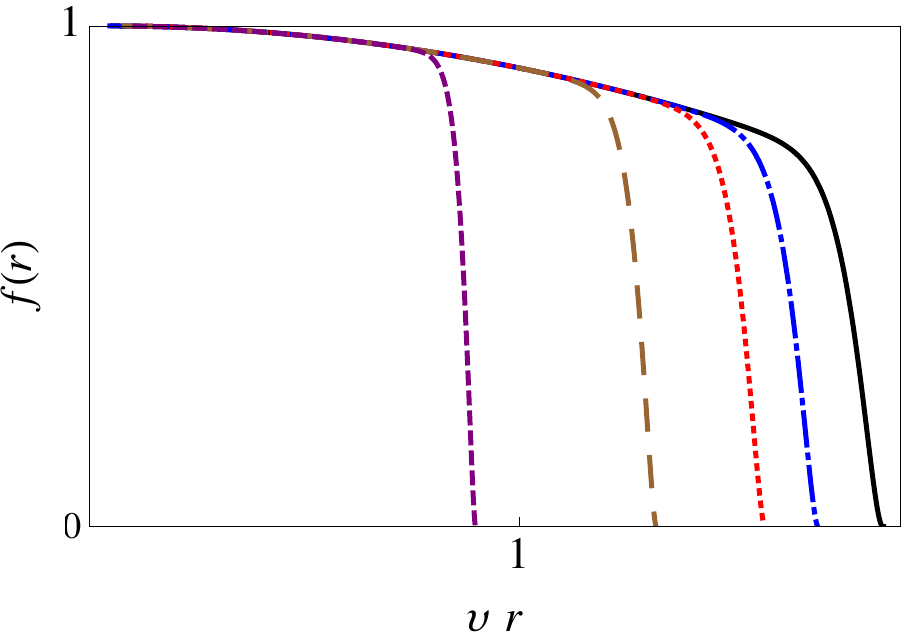}
\\   (c)
\end{tabular}
\caption{Variation of the bounce with barrier height for large $\alpha$, with the short-dashed purple, long-dashed brown, dotted red, dot-dashed blue, and solid black lines corresponding to $\alpha$ equal to 15, 20, 25, 40, and 100.  For all of these $U_0=2$,  $\epsilon=0.3$, and $8\pi G v^2 = 0.126$.  As in the previous figures, the black dot indicates the value of the field at the top  of the barrier.}
\label{LargeAlpha}
\end{figure}
The most interesting behavior was that  which corresponds to changing   the barrier height $\alpha$. Increasing the barrier height has two effects. First, it increases  the surface tension of the bubble which in turn (as in flat spacetime) increases the bubble radius. Second,it makes  the wall thinner. 
The situation for bubble nucleation in de Sitter spacetime is shown in Figs. \ref{SmallAlpha} and \ref{LargeAlpha}. For small values of $\alpha$, illustrated in Fig.\ref{SmallAlpha}, the field starts very close to the true vacuum at the center and approaches the false vacuum close to the horizon. However, in contrast with the flat spacetime case, the wall gets closer to the center of the bubble as $\alpha$ and therefore $\sigma$ increases. This is due to the shrinkage of the horizon radius which makes all of the evolution  happen on a shorter scale. However, when the field profiles are viewed as  functions of $r/r_H$, the results resemble the flat space case. In both cases, whether as a function of $r$ or $r/ r_H$, the walls get thinner when $\alpha$ gets larger and they get closer to the horizon. 

We may expect from this pattern that by increasing  $\alpha$ further we might make the wall arbitrarily narrow and close to the horizon. However, this trend does not continue when the wall starts reaching the horizon for larger values of $\alpha$. For very large values of $\alpha$, shown in Fig.\ref{LargeAlpha}, the wall is cut by horizon and there is no false vacuum region near the horizon. This means that the thin-wall approximation has already been broken, as proved in Sec.\ref{O(3)-sub-NoNariai}. $\phi(r_H)$ falls visibly short of the false vacuum value as $\alpha$ gets bigger. [For the largest $\alpha$ shown here, $\phi(r_H)=-0.85v$.] The wall does not get thinner and, as shown in panel b of Fig.\ref{LargeAlpha}, the shape of the wall does not change much. The field profiles as  functions of $r/r_H$ are almost indistinguishable. In this region the behavior of the field as a function of $r$ near the horizon is
\begin{equation}
	\phi(r) = \phi(r_H) + B \sqrt{r_H-r}~,
\end{equation}
for a constant $B$ that depends on the potential and therefore $\phi'$ diverges. But this divergence is a matter of a  bad coordinate system. Choosing the physical distance defined by
\begin{equation}
	r_{\rm phys} = \int_0^r dr' \sqrt{A(r')}~,
\end{equation}
as a coordinate  removes this divergence. Even when $\phi'(r)$ develops a singularity,  the boundary conditions  introduced in Eq.~\eqref{O(3)-eq-BCA} and Eq.~\eqref{O(3)-eq-BCA2} still hold. The actions of these bounces, which only depend on the horizon radius, remain finite and therefore we can neglect this coordinate singularity.

\section{Summary and conclusions}
\label{O(3)-sec-Summary}
In this chapter we studied bounces  with $O(3)\times O(2)$ symmetry. These bounce solutions show the thermal nucleation of a single bubble at the center of a horizon volume. The conventional interpretation of these bounces is the creation of two bubbles. However following the interpretation of \cite{Brown:2007sd} and taking a constant $\tau$ slice through the center of the bubble  (which is topologically a three-sphere)  gives the configuration of metric and the scalar field inside a horizon volume without any reference to quantities beyond the horizon. 

We used general quartic potentials  and scanned the parameter space of these potentials. We showed numerically and analytically that increasing the gravitational constant $G$ or adding a large positive constant to the field potential $V(\phi)$ drives the bounce towards the Hawking-Moss solution and calculated the threshold for $G$ at which the independent $O(3)\times O(2)$ solutions ceases to exist in Eq.~\eqref{O(3)-CriticalBeta}. 

The effects of decreasing $\epsilon$, the energy difference between the true and false vacua, is very different from their flat space counterparts. In flat space decreasing $\epsilon$ makes the bubble radius large. As $\epsilon$ tends to zero, this radius goes to infinity and there is no tunneling from false to true vacuum. However, in these $O(3)\times O(2)$-symmetric solutions, because of the existence of the horizon and  a nonzero temperature, very small $\epsilon$ is not a very special  point in the parameter space and the solutions for both transitions from true to false vacua and vice versa are possible. If $\epsilon$ gets too large (either  negative or positive), it loses its significance because after a while the field does not start close to the true vacuum and therefore does not see a very large $\epsilon$. 

The effect of increasing the barrier heights is quite  notable. In flat space, increasing the barrier height increases the surface tension and, as a result, pushes the bubble radius outwards and makes the wall thinner. But this is not completely true for our solutions. Increasing the  barrier height increases the surface tension and as a result decreases the horizon radius. Therefore the wall is pushed inwards (in $r$ coordinates). But if we look at the field profiles as  functions of $r/r_H$, the situation is more similar to the flat space case. The wall is pushed outwards and gets narrower. This behavior continues until the wall reaches the horizon and in this regime increasing the barrier heights further does not make the wall thinner. The scalar field rapidly  changes near the horizon and $\phi'$ develops a singularity near the horizon. However it is a coordinate singularity and none of the physical quantities blow up. 

For our solutions to be relevant to tunneling processes, it is not sufficient for them to satisfy the field equations. The fluctuations around these solutions must have one and only one negative mode. If there are  more than one negative mode, it is an indication that there is another path in field space that  has a lower Euclidean action and that dominates the decay. There are potentially three different types of negative modes for our $O(3)\times O(2)$-symmetric solutions. The first one is the usual $\tau$-independent negative mode corresponding to increasing or decreasing the radius of the bubble. This is what we expect for a critical bubble and we expect it to always be present. The second type of negative modes is  similar to the negative modes explained in Sec. \ref{Intro-sec-HM} which correspond to modulating the amplitude of the expansion-contraction mode with a sinusoidal variation in imaginary time. As explained in \ref{Intro-sec-HM}, these additional modes are present when the radius of the bubble is much smaller than the horizon radius and therefore we have to discard these very small bubbles.  The third type of negative modes is closely related to the four-sphere geometry of the bounce. The easiest way to visualize this mode is by looking at the shaded region in Fig.\ref{O3O2picture}. If this region gets too narrow, we can reduce its length and therefore its action by moving along the $\tau$ direction . This mode may be avoided if the radius of bubble is not too small and the the wall is not very thin. This is the same parameter region which we may need to be in to avoid the second type of negative modes.

\begin{figure}
\centering
\begin{tabular}{cc}
\includegraphics[height=1.95in]{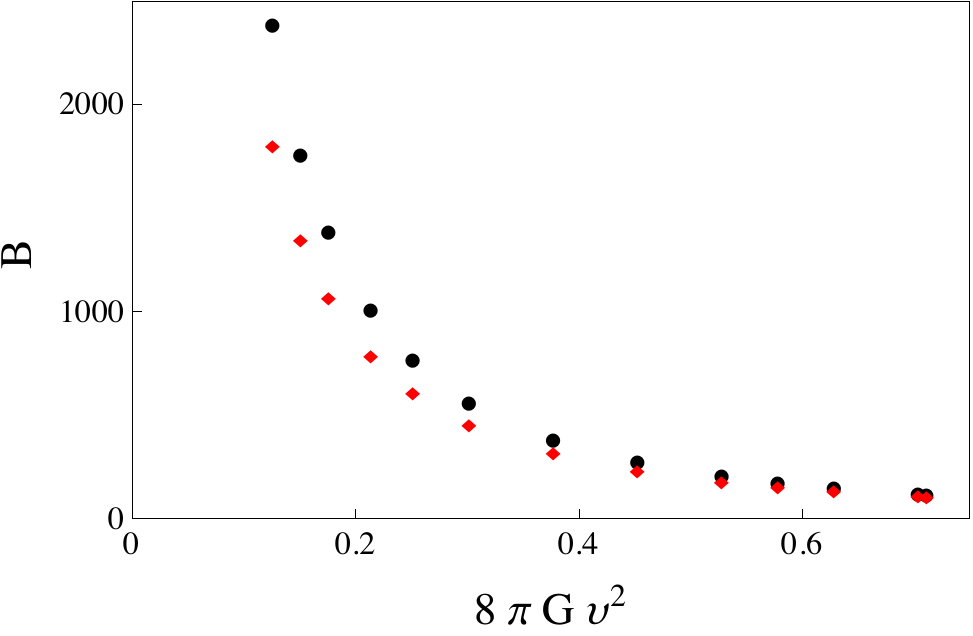}
\hskip .18in
\includegraphics[height=1.95in]{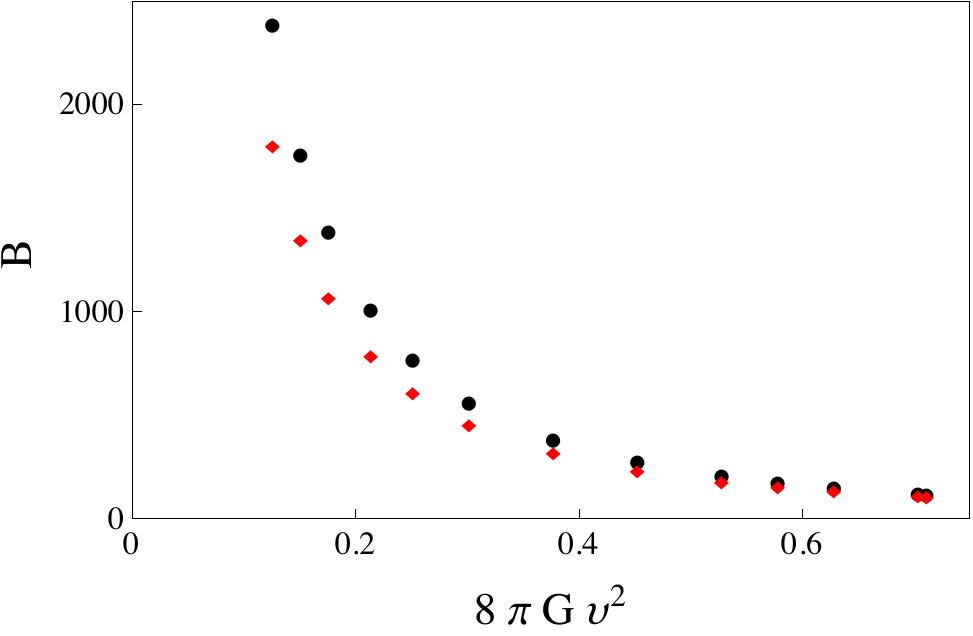}
\\   (a) \hskip 3.0in (b) \\
\includegraphics[height=1.95in]{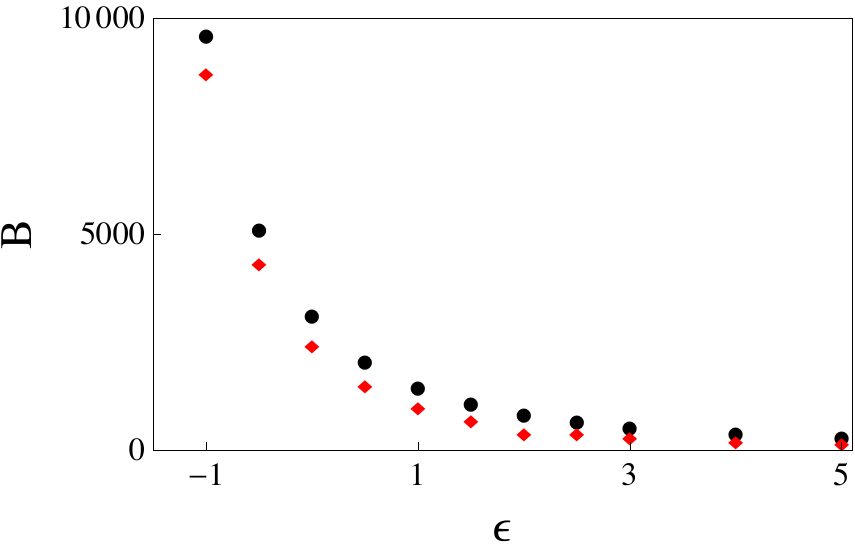}
\hskip .18in
\includegraphics[height=1.95in]{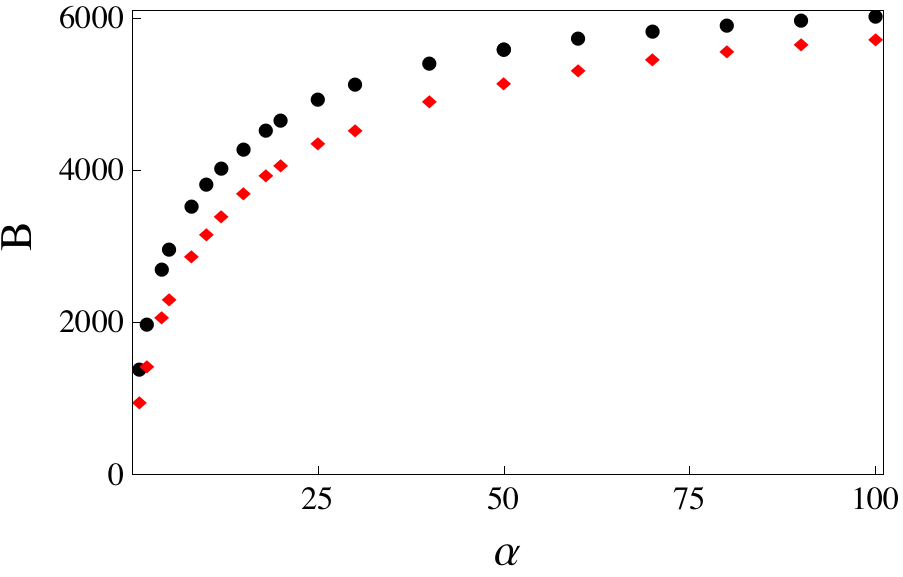}
\\   (c) \hskip 3.0in (d) 
\end{tabular}
\caption{Comparison of the tunneling exponents for the CDL bounce (red
  diamonds) and the O(3)$\times$O(2)-symmetric bounce (black circles)
  as one parameter is varied with the other three held fixed.  When
  held fixed, $U_0=2$, $\alpha=3$, $\epsilon=0.3$, and $8\pi G
  v^2=0.126$. }
\label{actionPlots}
\end{figure}

Even if these $O(3)\times O(2)$ symmetric bubbles have only one negative mode, they will not be very relevant if their action is higher than the CDL action. One of our main motivations for studying these solutions was checking  whether the $O(4)$-symmetric solutions  continue to be the dominant solutions in curved spacetime or not. In Fig.\ref{actionPlots} we plotted the action of our solutions compared to the CDL bounce.It is clear from the graphs that in all of these cases the CDL bounces have lower action and therefore  are the dominant path of tunneling. This is in agreement with the thin-wall approximation and brane nucleations presented in \cite{Garriga:2004nm}. However we did not restrict ourselves to the thin-wall limit and by scanning over a much larger portion of the parameter space, although not conclusively, we provided more evidence that the $O(4)$-symmetric bounces are the dominant tunneling paths even in curved spacetime.

\clearpage

\chapter{Bubble nucleation in spatial vector fields}
\label{VectorBubble}
In Chapter \ref{O(3)} we gave a plausible argument  that even in the presence of gravity the bubbles with ${\rm O}(3) \times {\rm O}(2)$ symmetry are subdominant and  O(4)-symmetric bubbles dominate. But all of the theories we studied were Lorentz invariant  field theories. In this chapter we study a three-component spatial   vector field theory which prefers  non-spherical bubbles as the dominant mode of the decay. The model we study has a field that transforms  under spatial rotations and  that has different longitudinal and transverse speeds of sound.  This chapter is based on \cite{Masoumi:2012aa}.
First we show that the domain wall tension depends on the orientation of the wall and describe analytical and numerical methods to calculate this orientation-dependent domain-wall tension. We then show how to calculate the shape of the critical bubble for thermal tunneling. The tunneling exponents get modified  by a factor of the ratio of the speeds of sound in the longitudinal and  transverse directions. If this ratio goes beyond (a model dependent) threshold, the flat wall becomes  unstable to the formation of zigzag segments of wall. This causes a kink in the shape of the critical bubble. The  scaling of  of the tunneling exponents with the speeds of sound  is very different from what we expect for smooth bubbles.  

\section{Motivations and outline }

As mentioned earlier, most  first-order phase transitions happen through  thermal or quantum  nucleation of a bubble  of true vacuum in the surrounding false vacuum. In this chapter we focus on thermal nucleation of  bubbles. If the bubble is too small, the surface tension makes it collapse to nothing. If it is too large, it  expands to complete the phase transition. In this sense, we need to create a critical bubble which corresponds to  the lowest saddle point of the energy functional with  only one negative eigenvalue, along the direction of expansion-contraction. For thermal tunneling, the dominant solutions are constant  in time. Therefore time variable is treated separately and in general easier to study. This makes the non-relativistic theories which have vectors  transforming under the spatial rotation group interesting. Studying them may shed light on many subtleties which are difficult to see in other contexts. For the case of scalar fields in N spatial dimensions, it has been shown \cite{Coleman:1987rm,Coleman:1977th} that the O(N) symmetric solutions have the lowest action. The probability $\Gamma$ of creating a bubble of radius $R$ is 
\begin{eqnarray} \label{eq-rate}
\log \Gamma=  -\frac{E_s}{k_b T}~,
\end{eqnarray}
where $E(R)$ is given by
\begin{equation} \label{VectorBubbleEnergy}
	E(R) = \sigma \Omega_{N-1} R^{N-1} - \Delta V \Omega_N R^N~,
\end{equation}
where  $\Delta V = V_{\rm fv} - V_{\rm tv}$ and  $\Omega_N$ is the area of a unit N-sphere. The surface tension $\sigma$ depends on the path $\gamma$ in the field space that takes the field from false vacuum to near the true vacuum 
\begin{equation} \label{VectorSurfaceTension}
\sigma = v_F \int_\gamma  d\,\phi \: \sqrt{2V}~.
\end{equation}
The scaling with the Fermi velocity $v_F$ follows from the equations of motion. For the critical bubble we choose the path that minimizes Eq.~\eqref{VectorSurfaceTension} and then we minimize Eq.~\eqref{VectorBubbleEnergy} with respect to $R$. This leads to the commonly used estimate for the tunneling rate. 
\begin{equation}
	\log \Gamma \sim \frac{\sigma^N}{(\Delta V)^{N-1}}\frac{1}{k_b T}~, \\
\end{equation}

In this chapter  we will generalize the theory to include vector fields. Our motivation arises  from condensed matter systems such as  liquid crystals, Helium 3 and Langmuir monolayers\cite{GalFou95,Leg75,Whe75,He3,Fou95,MacJia95,RudLoh99,SilPat06}.\footnote{In cases like the famous A-B transition in liquid $^3$He, the system shows different longitudinal and transverse speeds of sound. However, people used to study their nucleation properties using spherical bubbles.}   We study models which show non-spherical bubbles. To do so, we use spatial N-dimensional vector fields in an (N+1)-dimensional spacetime. These vectors  transform under the spatial rotation group and have different speeds of sound in the longitudinal and transverse directions. The Lagrangian for this model is
\begin{equation}
\mathcal{L}=\frac{1}{2}
\left(\dot{\phi}_i^2 - c_T^2 \partial_i\phi_j\partial_i\phi_j - (c_L^2-c_T^2) \partial_i\phi_i\partial_j\phi_j \right) - V(\phi_i)~.
\label{eq-L}
\end{equation}
The transitions occur between the two minima of the potential $V$, which we denote  by $\vec{\phi}_\pm$. To avoid ghost-like instabilities  we need $c_L \geq c_T$ . When $c_L\neq c_T$, the potential can minimally break  the spatial $SO(N)$ symmetry. We will consider this minimal breaking. In this set up, $(\vec{\phi}_+ - \vec \phi_-)$ specifies a direction which is the direction of the longitudinal wall and  causes non-spherical bubbles. 

We first study the planar domain-walls in \ref{sec-orient} and set up numerical and analytical results to determine the orientation dependence of the tension. We show a variety of different behaviors of $\sigma(\theta)$, the tension of the wall oriented at the polar angle $\theta$, in appendix \ref{sec-examples}. An instability in the flat wall is studied in \ref{sec-kink} and the corresponding effects on bubble shape are shown in \ref{sec-shape}. When $c_T$ is not much smaller than $c_L$, the bubble shape is close to a sphere. By increasing this ratio, the bubble shape gets more and more skewed and after reaching a critical value, it becomes a multivalued  function of angle. In \ref{sec-shape} we study these effects and give the correct interpretation of the singular shape of the bubble and find the corrections made to the tunneling exponents and transition rate. 
These techniques are identical to the one used for equilibrium bubbles, known as the Wulff construction\cite{RudBru95}.  It has been applied to ``soft matter'' systems like liquid crystals and Langmuir monolayers\cite{GalFou95,Fou95,MacJia95,RudLoh99,SilPat06}.  Our result agrees with the major conclusions in  these earlier works.  
\section{Orientation Dependence}
\label{sec-orient}

\begin{figure}
   \centering
   \includegraphics[width=12cm]{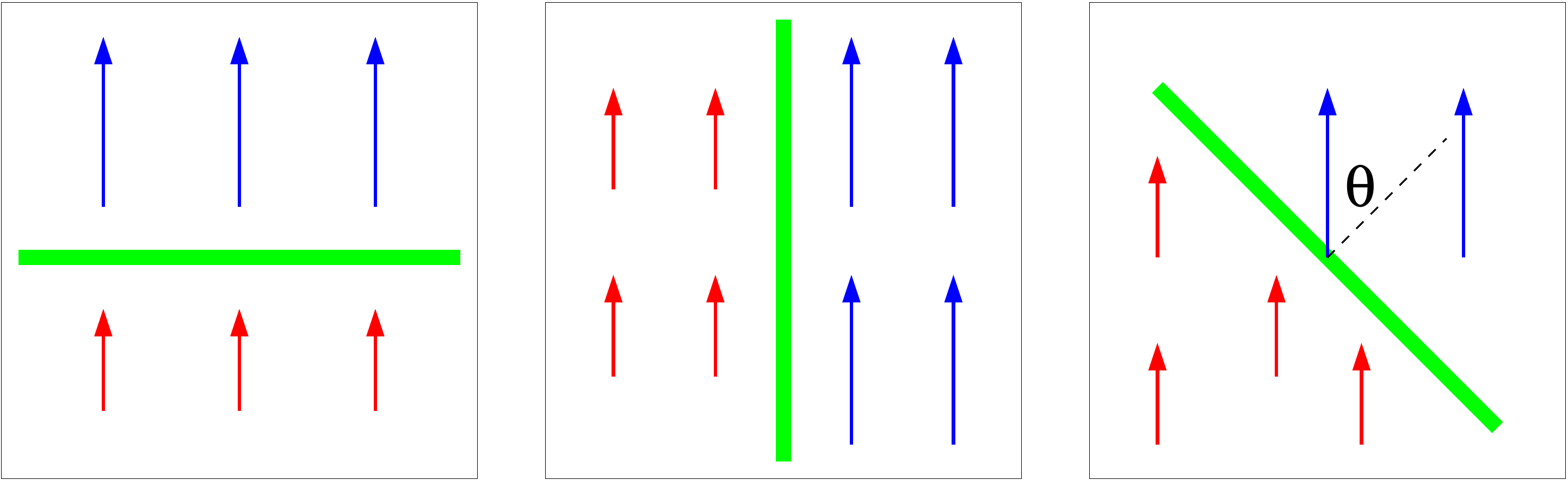} 
   \caption{The blue (longer) and red (shorter) arrows represent the vector field  at the two vacua.  The thick green line is the domain-wall.  From left to right, we show a longitudinal wall, a transverse wall, and a wall with orientation $\theta$.  The orientation is defined such that for a longitudinal wall $\theta=0$, and for a transverse wall $\theta=\pi/2$.
\label{fig-orientation}}
\end{figure}

The equation of motion derived from the Lagrangian in Eq.~(\ref{eq-L}) is
\begin{equation}
\ddot{\phi}_i - c_T^2\partial_j^2\phi_i
- (c_L^2-c_T^2) \partial_i \partial_j\phi_j = 
-\frac{\partial V}{\partial \phi_i}~.
\label{eq-eom}
\end{equation}
Here  it is more apparent that $c_T$ and $c_L$ correspond to the transverse and the longitudinal sound speeds.
To make the transitions possible, the potential $V$ must have at least two isolated local minima. There are many potentials which show this feature. For example, if a vector field $\vec \phi$ is put in an external field $\vec H$, the potential  to the second order in the external field $H$  is 
\begin{equation}
V(\vec{\phi}) = \frac{m^2}{2}|\vec{\phi}|^2 + 
\frac{\lambda}{4} |\vec{\phi}|^4 
+ a (\vec{H}\cdot\vec{\phi}) + b (\vec{H}\cdot\vec{\phi})^2 + c (\vec{H} \times \vec \phi)^2 + f (\vec H)^2 (\vec \phi)^2~.
\label{eq-potmot}
\end{equation}
Here we assume that $c=f=0$. Neglecting $a$ and assuming $b<0$ makes the direction parallel to $\vec H$ preferable. If $b|\vec{H}|^2 + m^2/2 <0$, we get two degenerate vacua at 
\begin{equation}
\vec{\phi}_\pm = \pm\sqrt{\frac{m^2+2b\vec{H}^2}{\lambda}}\frac{\vec{H}}{|\vec{H}|}~.
\end{equation}
The degeneracy in the minima can be removed by allowing a small $a$ to allow a first order phase transition. This example shows how generic this model is. Our further analysis in this chapter will be independent of the form  of the  potential in Eq.~(\ref{eq-potmot}). Later we show other examples. We start by considering a potential with  two almost degenerate vacua. This allows for a thin-wall approximation which simplifies the problem to that of  finding the tension for different orientations. Since the interpolation between the two vacua connects two vectors in the field space, it breaks the spatial rotational symmetry, as shown in Fig.~\ref{fig-orientation}. $\vec{\phi}_-$, which is the  field at the true vacuum (interior of the bubble), is shown as red arrows and $\vec{\phi}_+$, the vector field at the false vacuum (exterior of the bubble), is shown as blue vectors. The transition between them in different orientations corresponds to different contributions from gradient terms in the longitudinal and transverse directions  which in turn  makes the action along the walls (which per unit area is the tension) dependent on the orientation. The longitudinal ones are the most massive domain walls. The field wants to spend less time in those orientations. This skews the shape of bubble and creates non-spherical bubbles.  Here we will provide the general formalism to find $\sigma(\theta)$, and then in Sec.\ref{sec-shape} we will use this to find the bubble shape.

\subsection{Domain walls in two spatial dimensions}
The easiest cases to study are flat domain walls in a theory with vector fields in two spatial dimensions. The domain wall solutions for the case of degenerate vacua are static solutions of the equations of motion obtained in Eq.~\eqref{eq-eom} 
\begin{eqnarray}
-c_T^2(\partial_x^2+\partial_y^2)\phi_x-(c_L^2-c_T^2)\partial_x(\partial_x\phi_x+\partial_y\phi_y)&=&-\frac{\partial V}{\partial\phi_x}~,\nonumber \\
-c_T^2(\partial_x^2+\partial_y^2)\phi_y
-(c_L^2-c_T^2)\partial_y(\partial_x\phi_x+\partial_y\phi_y)
&=&-\frac{\partial V}{\partial\phi_y}~.
\end{eqnarray}
If the wall is along the $x$ direction, the boundary conditions can be written as
\begin{eqnarray}
	\partial_x \vec \phi &=& 0~, \cr
	\lim_{y\pm \rightarrow \infty}\vec{\phi}(x , y) &=& \vec{\phi}_\pm~,
\end{eqnarray}
where $\vec{\phi}_\pm$ are the true and false vacua.  Similarly if the wall is parallel to $\vec v$ and orthogonal to $\vec u$ the boundary conditions can be written as 
\begin{eqnarray}
(\vec{v}\cdot\nabla)\vec{\phi} &=& 0~, \label{eq-sym} \\
\lim_{\lambda\rightarrow\pm\infty}\vec{\phi}(\lambda\vec{u})
&=&\vec{\phi}_{\pm}~.
\end{eqnarray}
For a given wall, we can always rotate the coordinate systems (and at the same time $\vec \phi$) to set $\vec u$ to be along the $y$-axis. If the rotation is through an angle $\theta$ and the rotated vector field is $\tilde {\vec \phi}$, we can define the rotated potential $V_\theta(\phi_x,\phi_y)$ as 
\begin{equation}
V(\tilde \phi_x, \tilde \phi_y )=V(\phi_x\cos\theta+\phi_y\sin\theta,\phi_y\cos\theta-\phi_x\sin\theta) = V_\theta(\phi_x, \phi_y)~.
\end{equation}
In terms of this new field $\tilde{ \vec \phi}$, the wall is along the $x$ axis and therefore $x$-independent.  After this we will drop the tildes for simplicity. The field profile only depends on $y$.  This simplifies the equation of motion to
\begin{eqnarray}
- c_T^2 \partial_y^2\phi_x &=& 
-\frac{\partial V_\theta}{\partial\phi_x}~, \nonumber \\
- c_L^2 \partial_y^2\phi_y &=& 
-\frac{\partial V_\theta}{\partial\phi_y}~.
\label{eq-eom2}
\end{eqnarray}

The tension of the domain  wall is given by the total energy per unit $x$ (assuming the potential of the vacua is set to be zero).
\begin{eqnarray}
\sigma(\theta) = \int dy \left[ \frac{1}{2}\left(c_L^2\phi_y'^2+c_T^2\phi_x'^2\right)+V_\theta \right]=  2\int dyV_\theta ~.
\label{eq-tension}
\end{eqnarray}
The last step came from integrating the equations of motion in Eq.~\eqref{eq-eom2}. The practical way to find the domain-wall solution is to numerically minimize this tension\cite{AguJoh09a,GibLam10,AhlGre10}, which is what we do in Appendix~\ref{sec-examples}.

For a general potential, the orientation dependence of $\sigma$ can be complicated. Here we start from a simple and, in some sense,  the typical case.  Imagine the situation where at $\theta=0$ the interpolation is purely longitudinal,\footnote{Note that we talk about a particular solution, instead of imposing some symmetry on $V$.  This is necessary.  One might try a rotational  symmetry on $V$ along the vector $(\vec\phi_+-\vec\phi_-)$.  That turns out to be neither necessary nor sufficient to guarantee that $\phi_x$ is constant.} $\phi_x=const$.   Since a rotation of $\pi/2$ just exchanges $\phi_x$ and $\phi_y$, the interpolation will become purely transverse with $\phi_y=const$.  It is then easy to work out from Eq.~(\ref{eq-tension}) that
\begin{eqnarray}
\sigma(0) &=& c_L \int_{\rm path} \sqrt{2V} |d\vec\phi|~,\nonumber \\
\sigma(\frac{\pi}{2}) &=& c_T \int_{\rm path} \sqrt{2V} |d\vec\phi|~,
\label{eq-simpletension}
\end{eqnarray}
where the two integration paths are the same, so
\begin{equation}
\frac{\sigma(0)}{\sigma(\frac{\pi}{2})}=\frac{c_L}{c_T}~.
\label{eq-ratio}
\end{equation}
Potentials given by Eq.~(\ref{eq-potmot}) when $m^2>0$ satisfy  Eq.~(\ref{eq-ratio}). So do the simpler potentials we use in Appendix~\ref{sec-examples}. They not only show a good agreement with Eq.~(\ref{eq-ratio}), but also demonstrate an excellent fit to a na\"ive interpolation,
\begin{equation}
\sigma(\theta) = \sigma(0)\cos^2\theta + \sigma(\frac{\pi}{2})\sin^2\theta~,
\label{eq-sigma}
\end{equation}
in the regular range of parameters.  From the symmetry of the problem, it seems natural to expand $\sigma(\theta)$ as a polynomial of $\sin^2\theta$ and keep the lowest order terms.

We also analyze two extreme choices of parameters in Appendix~\ref{sec-examples}. One of them corresponds to Eq.\eqref{eq-sigma} and the other satisfies 
\begin{equation}
\sigma(\theta) = \sqrt{\sigma(0)^2\cos^2\theta+
\sigma(\frac{\pi}{2})^2\sin^2\theta}~.
\label{eq-sigmamotion}
\end{equation}

It turns out that Eq.~(\ref{eq-sigma}) and (\ref{eq-sigmamotion}) are quite representative for our further analysis.  Despite their simple forms which will simplify the calculation, they can actually have dramatically different behaviors.

\subsection{Flat Wall Instability}
\label{sec-kink}
Because $c_L > c_T$ and $\vec{\phi_+} - \vec{\phi_-}$ is along the $y$-axis, it is apparent from Eq.~\eqref{eq-ratio} that the wall in the longitudinal direction has the highest tension $\sigma(0)$~. Since the energy of a domain wall is proportional to its length, we would naively  expect  straight flat walls to have the lowest energy. However, if in some direction the wall gets too heavy, it may be favorable to create a wall which is longer (not straight-line) but with lower energy. 
Therefore, even if the boundary condition is set up to preserve the $x$ translational symmetry,  the minimum energy interpolation can spontaneously break that symmetry.
The full treatment of this problem is to remove the condition~(\ref{eq-sym}) and see if a symmetry breaking configuration can further minimize the total energy. However, this approach requires  a complicated  numerical work and we will not pursue it in this chapter. We will simply demonstrate this possibility in the thin-wall approximation.

\begin{figure}
   \centering
   \includegraphics[width=10cm]{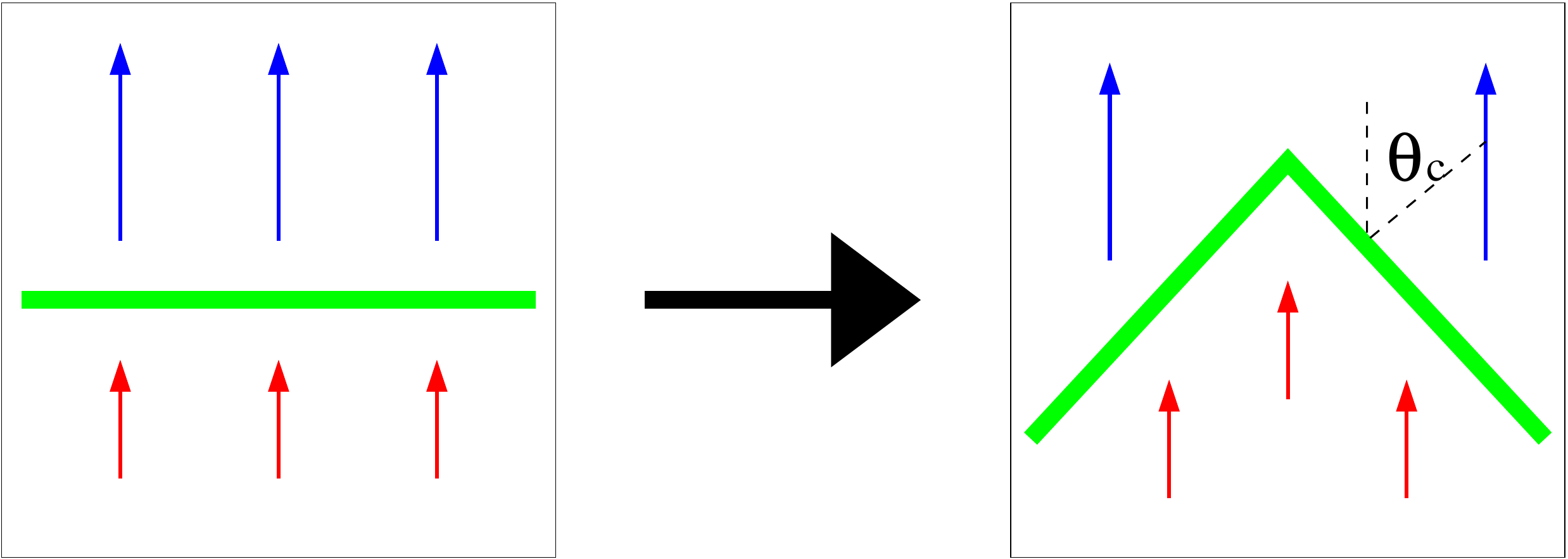} 
   \caption{When the flat, longitudinal wall on the left figure is too massive, it will spontaneously breaks into zigzag segments in the right figure.  Although the total wall area increases, the reduced tension still reduces total energy.
   \label{fig-zigzag}}
\end{figure}

The total energy is a functional of the domain-wall shape, $y(x)$.
\begin{equation} \label{eq-PureWall}
	E[y(x)] = \int_{x_1}^{x_2}
	 \sigma(-\tan^{-1} y') \sqrt{1+ y'^2} dx = 
	 \int_{x_1}^{x_2} \frac{\sigma(\theta)}{\cos\theta} dx ~.
\end{equation}
Here $y'$ denotes the derivative with respect to $x$. Now, given a symmetric boundary condition $y(x_1)=y(x_2)$, we can study whether the flat wall is stable to perturbations that deform the wall and, if it is unstable, we can look for other configurations which have the lowest possible energy.  Expanding $E$ near $\theta=0$ gives (assuming $\sigma(0)$ is a stationary point )
\begin{equation}
E \approx E_{\rm flat} + \int_{x_1}^{x_2} 
\left(\frac{1}{2}\frac{d^2\sigma}{d\theta^2}\bigg|_{\theta=0}  +\frac{\sigma(0)}{2}\right) \delta\theta^2 dx~.
\end{equation}
Thus, the perturbative stability condition is
\begin{equation}
\frac{1}{\sigma(0)}\frac{d^2\sigma}{d\theta^2}\bigg|_{\theta=0}>-1~.
\end{equation}
Next, the non-perturbative instability can occur if there is a $\theta\neq0$ such that
\begin{equation}
\sigma(\theta)<\sigma(0)\cos\theta~.
\end{equation}
When either or both instabilities  exists, there is a critical angle $\theta_c$ such  that $\sigma(\theta_c)/\cos\theta_c$ is the global minimum, and the wall prefers to settle into the zig-zag configuration in which every segment is oriented at $\theta_c$, as shown in Fig.\ref{fig-zigzag}.  It is favorable to not have  many corners which will cost more  energy. It is straightforward to see that domain walls more massive  than $\sigma(\theta_c)$ always break into zigzags, while  less massive walls are unaffected. 

The edges of the zigzag will contribute to the energy of the wall and the thin-wall approximation will break down at the sharp corners. A more careful treatment using thick-wall effects is necessary. However, if the distance between points $x_1$ and $x_2$ is large enough and there are not many kinks, the contribution from the corners will be negligible and the zigzag walls will  have lower energy and  still  be favorable. For the case of bubbles, we will not necessarily have the possibility of making the energy of the kinks negligible. However if the difference between the energy of the true and false vacuum is small, the bubble size will become very large and we can restore the results for the breakdown of flat walls \footnote{There are  two different  scales for nucleation of  kinky bubbles. One of them is the thin-wall approximation which demands that the physical length in which the transition happens be small compared to the radius of the bubble. We can achieve this approximation by making the energy difference of the two vacua small. The other hierarchy we need is the smallness of the energy of the kinks to the energy stored in the surface tension of the wall.  Fortunately this also can be achieved by creating a large bubble. Therefore, in order to make both of these approximations valid, we just need to  make the two vacua almost degenerate.}. 

\begin{figure}[htbp] 
   \centering
   \includegraphics[width=2.5in]{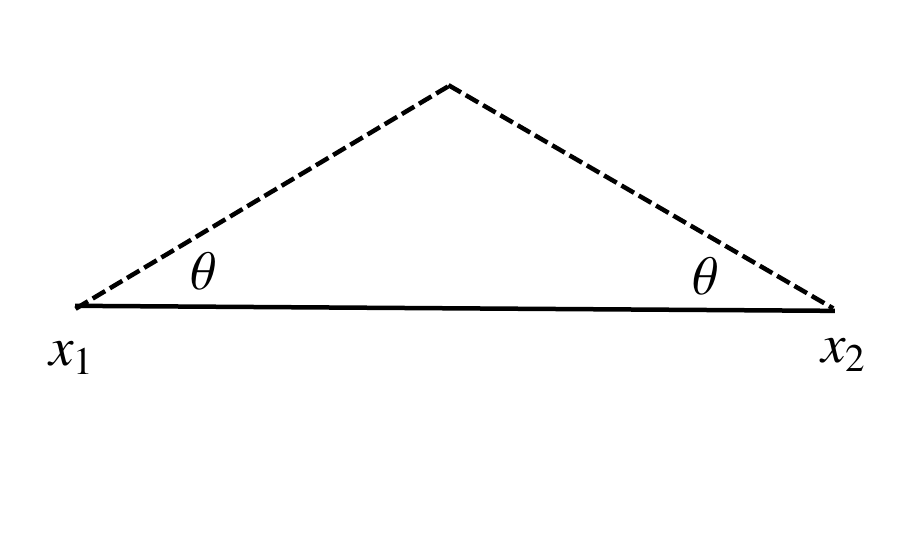}
   \includegraphics[width=2.5in]{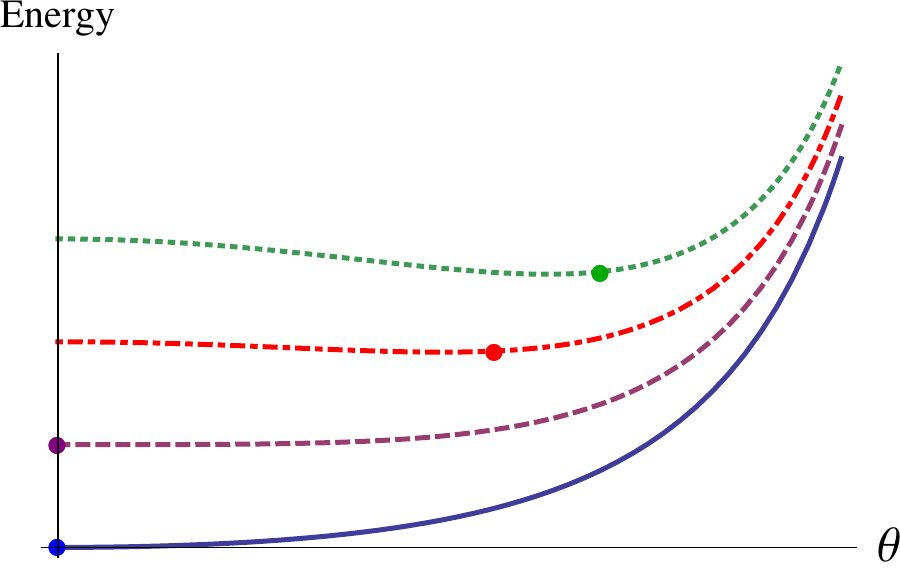} 
   \caption{The right figure is the energy stored in the wall connecting  points $x_1$ and $x_2$ (left figure) vs the angle $\theta$. The solid blue, dashed purple, dot-dashed red and dotted green lines correspond to $c_L/c_T = 1.5, 2, 2.5 $ and 3. The dots show the angle with minimum energy. Clearly, for high ratios of $c_L/c_T$, the zigzag walls ($\theta_{\rm min} \neq 0$) are energetically favorable. }
   \label{WallEVsTheta}
\end{figure}

The two examples whose $\sigma(\theta)$ is given by Eq.~\eqref{eq-sigma} and Eq.~\eqref{eq-sigmamotion} show very different  behaviors for different ratios of $c_L /c_T$. While the latter  always gives stable flat walls, the former  develops kinks once 

\begin{equation}
\frac{\sigma(0)}{\sigma(\pi/2)}=\frac{c_L}{c_T}>2~.
\end{equation}
A flat wall with $\theta=0$ would break and settle into zigzag segments with 
\begin{equation}
\theta_c = 
\sin^{-1}\sqrt{\frac{\sigma(0) - 2 \sigma(\frac{\pi}{2})}
{\sigma(0) - \sigma(\frac{\pi}{2})}} 
=\sin^{-1} \sqrt{\frac{c_L- 2 c_T}{c_L- c_T}}~.
\end{equation}
The wall shape and energy between two points $x_1$ and $x_2$ for the latter are shown in Fig. \ref{WallEVsTheta}.

\section{Bubble Shape}
\label{sec-shape}
\begin{figure}[htbp] 
   \centering
   \includegraphics[width=1.8in]{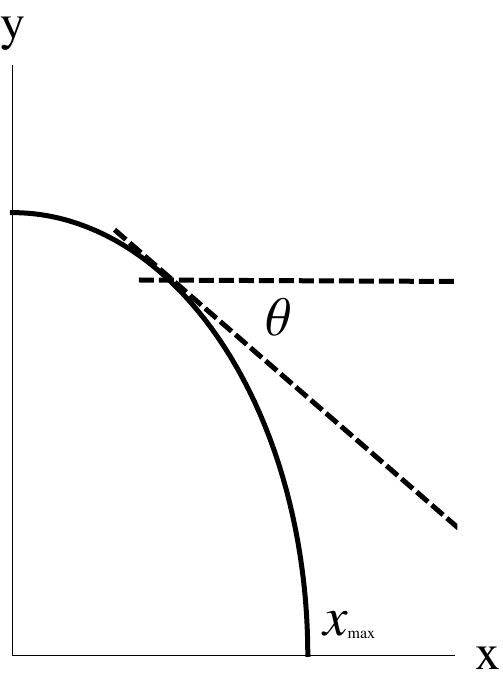} 
   \caption{Bubble shape in the first quadrant.}
   \label{OrdinaryWall}
\end{figure}

In order to allow a phase transition, we have to break the degeneracy between the two vacua and in order to keep the thin-wall approximation valid, the energy difference between the two vacua should be  kept small. The phase transition via  thermal bubble nucleation has a rate 
\begin{equation}
\Gamma \sim \exp\left[-\frac{E_s}{k_bT}\right]~,
\end{equation}
where $E_s$ is the saddle point energy of the bubble.  We can find this saddle point by treating $E$ as a functional of the bubble shape $y(x)$,
\begin{eqnarray}\label{EnergyFunctionalAug26}
E[y(x)] &=& (\text{surface integral of }\sigma)  - \Delta V ({\rm volume})
\nonumber \\
&=& 4 \int ~dx~ \left[ \sigma\left(-\tan^{-1}y'\right) \sqrt{1+y'^2} 
- y ~ \Delta V \right].
\end{eqnarray}
We put the origin at the center of the bubble. The bubble has a reflection symmetry along both the $x$- and $y$-axes. Therefore the contributions from different  quadrants to the bubble energy are equal and we can keep only the first quadrant. In this quadrant $y'= - \tan \theta$. This configuration is shown in Fig. \ref{OrdinaryWall}. To minimize the energy functional, we need to specify the boundary conditions. The ordinary boundary conditions are  $y'(0) = 0$ and  $y(x_{\rm max} )= 0$ at some unknown value of $x_{\rm max}$, which will be determined later by the energy difference between the two vacua. But we learned from  Sec.\ref{sec-kink}  to replace $y'(0)=0$ by $y'(0)=-\tan\theta_c$.  Despite that $y(x)$ is not smooth, it does eliminate the boundary variation and is indeed what we get from the Euler-Lagrange equation. A more formal argument is to write down $E[x(y)]$ instead, for which the standard choice, $x=0$ and $x'=0$ does not exclude the kink.  The resulting Euler-Lagrange  equation is essentially the same and we can just use it. 

We will keep it simple and solve the Euler-Lagrange equation for Eq.\eqref{EnergyFunctionalAug26}.
\begin{equation}
\Delta V = \frac{d}{dx} \left[\frac{d\sigma}{dy'}\sqrt{1+y'^2}+\sigma\frac{y'}{\sqrt{1+y'^2}}\right]
=\frac{d}{dx}\left(\sigma\sin\theta+\frac{d\sigma}{d\theta}\cos\theta\right)~.
\end{equation}
The general solution for $x$ is
\begin{equation}
x(\theta) = const. + \frac{1}{\Delta V}
\left(\sigma\sin\theta+\frac{d\sigma}{d\theta}\cos\theta\right)~.
\label{eq-x}
\end{equation}
The quantity in the parentheses is zero at both $\theta = \theta_c$ and $\theta=0$.  Therefore, solutions starting at either value will eliminate boundary variations, as promised, and also set that integration constant to zero. We can then integrate to find
\begin{equation}
y(\theta) = \frac{1}{\Delta V}
\left(\sigma\cos\theta-\frac{d\sigma}{d\theta}\sin\theta\right)~.
\end{equation}
For the case of higher dimensions, the structure of the potential can further break the ${\rm SO}(N-1)$ symmetry, since the exact interpolation between $\vec\phi_{\pm}$ may still involve nontrivial profiles of the transverse fields.   But if the  solutions retain ${\rm SO}(N-1)$ symmetry  the generalization to higher dimensions is straightforward. We can choose the $x_N$-axis along $(\vec{\phi}_+-\vec{\phi}_-)$ and denote  it by $x_L$ and $x_1$ to $x_{N-1}$ as the transverse directions $x_T$.  In these cases $\sigma$ is only a function of the polar angle $\theta$ and the energy is 
\begin{equation}
E[x_L(x_T)] = 2 \Omega_{N-2} \int  dx_T \,x_T^{N-2}
\left(\sigma(\tan^{-1}x'_L)\sqrt{1+x_L'^2}-\Delta V x_L \right)~,
\end{equation}
where $\Omega_{N-2}$ is the area of a unit  $(N-2)$-sphere and  $x_T=\sqrt{\sum_{i=1}^{N-1} x_i^2}$.

This leads to the general solution,
\begin{eqnarray}
x_T(\theta) &=& \frac{(N-1)}{\Delta V}
\left(\sigma\sin\theta+\frac{d\sigma}{d\theta}\cos\theta\right)~, \\
x_L(\theta) &=& \frac{(N-1)}{\Delta V}
\left(\sigma\cos\theta-\frac{d\sigma}{d\theta}\sin\theta\right)~.
\label{eq-xN} 
\end{eqnarray}

\begin{figure}
\begin{center}
\includegraphics[width=2in]{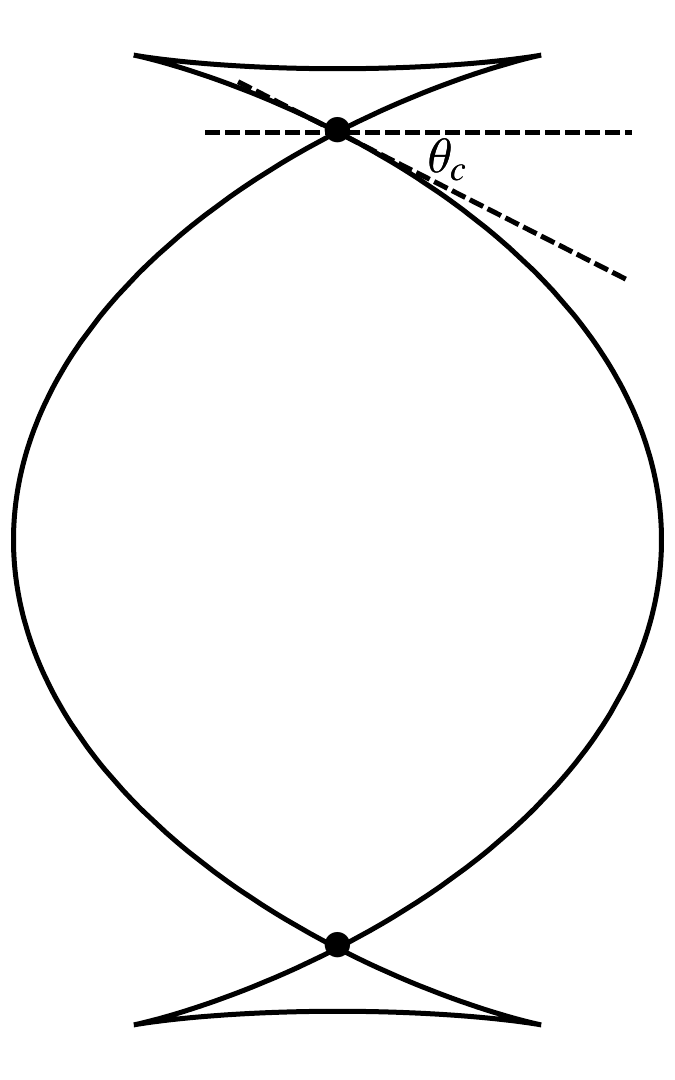} 
\end{center}
\caption{The bubble profile given by Eqs.~(\protect\ref{eq-xN}) and (\protect\ref{eq-sigma}) when $c_L>2c_T$.  The correct profile of the critical bubble is simply the middle portion.
}
\label{fig-candy}
\end{figure}

Although this is a naive generalization of \cite{RudBru95}, we should take a closer look.  Note that by symmetry we have $\frac{d\sigma}{d\theta}=0$ at $0$ and $\pi/2$.  Also, for simplicity we can treat $\sigma(\theta)$ as a monotonically decreasing function.  So we can see that $x_N$ is positive definite, but there is a risk of  $x_T$ being negative.  Since $x(\pi/2)$ is still always positive definite, and $x$ goes to zero exactly at $\theta_c$, if we na\"ively plot Eq.~(\ref{eq-xN}) for the range $\theta \in [0,2\pi]$, we may get something like a wrapped candy, as in Fig.\ref{fig-candy}. 
We will provide a simple argument using the number of negative modes  to prove the following statement:

{\bf Eq.~(\ref{eq-xN}) always gives the correct critical bubble profile.  When the flat longitudinal domain-wall is stable, this solution is valid for  $\boldsymbol{0 < \theta<\pi/2}$.  When the flat longitudinal domain-wall is unstable, we should take the largest $\boldsymbol{\theta_c}$ such that $\boldsymbol{x_T(\theta_c)=0}$ and use the portion $\boldsymbol{\theta_c<\theta<\pi/2}$. Therefore in Fig.\ref{fig-candy} we only keep the middle part of the candy shape region. } 

\begin{figure}[htbp] 
   \centering
   \includegraphics[width=2.5in]{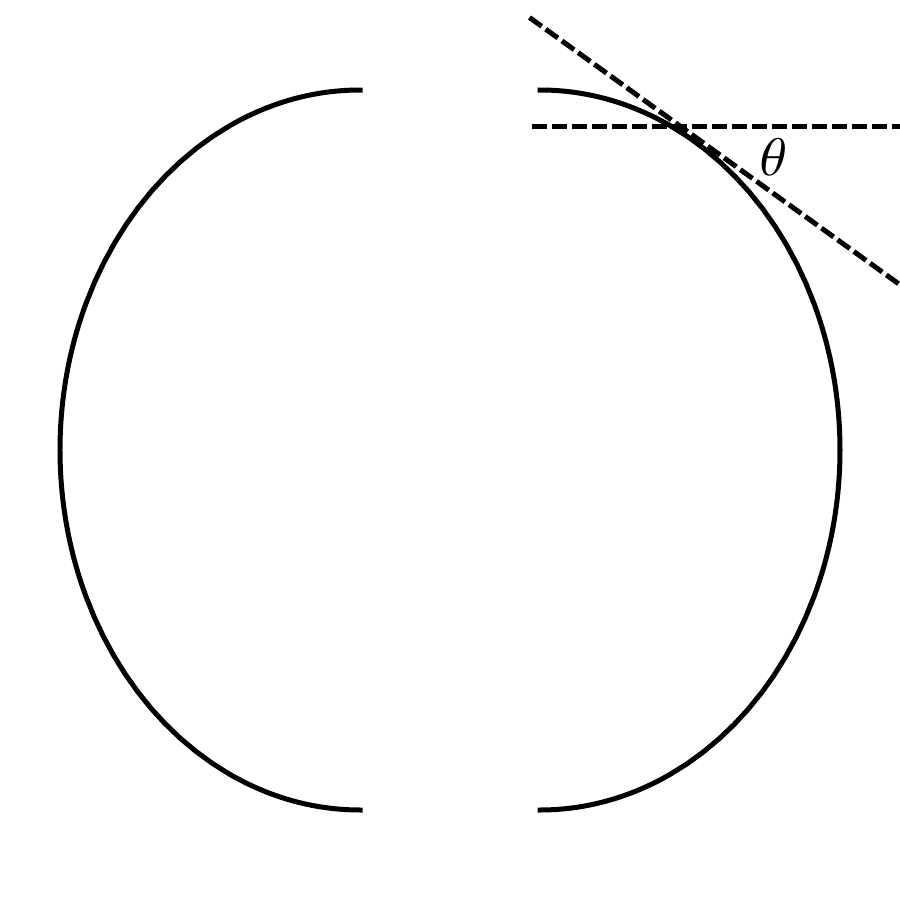} 
   \includegraphics[width=2.5in]{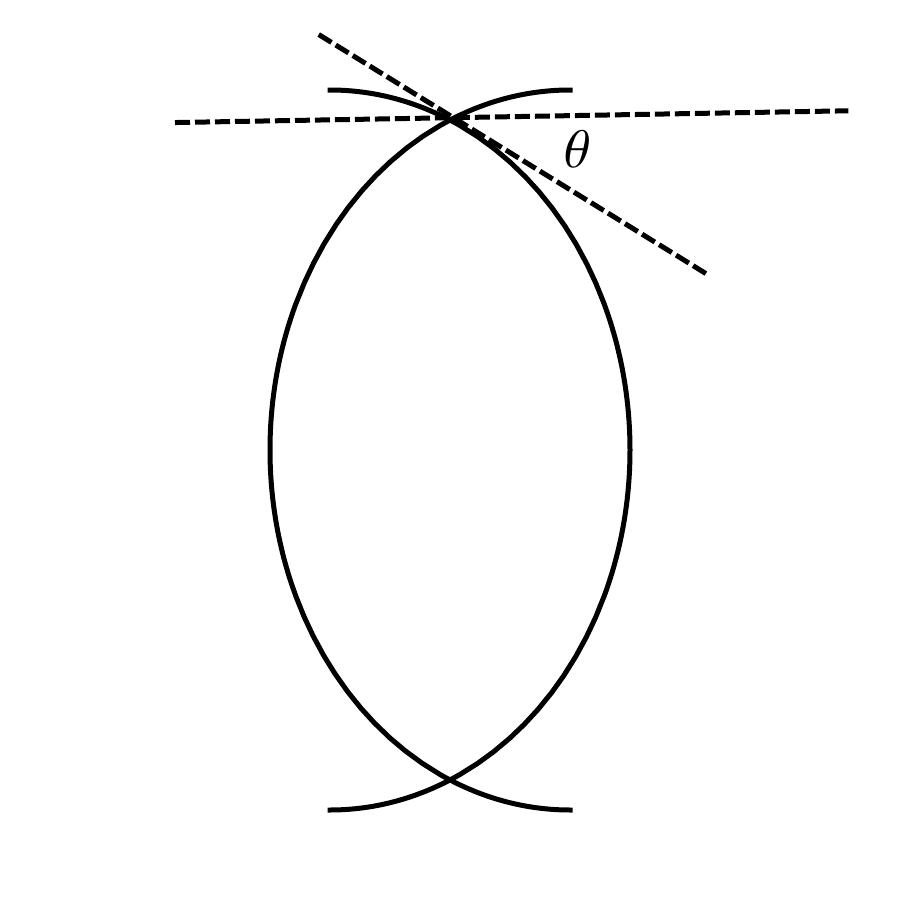} 
   \includegraphics[width=2.5in]{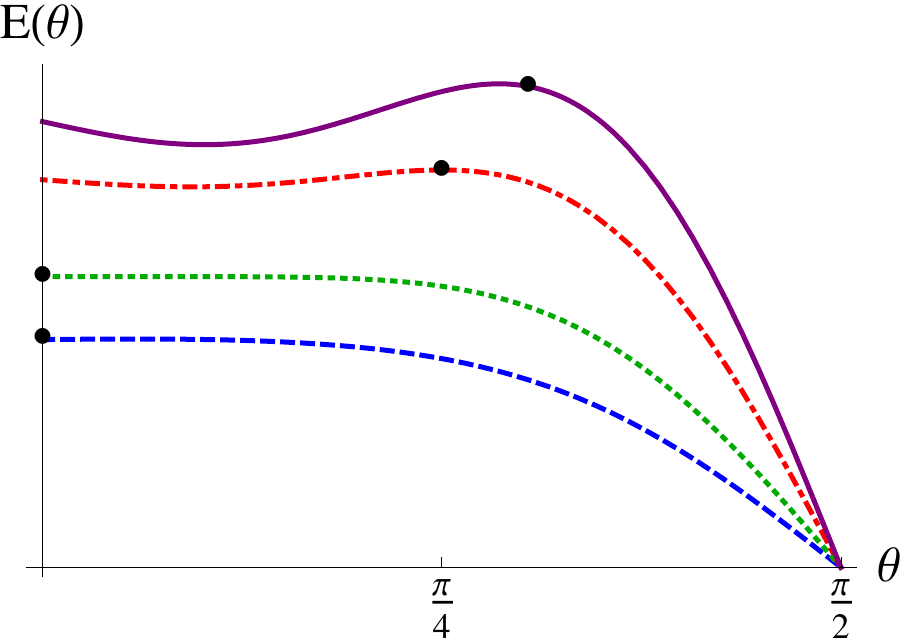} 
   \caption{Picking two pieces of the wall (top left panel) and gluing them at  angle $\theta$ (top right panel). Each piece locally satisfies Eq.~\eqref{eq-xN} and therefore gives the minimum energy for the fluctuations parallel to the wall. The energy of the bubble vs the gluing angle $\theta$ is shown in the lower panel for different values of $c_L/c_T$ for a tension that satisfies Eq.~\eqref{eq-sigma}. Blue dashes, green dots, red dot dashes and solid purple lines correspond to $c_L/c_T= 1.5, 2, 3$ and 4. The correct  spot for gluing the two is the angle which gives a local maximum (negative mode). For $c_L/c_T<2$ this gives a smooth bubble. For larger ratios it gives a kink. }
   \label{GluingAngleFig}
\end{figure}

\begin{figure}
\begin{center}
\includegraphics[width=10cm]{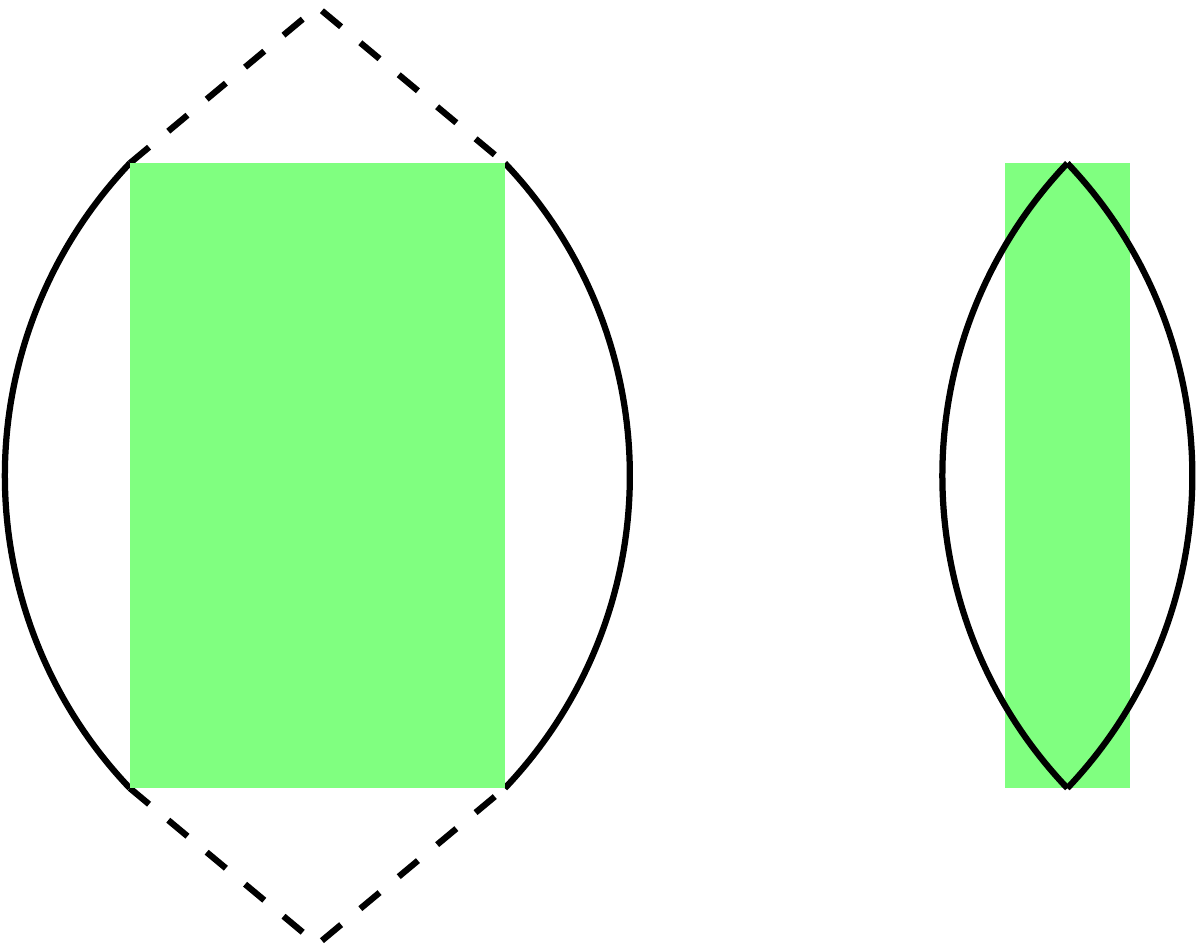}
\caption{The left figure depicts Eq.~(\protect\ref{eq-insert}), where we attempt to make a larger bubble by inserting true-vacuum regions and extra interpolation walls.  The right figure depicts Eq.~(\protect\ref{eq-remove}), where we try to make a smaller bubble by removing  part of the walls and part of the true-vacuum region.  Both result in smaller total energy, which shows that the kinky shape is indeed a saddle point.
\label{fig-saddle}}
\end{center}
\end{figure}

Any valid  tunneling configuration  must have one and only  one negative mode, in the expansion-contraction direction. The energy should increase for any other perturbation. This means that fluctuations of  the wall shape should still correspond to positive modes and every segment of the wall should still settle into a minimum energy configuration.   However, in the contraction-expansion direction  we must have a maximum energy.  Let's go back to the (2+1)-dimensional case and pick two pieces of  wall and glue them together. Since these pieces satisfy Eq.~\eqref{eq-xN}, they are indeed  local minima with respect to  fluctuations that change the shape of the wall. However, to get one negative mode, at the location of gluing, we must have a maximum of energy for the critical bubble. This situation is shown in Fig. \ref{GluingAngleFig}. We see that, at least for the case that satisfies Eq.~\eqref{eq-sigma}, for large values of  $c_L/c_T$  this angle is not at zero and therefore there are kinky bubbles. To make this more rigorous, let's look at Fig.\ref{fig-saddle}.
We can glue the two pieces, as in  the right panel,  using a smaller portion of the two shells.  Also, can we take the two shells further apart and interpolate between them with zigzag walls.  The former possibility is making the bubble smaller, while the later is making it bigger.  Through a pictorial argument, we can show that they both make the total energy smaller, establishing that Eq.~(\ref{eq-soln}) is really the saddle point with this unique negative mode. From Eq.~(\ref{eq-xN}) we get
\begin{equation}
\Delta V x_L(\theta_c) =(N-1) \frac{\sigma(\theta_c)}{\cos\theta_c}~,
\label{eq-crit}
\end{equation}
which means  that the energy difference due to volume for a cylinder,  an $(N-1)$-sphere times height $x_L(\theta_c)$, is equal to the energy in the domain wall that covers the $(N-1)$-sphere by a zigzag profile with orientation\footnote{There are two different   scales and approximations involved in this problem. The first  is using a flat-wall approximation to derive the orientation dependence of the tension. The second is the ratio of the contribution of the kink to the  wall in the bubble energy which we neglected here. Both of these approximations are valid if we choose a small energy difference between the two vacua [for example by choosing a very small $a$ in Eq.~(\ref{eq-potmot})]which makes the bubble bigger and flatter and, by increasing the area of the bubble and keeping the kink contribution unchanged, makes the kink contribution negligible.   }
 $\theta_c$. 

Then, as shown in the left portion of Fig. \ref{fig-saddle}, the energy lost due to the green (shaded) region is equal to the contribution from the dotted domain-wall.  
After these cancel each other, the two extra triangular regions still contribute $-\Delta V$, so the total energy is indeed less. 
\begin{equation}
({\rm zigzag \ wall})=\Delta V ({\rm rectangle})
<\Delta V (\rm extra \ false \ vacuum \ region)~.
\label{eq-insert}
\end{equation}

In the right portion of Fig.\ref{fig-saddle} we try to make a smaller bubble by removing the true vacuum region and domain walls covered by the green (shaded) rectangle and then patching the remaining two shells together as a smaller bubble, with a kink angle larger than $\theta_c$.  Since $\sigma(\theta_c)/\cos\theta_c$ is a minimum of $\sigma(\theta)/\cos\theta$, losing those wall segments overcompensates for the energy gain even if we remove $-\Delta V$ of the entire green (shaded) rectangle.
\subsection{Smooth Bubbles}

We can get more intuition by solving the exact bubble shape for a specific $\sigma(\theta)$.  In the first example we will use Eq.~(\ref{eq-sigmamotion}), where no spontaneous symmetry breaking should occur.  Thus, we are expecting a smooth bubble.  Plugging into Eq.~(\ref{eq-xN}), we get
\begin{eqnarray}
x_T(\theta) &=& \frac{N-1}{\Delta V} \frac{\sigma(\pi/2)^2\sin\theta}
{\sqrt{\sigma(0)^2\cos^2\theta + \sigma(\pi/2)^2\sin^2\theta}}~, \\
x_L(\theta) &=& \frac{N-1}{\Delta V} \frac{\sigma(0)^2\cos\theta}
{\sqrt{\sigma(0)^2\cos^2\theta + \sigma(\pi/2)^2\sin^2\theta}}~.
\end{eqnarray}

Obviously, the bubble takes the shape of an ellipsoid,
\begin{equation}
\frac{\sum_{i=1}^{N-1}x_i^2}{c_T^2}+\frac{x_L^2}{c_L^2} = (N-1)^2 r_0^2~,
\end{equation}
where 
\begin{equation}
r_0 = \frac{\sigma(0)}{c_L \,  \Delta V} = \frac{\sigma(\pi/2)}{c_T\,  \Delta V} 
= \frac{1}{\Delta V} 
\int_{\rm path} \sqrt{2V} |d\vec\phi|
\end{equation}
comes from Eq.~(\ref{eq-simpletension}).

It is then straightforward to calculate the saddle point energy,
\begin{equation}
E_s = S_{N-1}\frac{(N-1)^{N-1}}{N} 
\frac{\left(\int_{\rm path} \sqrt{2V} |d\vec\phi|\right)^N}{\Delta V^{N-1}}
c_L c_T^{N-1}~.
\end{equation}
Comparing this  to the usual form people use assuming a spherical bubble, Eq.~(\ref{eq-rate}), the difference can be characterized by an effective Fermi velocity, 
\begin{equation}
v_F \rightarrow (c_L c_T^{N-1})^{1/N}~,
\label{eq-Fermi}
\end{equation}
which is a weighted geometric average of sound speeds.

\subsection{Kinky Bubbles}

Now we turn our attention to Eq.~(\ref{eq-sigma}).  Plugging  it into Eq.~(\ref{eq-xN}), we get
\begin{eqnarray}
x_T(\theta) &=& \frac{(N-1)}{\Delta V}
\left(\bigg[2\sigma(\pi/2)-\sigma(0)\bigg]\sin\theta+\bigg[\sigma(0)-\sigma(\pi/2)\bigg]\sin^3\theta\right)~, \\
x_L(\theta) &=&  \frac{(N-1)}{\Delta V}
\left(\bigg[2\sigma(0)-\sigma(\pi/2)\bigg]\cos\theta-\bigg[\sigma(0)-\sigma(\pi/2)\bigg]\cos^3\theta\right)~.
\label{eq-soln}
\end{eqnarray}
As expected, when $\sigma(0)<2\sigma(\pi/2)$, we still have a smooth bubble profile.  When $\sigma(0)>2\sigma(\pi/2)$ we just use the portion $\theta_c<\theta<\pi/2$. This is proved in Sec.\ref{sec-shape}

The expression for $E_s$ is quite complicated in arbitrary dimensions, so we only present the results for  ``realistic'' dimensions.  For $N=2$, we have
\begin{eqnarray}
E^{N=2}_s &=& 
\frac{\left(\int_{\rm path}\sqrt{2V}|d\vec\phi|\right)^2}{4\Delta V}
(10c_Lc_T-c_L^2-c_T^2) \frac{\pi}{2}~, \ \ \ {\rm for} \ \ \ c_L<2c_T~, \\
&=& \frac{\left(\int_{\rm path}\sqrt{2V}|d\vec\phi|\right)^2}{4\Delta V}
\bigg( (10c_Lc_T-c_L^2-c_T^2) \cos^{-1}\sqrt{\frac{c_L-2c_T}{c_L-c_T}}
\nonumber \\ 
&+& (c_L+13c_T)\sqrt{(c_L-2c_T)c_T} \bigg)~, \ \ \ {\rm for} \ \ \ c_L>2c_T
\end{eqnarray}

This is quite complicated.  We should again compare it to the spherical bubble and think in terms of the effective Fermi velocity, especially in the limit $c_L\gg c_T$.
\begin{eqnarray}
v_F &\rightarrow& \left(\frac{10c_Lc_T-c_L^2-c_T^2}{8}\right)^{1/2}~,
\ \ \ {\rm for} \ \ \ c_L<2c_T~, \nonumber \\
v_F &\rightarrow& \frac{4}{\sqrt{3}}(c_Lc_T^3)^{1/4}~, 
\ \ \ {\rm for} \ \ \ c_L\gg c_T~.
\end{eqnarray}

For $N=3$, we have
\begin{eqnarray}
E^{N=3}_s &=& 4\pi 
\frac{\left(\int_{\rm path}\sqrt{2V}|d\vec\phi|\right)^3}{\Delta V^2}
\frac{4(c_L^2-10c_L^2c_T+52c_Lc_T^2-8c_T^3)}{105}~,
\ \ \ {\rm for} \ \ \ c_L<2c_T~, \nonumber \\
&=& 4\pi 
\frac{\left(\int_{\rm path}\sqrt{2V}|d\vec\phi|\right)^3}{\Delta V^2}
\frac{32c_T^2(7c_L-6c_T)\sqrt{c_T}}{105\sqrt{c_L-c_T}}~,
\ \ \ {\rm for} \ \ \ c_L>2c_T~.
\end{eqnarray}
Similarly we have
\begin{eqnarray}
v_F \rightarrow 
\left(\frac{8}{5}c_L^{1/2}c_T^{5/2}\right)^{1/3}~, 
\ \ \ {\rm for} \ \ \ c_L\gg c_T~.
\end{eqnarray}

Comparing these to Eq.~(\ref{eq-Fermi}), we find that the effective Fermi velocity is still a weighted geometric mean, but the weight on $c_L$ is always reduced by half.  This is quite understandable since the critical bubble approaches a thin slit.  A major portion of its domain-wall is aligned in the transverse direction.  It is straightforward to show that this limit generalizes to $N$ dimensions as
\begin{equation}
v_F \sim (c_L^{1/2}c_T^{N-1/2})^{1/N}~.
\end{equation}

\section{Summary}
\label{sec-dis}
In this chapter we studied the vacuum decay problem in a vector field theory which has different longitudinal and transverse speeds of sound and  constructed the critical bubbles for thermal tunneling. Because the spherical symmetry is broken due to  a combination of different sound speeds and having a singled out direction $(\vec{\phi}_+ - \vec{\phi}_-)$, the surface tension has a  dependence on the orientation of the wall. This makes the critical bubbles non-spherical. We derived the shape of the bubble using a simple formula, Eq.~\eqref{eq-xN}, and showed a modification in the bubble energy  by a factor of speeds of sound ratios. Since the tunneling rate depends exponentially on this energy, this implies a large modification to the rates. We showed that if the ratio of speeds of sound goes beyond a critical value, the flat walls become unstable and break spontaneously into zigzag segments. The same effect is seen for the nucleation of bubbles, and so the critical bubble develops kinks. We gave a recipe for making these non-smooth bubbles and  explained the correct way to interpret them according to the number of negative modes. We also developed numerical ways to calculate the orientation dependence of the tension and showed examples for different behaviors of the surface tension\footnote{Analysis of the bubble shape for nucleation is identical to that for  equilibrium bubbles, known as the Wulff construction.  Earlier works\cite{GalFou95,Leg75,Whe75,He3,Fou95,MacJia95,RudLoh99,SilPat06} have qualitatively similar results.  We further show that the flat wall settles into  zigzag segments with a corner angle of $\theta_c$.  The recognition and interpretation of the kinky bubble shape is also more transparent in our analysis.  The tension previously studied is often expanded as $\sigma(\theta) = \sigma_0 + a\cos\theta + b\cos 2\theta$, and most analysis focused on the effect of $a\neq0$.  In our model there is a reflection symmetry---the domain-wall tension does not change when you look at it from the other side.  Thus we always have $a=0$.  This makes our situation closer to a two dimensional  lattice model\cite{AleSur90}, where similar bubble shape was observed.}.

Our analytic and numerical study shows that the freedom to take different paths in a multi-dimensional field space is essential for the instability.  If we choose parameters that reduce the number of dynamical fields down to one, the longitudinal wall is always stable.  That is however an extreme choice.  For typical choices of parameters, at $c_L/c_T>2$ the longitudinal wall becomes unstable, and the critical bubble develops two kinks.  Such behavior can appear with an even smaller sound speed ratio, $c_L/c_T>\sqrt{2}$, if we tune the potential to the other extreme limit.  This range of sound speed ratios is not hard to find in real materials.

We picked two representative forms of $\sigma(\theta)$, given by Eq.~(\ref{eq-sigma}) and (\ref{eq-sigmamotion}), to calculate the exact shapes of critical bubbles.  This allowed us to observe the scaling property of tunneling rates.  When we further increase the sound speed ratio  the critical bubble is deformed but still smooth, we can modify the standard tunneling rate formula, Eq.~(\ref{eq-rate}), through an effective Fermi velocity,
\begin{equation}
v_F \rightarrow (c_L c_T^{(N-1)})^{1/N}~.
\end{equation}
This is quite intuitive, since one particular orientation is longitudinal  and all others are transverse.  They care about the sound speeds in their own orientations.  When we increase the sound speed ratio  the bubble starts to develop kinks and the scaling behavior changes to
\begin{equation}
v_F \rightarrow (c_L^{1/2} c_T^{(N-1/2)})^{1/N}~.
\end{equation}
This is because the kink development removes a large portion of the longitudinally oriented wall, so $c_L$ becomes less important.

On top of modifying the tunneling rate estimation, our result has a practical impact.  Typically,  experimental measurement of domain-wall tension involves measuring the bubble radius\cite{OshCro77}.  That is done by observing a domain wall popping through a partition with holes.  When it does, the radius of the hole is identified with the bubble radius.  Our result shows that for vector fields, the orientation of that partition is important. The popping radius can be identified with $x$ given by Eq.~(\ref{eq-xN}) only for a longitudinally oriented partition.  For other orientations  the hole and the bubble do not have common symmetries.  Therefore the exact relation between the popping radius and the critical radius requires further analysis.

We have only taken a small step toward a rich phenomenology.  Given the new insight here, many nontrivial questions arise.  How does the domain wall move (bubble expand) given this orientation dependence? When there is a kink, can we expect the tip to travel at $c_L$, leaving behind a Cherenkov-like tail of domain walls bounded by $c_T$?  How does the spontaneously broken planar symmetry interact with impurities or other external effects?  All these await future study, and may lead towards a more practical understanding of some exotic theories of phase transitions that rely on the properties of domain walls\cite{EasGib09,GibLam10,YanTye11}.

\clearpage

\chapter{Instabilities in large dimensional field spaces }
\label{Tumbling}
\begin{figure}[htbp] 
   \centering
   \includegraphics[width=2.5in]{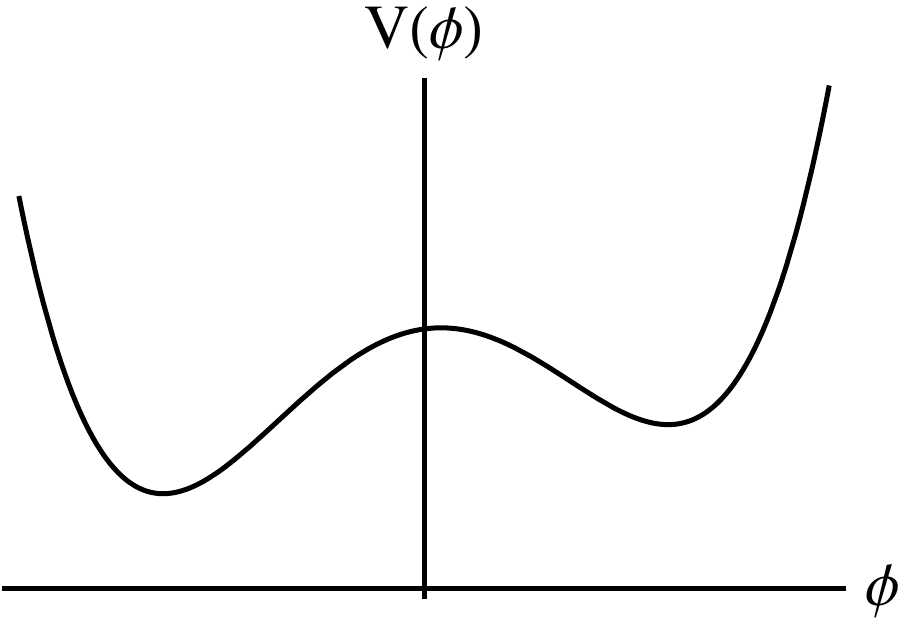} 
   \includegraphics[width=3in]{2DPotential.pdf} 
   \caption{In a theory with one scalar field (left picture), to go from the false vacuum (right minimum) to the true vacuum (left minimum) we have to cross the highest barrier. But for the case of a field theory with two fields (right picture), to go from the false vacuum to a truer one, we do not need to pass the highest barrier (central peak) and  can choose to go through many other lower barriers. This decreases the instanton action and enhances the tunneling rate.}
   \label{1Vs2DPots}
\end{figure}

There are  important qualitative and quantitative differences between tunneling in theories with one scalar field and  ones  which involve  many fields. We  studied one specific case in Chapter \ref{VectorBubble} that has more than one field and found dramatically different behavior. In general, finding  bounces solution in multi-field theories is not easy  and there is not much focus on it in the standard literature of tunneling. But there is one aspect of these theories which on very basic principles  may be worrisome and needs further exploration. From Fig. \ref{1Vs2DPots}, it is clear that, in the case with one field, to get from the false vacuum to the true vacuum  we must pass through the highest barrier that separates them, but that for  a theory with two fields  there are many other passages which do not require crossing the highest barrier. This may enhance the tunneling process. The number of saddle points and possible tunneling paths increases very rapidly with the number of fields. This may render most of the minima of these theories unstable and short-lived  due to quantum or thermal fluctuations. In this chapter, which is based on \cite{Greene:2013ida}, we show that, indeed, a generic minimum is not protected by high enough walls to ensure its stability. Specifically, we will study theories with multiple scalar fields and provide numerical evidence that  for a generic local minimum of the potential the semiclassical tunneling rate, $\Gamma = A e^{-B}$ increases rapidly as the number of fields increases. As a consequence, the fraction of vacua with rates low enough  to maintain metastability decreases exponentially with the number of fields. We also discuss possible implications for the landscape of string theory. If our results prove applicable to  string theory, the landscape of metastable vacua may not contain sufficient diversity to offer a natural explanation of dark energy.

The discovery that string theory admits a huge number of flux vacua \cite{Strominger:1986uh,Giddings:2001yu,Polchinski:1995mt} played a very important role in the development of the theory over the last decade. The landscape of vacua of theories with several hundred moduli fields  was considered to be a  good anthropic framework  to address the problem of naturalness of the cosmological constant. Many aspects of these vacua,  from their phenomenological and cosmological properties to their distribution and statistical features, have been extensively studied \cite{Grana:2005jc,Douglas:2006es}.  There have been many attempts to study some of the statistical features of these vacua in previous works \cite{Kachru:2003sx, Ceresole:2006iq, Dine:2007er,  Sarangi:2007jb,Tye:2007ja,Podolsky:2008du,Brown:2010bc,Brown:2007zzh}. But due to the complexity of this landscape of vacua, many important questions have not been answered. One  important one  is whether  the vacua obtained in this  way are long-lived or no. A direct analysis of this question brings formidable challenges in calculation. Therefore, we take a different approach and study  generic  field theoretic models of landscape and focus on how the stability of vacua varies as the dimension of the moduli space (the number of fields) increases. Our results suggest that tunneling rates, and hence vacuum instability grow so rapidly with the number of moduli fields that the probability of a given local minimum being metastable is exponentially small. At the semiclassical level, the transition rate is given by $\Gamma = A e^{-B}$ where $B$ is the instanton action and $A$ is given by the determinant of fluctuations around the classical path. We show that  $B$ drops very quickly with the number of moduli fields. For tunneling from de Sitter, we also have the possibility of Hawking-Moss solutions. The tunneling exponents in the latter  case in the thin-wall approximation are proportional to the difference between the saddle point energy density and the energy density of the false vacuum. We show that this difference drops very quickly with the number of moduli fields, so this tunneling channel  also gets a big enhancement. In string theory, the complexity of the geometry and the large number of moduli fields makes the situation much more complicated. Meanwhile, it opens up a new line of attack based on statistical analysis. For example, Douglas and Denef \cite{Denef:2004ze} developed a method of calculating the density of flux vacua in the string landscape in terms of the K\"ahler potential on the moduli space of a given Calabi-Yau compactification. Their work showed that the vacua tend to accumulate near the conifold locus in moduli space. Dine et al.\cite{Dine:2007er} used a scaling argument to show that the vacua with small cosmological constant in the landscape of string theory become unstable when the fluxes are large compared to their compactification volume. Chen et al.\cite{Chen:2011ac} studied the chance that a given stationary point in the landscape is a minimum. They found that the chance that a randomly chosen saddle point is a minimum is suppressed exponentially for large numbers of moduli fields. They showed that the suppression becomes more severe as the cosmological constant of the vacuum gets higher. In Sec. \ref{Land-sec-Background}, we discuss our approach for estimating tunneling rates in a field theoretical context. In Sec. \ref{Land-sec-Numeric}, we show the numerical methods and results. These results show a general feature of high-dimensional field theories and are independent of the application to  string theory. In Sec. \ref{Land-sec-String}, we discuss the possible implications of our results for the landscape of string theory and explain some of the considerations that would need to be resolved before applying this model to the string landscape. Finally, in Sec. \ref{Land-sec-Summary}, we estimate the maximum dimension of moduli spaces that allows enough diversity to explain the cosmological constant in a natural way, and we suggest future directions for studying these models.

\section{Background and approximation method}
\label{Land-sec-Background}
We consider the dynamics of a system composed of $N$ scalar fields $\phi_j$ with a Lagrangian 
\begin{equation}\label{Land-Lagrangian}
	{\mathcal L} = \frac{1}{2} \sum_{j=1}^N \partial_\mu \phi_j \partial^\mu \phi_j - V(\phi_1,\phi_2, \ldots, \phi_N)~.
\end{equation}
In general the potential $V$ has many local minima that represent  metastable false vacua. We are interested in the rate at which the transitions between these vacua occur. Let's consider one of these minima and use the freedom to shift the field by a constant value to bring this minimum to the origin. Since we are ignoring gravity at this point, we can shift the value of the potential by a constant number and set it to zero at the origin. If the potential is smooth enough around this minimum, we can expand the potential in a power series,
\begin{equation}
	\label{Land-potential}	
	V =  \lambda \left(\sum_i A_{ii}^{(2)} \phi_i^2 v^2 + \sum_{ijk} A_{ijk}^{(3)} \phi_i \phi_j \phi_k v + \sum_{ijkl}A_{ijkl}^{(4)} \phi_i \phi_j \phi_k \phi_l + \ldots\right)~.
\end{equation}
Here we used the freedom of using an $O(N)$ transformation to diagonalize the quadratic terms. Assuming that the origin is a minimum makes $A_{ii}^{(2)}$ positive definite. Here we have extracted a mass scale $v$ from the potential to make $A_{ijk}^{(3)}$ and $A_{ijkl}^{(4)}$ dimensionless. This  scale determines the typical distance between the stationary points of $V$ in the field space. Finally, we have extracted an overall factor $\lambda$ from the potential to make the quantities inside the parenthesis of order unity. Although the potential itself is bounded from below, if we keep a finite number of terms in the series, the resulting approximation may not be so. We are not concerned about making the potential in Eq.\eqref{Land-potential} bounded form below as long as we care about the immediate vicinity of the origin. Therefore the coefficients $A_{ijk}^{(3)}$ and $A_{ijkl}^{(4)}$ can be positive or negative without imposing an overall positivity condition. 

The tunneling exponent $B$ is the Euclidean action of the bounce solution. Since we are ignoring  gravitational effects, we can safely assume that the dominating bounce has an $O(4)$ symmetry and the fields are all functions of $s= \sqrt{{\bf x}^2 +x_4^2 }$. The equation for the bounce solution \cite{Coleman:1977py} is
\begin{equation}
	\label{Land-BounceEOM}
	\frac{d^2 \phi_j}{d s^2} + {3 \over s} \frac{d \phi_j }{d s}= \frac{\partial V}{ \partial \phi_j}~.
\end{equation}
This configuration will have a region on the true vacuum side at the center of the bubble and, in order to ensure it has a finite action, at large $s$ it should approach the false vacuum, i.e. $\lim_{s\rightarrow \infty}\phi(s) = \phi_{\rm fv}$. We do not specify $\phi(0)$, the field at the center of bubble,  a priori and since this is a second order differential equation, we would need another boundary condition.  We demand $\phi'(0)=0$ to avoid a singularity at $s=0$. Except for the thin-wall approximation, we do not expect the field at the origin be very close to the true vacuum. 

For  our study, we will generate a large ensemble of potentials with random coefficients. In order to calculate  the tunneling rates, we would have to solve Eq.~\eqref{Land-BounceEOM} for each sample. But these are $N$ coupled nonlinear  differential equations and solving them numerically in a reasonable time is not possible. Instead, we  develop an approximation to make this problem tractable. Our starting point is the thin-wall approximation for a single scalar field. In the thin-wall case, the field configuration can be thought of as a ball, filled  inside with  true vacuum while the outside is in false vacuum. The two phases are separated and patched smoothly by a transition region which is thin compared to the length scale in the problem (the radius of the ball). In this transition region, the potential barrier and the gradient terms have  an energy density per unit of area which is equivalent to a  tension. Therefore the net tunneling exponent is  \cite{WeinbergBook}
\begin{equation}\label{Land-TunnelingExponent}
	B = 2\pi^2 \int_0^\infty ds \,s^3 \left[ {1 \over 2} \phi'^2 + V(\phi) - V_{\rm fv}\right] = 2 \pi^2 R^3 \sigma - {1 \over 2} \pi^2 R^4 \epsilon~.
\end{equation}
where the factors of $\pi^2$ come from the volume elements in  four-dimensional spherical coordinates, $R$ is radius of the bubble, $\epsilon= V_{\rm fv} - V_{\rm tv}$ comes from the integration over the  inside of  the bubble and $\sigma$ represents the surface tension. By virtue of the equations of motions, $\sigma$ can be rewritten as
\begin{equation}\label{Land-tensionEq}
	\sigma = \left| \int_{\phi_{\rm fv}}^{\phi_*} d\phi \sqrt{2 \left[ V(\phi)- V_{\rm fv}\right]}\right| ~,
\end{equation}
and the integration is taken from the false vacuum outside the bubble to some point $\phi_*$ on the true vacuum side of the barrier which has the same potential as the false vacuum $V(\phi_*) = V_{\rm fv}$. For the case of multi-field theories, the possible $\phi_*$ constitute a co-dimension one subset in the field space. Depending on the endpoint, we will have different values for tension. 

To find the dominant path, we minimize this action with respect to $R$ (and also the endpoints ) to get
\begin{eqnarray} \label{Land-ThinWallB}
	R &=& {3 \sigma \over \epsilon} ~.\nonumber \\
	B &=&{\pi^2 \over 2}\sigma R^3= {27 \pi^2 \sigma^4 \over 2 \epsilon^3}~.
\end{eqnarray}
These results are only valid if $\epsilon$ is much smaller than the energy scales involved and therefore the tunneling rate given by Eq.~\eqref{Land-ThinWallB} is very slow. Also this result is very sensitive to $\epsilon$ which cannot be correct outside the thin-wall limit, because the field does not get close enough to the true vacuum to see the real energy difference. But we can extract some useful information for the thick wall case.  

\begin{figure}[h!] 
   \centering
   \includegraphics[width=2.7in]{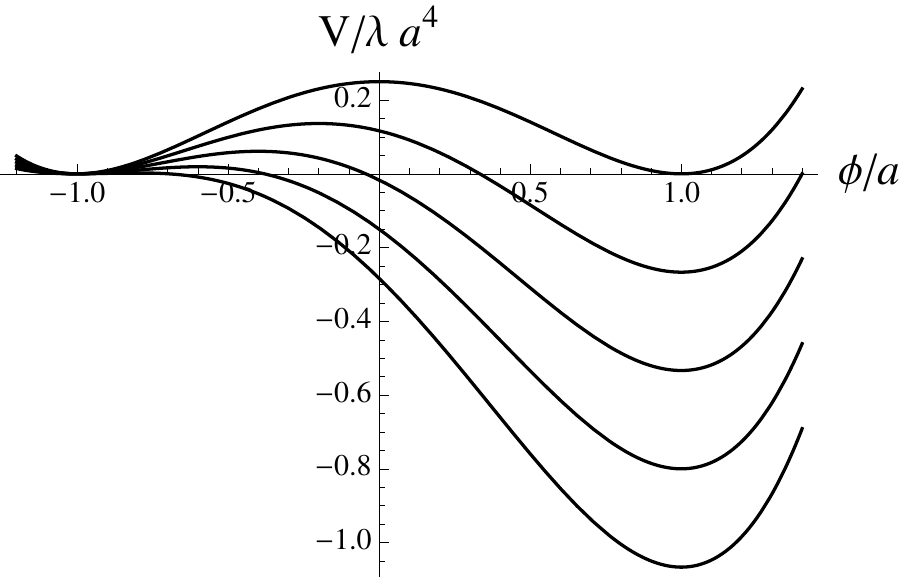} 
   \includegraphics[width=2.7in]{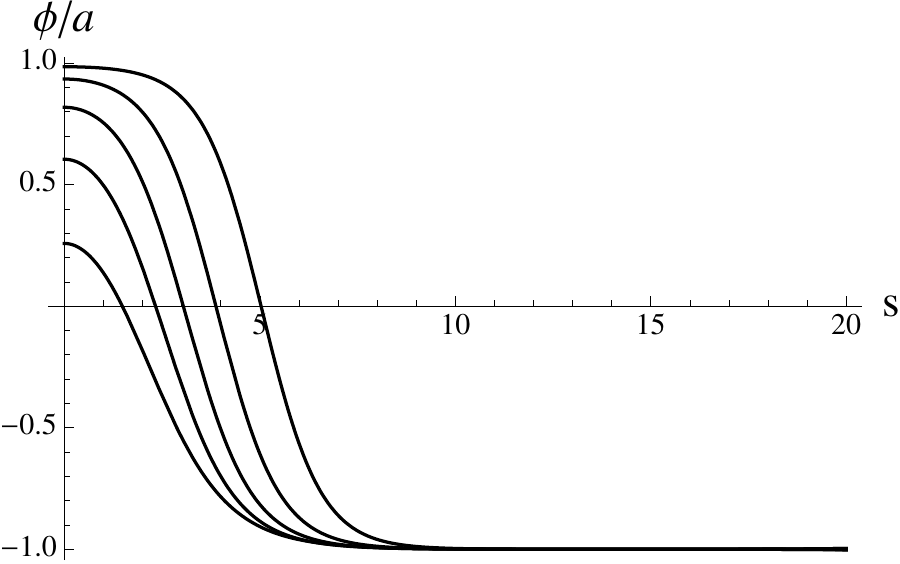} 
   \caption{Left panel shows the potential introduced in Eq.~\eqref{EricksPotentials}. The progression from  top to  bottom shows the potentials corresponding to $b=0, 0.2, 0.4, 0.6$ and 0.8. Case $b=0$ is a symmetric double well and the larger values show more and more deviation from the $\mathbb{Z}_2$ symmetry. The right panel shows the corresponding field profiles (top to bottom) of bounce solution for the values of $b= 0.4, 0.5, 0.6, 0.7$and 0.8. The solution deviates more and more from a thin wall and at $b=0.9$ it merges to a Hawking-Moss instanton. }
   \label{Land-FieldPotProfile}
\end{figure}

Here we follow chapter 12 of \cite{WeinbergBook} to get an approximation for the tunneling exponent in the thick-wall limit. Let's rewrite the potential in Eq.~\eqref{Land-potential} as (notice that in this case $V_{\rm fv}=0$)
\begin{equation}
	V(\phi) = \lambda v^4 \hat{V}(\phi/v)+ V_{\rm fv}~.
\end{equation}
We can define a dimensionless length and field as 
\begin{eqnarray}
	s&=& {\hat s \over v \sqrt \lambda} ~, \nonumber \\
	\phi &=& v \hat \phi~.
\end{eqnarray}
In terms of these new variables, the tunneling exponent in Eq.~\eqref{Land-TunnelingExponent} becomes
\begin{equation}
	 B = {2\pi^2 \over \lambda} \int_0^\infty d\hat s \; \hat s^3 \left[ {1 \over 2} \left(\frac{d \hat \phi}{d \hat s}\right)^2 + \hat{V} (\hat \phi) \right]~.
\end{equation}
Therefore the tunneling exponent is independent of $v$ and inversely proportional\footnote{Although $\mathcal B$ is independent of $v$, the value of $v$ effects the prefactor in the  tunneling rate.}  to $\lambda$. The bounce solution follows a path $\gamma_1$ in the field space starting  from some point on the true vacuum side of the barrier and evolving into the false vacuum. At some point $s_{\rm top}$, it passes halfway through the barrier  that separates the two vacua. Similarly to the case of a thin wall, we can assign a surface tension to this path. 
\begin{equation}\label{TensionApprox1}
	\tilde \sigma = \int_ \gamma d\phi \sqrt{2 \left[ V(\phi)- V_{\rm fv}\right]} =  \sqrt{\lambda} v^3\int_ \gamma d\hat \phi \sqrt{2 \left[ \hat V(\hat\phi)- V_{\rm fv}\right]} = \sqrt \lambda v^3 \tilde s~.
\end{equation}
Since the contribution from $s=0$ to $s=s_{\rm top}$ on average is the same as contribution from $s=s_{\rm top}$ to the point where  it reaches the false vacuum, we can evaluate the integral over a new path $\gamma'$ which is the portion of $\gamma$ that extends from $s=0$ to $s=s_{\rm top}$ . This causes a factor of 2 to appear in front of the integral to compensate for the part that we have neglected. 
\begin{equation}\label{TensionApprox2}
	\tilde \sigma =2 \int_{ \gamma'} d\phi \sqrt{2 \left[ V(\phi)- V_{\rm fv}\right]} =2  \sqrt{\lambda} v^3\int_{ \gamma'} d\tilde \phi \sqrt{2 \left[ \tilde V(\tilde\phi)- V_{\rm fv}\right]} = \sqrt \lambda v^3 \tilde s~.
\end{equation}
Outside the thin-wall limit, we cannot talk about the radius of the bubble, since there is no sharply defined boundary between the true and false vacuum. However we can introduce a notion of radius to be corresponding to $s_{top}$~. This  quantity which should be proportional to $R= s_{\rm top} = \tilde{s}_{\rm top} (\sqrt \lambda v)^{-1} \sim k (\sqrt \lambda v)^{-1}  $ where $k$ is a number of order unity. Now we can generalize  the tunneling exponent for the thin-wall solution calculated  in Eq.~\eqref{Land-ThinWallB} to the case of a thick-wall solution, 
\begin{eqnarray} \label{Land-ThickWallB2}
	&&R \sim {1 \over \sqrt \lambda} {1 \over \sqrt v} k ~, \cr \cr
	&&B \sim \pi^2 R^3 \tilde \sigma \sim {\pi^2 \over \lambda} k^3 \tilde{s}~.
\end{eqnarray}
We can check the reliability of this  approximation  using a one field example  which has thick-wall solutions. We start from a potential of the form \footnote{This example and the related graphs are taken from chapter 12 of \cite{WeinbergBook}.}
\begin{equation} \label{EricksPotentials}
	V(\phi) = \frac{\lambda}{4} \left[ (\phi^2-v^2)^2+ {4 b \over 3}(v \phi^3 - 2 v^3 \phi - 2 v^4)\right]~, \qquad 0\le b < 1~.~
\end{equation}
This potential has two local minima at $\phi=\pm v$ and a local maximum at $\phi = -b v$, while  $b$ is a measure of the breaking of ${\mathbb Z}_2$ symmetry between the two vacua. If $b\approx0$, the two vacua are almost degenerate and the thin-wall approximation works very well. The potential for different values of $b$ and the corresponding exact bounce solutions are shown in Fig.\ref{Land-FieldPotProfile}.  For large values of $b$, the wall gets thick and we can see how well the approximation in Eq.~\eqref{Land-ThickWallB2} works. To do so, first we calculate the exact tunneling exponent by solving the bounce equation and then calculate $\tilde{s}$ from Eq.~\eqref{TensionApprox2} by evaluating the integral in the field space from $\phi=-v$ to $\phi=-b v$. We then calculate the corresponding value for $k$. The values of $k$ are shown in Table.\ref{Land-Kvalues}. There is not  much change in  $k$ when we scan through the values of $b$. A typical value of $k$ is some number between 5 and 6. Therefore the approximation introduced in Eq.~\eqref{Land-ThickWallB2} is valid and we will use it in all of our numerical calculations.

\begin{table}[h!]
\caption{Different values of $k$ calculated over a wide range of potentials which have a thick-wall solution. Despite all the changes in the parameters of the potential, the values of $k$ do not change much when we scan potentials that give nearly thin-walls solutions to the ones that give Hawking-Moss instantons.}
\begin{center}
\begin{tabular}{|c|c|c|c|c|c|c|c|c|}
\hline
$b$ & 0.4 & 0.5  & 0.6 & 0.7 & 0.8\\
\hline
$\sigma $ &0.2746 &  0.1758 &  0.1017 &  0.0500 & 0.0183\\
\hline
$ B$ & 30.199 & 13.973 &  7.017 &  3.576 & 1.735\\
\hline
$k$ & 6.04 &  5.42 & 5.17 & 5.23&  5.75 \\ \hline
\end{tabular}
\end{center}
\label{Land-Kvalues}
\end{table}%
Now we can use a prescription to estimate the tunneling action $B$ in a simple form. We choose a number between 5 and 6 for $k$ and use the  $\tilde s$ introduced in Eq.~\eqref{TensionApprox2} as a measure of the tension for different paths. Minimizing the action is equivalent to minimizing the surface tension among all different paths $\gamma'$ in the field space. However, this is still too complicated if the dimensionality of field space, and hence the number of  possible choices for path $\gamma'$, are large. But we expect that the paths that give the minimum tension should pass through low barriers which are the saddle points of the potential. As an upper bound for the tension, we can look at all  the paths that go along  straight lines from the false vacuum to a  saddle point of the potential and evaluate the value of $\tilde {s}$ along these lines. We keep the lowest of these tensions $s= \tilde {s}_{\rm min}$. As a measure of tunneling exponent, we can use  Eq.~\eqref{Land-ThickWallB2} (setting $k\approx 5$)
\begin{equation} \label{Land-ThickWallB3}
	B \sim 10^3{ \tilde s \over \lambda}~.
\end{equation}
Of course this gives an upper bound for $B$. The actual value should be lower than this. If the lifetime of the metastable vacua calculated in this approximation is not long, it should be even shorter when we do not use the approximation. If the tunneling exponents get too small, let's say of order  unity or less, then too many bubbles pop out and this  make the dilute gas approximation,  which is a pillar  of the semiclassical approximation, invalid. In this case, the metastability of the false vacua is essentially gone.

\section{Numerical results}
\label{Land-sec-Numeric}
We numerically studied large ensembles of theories with potentials  given by Eq.~\eqref{Land-potential} with random coefficients. We use $s$ as a measure of stability of the vacua. Since the numerical calculations are very time consuming, we only keep terms up to quartic order in fields. We later use different orders in the expansion and show that our results do not change much. An ensemble is defined by choosing random numbers for the $A^{(n)}$ with uniform distributions over ranges defined by
\begin{eqnarray}\label{Land-Ranges}
	A_{ii}^{(2)} &\in& [0,a_2]~, \nonumber \\
	A_{ijk}^{(3)} &\in& [-a_3,a_3]~, \nonumber \\
	A_{iikl}^{(4)} &\in& [-a_4,a_4]~, \nonumber \\
\end{eqnarray}
where the $A^{(2)}$ are chosen to be positive to ensure that the origin of the field space is a minimum. The coefficients of the cubic and quartic terms  are not necessarily positive and therefore the potentials are  not in general bounded from below. However, because the important region for tunneling is  a small neighborhood of the origin, the lack of a lower bound at large field values does not concern us  and the boundedness of the $V$ at large distance is achieved by higher order terms. 

Because we are interested in the dependence of the tunneling exponent on the number of fields, it is important to have a prescription for how to change the ranges for $a_2, a_3$ and $a_4$ in Eq.~\eqref{Land-Ranges} when we change $N$. Our guiding principle is to keep  the variation of the potential inside an $N$-dimensional ball of radius $\phi_R$ independent of the number of fields. This has the  advantage that if we constraint ourselves to an $N_0\le N$-dimensional hypersurface of the field space, we can recover the same variation inside an $N_0$-dimensional ball of the same radius $\phi_R$. This ensures that our normalization does not have a peculiar dependence on the number of fields. 

With this assumption, it is an easy job to find the dependence of the $a_n$ on the number of fields. The typical variation of each field $\phi_i$ inside a ball of radius $\phi_R$ is $\phi_R/\sqrt N$. There are $N$ positive quadratic terms and their sum  is roughly $\phi_R^2$. Therefore the variation of the quadratic terms in the  potential is independent of $N$ if $a_2$ is. Similarly, there are of order $N^3$ cubic terms. Because they can be positive or negative, their sum is effectively $N^{3/2}$ times a typical value of $\phi_i^3$ which is $\phi_R^3 / N^{3/2}$. Therefore  we need to keep $a_3$ independent of $N$. Similarly there are $N^4$ quartic terms. Because they may be positive or negative, effectively we have $N^2$ terms. Each of them is roughly $\phi_R^4/ N^2$ and therefore $a_4$ should be independent of $N$. 

It is important to notice that the range of all the coefficients are not independent and they can have trivial effects. For example, if we multiply the potential by an overall factor, the ranges of the $a_i$ get rescaled by  the same overall factor. Also, by redefining the fields by a multiplicative factor, we can set the ratio $a_3/a_2$ to be any arbitrary number. Therefore if we keep only up to  quartic order, the only independent range will be the  ratio of $a_4/a_3$ (or the ratio of any chosen pair). Therefore, without losing generality, we can assume $a_2=a_3=1$. For our numerical calculations, we also set $a_4=1$. We get back to this shortly. 

\begin{figure}[htbp] 
   \centering
   \includegraphics[width=2.8in]{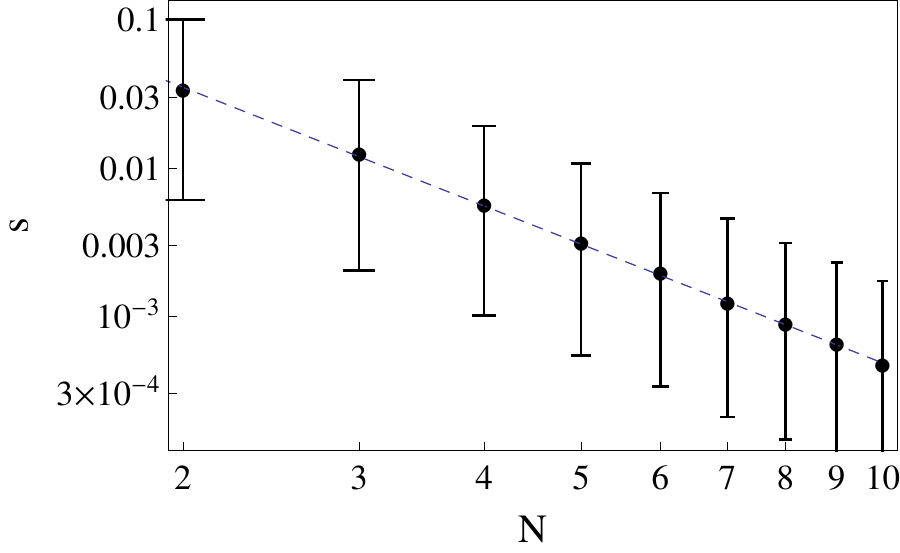} 
    \includegraphics[width=2.8in]{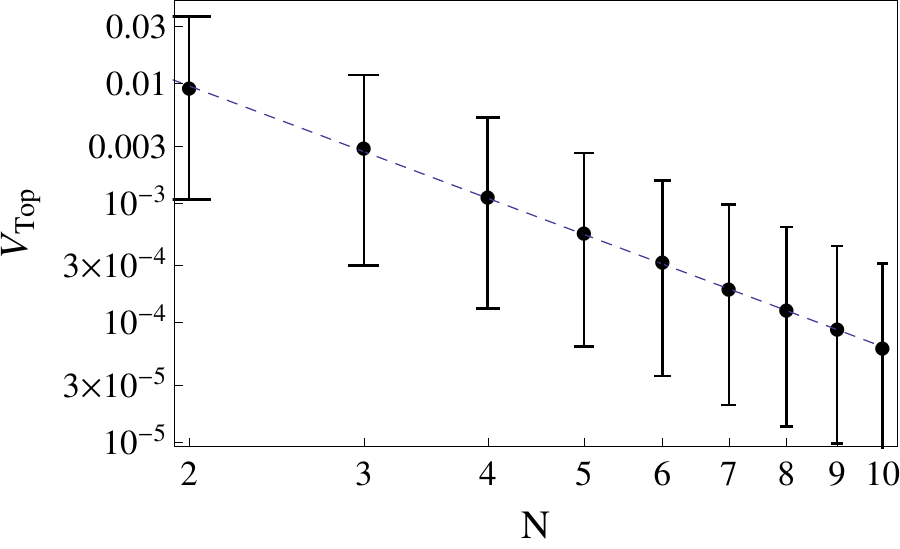} 
   \caption{Left panel, the median value of  $\tilde s$ for quartic potentials on a log-log plot. Right panel, the median value of the lowest barrier heights for quartic potential in units of $\lambda v^4$. The bars in both cases show the 25th and 75th  percentile.}
   \label{Land-QuarticActions}
\end{figure}

\begin{figure}[htbp] 
   \centering
   \includegraphics[width=2.8in]{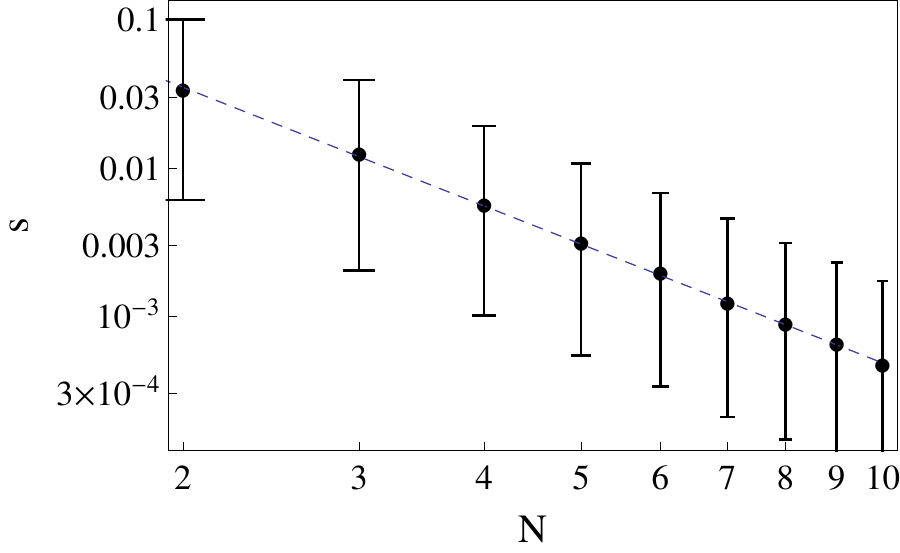} 
    \includegraphics[width=2.8in]{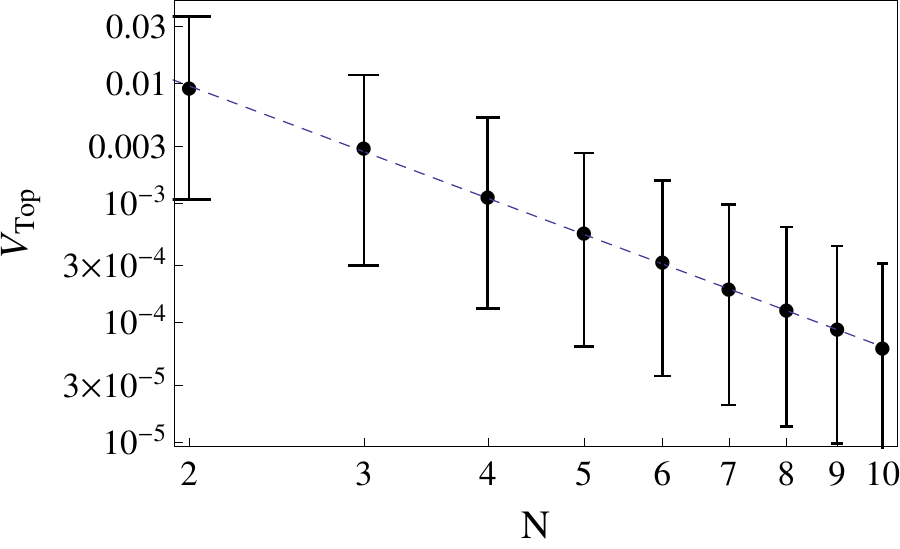} 
   \caption{Left panel, the median value of  the distances to the saddle point which has the smallest $s$  for quartic potentials on a log-log plot. Right panel, the median value of  distances to the saddle points with the  lowest barrier height for quartic potentials in units of $v^{-1}$. The bars in both cases show the 25th and 75th percentile.}
   \label{Land-QuarticDistances}
\end{figure}

For a given $N$, we generated an ensemble of 10000 random potentials and found all the saddle points.For each potential, we calculated the lowest $s$ (as explained in the Sec.\ref{Land-sec-Background}), the lowest barrier height, the distance to the lowest saddle point and the distance to the saddle point which has the smallest $s$. In Fig.\ref{Land-QuarticActions}, the median values of the lowest  $\tilde s$ and the lowest barrier heights are shown for quartic potentials. As the graphs indicate, the medians drop very quickly with the number of fields and they follow a power law relation. Not only do the barrier heights and the domain-wall tensions drop very quickly, these saddle points get closer to the origin (false vacuum) and this is a good indication that our truncation at the quartic level was the right idea. In Fig.\ref{Land-QuarticDistances}, the distance to the saddle points with the lowest $\tilde s$ and the saddle points with lowest barrier height for a quartic potential are plotted on a log-log plot. It shows very good agreement with a power law relation. The two graphs are almost identical.

To understand this behavior better, in Fig.\ref{Land-QuarticSaddleNum}, the median value for the number of saddle points around the origin are plotted on a log graph. Apparently this fits very well to an exponential. Although this is a numerical work, there are rigorous proofs that the number of stationary points of a polynomial of order $D$ in $N$ variables is an exponential in $N$ \cite{Blum:1997:CRC:265020,Dedieu200889}. Now let's  look at the lowest barrier heights. We kept the variations of the potential independent of $N$ (this was the guiding principle for choosing  the $a^{(n)}$s). If there are $K$ saddle points and they are randomly distributed (in fact being well scattered is enough and we do not really need randomness here), we expect the lowest of them to have a height proportional to $K^{-1}$. Therefore if the number of saddle points $K$, increases rapidly with the number of fields (in this case an exponential growth), we expect the lowest of them to fall very rapidly with $N$. A similar argument may be used to justify that the wall tensions should drop rapidly with the number of dimensions. 
\begin{figure}[htbp] 
   \centering
   \includegraphics[width=3.5in]{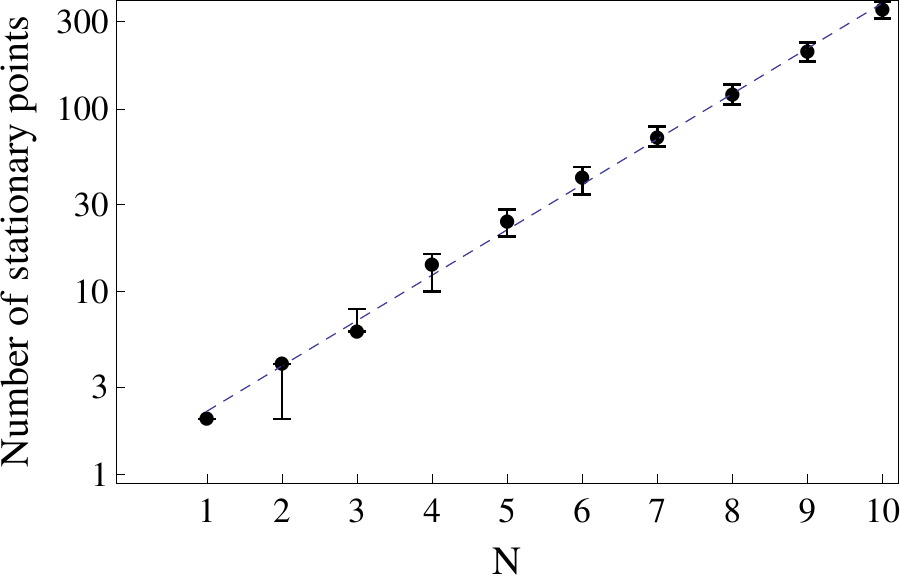} 
   \caption{The median value of the number of saddle points around the origin on a log plot for quartic potentials. As  is apparent from here, this number fits very well to an exponential. The bars show the 25th and 75th percentile.}
   \label{Land-QuarticSaddleNum}
\end{figure}

\begin{figure}[htbp] 
   \centering
      \includegraphics[width=2.8in]{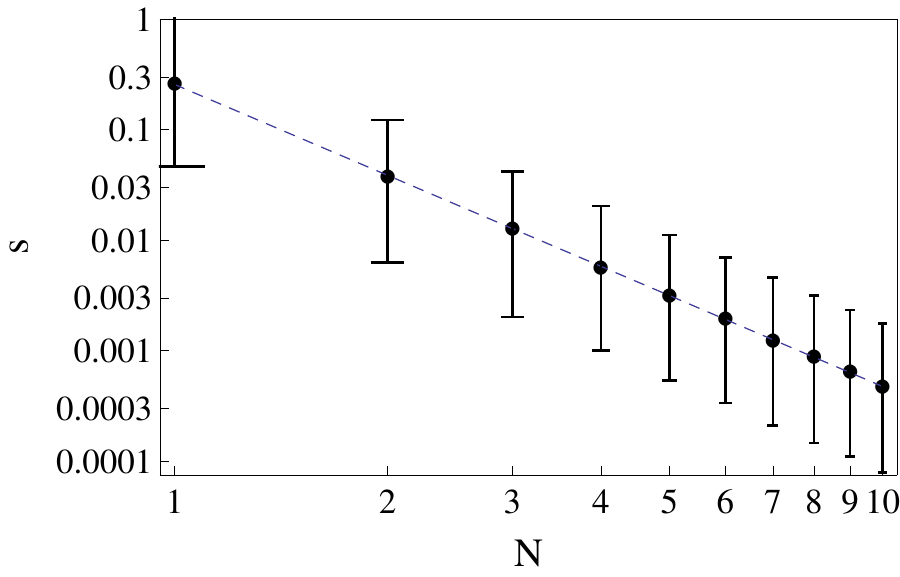} 
    \includegraphics[width=2.8in]{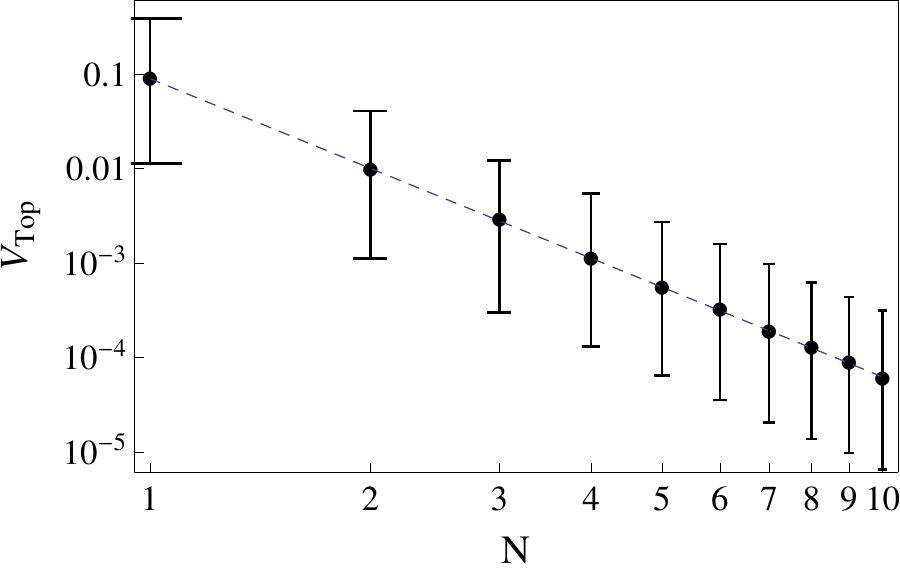} 
   \caption{Left panel, the median value of  $s$ for cubic potentials on a log-log plot. Right panel, the median value of the lowest barrier heights for cubic potentials in units of $\lambda v^4$. The bars in both cases show the 25th and 75th percentile.}
   \label{Land-CubicActions}
\end{figure}

\begin{figure}[htbp] 
   \centering
   \includegraphics[width=2.8in]{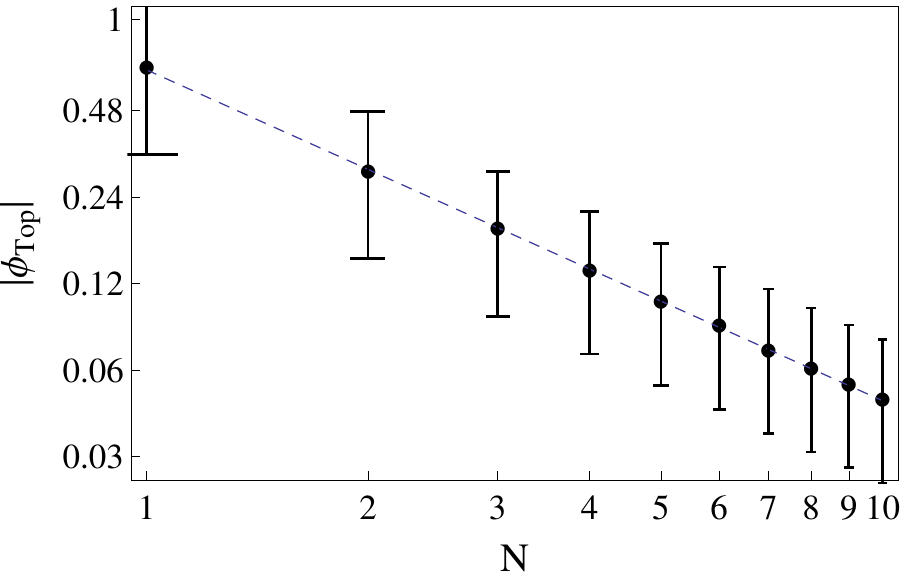} 
    \includegraphics[width=2.8in]{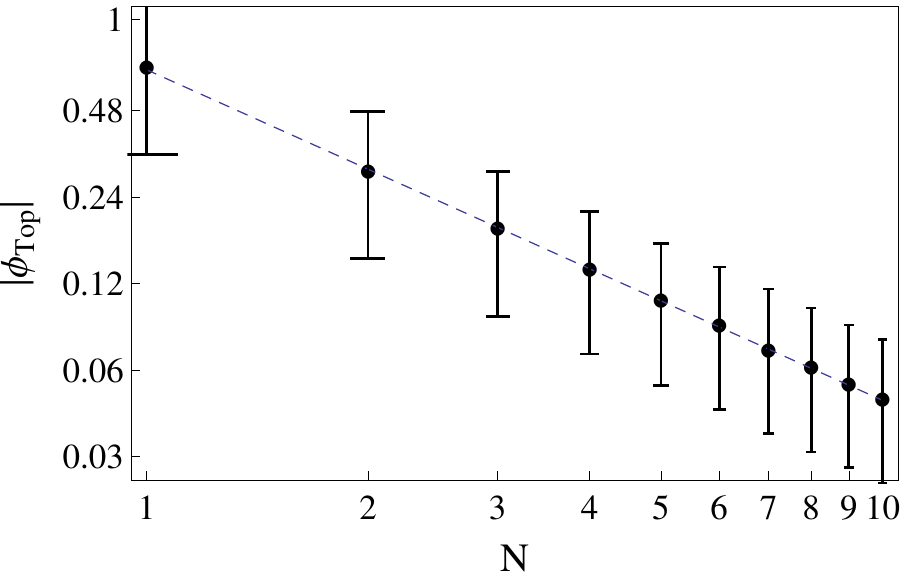} 
   \caption{Left panel, the median value of  the distances to the saddle points which have the smallest $s$  for cubic potentials on a log-log plot. Right panel, the median value of the distances to the saddle points with  lowest barrier heights for cubic potentials in units of $v^{-1}$. The bars in both cases show the 25th and 75th percentile.}
   \label{Land-CubicDistances}
\end{figure}

From these figures, we see that the median value of all the quantities we are interested in drops like a power law for  randomly chosen quartic potentials. Let's define this power to be $\alpha$. For example
\begin{eqnarray}\label{Land-Alphadef}
 	s_{\rm median}= C_{\rm tension} N^{-\alpha_{\rm tension}} \nonumber ~.
\end{eqnarray}
The values of $\alpha$ and $C$ for $s$, lowest barrier heights, and distance to the saddle point with lowest $s$ are shown in Table.\ref{Land-alpha} for the case of randomly chosen quartic, cubic and supersymmetric potentials (they are explained in the next few paragraphs). 
We may question the validity of cutting the potential at quartic order. To have a check, we repeated the same calculation for  randomly chosen cubic potentials. For each dimension, we chose ensemble of 10000 randomly chosen cubic potentials, and using the prescription of Sec. \ref{Land-sec-Background}, we calculated $\tilde s$, the lowest barrier heights and their distances to the false vacuum. The medians of the different quantities are shown in Figs.\ref{Land-CubicActions}, \ref{Land-CubicDistances} and \ref{Land-QuarticSaddleNum}. The same power law decrease is observed in the median values of the different quantities of interest. 
\begin{figure}[htbp] 
   \centering
   \includegraphics[width=3.5in]{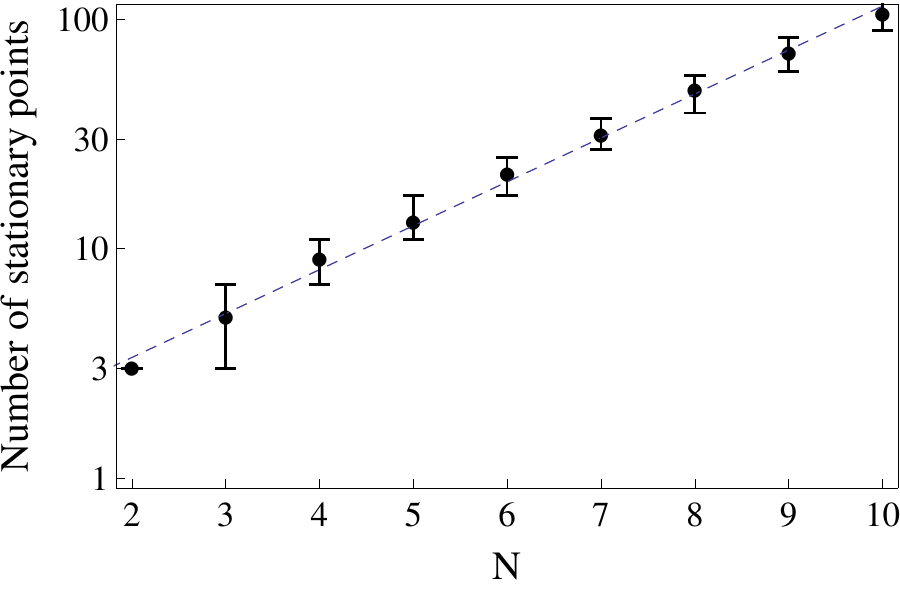} 
   \caption{The median value of the number of saddle points around the origin on a log plot for cubic potentials. As is apparent from here, this number fits very well to an exponential. The bars show the 25th and 75th percentile.}
   \label{Land-QuarticSaddleNum}
\end{figure}

\begin{figure}[htbp] 
   \centering
   \includegraphics[width=2.8in]{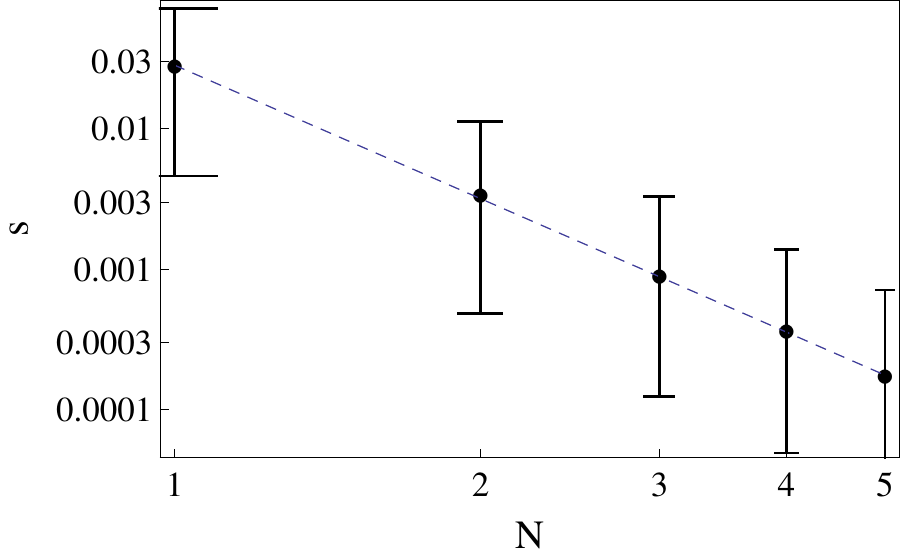} 
    \includegraphics[width=2.8in]{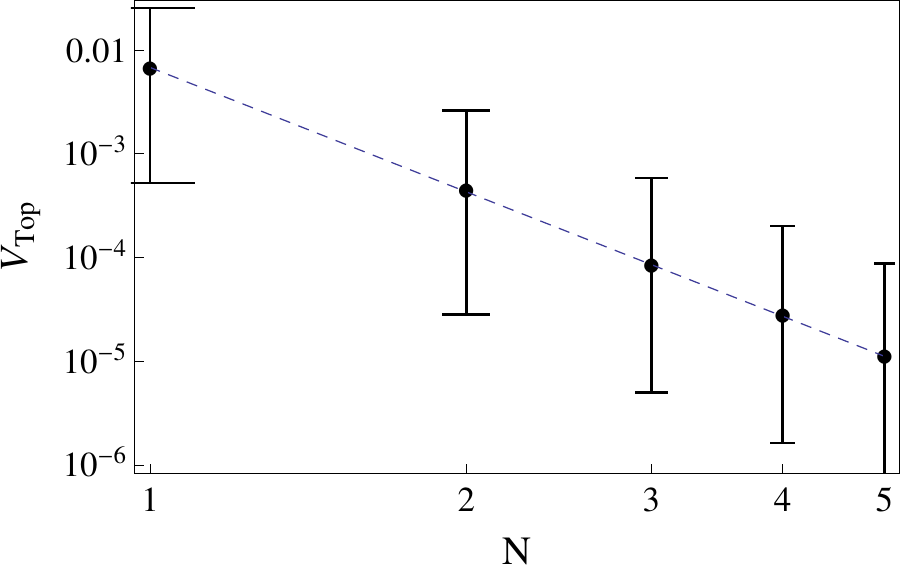} 
   \caption{Left panel, the median value of  $s$ for supersymmetric potentials on a log-log plot. Right panel, median value of the lowest barrier height for supersymmetric potentials in units of $\lambda v^4$. The bars in both cases show the 25th and 75th percentile.}
   \label{Land-SUSYActions}
\end{figure}
\begin{figure}[htbp] 
   \centering
   \includegraphics[width=2.8in]{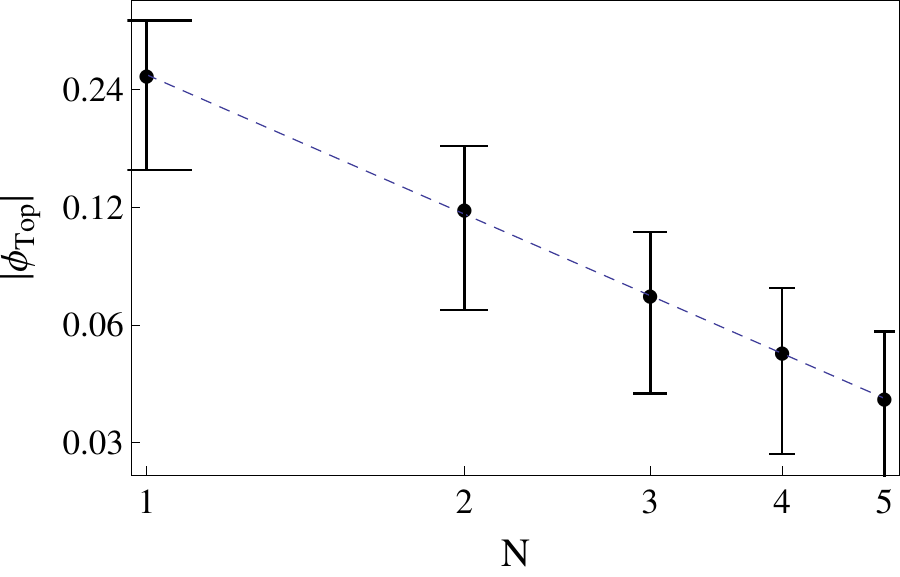} 
    \includegraphics[width=2.8in]{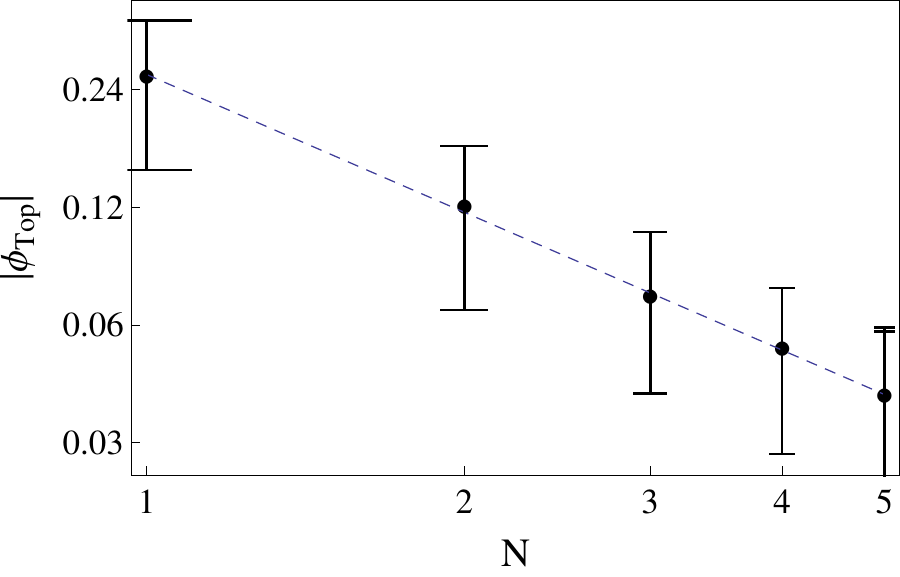} 
   \caption{Left panel, the median value of  the distances to the saddle points which have the smallest $s$  for supersymmetric potentials on a log-log plot. Right panel, the median value of  distances to the saddle points with  lowest barrier height for supersymmetric potentials in units of $v$. The bars in both cases show the 25th and 75th percentile.}
   \label{Land-SUSYDistances}
\end{figure}
\begin{figure}[htbp] 
   \centering
   \includegraphics[width=3.5in]{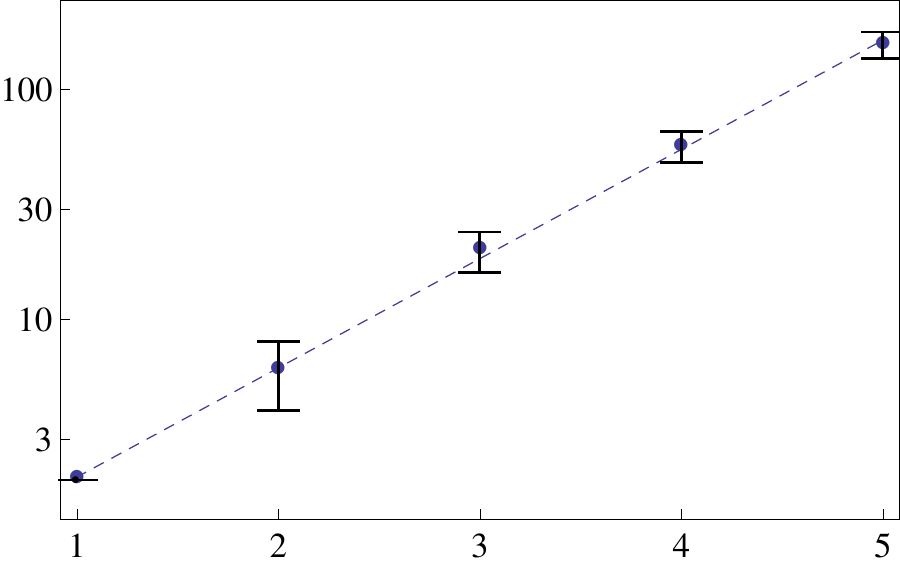} 
   \caption{The median value of the number of saddle points around the origin on a log plot for supersymmetric potentials. As  is apparent from here, this number fits very well to an exponential. The bars show the 25th and 75th percentile.}
   \label{Land-SUSYSaddleNum}
\end{figure}

Another issue to address is the correlation between the coefficients. We chose all the coefficients in Eq.~\eqref{Land-potential} as uncorrelated random numbers. We may worry that the results we found about the rapid drop of different quantities versus the number of fields could be merely an artifact  of choosing so many random uncorrelated numbers. To show that this is not the case, we started from the superpotential of a set of $N$ scalar superfields. This potential is described in details in Appendix. \ref{SUSYAppendix}. We chose the coefficients $m_{ij}, g_{ijk}$ and $h_{ijmk}$ to be randomly chosen numbers. Using an $SU(N)$ rotation in the field space, we can diagonalize the positive definite matrix $m_{ij}$.  We chose these coefficients to be uniformly distributed over the following intervals
\begin{eqnarray}
	m_{ij} &\in& [0, a_2] \nonumber ~, \\
	\text{ Re } g_{ijk} &\in& [0, a_3] \nonumber ~, \\
	\text{Im }g_{ijk} &\in& [0, a_3] \nonumber ~, \\
	\text{ Re }h_{ijkl} &\in& [0, a_4] \nonumber ~, \\
	\text{ Im } h_{ijkl} &\in& [0, a_4] ~. \\
\end{eqnarray}
 Although we chose uncorrelated numbers for the $m_{ij}, g_{ijk}$ and $h_{ijkl}$,  we can see from Eq.~\eqref{SUSYPotentialExpansion}, that the coefficients in the expansion are highly correlated. Please notice that the supersymmetry is broken after we truncated Eq.~\eqref{Potential2} at the quartic order and this example only addresses the effect of correlation between coefficients on the median of different quantities. The median values of the quantities of interest are shown in Figs. \ref{Land-SUSYActions}, \ref{Land-SUSYDistances}, \ref{Land-SUSYSaddleNum}. Because the $N$ chiral superfields correspond to $2N$ real scalar fields, instead of writing the fits to
the data as in Eq.~\eqref{Land-Alphadef} we write, e.g.,
\begin{equation}
     s_{\rm median} = C_{\rm tension} (2N)^{-\alpha} \, . 
\end{equation}
Again we get a power law decrease for all the median quantities, but the powers are different. The   drop with the number of fields is even more severe in the case of supersymmetric potentials which have correlated coefficients.

\begin{table}[htdp]
\begin{center}
\begin{tabular}{|l|c|c|c|c|} \hline
                                &       ~$\alpha_{\rm tension}$~ &
  ~$\alpha_{\rm height}$~& ~$\alpha_{\rm distance}$~
    \\ \hline
        Cubic potentials     &        2.73   &   3.16       &  1.15  \\ \hline
        Quartic potentials   &        2.66   &   3.12  &  1.10    \\ \hline
        SUSY                 &        3.16   &   3.99  &  1.19    \\ \hline
\end{tabular}
\end{center}

\begin{center}
\begin{tabular}{|l|c|c|c|c|} \hline
                &       ~$C_{\rm tension}$~ & ~$C_{\rm height}$~ &
   ~ $C_{\rm distance}$~  \\ \hline
        Cubic potentials     & 0.26  &  0.090    & 0.67      \\ \hline
        Quartic potentials   & 0.22   & 0.083    & 0.60    \\ \hline
        SUSY                 & 0.25   & 0.11   &  0.60 \\ \hline
\end{tabular}
\end{center}
\caption{Best fit parameters, defined as in Eq.~(\ref{Land-Alphadef}), for
power law fits to the data in Figs.~\ref{Land-QuarticActions}, \ref{Land-QuarticDistances},
\ref{Land-CubicActions},  \ref{Land-CubicDistances}, \ref{Land-SUSYActions} and \ref{Land-SUSYDistances},.}
\label{Land-alpha}
\end{table}

\begin{figure}[htbp] 
   \centering
   \includegraphics[width=3in]{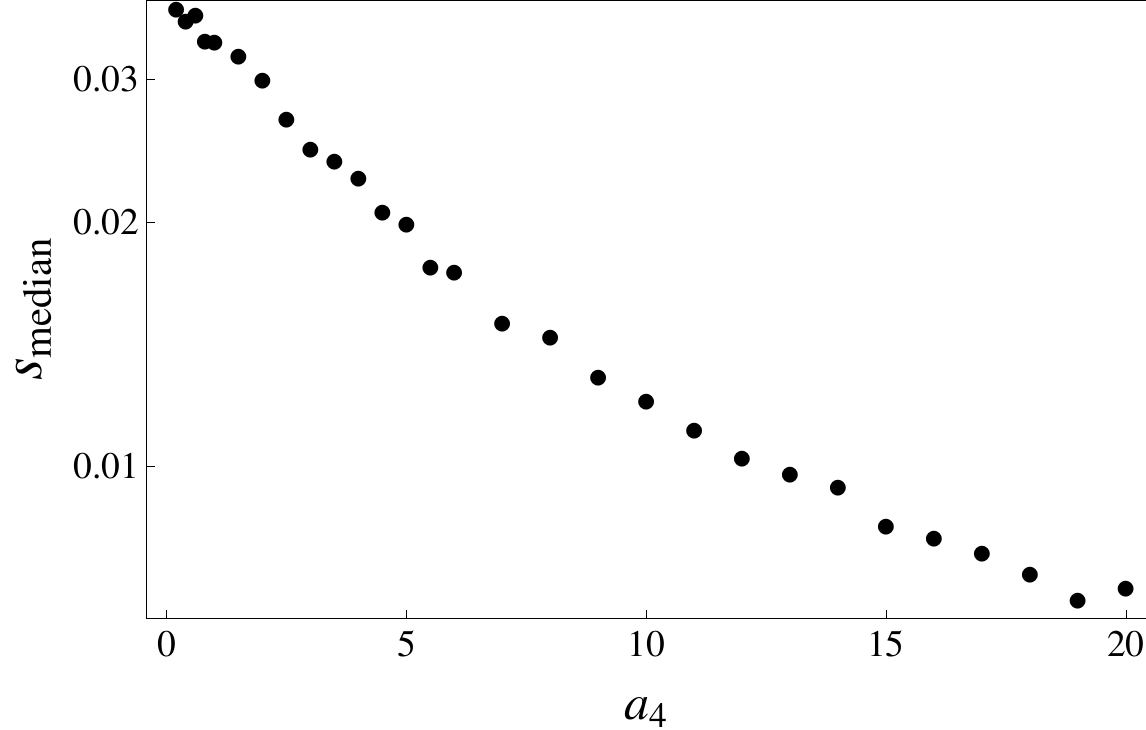} 
   \caption{The dependence of the median value of $s$ on the range of parameter $a_4$. The data shown are for $N=2$.}
   \label{MedianVsa4}
\end{figure}

Now we address one more issue regarding the numerical calculations. We chose the  range of the coefficients $a_i$ to be all  equal to one. As mentioned earlier, the values  $a_2$ and $a_3$ can be arbitrarily set by an overall factor for the potential and rescaling the fields. We studied the effects of changing $a_4$ on the median value of $s$ in Fig.\ref{MedianVsa4}. Making the value of $a_4$ larger makes the tunneling exponents and barrier heights smaller and the limit of $a_4\rightarrow0$ simply reduces to the results we had for cubic potentials as expected. Therefore we cannot compensate for the drop of median values by changing the value of $a_4$ (or $a_2$ and $a_3$).

\begin{figure}[htbp] 
   \centering
   \includegraphics[width=4.5in]{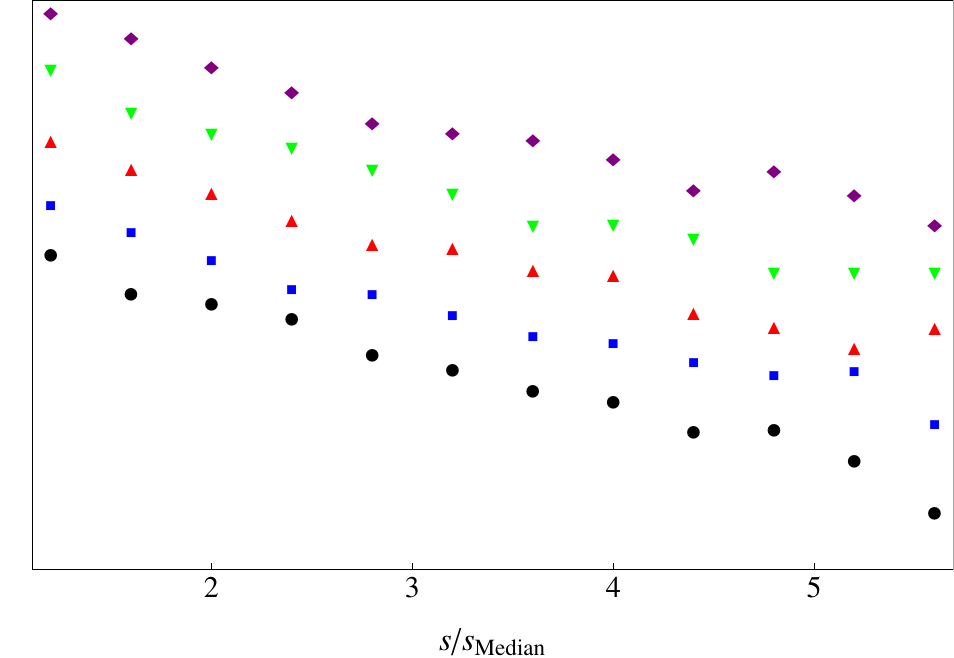} 
   \includegraphics[width=4.5in]{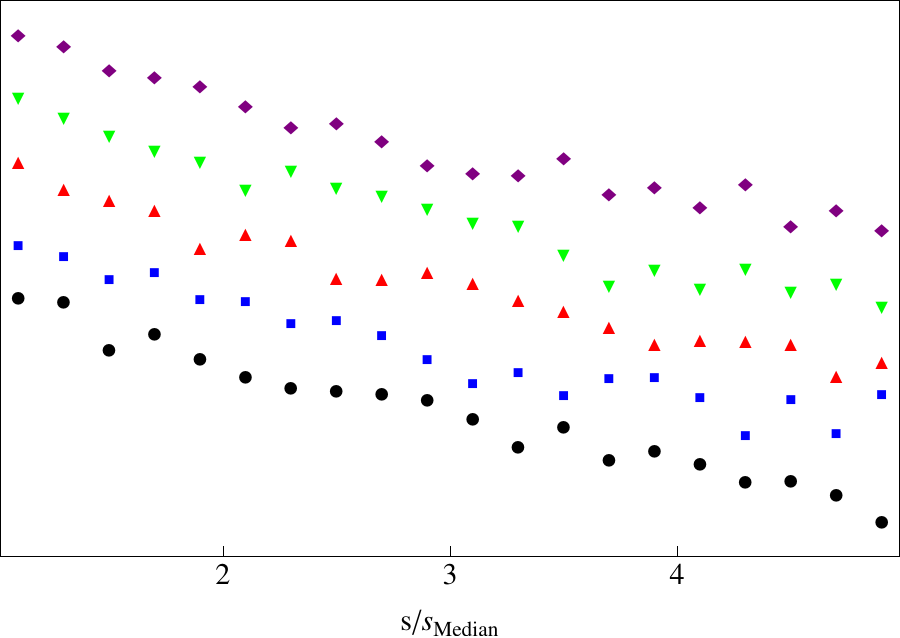}
   \caption{The distribution of values of $\tilde s$ in the ensembles for  quartic (top) and cubic (bottom) random potentials for various values of $N$. The vertical axis represents the natural logarithm of the number of values in each bin. For the sake of clarity, the data for different values of $N$ have been offset by constants, so only the slopes are meaningful. Purple diamonds correspond to $N=2$, green down-pointing triangles, $N=4$ ; red up-pointing triangles, $N=6$ ; blue squares, $N=8$ and black circles, $N=10$.}
   \label{Land-s-Distributions}
\end{figure}

\begin{figure}[htbp] 
   \centering
   \includegraphics[width=4.5in]{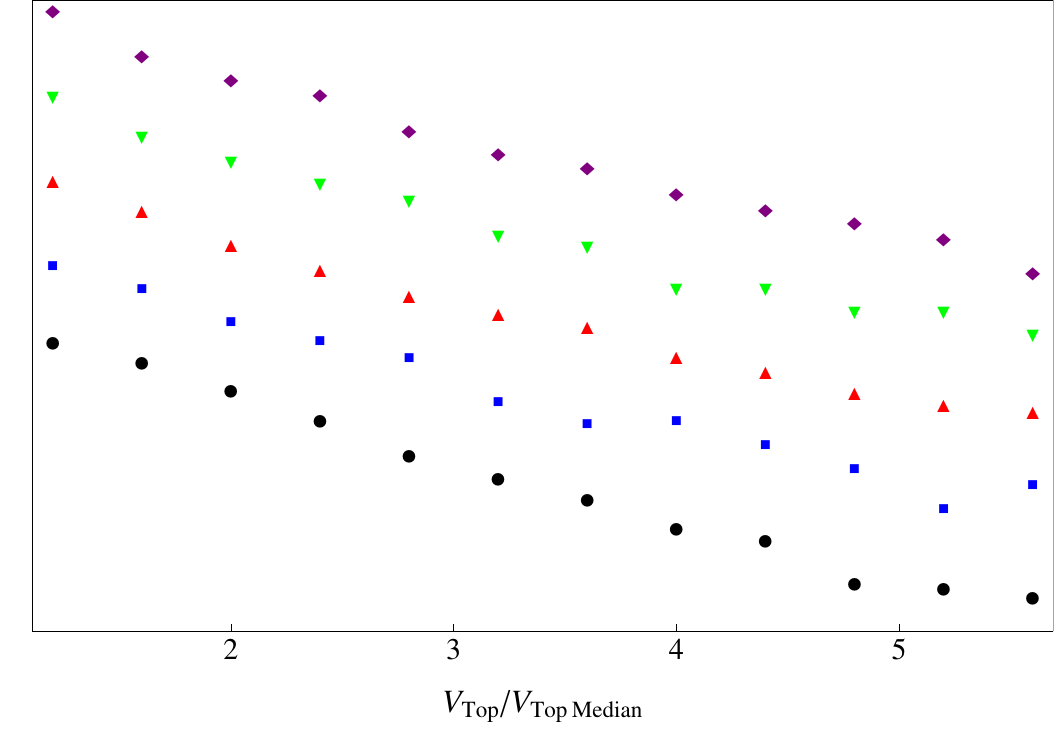} 
   \includegraphics[width=4.5in]{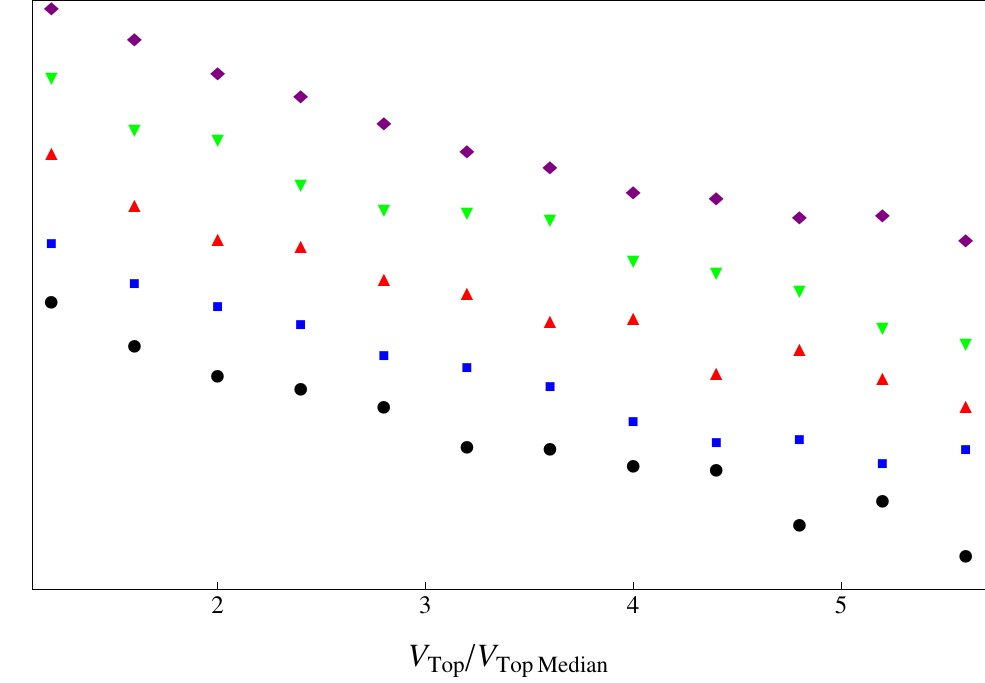}
   \caption{The distribution of values of lowest barrier heights in the ensembles for  quartic (top) and cubic (bottom) random potentials for various values of $N$. The vertical axis represents the natural logarithm of the number of values in each bin. For the sake of clarity, the data for different values of $N$ have been offset by constants, so only the slopes are meaningful. Purple diamonds correspond to $N=2$, green down-pointing triangles, $N=4$ ; red up-pointing triangles, $N=6$ ; blue squares, $N=8$ and black circles, $N=10$.}
   \label{Land-V-Distributions}
\end{figure}

Until now, we have seen that the median values of tensions, barrier heights and tunneling exponents drop with a power law of the  number of dimensions. Therefore if we choose a generic vacuum in these theories, it will not have a long lifetime. This by itself does not tell us about the lifetime of the  ``non-generic" points and the number of stable points. To find the population of these outliers, we studied the distribution of  tunneling exponents, barrier heights and the distances to the false vacuum around their median values  for different dimensions. In Figs. \ref{Land-s-Distributions}, \ref{Land-V-Distributions} and \ref{Land-D-Distributions} we show the distribution of values of $s$, $V_{\rm top}$ and saddle point distances for several choices of $N$ with  quartic, cubic and supersymmetric potentials. They roughly coincide when plotted as a functions of $s/s_{\rm median}$. Our data suggest that the frequency of finding a vacuum with large $s$ has an approximately exponential falloff which we can describe by
\begin{equation} \label{Land-eq-Distributionfit}
	n(s)\approx n_0 \exp{\left(- \gamma s /s_{\rm median}\right)}~.
\end{equation}
Similar results are found for the barrier heights and distances of the lowest saddle points to the false vacuum. The values of $\gamma$ for various $N$ are shown in Tables.\ref{GammasFit}, \ref{GammaVFit} and \ref{GammaDistFit}. Now using the equation for medians from Eq.~\eqref{Land-Alphadef} in  Eq.~\eqref{Land-eq-Distributionfit}
\begin{equation}
	n(s) \approx n_0 \exp{\left( - {\gamma \over C_{\rm tension}} N^{\alpha_{\rm tension}} s\right)}~.
\end{equation}
We can use this result to find an approximation for the distribution of tunneling exponents using Eq.~\eqref{Land-ThickWallB3}. This suggests that the fraction of vacua with a tunneling exponent greater than some value $B$ is roughly
\begin{equation}\label{Land-SurvivingFraction}
	f( B) \sim \exp{\left( - \beta N^{\alpha_{\rm tension}}  B\right)}~,
\end{equation}
where 
\begin{equation}
	\beta = 10^{-3} \frac{\gamma \lambda}{C_{\rm tension}}~.
\end{equation}
From the numerical analysis of this section, we see that $\gamma/ C_{\rm tension}$ is of order unity. We extracted an overall factor $\lambda$ from the shape of the landscape potential. There is no reason for it to be a small coupling constant. It could be well of order unity, which makes $\beta \sim 10^{-3}$. Since $\alpha_{\rm tension}\sim 3$, Eq.~\eqref{Land-SurvivingFraction} shows  a huge suppression in the number of long-lived vacua in high dimensional landscapes. 

We also show that the barrier heights that protect the false vacuum got very low, with a very similar relation 
\begin{equation}
	 f( V_{\rm barrier}) \sim \exp{\left( - \frac{\gamma}{C_{\rm barrier} } N^{\alpha_{\rm barrier}} V_{\rm barrier}\right)}~.
\end{equation}
Therefore the fraction of barrier heights which are high enough to support stable vacua drop with an exponential of some power of $N$. This huge suppression means that only an extremely small portion of barrier heights are high. Until now we have neglected gravity. But in the presence of gravity and a de Sitter space with a relatively flat barrier, the Hawking-Moss bounce provides an alternative mode of decay. The corresponding decay exponent is
\begin{equation} \label{Land-HMRate}
	 B_{\rm HM} = {3 \over 8 G_{\rm N}^2} { V_{\rm barrier} - V_{\rm fv} \over V_{\rm barrier} V_{\rm fv}}~,
\end{equation}
where $V_{\rm barrier}$ is the potential on top of the barrier and $G_{\rm N}$ is the Newton's constant. Our numerical solutions suggest that the barrier heights relative to the false vacuum, (i.e., $V_{\rm barrier} - V_{\rm fv}$) drop like an exponential of a power and therefore this mode of  decay also gets  enhanced for large number of fields and  adds to the instability of the false vacuum. 
\begin{table}[htdp]
\begin{center}
\begin{tabular}{|c|c|c|c|} \hline
    ~$N$~    &      ~ Cubic~ &  ~Quartic~ & ~SUSY~  \\ \hline
        1       &       0.58   &  0.46  & 0.45 \\ \hline
        2       &       0.38   &  0.39  & 0.33\\ \hline
        3       &       0.40   &  0.35  & 0.33 \\ \hline
        4       &       0.34   &  0.33  & 0.32  \\ \hline
        5       &       0.37   &  0.35  &  0.32 \\ \hline
        6       &       0.37   &  0.34  &  \\ \hline
        7       &       0.38   &  0.34  &  \\ \hline
        8       &       0.38   &  0.35  &   \\ \hline
        9       &       0.38   &  0.37  &   \\ \hline
        10       &      0.35   &  0.34  &   \\ \hline
\end{tabular}
\end{center}
\caption{Best fit exponents for the $s$ distributions of results for cubic
  and quartic non-supersymmetric and quartic supersymmetric potentials
  with various $N$.}
\label{GammasFit}
\end{table}
\begin{table}[htdp]
\begin{center}
\begin{tabular}{|c|c|c|c|} \hline
    ~$N$~    &      ~ Cubic~ &  ~Quartic~ & ~SUSY~  \\ \hline
        1       &       0.49   &  0.75  & 0.78 \\ \hline
        2       &       0.32   &  0.69  & 0.64\\ \hline
        3       &       0.34   &  0.65  & 0.50 \\ \hline
        4       &       0.31   &  0.60  & 0.55  \\ \hline
        5       &       0.29   &  0.60  &  0.54 \\ \hline
        6       &       0.32   &  0.59  &  \\ \hline
        7       &       0.32   &  0.58  &  \\ \hline
        8       &       0.35   &  0.59  &   \\ \hline
        9       &       0.34   &  0.61  &   \\ \hline
        10       &      0.32   &  0.62  &   \\ \hline
\end{tabular}
\end{center}
\caption{Best fit exponents for the  distributions of the lowest barrier heights of results for cubic
  and quartic non-supersymmetric and quartic supersymmetric potentials
  with various $N$.}
\label{GammaVFit}
\end{table}
\begin{table}[htdp]
\begin{center}
\begin{tabular}{|c|c|c|c|} \hline
    ~$N$~    &      ~ Cubic~ &  ~Quartic~ & ~SUSY~  \\ \hline
        1       &       0.85   &  1.42 & 1.93 \\ \hline
        2       &       1.28   &  1.52  & 1.89\\ \hline
        3       &       1.52   &  1.98  & 3.17 \\ \hline
        4       &       1.72   &  2.04  & 2.2  \\ \hline
        5       &       1.78   &  2.34&   2.5\\ \hline
        6       &       1.79   &  1.95  &  \\ \hline
        7       &       1.53   &  1.75  &  \\ \hline
        8       &       1.54   &  1.68  &   \\ \hline
        9       &       1.52  &   1.67  &   \\ \hline
        10       &      1.49   &  1.54  &   \\ \hline
\end{tabular}
\end{center}
\caption{Best fit exponents for the  distributions of the distances of the lowest saddle points to the false vacuum for  results for cubic
  and quartic non-supersymmetric and quartic supersymmetric potentials
  with various $N$.}
\label{GammaDistFit}
\end{table}
\begin{figure}[htbp] 
   \centering
   \includegraphics[width=4.5in]{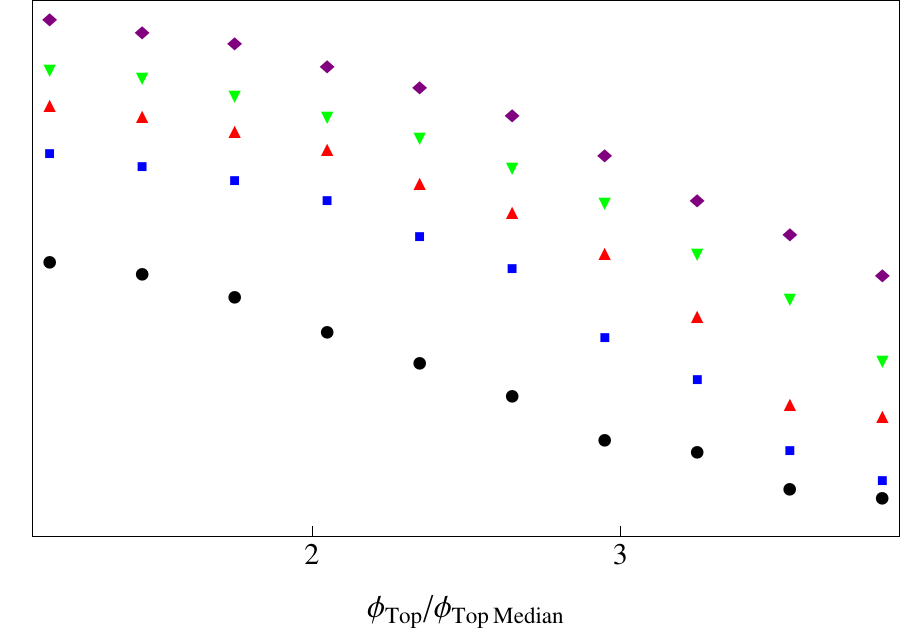} 
   \includegraphics[width=4.5in]{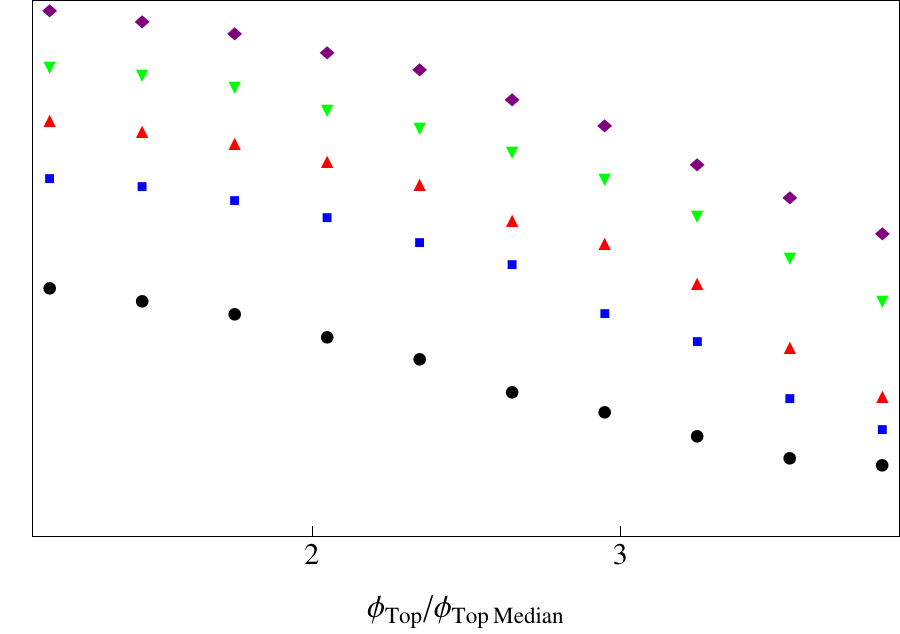}
\caption{The distribution of values of the distance of the lowest saddle point to the false vacuum in the ensembles for  quartic (top) and  cubic (bottom)   random potentials for various values of $N$. The vertical axis represents the natural logarithm of the number of values in each bin. For the sake of clarity, the data for different values of $N$ have been offset by constants, so only the slopes are meaningful. Purple diamonds correspond to $N=2$, green down-pointing triangles, $N=4$ ; red up-pointing triangles, $N=6$ ; blue squares, $N=8$ and black circles, $N=10$.}
\label{Land-D-Distributions}
\end{figure}
\begin{figure}[htbp] 
   \centering
   \includegraphics[width=4.5in]{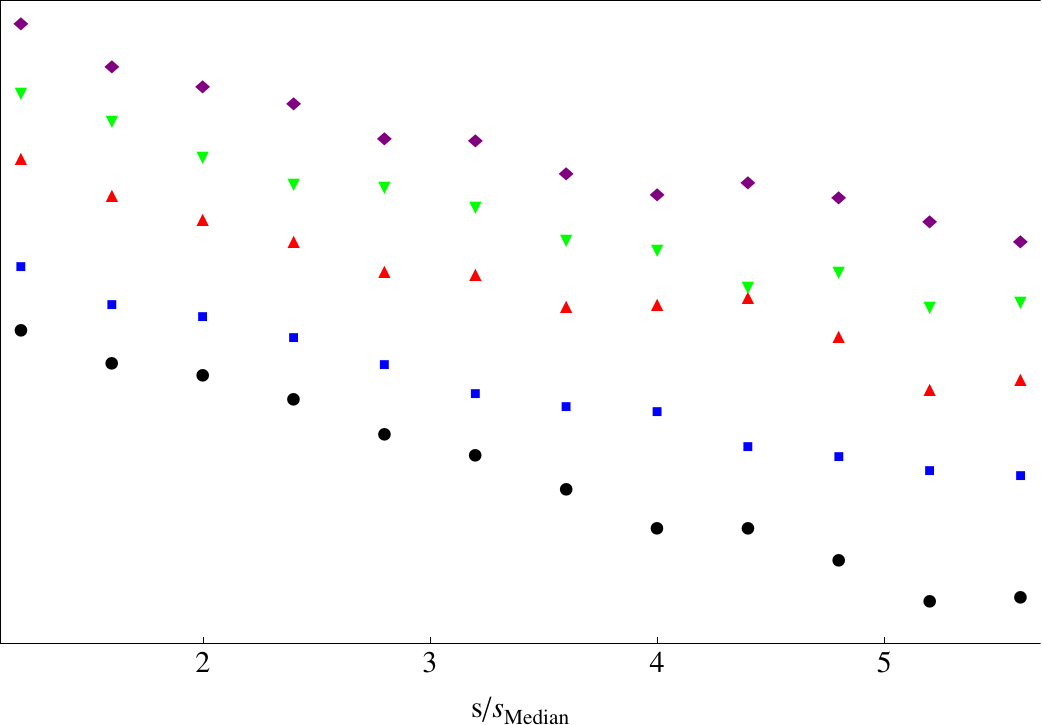} 
   \includegraphics[width=4.5in]{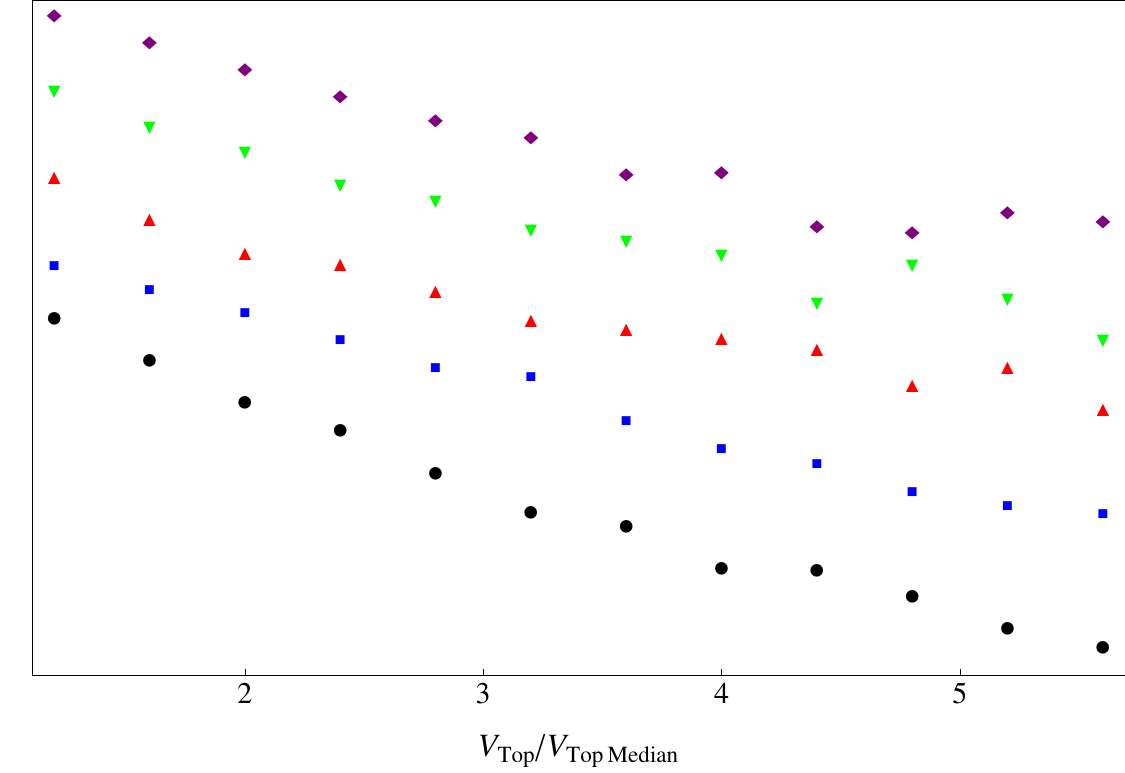}
\caption{The distribution of values of $\tilde s$ (top) and lowest saddle point heights (bottom)  in the ensembles for  supersymmetric  random potentials for various values of $N$. The vertical axis represents the natural logarithm of the number of values in each bin. For the sake of clarity, the data for different values of $N$ have been offset by constants, so only the slopes are meaningful. Purple diamond correspond to $N=1$, green down-pointing triangles, $N=2$ ; red up-pointing triangles, $N=3$ ; blue squares, $N=4$ and black circles, $N=5$.}
\label{Land-D-Distributions}
\end{figure}

\section{Possible implications for string landscape}
\label{Land-sec-String}
The results we showed so far are general features of  quantum field theories of scalar fields. However our main motivation in this work is understanding  the implications of large dimensions for the tunneling rates of metastable vacua of the string landscape. It this section, we contemplate  the applicability of our assumptions and results for the string landscape.

Although the large number of possible vacua in the landscape of moduli of the underlying Calabi-Yau manifolds  makes the string landscape a very intriguing place to start anthropic arguments for almost everything, it may contain the seeds of its demise within by destabilizing such an enormous number of vacua. Not only does the huge number of minima in  high dimensional string theory landscapes (which can be the endpoints of tunneling events)\footnote{The density of vacua  near the conifold point, $\rho_{\rm conifold}$, in a one-dimensional   moduli space is described by $1/r^2(C+\log r)^2]$, where $r$ is the  distance from the conifold point \cite{Denef:2004ze}; applying this  result near a generic point along the conifold locus an  $n$-dimensional moduli space, shows he rapid growth in the number   of vacua, $\int d^nr \,\rho_{\rm conifold}$ with $n$.} offer the possibility of a significant enhancement of quantum tunneling processes, but we also have seen that the huge growth of the number of saddle points makes the lowest of the saddle points  exponentially low and enhances  thermal tunnelings. The main question of this section is to what extent the features of the landscape of random potentials that we have used in the previous sections provide an accurate insight into the properties of the string landscape. 

First we used a straight line that connects the false and true vacuum (to be more accurate, twice the contribution from a straight line that connects the false vacuum to a  saddle points on the surrounding barrier)  as an upper bound for the tension. But this very simple approximation may not be correct in the string landscape or in other high-dimensional spaces. The correct tunneling trajectories may be very tricky and complicated. Explicit examples in the landscape are the conifunneling trajectories between monodromy-related flux vacua found in \cite{Ahlqvist:2010ki} (see also \cite{ Danielsson:2006xw,  Johnson:2008kc}). An additional, and potentially pivotal, complication is that the different flux vacua that are not monodromy related are generally minima of distinct potentials. Physically, such transitions invoke features not captured by our local field theoretic model, including for example the nucleation of branes to absorb changes in flux \cite{Brown:1988kg,deAlwis:2006cb}. These effects might significantly affect the tunneling action, and possibly mitigate the field theory instabilities we have identified.

Second, we have assumed that as the stabilizing contributions to a given model are varied, the effective potentials around local minima will have expansions that are well modeled by random polynomials. In Eq.~\eqref{Land-potential}, if we keep to order $D$, there will be roughly $D^N$ random coefficients. But are there that many random free parameters to choose in a string landscape?  In the landscape, the ``random" parameters of the potential are the fluxes. There are far fewer fluxes in this system than the number of random coefficients we chose in our model. However, the associated minima of $V$ will occur at different locations in the moduli space of a Calabi-Yau.  As the location of these minima changes in the moduli space, it may help to pseudo-randomize the potential further. It seems reasonable to us that this will result in local expansions well modeled by the random potentials invoked in Sec.~\ref{Land-sec-Background}, but we do not have a firm argument.  One important issue that we need to point out is that the decrease in barrier heights and tensions is merely a result of the fact  that there are many positive and negative terms in the potential and that the number of coefficients grows exponentially with the number of fields. As long as these numbers are well scattered over positive and negative numbers, we expect the same results to hold. We probably do not need a perfect randomness as long as we have scattered  the coefficients enough over positive and negative numbers. As mentioned, because the location of minimum can be anywhere in the landscape, that helps spreading these numbers, even if they are not random or freely chosen.

Third,  we used the results achieved from theories with canonically normalized kinetic terms to calculate the tunneling exponents. It is well known, however, that string vacua are densest in the vicinity of the conifold locus, where the classical moduli space metric suffers from a curvature singularity. In particular, near a generic point on the conifold locus we can choose local coordinates $(Z^1, Z^2, \dots ,Z^P)$ on the moduli space such that $Z^1 = 0$ labels the conifold. Near $Z^1 = 0,$ the moduli space metric $G$ behaves as:
\begin{equation}
G^{11}(Z) \sim {\rm ln} (|Z^1|^2)  \, .
\end{equation}
The local form of the action then takes the form
\begin{equation}
\int \sqrt {-g} [g^{\mu \nu} G_{i j} \partial_\mu Z^i \partial_{\nu} Z^j - V(Z)],
\end{equation}
where $g$ is the space-time metric. In this expression the coordinates $Z$ are the moduli space representation of the scalar fields $\phi$ and $V$ is their flux potential.  At any non-singular point we can, of course, use a local change of field variables to absorb $G$ into the $Z$, yielding a canonical kinetic term. But as we approach a conifold point, this change of variables corresponds to reducing the barrier heights in $V$ (assuming $V$ is continuous and is being expanded about a local minimum) and thus increases tunneling rates. In the regions of moduli space that are most densely populated with string vacua, we therefore expect the non-canonical kinetic terms to augment the destabilization we have found.

Fourth, since the flux potential $V$ is derived from the superpotential $W$, and it is $W$ that directly incorporates flux values, a more accurate representation of the landscape arises from randomly varying coefficients in a local expansion of $W$, and then using the result to calculate the random potentials. In principle, the relationships between the coefficients in $V$, which reflect its origin in $W$, could alter our findings. For this reason we have repeated the calculations for a supersymmetric potential which is described in Appendix \ref{SUSYAppendix}. The results, which were very similar to the random potentials, were presented in Sec. \ref{Land-sec-Numeric}.

The examples we worked out in this chapter  illustrate the challenges of investigating random potentials for large numbers of fields. In all of our studies, supersymmetric or not, computational considerations have forced us to only probe a limited range of values of $N$, the number of fields, and work with truncated potentials and use various  approximations for tunneling rates. We are assuming that the pattern we have found, as evidenced in Figs.~\ref{Land-QuarticActions}-\ref{Land-SUSYSaddleNum}, will continue to hold as these constraints are relaxed. 

\section{A multiverse explanation of the cosmological constant? }
\label{Land-sec-Mutliverse}
Our results have important and immediate implications for attempts to use multiverse scenarios as an explanation for the smallness  of the cosmological constant. These argument are based on  simple counting. Even if the natural scale of the underlying theory for the cosmological constant is Planckian, one might naturally find vacua with $\Lambda$ as low as $10^{-120}$ if there are many more than $10^{120}$ vacua. However, a low cosmological constant is not the only  feature of the observed universe. Our universe is several billion years old and whatever vacuum it is living in  must be very long-lived. Therefore what we really need for an anthropic and natural explanation of the cosmological constant is an abundance (much more than $10^{120}$) of truly metastable and long-lived vacua. 

The generally accepted counting is that for a case of $N$ fields, the number of vacua $\mathcal N_{\rm vac}$ is of order $g^N$ where $g$ is some number of order unity, let's say 10. However our results suggest that the chance of finding a stable vacuum decreases like $e^{-\beta N^\alpha}$ where $\alpha$ is a model dependent number (in the cases we studied it was roughly 2.7 for quartic random potentials and 3.2 for supersymmetric potentials). Therefore, for very large $N$  the chance of finding even one stable vacuum is exponentially small.
\begin{figure}[htbp] 
   \centering
	\includegraphics[width=3.5in]{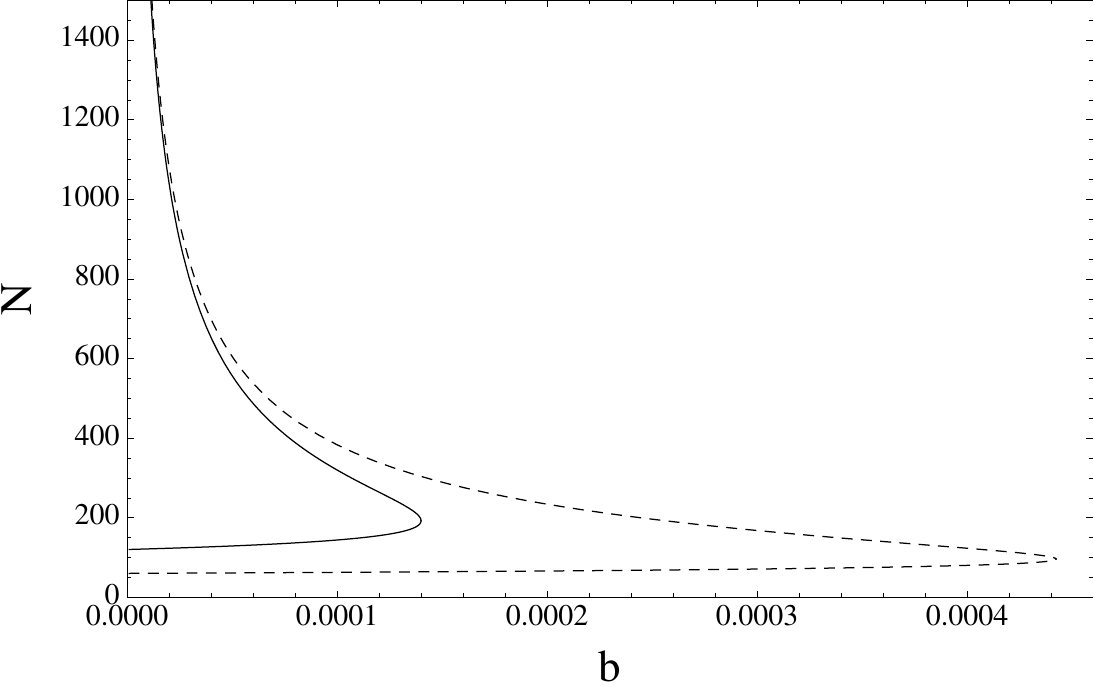}
   \caption{Parameter ranges allowing multiverse explanations of the cosmological constant. With $\alpha = 2.66$ and $g=10$, a sufficient number of metastable vacua is only possible for parameters in the region to the left of the solid line. This region is extended to the dashed line if $g=100$. }
   \label{Land-StabilityRegion1}
\end{figure}
\begin{figure}[htbp] 
   \centering
	\includegraphics[width=3.5in]{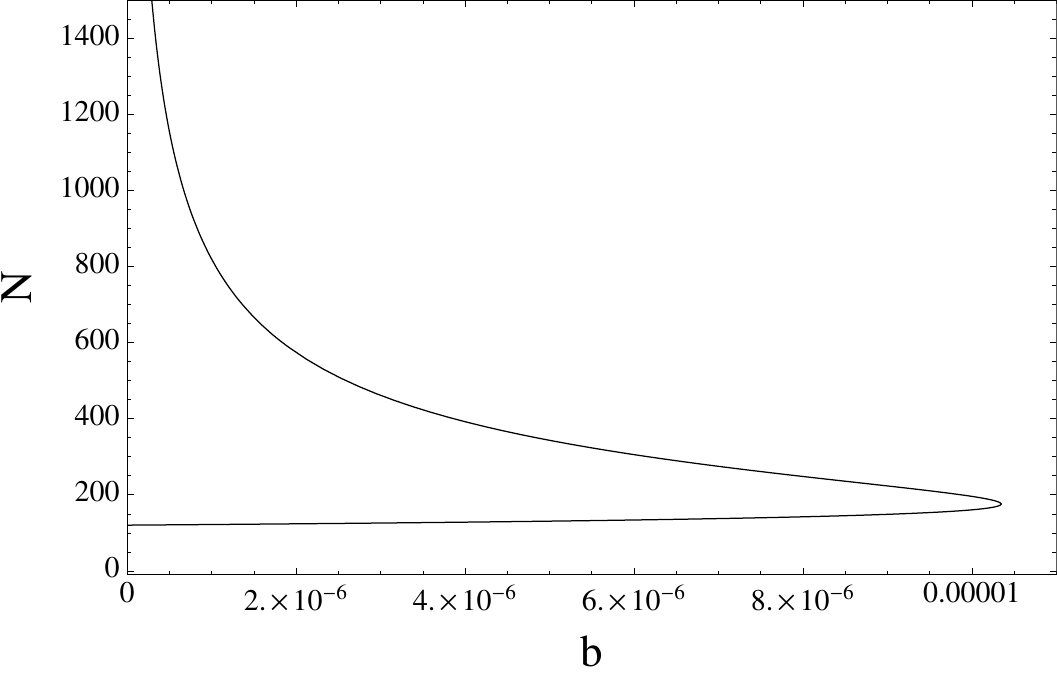}
   \caption{Like Fig. \ref{Land-StabilityRegion1}, but for $\alpha = 3.16$ and $g=10$. }
   \label{Land-StabilityRegion2}
\end{figure}
If the number of vacua is $g^N$, the number of metastable ones is
\begin{equation}
	\mathcal N_{\rm vac} f( B_{\rm min}) \sim g^N e^{-\beta  B_{\rm min} N^\alpha}~,
\end{equation}
where $f( B)$ given by Eq.~\eqref{Land-SurvivingFraction}. The requirement that this number be greater than $10^{120}$ is
\begin{equation} \label{Land-Inequality}
	N- {b \over \ln g} N^{\alpha} > 120 \left( \ln 10 \over \ln g\right),
\end{equation}
with 
\begin{equation}
	b = \beta  B_{\rm min} = \left( {10^{-3} \gamma \over C_{\rm tension}} \right) \lambda  B_{\rm min}\sim 10^{-3} \lambda ~.
\end{equation}
$N$ must be a large number to satisfy Eq.~\eqref{Land-Inequality}, at least more than 120 if $g=10$. The new feature we have found is that $N$ cannot be too large, since the enhancement in the tunneling rate at large $N$ will make the number of available stable vacua smaller rather larger. Therefore there can only  be a window in $N$ values, larger than let's say 120, but small enough that the tunnelings  leave us enough stable vacua to explain the cosmological constant. In fact there may be no such a window which satisfy this inequality, as illustrated in Fig. \ref{Land-StabilityRegion1} where we used $\alpha=2.66$ from our results for non-supersymmetric potentials in Sec.\ref{Land-sec-Numeric}. The allowed region for $b$ and $N$ is on the left of the solid line for $g=10$ and  dashed line for $g=100$. As we can see, there is no value of $N$ which satisfy this criteria if $b>1.4 \times 10^{-3}$ for $g=10$  and the range of allowed $N$ is  small until $b$ gets much smaller than this threshold. Changing the value of $g$ does not help much. For example for $g=100$, the allowed values for $b$  only increases by a factor of 4. We need to increase $g$ exponentially to change the range of $b$ significantly. 

In Eq.~\eqref{Land-Inequality} $\alpha$  is model dependent. The effect of varying it is shown in Fig. \ref{Land-StabilityRegion2} where we used $\alpha=3.16$ from our supersymmetric potentials in Sec. \ref{Land-sec-Numeric}. Although the curve looks similar to the one of the  previous case, the allowed values for $b$ have shrunk by more than one order of magnitude. 

\section{Summary}
\label{Land-sec-Summary}
In this chapter we studied the effect of increasing the number of fields on the stability of vacua in multi-field quantum theories. Our motivation arises from the landscape of string theory and whether or not it can furnish a natural framework to resolve the cosmological constant problem. We used random polynomials of quartic and cubic order with $N$ fields ($N\le10$)  as an approximation for the potential near its minimum. To make the computations tractable, we used some approximations  explained in Sec.\ref{Land-sec-Background}, and kept the number of fields relatively small. Even with these constraints, our data provide evidence that there is an exponentially large enhancement in the transition rates due to quantum and thermal fluctuations and that the rates increase so rapidly as a function of $N$ that all but an exponentially small fraction of the generic vacua become unstable. Therefore the range of parameters that give a sufficiently large number of stable vacua to provide enough diversity to explain the cosmological constant problem is severely restricted.

The assumptions we made were quite generic and potentially relevant to any model invoking anthropic explanation for the cosmological constant problem in which the required diversity of vacua is a consequence of involving a large number of fields. However, since our results are based on models of field theory, the impact of these considerations on the landscape of string theory and its ability to provide a natural explanation of cosmological constant problem needs further study.

\clearpage

\chapter{Higher dimensional Einstein-Maxwell landscapes }
\label{EMLandscape}
Understanding the statistics, phenomenology, stability analysis and other features of the landscape of string theory is one of the most important and urgent problems in theoretical physics. Unfortunately, because of the complexity of this landscape, many of these features are not understood very well.  As a way of understanding these features, different toy models have gotten attention and have helped us to uncover some of these features. Among these, the landscape of Einstein-Maxwell theory (for brevity we call it EM after this) is a very popular and interesting one. 

This is a model of electromagnetism in curved spacetime with a minimal coupling between  electromagnetism and gravity. The magnetic flux can help to stabilize the compactification of some of the dimensions. For example, if we start from a six-dimensional spacetime and wrap magnetic fluxes around two of these dimensions, we end up with four non-compact dimensions. The remaining two dimensions have the geometry of a two-sphere. This compactification was first introduced in \cite{Freund:1980xh,RandjbarDaemi:1982hi} and later on many aspects of this theory were understood. 

This model that we will  describe later provides an excellent lab to test whether the field theoretical approach we used in Chapter \ref{Tumbling} is applicable to potentials obtained from compactification because

\begin{enumerate}
	\item The compact extra dimensions are stabilized by fluxes wrapping around them.
	\item  The kinetic terms  are not  canonical and even not diagonal. Nicely enough, there is a general formula for making them diagonal and canonical.
	\item Although there are  fluxes which  corresponding to magnetic charges, we can go to a decompactified phase without nucleation of branes.
	\item There are inherently random numbers (the fluxes for different spheres). This resembles the situation in the string landscape.
	\item The effective potential obtained by compactification is exactly of cubic order. This makes the Taylor expansions we used in Chapter \ref{Tumbling} legitimate. 
\end{enumerate}
The work we present in this Chapter is mainly based on a work in progress with Adam Brown and Alex Dahlen \cite{BrownA}. It will appear soon. We describe compactification of (4+2$N$)-dimensional space in EM theory over $\mathbb{R}^4\otimes \mathbb{S}^2\otimes \mathbb{S}^2\otimes \ldots \otimes \mathbb{S}^2.$
In Sec. \ref{EM-sec-6DReview} we review some of the features of the 6-dimensional EM theory in which two of the dimensions are compactified over a two-sphere. This corresponds to $N=1$. In Sec. \ref{EM-sec-Compactification} we describe the generalization to higher dimensions and calculate the effective potential. We show some of the main features of the $N=2$ case in Sec. \ref{EM-sec-N=2} and present some of the results and conjectures for higher dimensions in Sec. \ref{EM-se-HighN}. Finally  we summarize and discuss the future directions we will take in Sec. \ref{EM-sec-Summary}.
\section{Review of the six-dimensional model}
\label{EM-sec-6DReview}
\begin{figure}[htbp] 
   \centering
   \includegraphics[width=2.5in]{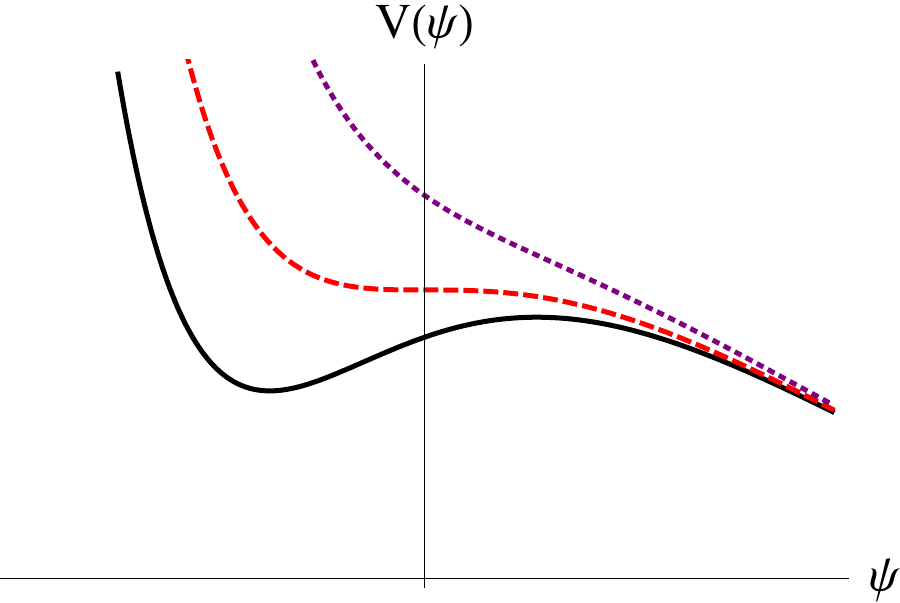} 
   \caption{Different behaviors of the potential as we change the number of fluxes. For very large $n$, there is no stationary point (dotted purple line). By lowering $n$, the potential gets a single stationary point (dashed red line) and for smaller $n$ (black line), it gets one maximum at a larger value of $\psi$ and one minimum at a smaller $\psi$. In the last case there is a stable compactified phase separated by a barrier from the decompactified phase. We are only interested in the last case.}
   \label{EM-Fig-DifferentVBehavior}
\end{figure}

In this section, we mainly follow the explanation and conventions in \cite{BlancoPillado:2009di}. The action of six-dimensional EM is 
\begin{equation}\label{EM-eq-6DLagrangian}
	\tilde S = \int d^6 \tilde{x} \sqrt{-\tilde g}\left(- {M_{(6)}^4 \over 2} \tilde{\mathcal R}^{(6)} +{1 \over 4 } F_{MN} F^{MN} + \tilde \Lambda \right)~,
\end{equation}
where $\mathcal{R}_{(6)}$ and $\tilde \Lambda$ denote the six-dimensional scalar curvature and cosmological constant. We use capital Latin letters to  denote  six-dimensional indices  and Greek letters  for four-dimensional indices.  
The equations of motion arising from this action are
\begin{eqnarray}
	\tilde{R}_{MN} - \frac{1}{2} \tilde{g}_{MN}\tilde{\mathcal R} &=& \frac{1}{M_{(6)}^4} T_{MN} , \cr \cr
	\frac{1}{\sqrt{-\tilde g} } \partial_M\left( \sqrt{-\tilde g} F^{MN}\right)&=&0~,
\end{eqnarray}
where the energy-momentum tensor  is
\begin{equation} \label{EM-6DEMTensor}
	T_{MN} = \tilde{g}^{PQ} F_{MP}F_{NQ} - {1\over 4} \tilde{g}_{MN} - \tilde{g}_{MN} \tilde \Lambda~.
\end{equation}
We can compactify two of the dimensions over $\mathbb{S}^2$ using a metric of the form 
\begin{equation} \label{EM-6DEMMetric}
	ds^2 = \tilde {g}_{MN}  dx^M dx^N = e^{-\psi(x)/M_P} g_{\mu\nu} dx^\mu dx^\nu + e^{\psi(x)/M_P} R^2 d\Omega_2^2~.
\end{equation}
Here we defined the four-dimensional Planck mass as
\begin{equation}
	M_P^2 =4 \pi R^2 M_{(6)}^4~.
\end{equation}
The only  ansatz for the electromagnetic equations consistent with the symmetries of the extra dimension and independent of the four-dimensional coordinates  \cite{RandjbarDaemi:1982hi} is
\begin{equation}
	A_{\phi} = - {n \over 2  g } (\cos \theta \pm 1 )~,
\end{equation}
which is a monopole-like configuration in the extra-dimensional spheres. Here $n$ is an integer and the signs correspond to the two patches necessary to describe a monopole configuration.  Assuming that the two representations for the electromagnetic field are  related by a single-valued gauge transformation quantizes $n$. The only nonzero components of $F_{MN}$ are
\begin{equation}
	F_{\theta \phi } = - F_{\phi \theta} = {n \over 2 g}\sin \theta.
\end{equation}
Using this ansatz for the electromagnetic sector, and the metric in  Eq.~\eqref{EM-6DEMMetric}, we can integrate out the two extra dimensions and get the four-dimensional effective action from Eq.~\eqref{EM-eq-6DLagrangian}  
\begin{equation}
S = \int d^4 x \sqrt{-g} \left(- {1 \over 2} M_P^2 {\mathcal R}^{(4)} + {1\over2 } \partial_\mu \psi \partial^\mu \psi +V(\psi)\right),
\end{equation}
where 
\begin{equation} \label{EM-eq-6DEffPotA}
	V(\psi) = {M_P^2 \over R^2 } \left( { \pi n^2 \over 2 g ^2  M_P^2} e^{-3 \psi /M_P}  - e^{-2 \psi/M_P} + {4 \pi R^4 \tilde \Lambda \over M_P^2}e^{-\psi/{M_P}} \right)~.
\end{equation}
For any nonzero value of flux, this potential gets large and positive for large negative $\psi$ (small extra dimensions) and tends to zero for large positive $\psi$ (decompactifid phase). If there is no flux, the potential is unbounded from below and  the extra-dimensional spheres collapse. The possible behaviors of the potential for nonzero fluxes is shown in Fig.\ref{EM-Fig-DifferentVBehavior}.

\begin{figure}[htbp] 
   \centering
   \includegraphics[width=3.5in]{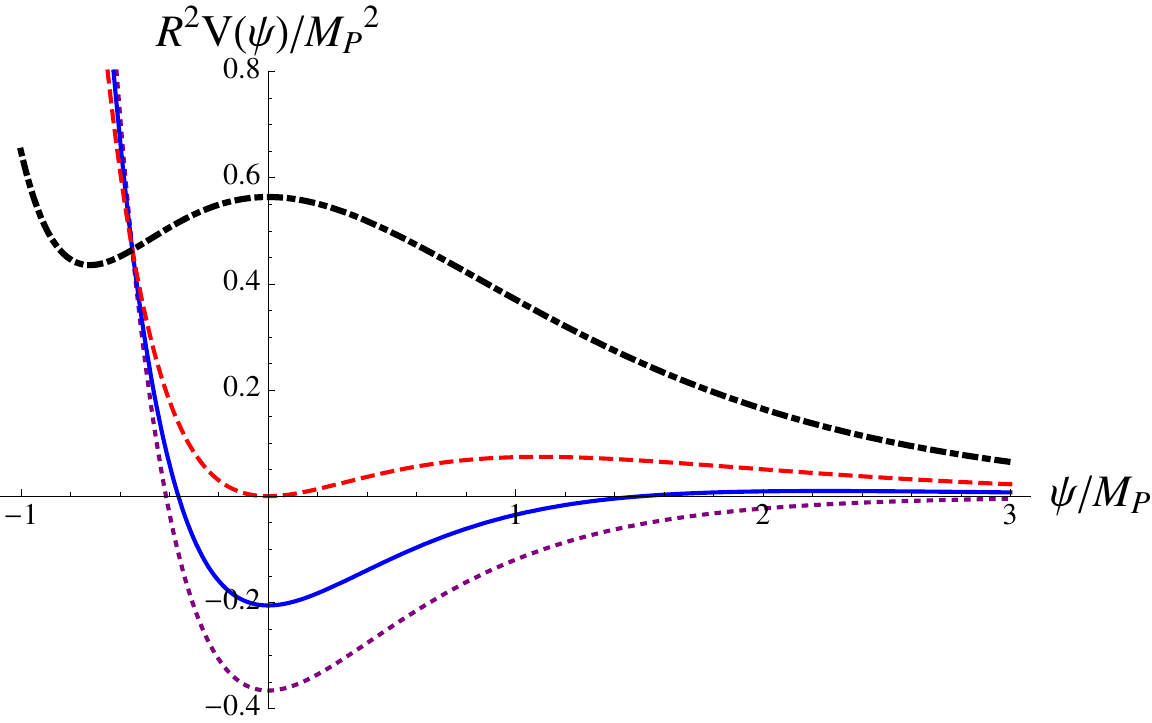} 
   \caption{Different possibilities for the minimum of the potential introduced in Eq.~\eqref{EM-eq-6DEffPotB}. As we change $\tilde n$ from very small values, the minimum changes from being de Sitter (black dotted-dashed line), to Minkowski (red dashed), to an AdS with a positive maximum (blue line) to an AdS space  which is never positive(dotted purple). These graphs  correspond to $\tilde n = 0.467, 1/\sqrt{2}, 0.776$  and $0.826.$  }
   \label{EM-Fig-DifferentVBehavior2}
\end{figure}

We are only interested in the  case where there is a (classically) stable  minimum. In this case, we can choose $R$ in such a way that one of the stationary points appears at $\psi =0$. (Our convention is different from the one used in \cite{BlancoPillado:2009di}.) This choice of $R$ is merely a shift in the values of $\psi$ or, equivalently, a rescaling of $R$. Also,  we define a dimensionless quantity $\tilde n$ as  $\tilde n^2=\pi n^2 /(2 g^2 M_P^2) $ (Please notice that the electric charge  in six-dimensional spacetime is no longer dimensionless). In terms of the rescaled $R$ and $\tilde n$, the potential takes a very simple form,
\begin{equation}\label{EM-eq-6DEffPotB}
	{R^2 V(\psi) \over M_P^2} = \tilde{n}^2 e^{-3\psi /M_P} - e^{-2\psi /M_P} + (2 - 3 \tilde n^2) e^{- \psi/M_P}~.
\end{equation}
Depending on $\tilde n$, the minima can be  de Sitter, anti-de Sitter or Minkowski. For $0< \tilde n< 1/\sqrt2$, this minimum is a de Sitter space. If $n=1/\sqrt2$, the minimum is Minkowski. For larger $\tilde n$, the minimum is  anti-de Sitter. But there are two possibilities. If $1/\sqrt{2} < n < \sqrt{2/3}$ there is a de Sitter minimum which is separated by a barrier with positive energy from the decompactification. For larger $\tilde n$'s there is no stationary point besides the anti-de Sitter minimum and this minimum is completely stable. Choosing a very large $\tilde n$ creates a minimum with arbitrarily negative potential. These possibilities are shown in Fig. \ref{EM-Fig-DifferentVBehavior2}.
This potential can tunnel to different values of $\tilde n$ by creating a charged brane or by decompactifying. These instantons are explained in details in \cite{BlancoPillado:2009di}.

\section{Compactification and effective potential}
\label{EM-sec-Compactification}
\subsection{Calculation of effective potential}
In this section, we describe a compactification of $(4+2N)$-dimensional Einstein-Maxwell (EM) theory over $\mathbb{R}^4 \otimes \underbrace{\mathbb{S}^2 \otimes \mathbb{S}^2 \ldots \otimes \mathbb{S}^2}_{\text{N times}}$. As explained in the beginning of this chapter this model has very interesting  features. 
Let's start from the action of the EM theory in $(4+2N)$-dimensional spacetime. As in the previous section, all the tilded quantities and also the capital Latin indices refer to the $(N+2)$-dimensional quantities and the Greek indices denote the four-dimensional non-compact spacetime. The action is 
\begin{equation} \label{EM-HighDActionA}
	S_{\text{EM}} = \int d \,\tilde x^{(4 + 2N)} \sqrt{-\tilde G} \left[ -\frac{M^4_{(4+2N)}}{2} \mathcal{\tilde R}^{(4+2n)} + \frac{1}{4} F_{AB} \,F^{AB} + \tilde \Lambda\right]~,
\end{equation}
We compactify the spacetime  over the product of $N$ two-spheres and $\mathbb{R}^4$  and write the metric as
\begin{equation} \label{EM-HighDMetricA}
	d\tilde s^2 = \tilde G_{MN}(x_K) d\tilde x^M d \tilde x^N=e^{-B \sum \psi_i(x_\mu)} g_{\mu\nu} dx^\mu dx^\nu + e^{A \psi_i(x_\mu)} R^2 d \Omega^2_{2(i)}~.
\end{equation}
Here $ d \Omega^2_{2(i)}$ is the length element of the $i$'th two-sphere. $B$ and  $A$ are constant numbers  and the $\psi_i$'s only depend on the four-dimensional spacetime coordinates.  $R$ is a length scale associated with the size of the extra-dimensional spheres that  keeps the dimensions correct. When we integrate out the extra dimensions, the action is in the   Einstein frame only when $A=B$. Here we set them equal, and by a rescaling of the fields we set both to be equal to one.  We are looking for solutions for the electromagnetic sector which are independent of the underlying four-dimensional spacetime. We want to choose these solutions in such a way that they stabilize the radius of the spheres.  An ansatz which is compatible with the symmetries of the extra dimensions and independent of the four-dimensional coordinates $x_\mu$ is \footnote{It was argued in \cite{BlancoPillado:2009di} based on \cite{RandjbarDaemi:1982hi} that this is the only compatible answer for the case of the six-dimensional  EM theory. We do not know whether or not this is the unique solution in the higher dimensional case.}
\begin{equation}
	F_{\theta_i \phi_i} = \frac{g N_i}{4 \pi}\sin \theta_i~,
\end{equation}
The other components are zero. The Lagrangian density of the electromagnetic tensor is
\begin{equation}
	F_{CD}F^{CD} = 2 \sum_i F_{\theta_i \phi_i} F^{\theta_i \phi_i} = 2 \sum_i g^{\theta_i \theta_i} g^{\phi_i\phi_i}F_{\theta_i \phi_i}^2 =2\sum_{i} e^{-2 B \psi_i} \left({g N_ i\over 4 \pi R^2}\right)^2~.
\end{equation}
Now we need to express the scalar curvature of the $(4+2N)$-dimensional  metric in Eq.~\eqref{EM-HighDMetricA} in terms of the associated scalar curvature of the four-dimensional space time. We have undertaken this calculation for the warped product of an arbitrary number of spaces in Appendix \ref{Appendix-WarpedMetric}. The result for this special case is
\begin{align}\label{EM-eq-HighDCurvatureA}
	\mathcal{R}^{(4+2N)} = e^{\Psi} \left[ \tilde{\mathcal{R}}^{(4)} +  \nabla^2 \Psi -\frac{1}{2} \nabla \Psi \cdot \nabla \Psi- \frac{1}{2}  \sum_{i=1}^N \nabla \psi_i \cdot \nabla \psi^i\right]+ \frac{2}{R^2} \sum_{i=1}^Ne^{-\psi_i}~,
\end{align}
where $\Psi=\sum_{i=1}^N\psi_i$ and all the covariant derivatives are calculated using the  the connections obtained from the four-dimensional metric ($g_{\mu\nu}$), so the inner product
\begin{equation}
 \nabla \Psi \cdot \nabla \Psi = \nabla_\mu \left(\sum_{i=1}^N \psi_i \right) \nabla^\mu\left(\sum_{j=1}^N \psi_j\right)~.
\end{equation}
We can readily integrate  the action in Eq.~\eqref{EM-HighDActionA} over the extra dimensions. We set the $(4+2N)$-dimensional Planck mass  $M_{(4+2N)}=1$. The effective actions of the different sectors are
\begin{eqnarray}
	S_{\text{GR}}&=&-\int d^4x \sqrt{-g }\; \frac{1}{2}(4 \pi R^2)^N\left[ \tilde{\mathcal{R}}^{(4)} +  \nabla^2 \Psi - \frac{1}{2}  \left| \nabla \Psi\right|^2 \right. \left. - \frac{1}{2}  \sum_{i=1}^N\left| \nabla\psi_i\right|^2 + \frac{2}{R^2}\sum_{i=1}^Ne^{- \Psi -  \psi_i}\right]  ~, \cr\cr
	S_{\text{E-M}} &=& \int d^4x \; (4 \pi R^2)^N\sqrt{-g} \sum_{i=1}^N 2 \left( {g N_i \over 8 \pi R^2}\right)^2 e^{- \Psi-2  \psi_i} ~, \cr\cr
	S_\Lambda &=&  \int d^4x \sqrt{-g} (4 \pi R^2)^N  \Lambda e^{-\Psi} ~.
\end{eqnarray}	
Let's define $M = (4\pi R^2)^{N/2}$.  We can drop the second term in the first equation, because it is a total derivative. Putting  everything together we get 
\begin{eqnarray} 
	&&S=- \int d^4x  \sqrt{-g}  M^2 \left[ {1 \over 2}  \mathcal{R}^{(4)} - \frac{1}{4} \left| \nabla \Psi\right|^2 - \sum_{i=1}^N \frac{1}{4}  \left| \nabla\psi_i\right|^2 +{1 \over R^2}\sum_{i=1}^N e^{- \Psi -  \psi_i}\right. \cr\cr
	&& \qquad \qquad \left. - 2 \sum_{i=1}^N \left( {g N_i  \over 8 \pi R^2}\right)^2 \, e^{-\Psi-2 \psi_i}- \tilde\Lambda e^{-\Psi}\right]~.
\end{eqnarray}	
We have used the freedom to set the $(4+2N)$-dimensional Planck mass to one, but still we can set  $R^2 \Lambda=1$. This is merely a shift in the $\psi_i$'s and is not setting a mass scale. This gives 
\begin{align} \label{EM-EffectiveActionC}
	&{ S \over M^2} =- \int d^4x  \sqrt{-g} \;   \left[ {1 \over 2}  \mathcal{R}^{(4)} - \frac{1}{4} \left| \nabla \Psi\right|^2 - \sum_{i=1}^N \frac{1}{4}  \left| \nabla\psi_i\right|^2 - {1 \over R^2} V_{\text{eff}} \right]~,
\end{align}	
where 
\begin{align}\label{EM-EffectivePotentialC}
	V_{\text{eff}}= \sum \left( n_i^2 \, e^{-\Psi-2 \psi_i}- e^{-\Psi - \psi_i}\right)+e^{-\Psi}~.
\end{align}
Here  $n_i = \frac{g N_i \sqrt{2}}{8 \pi R}$.  If we define $\alpha_i = e^{-\psi_i}$, the effective potential becomes a polynomial of order three. This verifies one of the claims we made in the beginning of this chapter .

\subsection{Bringing the kinetic terms into canonical form}
\label{EM-sub-Canonical}
The kinetic terms for the scalar fields in Eq.~\eqref{EM-EffectivePotentialC} are not canonical. To diagonalize them, we need a field redefinition. There are many different ways to do this.We present two of them here. The first one is a symmetric field redefinition which treats all the fields on the same footing. The second one is not symmetric, but it is more useful  for explicit calculations.

Let's start from the symmetric redefinition. 
Here  $n_i = \frac{g N_i \sqrt{2}}{8 \pi R}$. The kinetic terms of the scalar fields are not diagonal. To diagonalize them, we need a field redefinition. 
\begin{align}\label{EM-eq-SymRedef}
	 \psi_i = \sqrt 2  \phi_i + \frac{\sqrt 2}{A}\sum_j \phi_j~,
\end{align} 
where
\begin{equation}\label{EM-ConstantA}
A= -( N +1) \pm \sqrt{N+1} ~.
\end{equation}
To fix the notation, let's choose the plus sign in Eq.~\eqref{EM-ConstantA}. The inverse transformation is 
\begin{align}\label{EM-SymInvRedef}
	\sqrt 2\phi_i = \psi_i - \frac{1}{A+N}\sum_j \psi_j~.
\end{align}
In this notation, the kinetic terms are canonically normalized.
\begin{align}
	 \left| \nabla \Psi\right|^2+ \sum \left| \nabla\psi_i\right|^2 = 2 \sum \left| \nabla\phi_i\right|^2~.
\end{align}
The effective potential in terms of the new fields is
\begin{align} \label{EM-SYMVA}
	&V_{\text{eff}} =  e^{- \frac{A+ N}{A} \sqrt2\Phi} + \sum \left[ n_i^2 e^{-\frac{A+2+N}{A} \sqrt 2 \Phi-2 \sqrt 2\phi_i} -  e^{-\frac{A+1+N}{A} \sqrt 2\Phi- \sqrt 2\phi_i}\right]~,
\end{align}
where $\Phi = \sum \phi_i$~.
Although this field redefinition is very useful  for the case that all the $n_i$'s are the same, for other cases it is more  convenient to use a field redefinitions which does not treat all the fields the same. 
\begin{eqnarray}\label{EM-AsymRedef}
	\psi_1 &=& \tilde{\phi}_1 +\tilde{\phi}_2+  \dots+ \tilde{\phi}_N ~, \cr
	\psi_2&=& \tilde{\phi}_1 - (N-1) \tilde{\phi}_2~, \cr
	\vdots \cr
	\psi_j&=& \tilde{\phi}_1 + \dots+ \tilde{\phi}_{j-1} - (N-j+1) \tilde{\phi}_j  \cr
	\vdots \cr
	\psi_N&=&\tilde{\phi}_1 + \tilde{\phi}_2+ \dots +\tilde{\phi}_{N-1} -\tilde{\phi}_N~.
\end{eqnarray}
We can write this in a slicker way, 
\begin{equation}
	\psi_i = \sum_{j=1}^i \phi_j - (N-i) \phi_i  \qquad i \ge 2~.
\end{equation}
 The kinetic terms become
\begin{align}
	&\left| \nabla \Psi\right|^2 + \left| \nabla\psi_i\right|^2 = \sum_{i=1}^N (N+1-i)(N+2-i) \left| \nabla\tilde{\phi}_i\right|^2~.
\end{align}
The inverse of this transformation is 
\begin{eqnarray} \label{EM-AsymInvRedef}
	\tilde{\phi}_1 &=& \frac{1}{N}(\psi_1 + \psi_2+ \ldots + \psi_n )~,\cr
	\tilde{\phi}_2 &=&\frac{1}{(N-1)N}\left[\psi_1 - (N-1)\psi_2+\psi_3+\ldots + \psi_N\right] ~, \cr
	\vdots	\cr
	\tilde{\phi_j} &=& \frac{1}{(N+1-j)(N+2-j)} \left[\psi_1 - (N+1-j)\psi_j + \psi_{j+1}+\ldots+\psi_N \right]~,  \cr
	\vdots	\cr
	\tilde{\phi}_n&=& \frac{1}{2} (\psi_1- \psi_n)~.
\end{eqnarray}
Again we can write this in a closed form
\begin{equation}
	\tilde{\phi}_j =\frac{1}{(N+1-j)(N+2-j)} \left[ \psi_1 - (N+1-j) \psi_j + \sum_{i=j+1}^N \psi_i\right] \qquad j\ge 2 ~.
\end{equation}
This is diagonal. However, it is not canonically normalized and needs a  trivial rescaling of the field.In order to go to a canonically normalized frame, we define
\begin{align}
	& \phi_i = \sqrt{(N+1-i)(N+2-i) \over 2} \tilde{\phi}_i~.
\end{align}
The effective potential in terms of these fields is
\begin{align}
	&V_{\text{eff}} =  e^{-\sqrt{2 N \over N+1} \phi_1} + \sum \left[ n_i^2 e^{-\sqrt{2 N \over N+1} \phi_1-2 \psi_i} - e^{-\sqrt{2 N \over N+1} \phi_1 - \psi_i}\right]~.
\end{align}
Here we present the field redefinition and effective potential for small values $N$.\\ \\
{\bf N=2 two-spheres}

The new fields are given by 
\begin{equation} \label{EM-RedefN2Asym}
	\left( \begin{array}{c}  
		X \\ 
		\phi
	\end{array}\right) = \frac{1}{\sqrt2} 
	\left( 
	\begin{array}{cc} 
		\sqrt{\frac32}  &\sqrt{\frac32} \\
		\frac{1}{\sqrt2}& -\frac{1}{\sqrt2}
	\end{array}\right)
	\left( \begin{array}{c}  
		\psi_1 \\ 
		\psi_2
	\end{array}\right)~,
\end{equation}
and the effective potential Eq.~\eqref{EM-EffectivePotentialC} is
\begin{equation}\label{EM-ASYPotN2}
	V_{\rm eff} (X,\phi ) = e^{-4 X \sqrt3} \left(n_1^2 e^{-2\phi} + n_2^2 e^{2\phi}\right) - e^{-X/\sqrt{3}} \left( e^{-\phi}+ e^{\phi}\right)+ e^{-2X/\sqrt3}~.
\end{equation}

{\bf N=3 two-spheres}

The new fields are given by 
\begin{equation} \label{EM-RedefN3Asym}
	\left( \begin{array}{c}  
		X \\ 
		\phi_1 \\
		\phi_2
	\end{array}\right) = \frac{1}{\sqrt2} 
	\left( 
	\begin{array}{ccc} 
		{2 \over \sqrt3}  & {2\over\sqrt3 } & {2 \over \sqrt 3} \\
		\frac{1}{\sqrt6}& \frac{1}{\sqrt6} & {-2 \over \sqrt{6}} \\
		{1\over \sqrt2} & 0 & {-1 \over \sqrt2}
	\end{array}\right)
	\left( \begin{array}{c}  
		\psi_1 \\ 
		\psi_2 \\
		\psi_3
	\end{array}\right)~,
\end{equation}
and the effective potential is
\begin{eqnarray}\label{EM-ASYPotN3}
	V_{\rm eff} (X,\phi ) &=& e^{-5 X \sqrt6} \left[n_1^2 e^{-2\phi} + n_2^2 e^{2\phi}\right] \cr 
	&&- e^{-X/\sqrt{3}} \left( e^{-\phi}+ e^{\phi}\right)+ e^{-2X/\sqrt3}~.
\end{eqnarray}

{\bf N=4 two-spheres}

The new fields are given by 
\begin{equation} \label{EM-RedefN4Asym}
	\left( \begin{array}{c}  
		X \\ 
		\phi_1 \\
		\phi_2 \\
		\phi_3 
	\end{array}\right) = \frac{1}{\sqrt2} 
	\left( 
	\begin{array}{cccc} 
		{\sqrt5\over2}  	      & 	{\sqrt5\over2}				&{\sqrt5\over2}  		& {\sqrt5\over2}\\
		{1\over \sqrt {12}}      & 	{-3\over \sqrt{12}} 		&{1\over \sqrt {12}}   &	{1\over \sqrt {12}} \\
		\frac{1}{\sqrt6}		& 	0  							& {-2 \over \sqrt{6}}   &\frac{1}{\sqrt6}  \\
		{1\over \sqrt2}            &  0 							& 0 					& {-1 \over \sqrt2}
	\end{array}\right)
	\left( \begin{array}{c}  
		\psi_1 \\ 
		\psi_2 \\
		\psi_3 \\
		\psi_4
	\end{array}\right)~,
\end{equation}
and the effective potential is
\begin{eqnarray}\label{EM-ASYPotN4}
	&&V_{\rm eff} (X,\phi_1,\phi_2,\phi_3 ) = e^{-3\sqrt{5/2 }X } \left\{\left[\left(n_1^2 e^{-2\phi_1} + n_2^2 e^{2\phi_1}\right)e^{-2\phi_2/\sqrt3}+n_3^2 e^{4\phi_2/\sqrt 3}\right]e^{-\sqrt{2\over3} \phi_3}+n_4^2 e^{\sqrt6 \phi_3}\right\} \cr \cr
	&&-e^{-\sqrt{5/2} X} \left\{ \left[      \left( e^{-\phi_1} + e^{\phi_1}\right)e^{-\phi_2 /\sqrt3} + e^{2 \phi_2/\sqrt3}   \right] e^{-\phi_3/\sqrt 6} + e^{3\phi_3/\sqrt6}\right\} + e^{-2\sqrt{2/5}X}~.
\end{eqnarray}
This explicit form of potential is used in the calculations in the next few sections.

\section{Case of equal fluxes}
\label{EM-sec-Equaln}
When all the fluxes are equal, it is easy to get some understanding of the potential in Eq.~\eqref{EM-SYMVA}. Let's assume that all the $n_i's$ take a common value $n$. We expect that at least some of the stationary points of the potential lie  on the line where all the fields $\phi_i$ take a common value $\tilde \phi$. 
The effective potential Eq.~\eqref{EM-SYMVA} along this line is
\begin{equation}\label{Veff2}
	V_{\rm eff} = N n^2 e^{-\sqrt{2 \over 1+N}(2+N) \tilde{\phi}} - N e^{\sqrt{2(1+N)} \tilde{\phi}} + e^{\sqrt{2 \over 1+N} N \tilde{\phi}}~.
\end{equation}
To make Eq.~\eqref{Veff2} look better, let's define  $X=e^{-\sqrt2\tilde{\phi} (1+N/A )}$.
\begin{equation}\label{VAlongTheLine}
	V_{\text{eff}} =  X^{N}+ N n^2 X^{N+2}-  N X^{N+1}~,
\end{equation}
 The gradient vanishes at the location of critical points
\begin{align} \label{EM-GradientEqA}
	& \frac{\partial V}{\partial \phi_i}=-\frac{A+N}{A} \sqrt2 X^N  \left\{ n^2 X^2 (N+2)  - (1+N) X +1 \right\}~.
\end{align}
 Equation Eq.~\eqref{EM-GradientEqA} may admit at most two solutions,
\begin{equation} \label{EM-ExtremaA}
	\tilde X = \frac{N+1\pm \sqrt{(N+1)^2 -4 n^2  (N+2)}}{2 n^2 (N+2)}~.
\end{equation}
If both roots are positive numbers, the larger one, which corresponds to smaller spheres, will be a minimum and the smaller one will be a maximum along the line $\phi=\tilde \phi$.  To understand whether these points are minima, maxima or saddle points, we need to calculate  the other eigenvalues of the Hessian which correspond to fluctuations of potential in directions not parallel to $\Phi$. After some simplification, the Hessian matrix at the critical points is
\begin{equation}
	H_{ij} = 2 \tilde X^N \left[(4 n^2\tilde X^2 -\tilde X) (y^2 + \frac{1}{A}(y+1) +  \delta_{ij} ) -2 n^2 \tilde X^2 y^2\right] = 2 \tilde  X^N \left[U_1\delta_{ij} + U_2 \right]~,
\end{equation}
where $y = 1 + \frac{N}{A}$. Please notice that the terms which do not have indices on the right side are independent of $i$ and $j$. The eigenvalues of this matrix are $2\tilde X^N(U_1 + N U_2)$ and an  $(N-1)$-fold degenerate eigenvalue $2\tilde X^N U_1$. By looking at the behavior of the potential in Eq.~\eqref{VAlongTheLine} along the line $\phi_i=\tilde \phi$, it is clear that the non-degenerate eigenvalue is negative  at the smaller root and positive at the larger root. The other eigenvalues are all equal to $U_1= 4 n^2 \tilde X^2 - \tilde X$. The value of $U_1$ at the larger root (the one with positive sign in Eq.~\eqref{EM-ExtremaA}) is always positive. Therefore if this point exists, it is a minimum. 
\begin{equation}
	U_1 =
		  \frac{N+2 \sqrt{(1+N)^2-4 n^2 (2+N)}}{2+N}  \tilde X~,
\end{equation}
The value $U_1$ at the smaller root (corresponding to bigger spheres) is
\begin{equation}
	U_1=\frac{N-2 \sqrt{(1+N)^2-4 n^2 (2+N)}}{2+N} \tilde X~.
\end{equation}
The sign of $U_1$ can be easily expressed  in terms of $n_c$ defined as
\begin{equation}
	n_c = {1 \over 4} \sqrt{2+3N}~.
\end{equation}
If $n<n_c$, $U_1$ is negative and therefore this point corresponds to a maximum. For $n_c<n$, $U_1$ is positive and therefore this point is a saddle point with one negative mode and $N-1$ positive modes.
The vacua can be de Sitter, Minkowski and AdS. The Minkowski vacua happen at 
\begin{equation}
	n_{\rm M} = {\sqrt{N}\over2}~.
\end{equation}
For $n<n_{\rm M}$, the vacuum will be an AdS space and for $n_{\rm M}<n$, it is a de Sitter space. 

\begin{figure}[htbp] 
   \centering
   \includegraphics[width=3in]{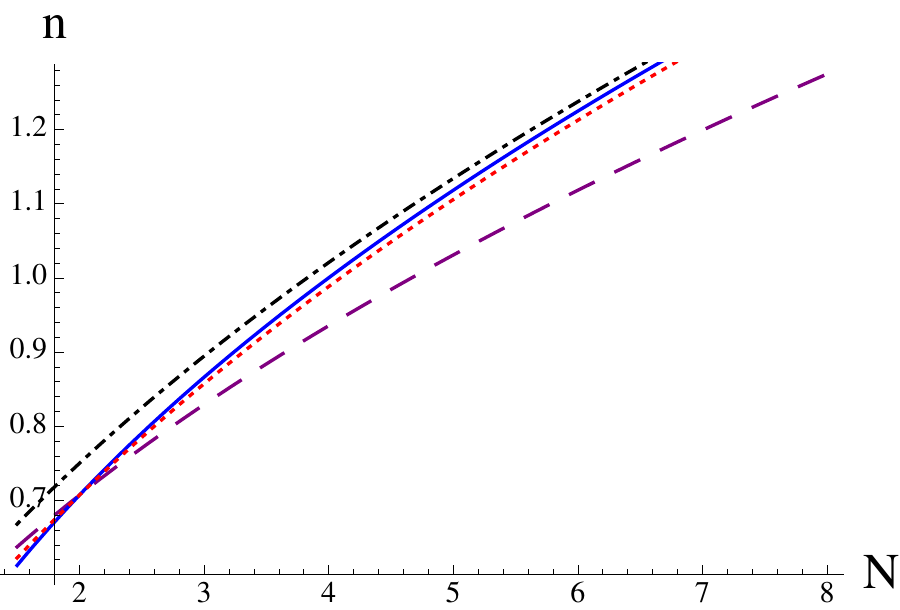} 
   \caption{The possible situations for the critical points with all the spheres of the same size for the effective potential of Eq.~\eqref{EM-SYMVA}. There is no minimum (vacuum) above the black dot-dashed line. The blue solid  line specifies the Minkowski vacuum, below it are the AdS and above the dS vacua. The transition between the CDL and HM bounces happens at the red dotted line. Above the purple dashed  line, there is a saddle point and a minimum and under it a maximum and a minimum.}
   \label{EM-EqualNPhaseSpace}
\end{figure}

There is no minimum for 
\begin{equation}
	n> n_{\rm max}= \frac{N+1}{2\sqrt{N+2}}~.
\end{equation}
We know that when the top of the barrier gets too flat, the CDL instantons cease to exist. The criteria for the existence  of CDL bounces is \cite{Hackworth:2004xb,Jensen:1988zx,Batra:2006rz}
\begin{equation}
	{V''(\phi_{\rm top})\over H_{\rm top}^2} >4.
\end{equation}
When this criteria is met, the dominating bounce is the CDL one. However if the CDL does not exist, the Hawking-Moss bounce dominates. Therefore the tunneling mode has a transition from CDL to HM at  $V''(\phi_{\rm top})= 4H_{\rm top}$ . This happens at 
\begin{equation}
	n_{\rm CDLHM}= {(1+N) \sqrt{6+N} \over 2(4+N)}~.
\end{equation}
Interestingly enough, $n_c$, $n_M$ and $n_{\rm CDLHM}$ all coincide for $N=2$. But this is only a coincidence. For all other $N$'s this does not happen. 
These lines are shown in Fig.\ref{EM-EqualNPhaseSpace}.

Unfortunately, the case where the fluxes are not equal is not as easy as this. In the next  section,  we derive many features of the potentials for $N=2$ and in Sec.\ref{EM-se-HighN} we derive some general results and also state some conjectures about the statistics and the Hubble parameter  of the landscape of these vacua. 

\section{Case of $N=2$}
\label{EM-sec-N=2}
In this section we study the features of the $N=2$ case, which is the compactification of eight-dimensional Einstein-Maxwell theory on two two-spheres.
\subsection{Phase diagram of the $N=2$ case}
\label{EM-sub-PhaseD}
\begin{figure}[htbp] 
   \centering
   \includegraphics[width=4in]{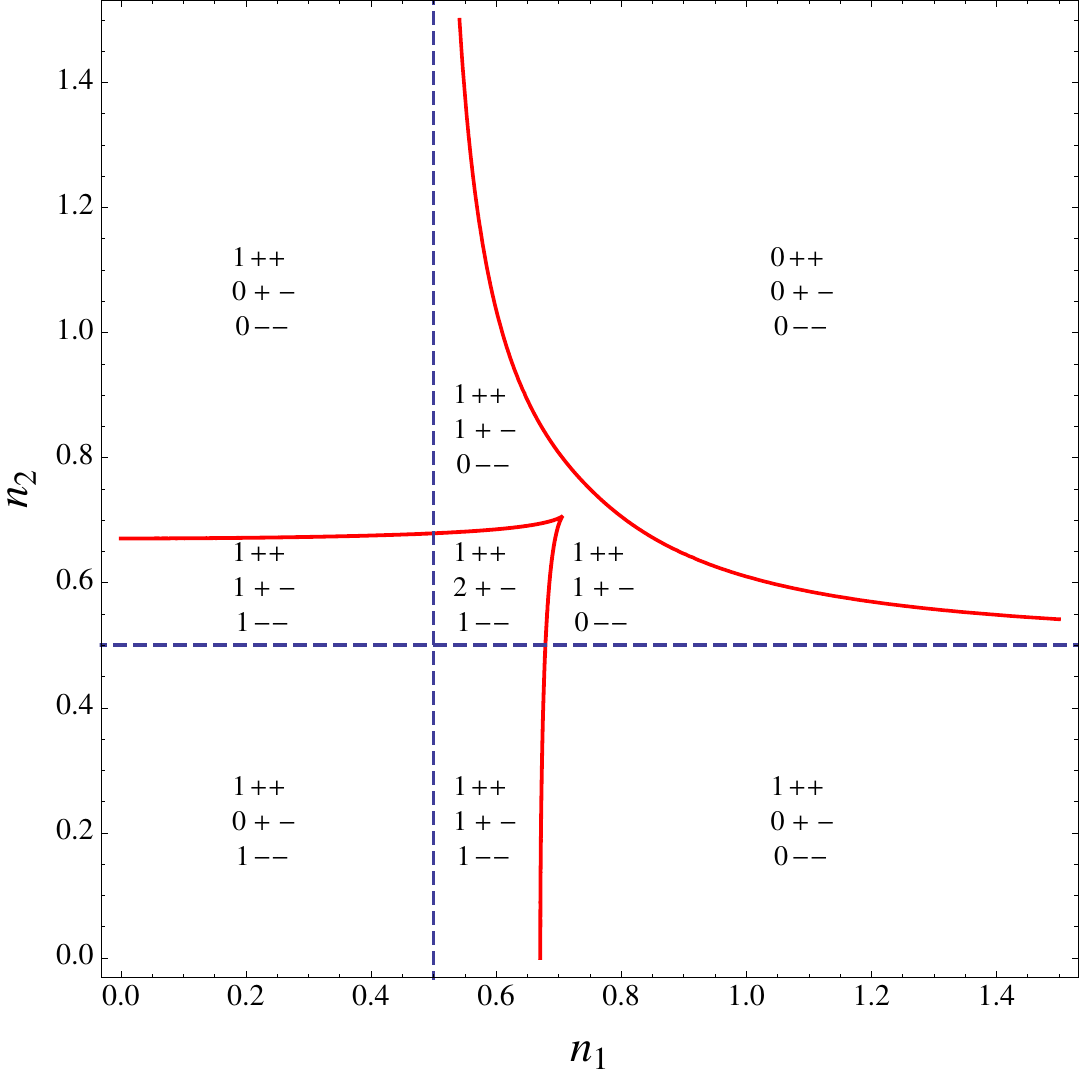} 
   \caption{The phase diagram of the number of minima (++), maxima$(- -)$ and saddle points$(+-)$ for the N=2 case. The two curves satisfy Eqs. \eqref{EM-Castastrophe} and \eqref{EM-HMCDL}. There is no stationary point above the red line. On the left of the vertical dashed blue line (and similarly under the  horizontal dashed blue line), a maximum and saddle point annihilate. The derivation of the curves and also the annihilations on the left of these lines are derived in the main text.  }
   \label{EM2DPhase}
\end{figure}

Let's assume that there are $n_1$ fluxes on the first two-sphere and $n_2$ on the second one.  The effective potential for this case in canonically normalize fields is given in Eq.~\eqref{EM-ASYPotN2} . We want to know the number of minima, maxima and saddle points for different numbers of fluxes.  A summary is shown in Fig. \ref{EM2DPhase}. The region above the top red line does not have any minimum. This resembles the $N=1$ case where increasing the flux causes the minimum to vanish. Along this line the minimum and the saddle point annihilate each other. To find the equation that the red line satisfies, let's rewrite the potential in Eq.~\eqref{EM-EffectivePotentialC} as
\begin{equation}
	V= e^{-(\psi_1+ \psi_2)} \tilde V~, 
\end{equation}
where 
\begin{equation} \label{EM-EffectivePotentialF}
	\tilde V= 1+ e^{-\psi_1} + e^{-\psi_2} -2 n_1^2 e^{-2\psi_1} - 2n_2^2 e^{-\psi_2}
\end{equation}
At any stationary point of the potential
\begin{equation}
	0={\partial V\over \partial \psi_i} = -V + \frac{\partial \tilde V }{\partial \psi_i} {V \over \tilde V}= V \left( -1 + \frac{1}{V} \frac{\partial \tilde V}{\partial \psi_i}\right)~.
\end{equation}
This leads to 
\begin{equation} \label{EM-Const1A}
	\tilde V = e^{-\psi_i} - 2 n_i^2 e^{-2\psi_i}~.
\end{equation}
At the red line, the minimum merges with the saddle point. Therefore one of the eigenvalues of the Hessian should vanish. The Hessian matrix at the extremum is
\begin{eqnarray}\label{EM-Const1B}
	\frac{\partial^2 V}{\partial \psi_i \partial \psi_j} &=& V \frac{\partial }{\partial \psi_j}	 \left(-1 + {1 \over \tilde V} {\partial \tilde V \over \partial \psi_i} \right) 
	= V\left(\frac{\partial^2 \tilde V}{\tilde V\partial \psi_i\partial \psi_j}-{1 \over \tilde V^2 } \frac{\partial \tilde V}{\partial \psi_i} \frac{\partial \tilde V}{\partial \psi_j}\right) \cr\cr
	&=& V\left( {\partial \tilde V \over \tilde V \partial \psi_i \partial \psi_j} -1\right) = {V \over \tilde V} \left\{ \left[ 4 n_i^2 e^{-2\psi_i} - e^{-\psi_i}\right]\delta_{ij} - \tilde V\right\}~.
\end{eqnarray}
Here we use a simple theorem about the determinant of matrices. For a matrix $M$ defined by $M_{ij} = a_i \delta_{ij} - b$, the determinant of $M$ vanishes if $\sum_i a_i^{-1} = b^{-1}$. Therefore the vanishing of one of the eigenvalue of the Hessian translates into 
\begin{equation}\label{EM-Const1C}
	\sum_i {1 \over 4n_i^2 e^{-2\psi_i} - e^{-\psi_i}} = {1 \over \tilde V}.
\end{equation}
This leads to 
\begin{equation}\label{EM-Const1D}
	\sum_i {-2 n_i^2 e^{-\psi_i} +1 \over 4n_i^2 e^{-\psi_i} -1} = 1~.
\end{equation}
After we combine  Eq.~\eqref{EM-Const1A} and Eq.~\eqref{EM-Const1D} and perform some manipulation, the equation for the red line in Fig. \ref{EM2DPhase} is given by
 \begin{eqnarray} \label{EM-Castastrophe}
 &2358 n_1^2 n_2^2+58848 n_1^4 n_2^4+65536 n_1^6 n_2^6+1125 \left(n_1^4+n_2^4\right)-15420 \left(n_1^4 n_2^2+n_1^2 n_2^4\right)\cr
 &-2500 \left(n_1^6+n_2^6\right)+22800 \left(n_1^6 n_2^2+n_1^2 n_2^6\right)-67584 \left(n_1^6 n_2^4+n_1^4 n_2^6\right)=0~.
 \end{eqnarray}
 The same equation also describes the lower part of the red line where a maximum and saddle point merge and disappear. 
 On the left side of the vertical dashed blue line and also under the horizontal blue dashed line in Fig. \ref{EM2DPhase}, one of the saddle points hits the infinity and disappears. We give a proof for this in Sec.\ref{EM-se-HighN}, where we prove a generalized version of this statement for arbitrary $N$.  
 
\begin{figure}[htbp] 
   \centering
   \includegraphics[width=3in]{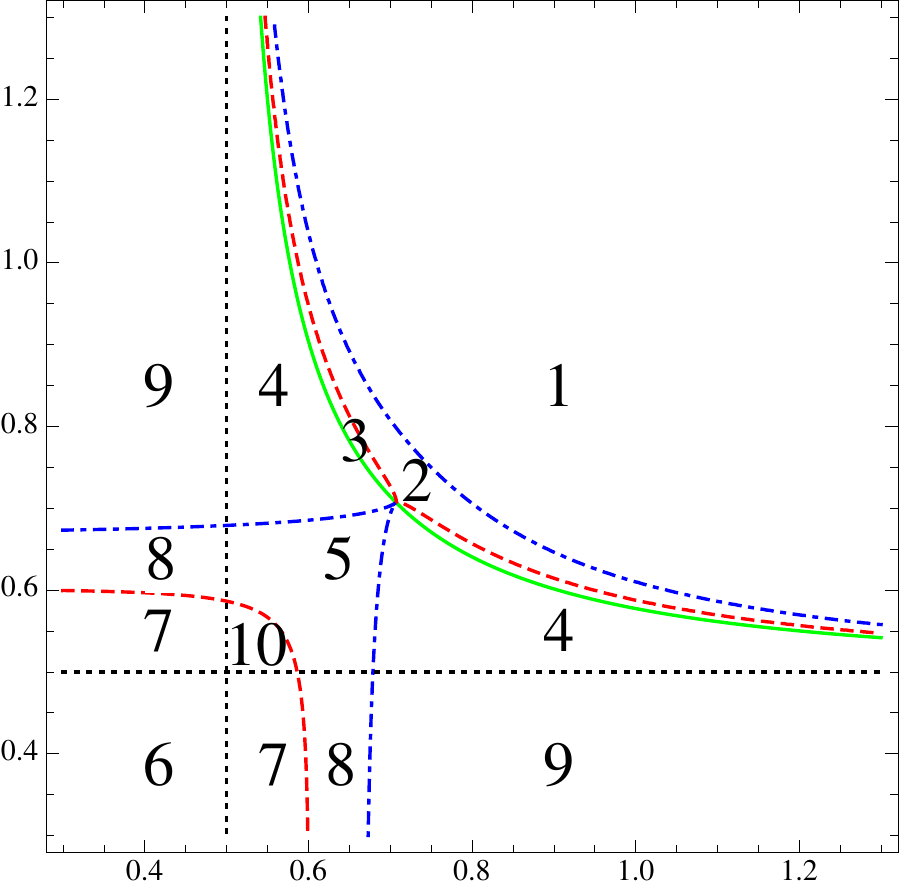} 
   \caption{The phase diagram for the $N=2$ case. The green line given by Eq.~\eqref{EM-2DMinkowski} shows the location of the Minkowski vacua, above it is the dS and below it the AdS minima. There is no minimum above the top dotted-dashed blue line (region 1) which is given by Eq.~\eqref{EM-Castastrophe}. The red dashed line, which is given by Eq.~\eqref{EM-HMCDL}, shows the boundary where the CDL bounces vanish and the HM bounces dominate .  The most interesting part of this landscape is the region between the top red line and the green line where all the dS minima live in (regions 2 and 3). As proven in the text, its area, and therefore the number of dS vacua, is finite.  }
   \label{SecondEM2DPhase}
\end{figure}

 The vacua that we get from the minimum of these potentials can decay through decompactification. This decompactification can be carried either by CDL or HM bounces. But if the top of the barrier gets too flat, the CDL disappears, and using a similar technique, we find that the line where the transition between CDL-dominated decay to HM-dominated decay occurs  along a curve [the criteria for this transition is $V''(\phi_{\rm top}) = 4 H_{\rm top}^2$]
 \begin{eqnarray} \label{EM-HMCDL}		
 	&1286 n_1^2 n_2^2+44256 n_1^4 n_2^4+65536 n_1^6 n_2^6+637 \left(n_1^4+n_2^4\right)-10012 \left(n_1^4 n_2^2+n_1^2 n_2^4\right)-1764 \left(n_1^6+n_2^6\right) \cr
	&+17808 \left(n_1^6 n_2^2+n_1^2 n_2^6\right)-59392 \left(n_1^6 n_2^4+n_1^4 n_2^6\right)=0~.
  \end{eqnarray}
This is the red dashed curve in Fig.\ref{SecondEM2DPhase}. The Minkowski vacua occur along the green solid curve in Fig.\ref{SecondEM2DPhase} which is described by

\begin{eqnarray}\label{EM-2DMinkowski}
	\frac{1}{n_1^2}+\frac{1}{n_2^2}=4~.
\end{eqnarray}
This is a special case of the result we present for arbitrary $N$ in Sec.\ref{EM-se-HighN}. 

Now we can give a full description of the phase diagram presented in Fig. \ref{SecondEM2DPhase}. In region 1 there is no stable minimum.  Region 2 has a single dS vacuum and the transitions are dominated by the HM bounces. Region 3, the tiny area between the green solid line and the dashed red line, has a single dS vacuum and the transitions are dominated by CDL bounces. Region  4 has a AdS vacuum and a saddle point and the CDL bounces are dominating. Region 5 has a AdS vacuum, two saddle points and one maximum and the CDL bounces dominate. Region 7, 8 and 9 have a AdS minimum and the potential  approaches zero from below for large $\psi$'s and therefore there is no tunneling in this region. They are only different in the number of saddle points and maxima. Region 10 has a AdS vacuum, one maximum and two saddle points and the HM bounces are dominating.

\subsection{Statistics of vacua}
\label{EM-sub-Statistics}  
It is not possible in general to find an analytic expression for the extrema of  the potential Eq.~\eqref{EM-EffectivePotentialC} or, equivalently, Eq.~\eqref{EM-EffectivePotentialF}. But we can infer some useful information about them. The area of the regions to the left or under the green line is infinite which means that there are an infinite number of AdS vacua. Also, by making the $n$'s small enough, these minima can have arbitrarily negative cosmological constants. The story for the more interesting dS vacua is different. First we show that there are a finite number of them and then we find approximately   the location of the minima and their cosmological constants for  large values of fluxes.  The dS vacua are bound between the two lines defined by  Eqs.~\eqref{EM-Castastrophe} and ~\eqref{EM-2DMinkowski}.  For large values of $n_1$, the tail of the green line is
\begin{equation} \label{EM-GreenTail}
	n_{2g} = {1 \over 2} + {1 \over 4 n_1^2}~,
\end{equation} 
and the tail of the green blue dotted-dashed line is 
\begin{equation}\label{EM-BlueTail}
	n_{2b} = {1 \over 2} + {1 \over 12 n_1^2}~.
\end{equation}
The area between these two lines is 
\begin{equation} \label{EM-WhyFiniteForN2}
	\int_{n_1^*} dn_1 \left[ n_{2g} - n_{2b}\right]
\end{equation}
which is finite. From Eq.~\eqref{EM-GreenTail} and Eq.~\eqref{EM-BlueTail} it is apparent that the right parameterization for the large values of $n_2$ is
\begin{eqnarray}
	n_1 &=& {1 \over 2 } + \epsilon^2 ~, \cr \cr
	n_2 &=& {K \over \epsilon}~,
\end{eqnarray}
where $\epsilon$ can be arbitrarily small. In terms of this parameterization, the minimum is located at
\begin{eqnarray}
	e^{-\psi_1} &=& 2+ \frac{1}{9} \left(-96+\frac{1}{K^2}+\frac{\sqrt{1-12 K^2}}{K^2}\right) \epsilon^2~, \nonumber \\
	e^{-\psi_2} &=& \frac{1+\sqrt{1-12 K^2}}{3 K^2} \epsilon^2~, \nonumber \\
	V_{\rm min} &=& \frac{2 \left(1+\sqrt{1-12 K^2}\right) \left(24 K^2-1-\sqrt{1-12 K^2}\right)\epsilon ^4}{27 K^4}~.
\end{eqnarray}
It is easy to see that $ K \le {1 \over \sqrt 12}$~. Now we can express $V_{\rm min}$ in terms of $n_1$ and $n_2$
\begin{equation}
	V_{\rm min}=\frac{2 \left(1+\sqrt{1+(6-12 n_1)n_2^2}\right) \left(1+(12-24n_1) n_2^2+\sqrt{1+(6-12 n_1) n_2^2}\right)}{27n_2^4}~.
\end{equation}
The eigenvalues of the Hessian in a canonically redefined field  at the minimum are 
\begin{eqnarray}
	\lambda_1&=&  {4\over 9 K^4}\sqrt{1-12 K^2}  \left(1+ \sqrt{1-12 K^2}\right)^2 \epsilon^4 ~, \nonumber \\
	\lambda_2 &=&\frac{16 \left(1+\sqrt{1-12 K^2}\right)}{9 K^2} \epsilon^2~.
\end{eqnarray}
Therefore the potential energy at the minimum is of order $\epsilon^2$. One of the eigenvalues is of order of $\epsilon^2$. However, the other one is of order $\epsilon^4$. Therefore one of the mass scales of excitations  around the minimum is much smaller than the energy scale of the vacuum.

\section{Higher $N$'s}
\label{EM-se-HighN}
It is very difficult to completely map out the landscapes for $N>3$. However we can infer some important and general features of them.

\subsection{Criteria for Minkowski vacua}
We can rewrite the effective potential in Eq.~\eqref{EM-EffectivePotentialC} as
\begin{align}\label{EM-EffectivePotentialG}
	V_{\text{eff}}= e^{-\sum \psi_i} \tilde V~,
\end{align}
where 
\begin{align}\label{EM-EffectivePotentialK}
	V_{\text{eff}}= 1+\sum \left( n_i^2 \, e^{-2 \psi_i}- e^{- \psi_i}\right)~.
\end{align}
At a Minkowski vacuum, $V=0$ and $\partial V /\partial \psi_i =0$. This leads to 
\begin{eqnarray}
	\tilde V&=& 0 ~,  \cr \cr
	{\partial \tilde V \over \partial \psi_i}&=&  -2 n_i^2 e^{-2\psi_i} + e^{-\psi_i}=0~.
\end{eqnarray}
Form the second equation we can find all the $\psi_i $'s and plugging it back into $\tilde V=0$ we get the criteria for Minkowski vacua. 
\begin{equation} \label{EM-HighNMinkowski}
	\sum_{i=1}^N {1 \over n_i^2}= 4~.
\end{equation}
The dS vacua correspond to $\sum_{i=1}^N {1 \over n_i^2}<4$ and AdS vacua to   $\sum_{i=1}^N {1 \over n_i^2}>4$. Equation Eq.~\eqref{EM-2DMinkowski} is a special case of this result.

\subsection{Bounds on values of $n_i$'s}
\label{EM-sub-BoundsOnFlux}

As we have seen in the previous sections,  a stationary point  does not exist for every choice of fluxes. Here we infer some bounds on the value of fluxes which is necessary for existence of a stationary point. It is more convenient to write the potential in Eq.~\eqref{EM-EffectivePotentialC} in terms of $\rho_i = e^{-\psi_i}$,
\begin{equation}\label{EffectiveV}
V_{\rm eff} = \rho_1 \ldots \rho_N \; f(\rho_1,\ldots,\rho_N)~,
\end{equation}
where 
\begin{equation}\label{EffectiveV2}
	f(\rho_1,\ldots , \rho_N) = 1+ \sum \left( n_i^2 \rho_i^2 - \rho_i\right)~.
\end{equation}
The critical points must satisfy
\begin{equation} \label{Critical}
	f+ \rho_i \frac{\partial f}{\partial \rho_i} = 1+ \sum_{j=1}^N \left( n_j^2 \rho_j^2 - \rho_j\right) + 2n_i^2 \rho_i^2 -\rho_i = 0
\end{equation} 
for every value of $i$. Therefore, at the critical point, $\rho_i(2n_i^2 \rho_i -1) $ should be independent of $i$. By completing the squares we can rewrite Eq.~\eqref{Critical} as
\begin{equation} \label{Critical2}
	\sum_{j=1}^N n_j^2 (\rho_j-{1 \over 2 n_j^2})^2 + 2n_i^2 (\rho_i-\frac{1}{4 n_i^2})^2 = \frac{1}{8 n_i^2} +  \sum_j \frac{1}{ 4n_j^2} -1 ~.
\end{equation}
Therefore 
\begin{equation}
	\forall i \qquad   \frac{1}{8 n_i^2} +  \sum_j \frac{1}{ 4n_j^2} \ge1
\end{equation}
This leads to
\begin{equation} \label{bound}
	 \sum_j \frac{1}{ n_j^2}  \ge \frac{8N}{2N+1}~.
\end{equation}
Because  we cannot set the left-hand side of Eq.~\eqref{Critical2} to zero for all $i$'s, this is not the strictest bound we can set on $n_i$'s. For large values of $N$, even this non-strict bound converges to the bound we found for the dS to AdS transition which means that the window in which we get a de Sitter space gets smaller and smaller by going to higher dimensions. However we will see soon that the number of vacua diverges for $N>2$.
\subsection{Location of the critical points}
\label{EM-sub-LocationOfExtrema}

We can rewrite  Eq.~\eqref{Critical2}  as
\begin{equation} \label{Critical3}
	2 n_i^2 \left(\rho_i - \frac{1}{4n_i^2}\right)^2 = \frac{1}{8 n_i^2} + \left[ \sum \frac{1}{4n_j^2}-1- \sum n_j^2 \left( \rho_j- \frac{1}{2n_j^2}\right)^2\right] = \frac{1}{8 n_i^2} + A~,
\end{equation}
where $A$ is defined as
\begin{equation} \label{ADefinition}
	A= \sum \frac{1}{4n_j^2}-1- \sum n_j^2 \left( \rho_j- \frac{1}{2n_j^2}\right)^2~.
\end{equation}
Solving  $\rho_i$ in terms of $A$ leads to
\begin{equation} \label{Solution}
	\rho_i =\frac{1}{4n_i^2} \left[ 1\pm \sqrt{1+ 8 n_i^2 A}\right]~.
\end{equation}
If $A$ is positive, only the positive sign applies to this problem. Now plugging Eq.~\eqref{Solution} into Eq.~\eqref{ADefinition}, we get an equation for $A$,
\begin{equation} \label{Solution2}
	 \frac{N+2}{2} A= -1 + \sum \frac{1}{8 n_i^2} \left(1\pm \sqrt{1+8 n_i^2 A} \right) ~. 
\end{equation}
 For positive $A$, only positive signs apply here. However, for negative $A$'s, the situation is not that clear and both positive and negative signs are applicable. The possibility of plus or minus in each term of the sum makes this equation less useful, but we will still be able to  use this equation later when we expand it in terms of small $\rho_i$'s.
 
\subsection{Case of small fluxes}
\label{EM-sub-SmallFluxes}
We saw in Fig.\ref{EM2DPhase} that for small values of the $n_i$'s, there always exists a stationary point which is a maximum(the diamond shape under the red line). In this section we show that this is indeed true for general $N$.

Starting from Eq.~\eqref{Critical} and expanding to 2nd order in $n_i$, we get
\begin{equation} \label{Solution3}
\rho_i =\frac{1}{N+1} - \frac{1}{(N+1)^3} \sum_{j} n_j^2 + \frac{2 n_i^2}{(N+1)^2}~,
\end{equation}
and, using the definition of $f$ in Eq.~\eqref{EffectiveV2}
\begin{equation}
	f(\rho_1, \ldots, \rho_N)|_{\rm Critical Point} = \frac{1}{N+1} \left( 1- \sum \frac{n_i^2}{(N+1)^2}\right)+\mathcal{O}(n_i^4)~.
\end{equation}
It is easy to show that 
\begin{equation}
		\left.\frac{\partial^2 V_{\rm eff}}{\partial \rho_i \partial \rho_j }\right|_{\rm critical \; point}= \rho_1 \ldots \rho_N \left[ 2n_i^2 \delta_{ij}- \frac{f}{x_i^2}\delta_{ij} - {f \over x_i x_j}\right] = -\rho_1 \ldots \rho_N  (N+1)\left[\delta_{ij}+1+ \mathcal{O}(n_i^2) \right]~.
\end{equation}
The eigenvalues of this matrix to zeroth order in $n_i$ are proportional to $(N+1)$ and $(N+1)^2$ which confirms that this extremum is indeed a maximum.

\subsection{Vanishing critical points because of hitting infinity}
\label{EM-sub-Vanishing}

In this section we  find the criteria for changing the number of critical points  due to the approach of some of the fields $\psi_i$ to infinity for general $N$ and find the location of these saddle points.  It will be a straightforward calculation to determine whether they are minima, maxima or saddle points. If the $i$'th radion $\psi_i$ approaches infinity, then $\rho_i = e^{-\psi_i}$ will approach zero. Let's separate the $\rho_i$'s that remain finite  by labeling them as $\rho_1, \ldots, \rho_K$ and the ones that approach zero  as $\rho_{K+1} \ldots \rho_N$. From Eq.~\eqref{Solution} we see that this behavior is  possible if and only if  $A$ is an infinitesimally small negative number and we have to choose the positive  root for $i=1\ldots K$ and the negative  root for $i=K+1,\ldots N$. Now expanding Eq.~\eqref{Solution2} to first order in $A$, we get
\begin{equation} \label{TaylorSolution}
\sum_{i=1}^K \frac{1}{4 n_i^2} =1+  (N+1-K) A= 1- \epsilon~,
 \end{equation}
where $\epsilon$ is an infinitesimally small positive number. Therefore, for any subset $n_1\ldots n_K$, the number of stationary points of the potential changes at a hypersurface defined by $\sum_{i=1}^K \frac{1}{4 n_i^2} =1$. The location of this critical point will be
\begin{equation} \label{EM-CriticalPointsLocationA}
	\rho_i = \left\{ 
	\begin{array} {ll}
		\frac{1}{2n_i^2}  & i\le K~, \\
		0	& i >K~.
	\end{array}
	\right.
\end{equation}
The two blue dashed lines we found  in Fig.\ref{EM2DPhase} ($ n_1= \frac{1}{2}$ or $n_2 = \frac{1}{2}$)  for $N=2$ are special cases of this argument and we  expect that a saddle point vanishes on the left (bottom) of these lines. Using   the value of $\rho_i$ we found in Eq.~\eqref{EM-CriticalPointsLocationA} we can easily  determine whether these points are minima, maxima or saddle points .

\subsection{Divergence of the number of dS vacua for $N\ge3$}
\label{EM-sub-InfiniteVacua}
One of the most interesting features of this landscape is that not also is the possible number of AdS vacua  infinite for $N=2$, but also for larger $N$  they accommodate an infinite number of dS vacua. In this section we prove this result for the case $N=3$.  It will be clear from this proof that the larger $N$'s also have an infinite number of dS minima. Clearly, for any finite value for the fluxes $n_1$, $n_2$ and $n_3$ there is a finite volume in  parameter space. Therefore if we want to find an infinite volume, these numbers should get very large. However,  from Eq.~\eqref{bound}, it is not possible for all of them to get large. Let's assume that $n_1$ remains finite and $n_2$ and $n_3$ get large. We are looking for dS minima. Using the argument in Sec.\ref{EM-sub-BoundsOnFlux} and Eqs.~\eqref{bound} and \eqref{EM-HighNMinkowski}, $n_1$ should remain very close to $1/2$. We use the following parameterization for $n_1$ and $n_2$ and $n_3$:
\begin{eqnarray} \label{KDefinition}
	n_1 &=& \frac{1}{2}+  \epsilon^2~,  \nonumber\\
	n_2&=& \frac{K}{\epsilon}~,\nonumber \\
	n_3&=& \frac{L}{\epsilon}~,
\end{eqnarray}
where again $\epsilon$ is a small positive number and $K$ and $L$ are positive numbers that should be chosen in such a way that   ${1 \over n_1^2} + {1 \over n_2^2} + \frac{1}{n_3^2}\le 4$. To the first non-zero approximation in $\epsilon$, the location of the minima is given by
\begin{eqnarray} \label{EM-N3Rhos}
	\rho_1 &=& 2 + A \epsilon^2 ~, \cr
	\rho_2 &=& B \epsilon^2 ~, \cr
	\rho_3 &=& C \epsilon^2 ~.
\end{eqnarray}
In order to satisfy Eq.~\eqref{Critical},  $A$, $B$ and $C$ satisfy 
\begin{eqnarray} \label{EM-EquationImpossible}
	12+A-B-C+B^2 K^2+C^2 L^2&=&0 \nonumber ~,\\
	4 - 2 B - C + 3 B^2 K^2 + C^2 L^2&=&0 \nonumber ~,\\
	4-B-2 C+B^2 K^2+3 C^2 L^2&=&0 ~.
\end{eqnarray}

Unfortunately, it is not very easy to solve these equations analytically and analyze the minima as we could do in the case of two two-spheres. However, we can prove that the volume in the $n$ space that creates dS vacua is infinite in contrast to the $N=2$ case. In the case of two two-spheres, we found a finite number of de Sitter vacua because the area between the graphs in Eqs.~\eqref{EM-GreenTail} and \eqref{EM-BlueTail} was finite. However, if we rotate  these two graphs around the $n_1$ axis, the integral in Eq.~\eqref{EM-WhyFiniteForN2} diverges logarithmically. To make this argument more precise, let's adopt a cylindrical coordinate system as follows:
\begin{equation} \label{Cylinder2}
	n_1 = n_1,  \qquad n_2= r \cos \theta,  \qquad n_3 = r \sin \theta~.
\end{equation}
The two graphs that specify the boundary of the region containing the dS vacua are 
\begin{eqnarray}
	n_1 &=& \frac{1}{2} \left[ 1 +\frac{1}{8 \sin^2 \theta \cos^2 \theta}  \frac{1}{r^2} \right]~,  \cr\cr
	n_1 &=& \frac{1}{2} \left[ 1 + \frac{G(\theta)}{r^2}\right]~,
\end{eqnarray}
where the first one is a result of Eq.~\eqref{EM-HighNMinkowski} and the second one comes from Eq.~\eqref{EM-EquationImpossible}. Our  numerical solutions show that 

$$F(\theta) = G(\theta) - \frac{1}{8 \sin^2 \theta \cos^2 \theta} >0. $$
Therefore the volume of phase space for the de Sitter vacua is 
\begin{equation}
	\int_{r_*}^\infty \int_0^{\pi/2} r dr d\theta \frac{F(\theta)}{r^2} =\int_{r_*}^\infty dr \frac{1}{r}\int_0^{\pi/2}d\theta F(\theta)~.
\end{equation}
This integral is logarithmically divergent and therefore the number of dS vacua is infinite. A similar argument is correct for higher $N$ and our conjecture is that this integral for different $N$'s gets a modification of the form 
\begin{equation}
	\int_{r_*}^\infty \int_0^{\pi/2} r^{N-1} dr d\Omega_{N-2}\frac{F(\theta)}{r^2} =\int_{r_*}^\infty dr \frac{1}{r^{N-3}}\int_0^{\pi/2}d\theta F(\theta)~.
\end{equation}
This expression is more divergent for larger $N$'s. Therefore in all of them we will have an infinite number of dS vacua and also an infinite number of AdS vacua with arbitrarily large negative cosmological constants. 

\section{Summary and future directions}
\label{EM-sec-Summary}
In this chapter  we studied compactifications of the $(4+2N)$-dimensional Einstein-Maxwell theory over a product of a four-dimensional Lorentzian space and   two-spheres.  We calculated the four-dimensional effective potential. The landscape of vacua for $N=2$ is completely  mapped out. We showed that the number of dS vacua in this case is finite and that there are an infinite number of AdS vacua. We studied the minima of this theory in the approximation that one of the fluxes gets very large. Although we could not completely study the landscape for higher $N$'s, we still could derive some of its important features. Especially we showed that there are an infinite number of dS vacua for higher $N$'s.

There are many things left here that we will pursue in future works. One of them is the stability of these spheres for higher $\ell$ modes. We also will check whether this model of compactification becomes  unstable for large $N$, as we could show for field theoretical models in Chapter  \ref{Tumbling}.

\clearpage

\chapter{Conclusion}
\label{section:conclusion}
The Universe can undergo large-scale phase transitions. When the hot and dense Universe got colder, there was a  chance that it did not land  on the true global minimum of the potential describing it. The way to move towards a more stable vacuum is a phase transition which is carried by nucleation of  bubbles of the new phase.  The  subsequent expansion may complete the phase transition. 
In this thesis I explained aspects of this decay. 

In Chapter \ref{section:intro} I laid the foundation for the subsequent chapters. I described the vacuum decay in a field theory.

In Chapter \ref{O(3)} I explained the bubbles which have an O(3) $\times$ O(2) symmetry, which is different from the conventionally used O(4) symmetry. I showed that these bubbles have higher actions and therefore are subdominant in the decay process.

 In Chapter \ref{VectorBubble} I showed the results for the decay of a spatial vector field theory with different transverse and longitudinal speeds of sound. I showed that in some limits the flat walls get unstable and the bubbles develop kinks.
 
 In Chapter \ref{Tumbling} I studied the effects of large numbers of fields on the decay rate. I  showed that the decay rates grow so quickly as to render the overwhelming majority of the vacua unstable, and therefore not  good candidates for our observed Universe.

 In Chapter \ref{EMLandscape} I presented a model which shares many of the features of the string landscape and makes a good lab for testing the assumptions made in Chapter \ref{Tumbling}. We will investigate the consequences of large number of fields in this model in  upcoming work.
\clearpage

 \appendix
 \chapter{de Sitter space}
\label{Appendix-DeSitter}
\begin{figure}[htbp] 
   \centering
   \includegraphics[width=2in]{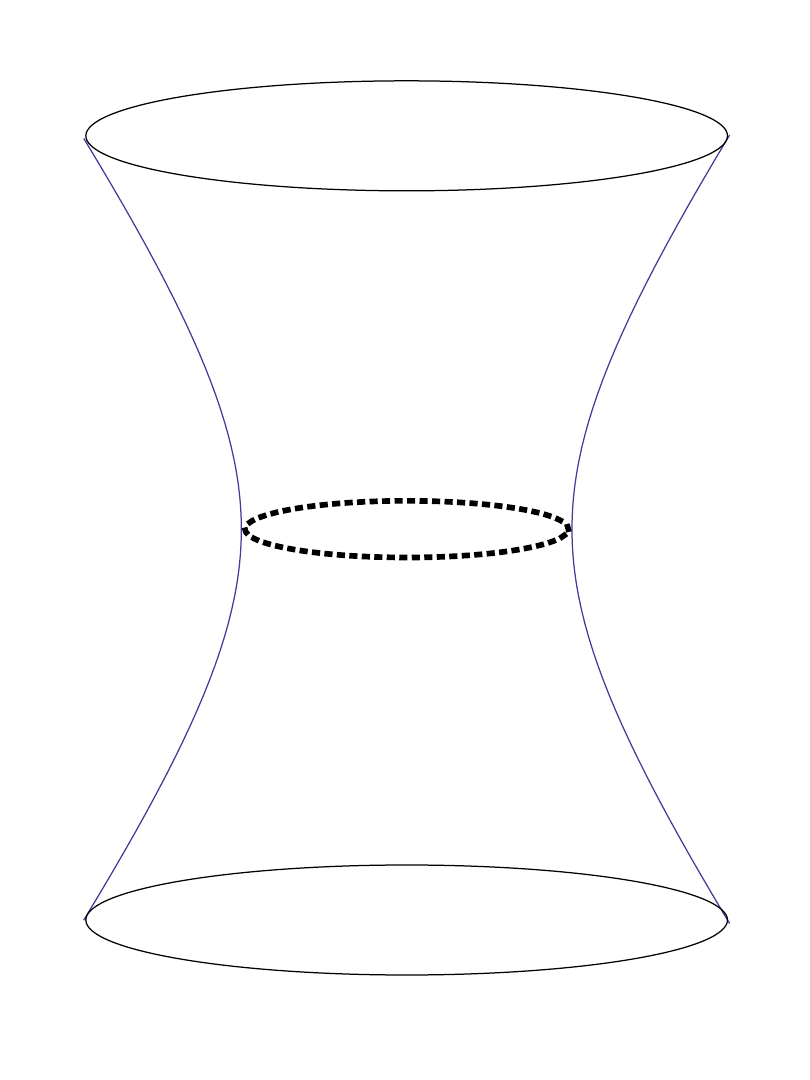} 
   \caption{A schematic view of the de Sitter space. Each of the points here is a (d-2)-sphere. The dashed line shows a (d-1)-sphere at the end of the contracting and beginning of the expanding phase.}
   \label{Appendix-ds-FirstPic}
\end{figure}

The d-dimensional de Sitter space ${\rm dS}_d$ can be thought as a hypersurface in a d+1-dimensional Minkowski space defined by \cite{Spradlin:2001pw}

\begin{equation}\label{AppDsAug27}
	-X_0^2 + X_1 + \ldots + X_d^2 = \ell^2~,
\end{equation}
where $\ell$ is the de Sitter radius.   This space satisfies the Einstein equation with a cosmological constant related to $\ell$ by
\begin{equation}
	\Lambda = {(d-2) (d-1) \over 2 \ell^2}~.
\end{equation}
Because de Sitter space has a horizon, it has a temperature \cite{Bekenstein:1973ur,Hawking:1974sw} which is given by
\begin{equation}
	T_{\rm ds}  = {1 \over 2 \pi \ell}~.
\end{equation}
We now set $\ell=1$. If we assume $X_0$ is the time direction, then we can foliate the spacetime with constant time hypersurfaces. Each of these hypersurfaces has the geometry of a (d-1)-sphere of radius $\sqrt{1 + X_0^2}$. This  foliates the space with    first contracting and then expanding spheres. This space has a $SO(d,1)$ symmetry group.  A simple picture of this space is shown in Fig.\ref{Appendix-ds-FirstPic}
There are many different coordinates systems used for de Sitter space. Here we introduce the two which are needed in the thesis. The first one is the global coordinate system which describes the above embedding. Let's devise a spherical coordinate system for the space-like  directions 
\begin{eqnarray}
	&&\omega^1 = \cos \theta_1 ~, \cr
	&&\omega^2 = \sin \theta_1 \cos \theta_2 ~, \cr 
	&&\vdots \cr
	&&\omega^{d-1} = \sin \theta_1 \ldots \sin \theta_{d-2} \cos \theta_{d-1} ~, \cr
	&&\omega^{d} = \sin \theta_1 \ldots \sin \theta_{d-2} \sin \theta_{d-1} ~.
\end{eqnarray}
Here $0 \le \theta_i \le \pi$ for $1\le i <d-1$ and $0\le \theta_{d-1}<2\pi$. In terms of these variables, we can put a global  coordinate system on the de Sitter space 
\begin{eqnarray}
	X_0 = \sinh \tau ~, \cr
	X_i = \omega_i \cosh \tau~.
\end{eqnarray}
The metric takes a very simple form
\begin{equation}
	ds^2 = - d \tau^2 + \cosh^2 \tau d\Omega_{d-1}^2~.
\end{equation}

\begin{figure}[htbp] 
   \centering
   \includegraphics[width=2in]{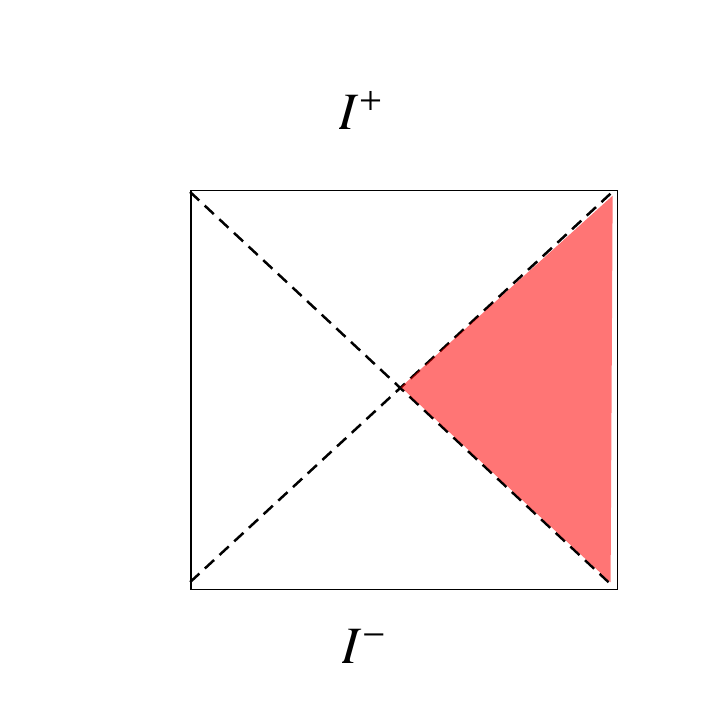} 
   \caption{Penrose diagram for the de Sitter space. The vertical solid lines are time-like points corresponding to the points on the north and south poles. As seen, it takes an infinite amount of time for a signal to travel between the poles and the north pole is only causally connected to the shaded region which is called the static patch.}
   \label{Appendix-ds-SecondPic}
\end{figure}

The Penrose diagram for  de Sitter space is shown in Fig.\ref{Appendix-ds-SecondPic}. Not all the points are causally connected and for any point there is a horizon. The shaded area in this picture shows all the points that can send and receive signals from the north pole (the solid vertical line on the right).  We can choose a coordinate system which only covers this shaded region
\begin{eqnarray}
	X_0 &=& \sqrt{1-r^2 } \sinh t ~, \cr
	X^a &=& r \omega^a ~, \qquad a=1, \dots , d-1 ~, \cr
	X_d &=& \sqrt{1-r^2 }\cosh t ~.
\end{eqnarray}
In this coordinate system the metric takes a very simple form
\begin{equation}
	ds^2 = -(1-r^2) dt^2 + {dr^2 \over 1-r^2} + r^2 d \Omega_{d-2}^2~.
\end{equation}
This is a static metric and $r=\pm1$ denotes the boundaries of the shaded area. Clearly the Euclidean version of the space given  in Eq.\eqref{AppDsAug27} is a d-sphere.

\chapter{Examples of potentials with various  orientation dependence of the tension. }
\label{sec-examples}

Here we present some analytic approaches and the numerical evaluation of the domain-wall tension given by the relaxation method\cite{AguJoh09a,GibLam10,AhlGre10}.  The first potential we study is a double well in one direction and a quadratic in the other direction.
\begin{equation} \label{eq-DoubleWell}
V(\phi_x, \phi_y)=  - \frac{1}{2}\mu^2  \phi_y^2 + \frac{1}{4} \lambda \phi_y^4 + \frac{1}{2} \beta \phi_x^2~.
\end{equation}
This is qualitatively similar to Eq.~(\ref{eq-potmot}) when $m^2>0$.  The differences are some 4th order terms involving $\phi_x$, which is not very important when $\beta>0$ stabilizes a trajectory near $\phi_x=0$.  The two potentials can be roughly related by
\begin{eqnarray}
-\frac{\mu^2}{2} &=& \frac{m^2}{2}+b|\vec{H}|^2~, \nonumber \\
\frac{\beta^2}{2} &=& \frac{m^2}{2}~.
\end{eqnarray}
The two degenerate minima sit at $(0,\pm\sqrt{\frac{\mu^2}{\lambda}})$.
For the purely longitudinal (or transverse) wall oriented along the $\vec{y}$ (or $\vec{x}$), we can solve the problem analytically and get the exact value of the tension. 

Longitudinal wall ($\theta = 0$):
\begin{align}
	& \phi_y(x,y) = \sqrt{\frac{\mu}{\lambda}} \tanh \left(\frac{\mu y}{c_L \sqrt{2}}\right) \\
	& \phi_x(x,y) = 0 \\
	& \sigma = \sigma(0) = \frac{2 \sqrt{2} \mu^3 c_L}{3 \lambda}~.
\end{align}
Transverse wall ($\theta = \frac{\pi}{2}$):
\begin{align}
	& \phi_y(x,y) = \sqrt{\frac{\mu}{\lambda}} \tanh \left(\frac{\mu x}{c_T \sqrt{2}}\right) \\
	& \phi_x(x,y) = 0 \\
	&\sigma = \sigma(\pi/2) = \frac{2 \sqrt{2} \mu^3 c_T}{3 \lambda}~.
\end{align}

For other orientations of the wall, we can evaluate the tension numerically.  
Before that, we can analyze two extreme cases.  Using the method of rotating the potential as described in Sec.\ref{sec-orient}, we have
\begin{eqnarray}
\sigma(\theta) &=& 
\int dx ~ \frac{c_T^2}{2}\phi_x'^2 + \frac{c_L^2}{2}\phi_y'^2
+V_\theta(\phi_x,\phi_y) \nonumber \\
&=& \int dx ~ \frac{c_T^2}{2}\phi_x'^2 + \frac{c_L^2}{2}\phi_y'^2
+V(\phi_x\cos\theta+\phi_y\sin\theta,\phi_y\cos\theta-\phi_x\sin\theta)~.
\end{eqnarray}
When $\beta\rightarrow\infty$, we effectively have a single field problem with
\begin{equation}
\bar\phi = \frac{\phi_x}{\cos\theta} = \frac{\phi_y}{\sin\theta}~,
\end{equation}
such that
\begin{equation}
\sigma(\theta) = 
\int dx ~ \frac{c_L^2\cos^2\theta + c_T^2\sin^2\theta}{2}\bar\phi'^2
+V(\bar\phi,0)~.
\end{equation}
Clearly, this gives us Eq.~(\ref{eq-sigmamotion}).

\begin{figure}
   \centering
   \includegraphics[width=3in]{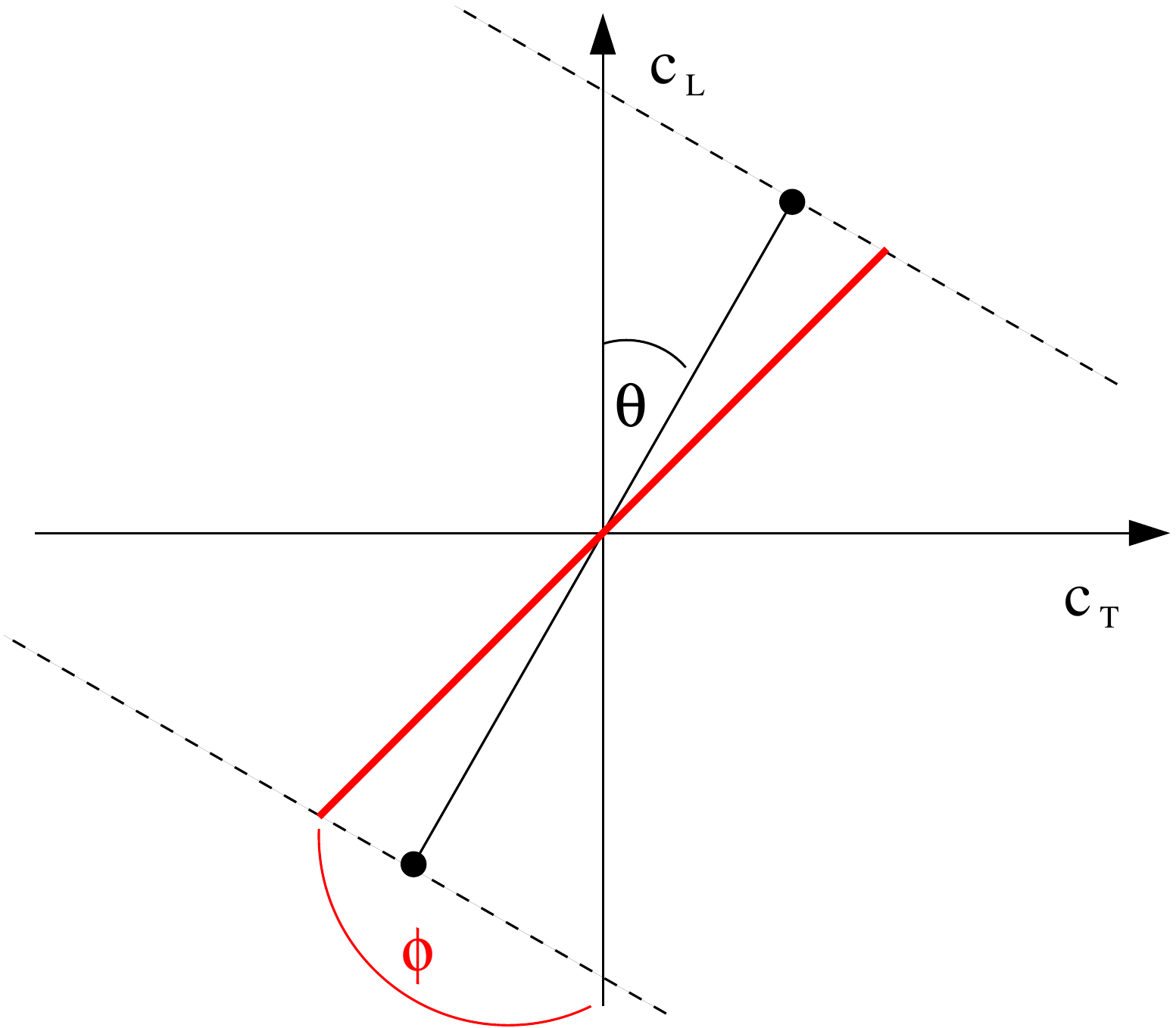} 
   \caption{The $c_L$ and $c_T$ axes are the directions in which the field has purely longitudinal and transverse sound speeds.  The two dots represent the two discrete vacua.  In the limit $\beta\rightarrow0$, the dashed lines through them are almost in the vacuum, too.  The important portion of the domain-wall is the red (thick) path from one line to the other, which is free to pick the best orientation $\phi$.}
   \label{fig-path}
\end{figure}

The other extreme limit is $\beta\rightarrow0$, in which the potential is flat in the $\phi_x$ direction.  The two degenerate vacua approach two separated lines.  Moving along these lines contributes nothing to the tension.  As shown in Fig.\ref{fig-path}, the path that minimizes the tension involves first moving along these lines to an appropriate angle $\phi$, then connecting them through a straight line.  The tension of this path is a function of both $\theta$ and $\phi$ through the orientation dependence in Eq.~(\ref{eq-sigmamotion}), and a simple projection of the length.
\begin{equation}
\sigma(\theta,\phi) = \frac{\sigma(0)}{c_L}\frac{1}{\cos(\theta-\phi)}
\sqrt{c_L^2\cos^2\phi + c_T^2\sin^2\phi}~.
\end{equation}
Minimizing this with $\phi$, we have
\begin{equation}
\phi_m(\theta) = \arccos \frac{c_T^2\cos\theta}
{\sqrt{c_T^4\cos^2\theta+c_L^4\sin^2\theta}}~,
\end{equation}
so the tension in this case should be
\begin{equation}
\sigma(\theta) = \sigma[\theta,\phi_m(\theta)]~.
\label{eq-sigmaextreme}
\end{equation}

We can apply the analysis in Sec.\ref{sec-kink} and calculate the stability condition for the flat longitudinal wall:
\begin{equation}
\frac{1}{\sigma(0)}\frac{d^2\sigma}{d\theta^2}
=\left(1-\frac{c_L^2}{c_T^2}\right)>-1~.
\end{equation}
We can see that the wall becomes unstable as soon as $c_L>\sqrt{2}c_T$.

\begin{figure}
   \centering
   \includegraphics[width=3in]{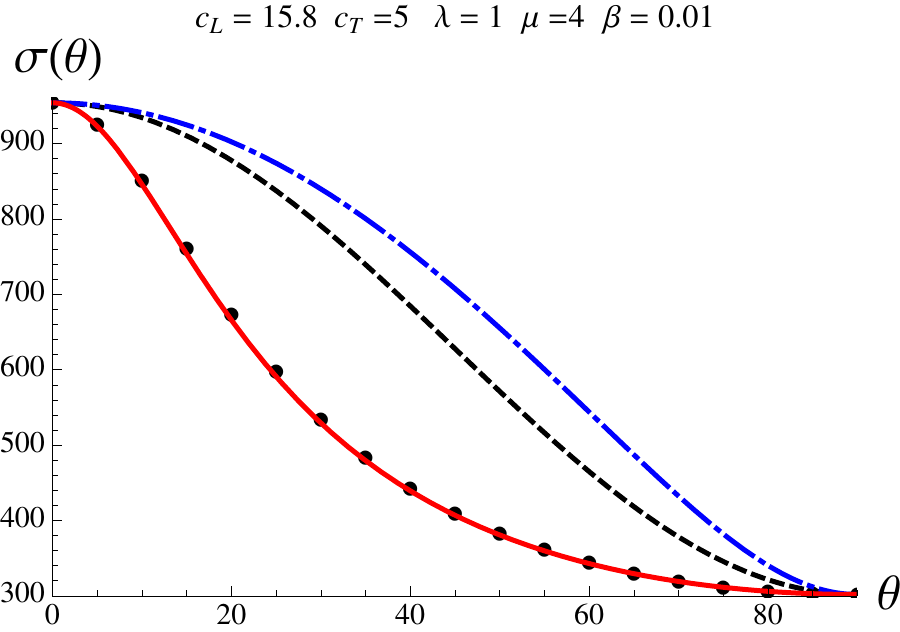} 
   \includegraphics[width=3in]{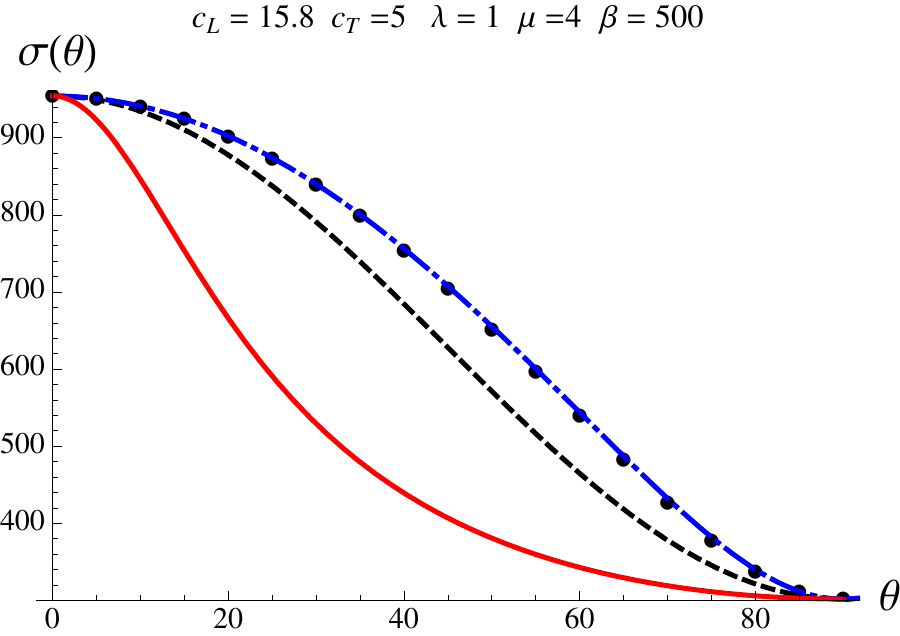} 
   \caption{The numerically calculated values of the tension for a double-well potential are shown in dots.  The three analytical fits: Eq~.(\protect\ref{eq-sigma}) is the dashed line, Eq.~(\protect\ref{eq-sigmamotion}) is the dot-dashed (blue) line, and Eq.~(\protect\ref{eq-sigmaextreme}) is the solid (red) line.  We can see that  for large $\beta$, Eq.~(\protect\ref{eq-sigmamotion}) is a good fit, and for small $\beta$, Eq.~(\protect\ref{eq-sigmaextreme}) is a good fit.}
   \label{fig-extreme}
\end{figure}

\begin{figure}
   \centering
   \includegraphics[width=3in]{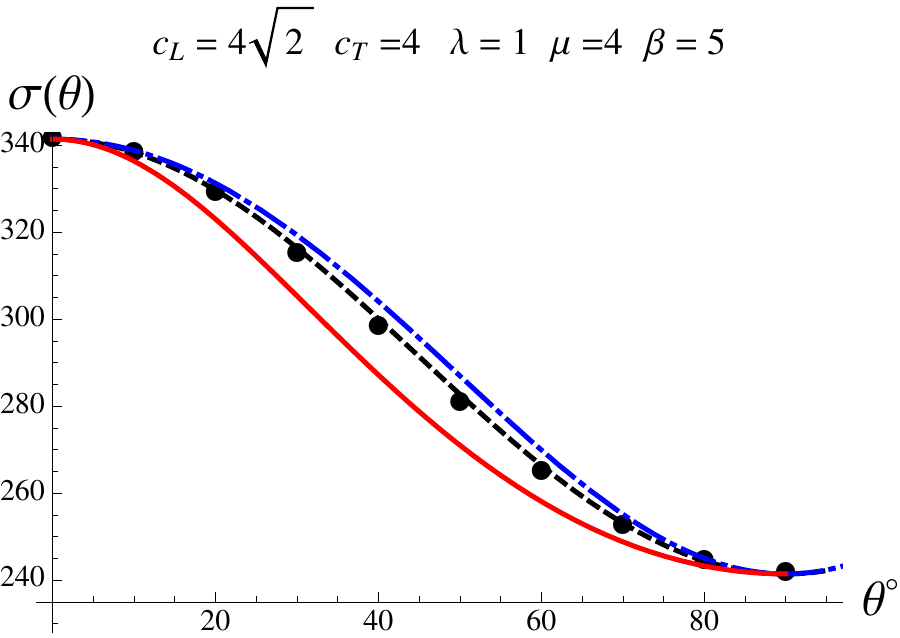} 
   \includegraphics[width=3in]{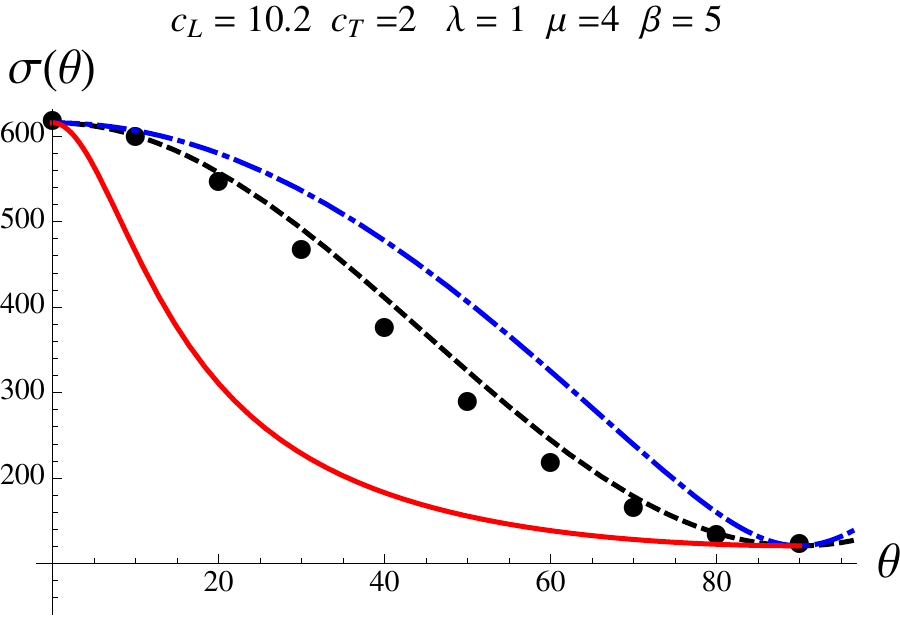} 
   \caption{The numerically calculated values of the tension for a double-well potential are shown in dots.  Again the three analytical fits: Eq.~(\protect\ref{eq-sigma}) is the dashed line, Eq.~(\protect\ref{eq-sigmamotion}) is the dot-dashed (blue) line, and Eq.~(\protect\ref{eq-sigmaextreme}) is the solid (red) line.  The two figures use the same potential but different sound speed ratios.  }
   \label{DoubleWellNumericalTension}
\end{figure}

We next provide several plots with the numerical values on top of the three possible fits, Eqs.~(\ref{eq-sigma}), (\ref{eq-sigmamotion}) and (\ref{eq-sigmaextreme}).  Fig.\ref{fig-extreme} shows that the two extreme limits indeed fit very well with our analysis.  Fig.(\ref{DoubleWellNumericalTension}) shows that with a more moderate choice of parameters, Eq.~(\ref{eq-sigma}) is quite reliable, independent of the  sound speed.

\begin{figure}
   \centering
   \includegraphics[width=10cm]{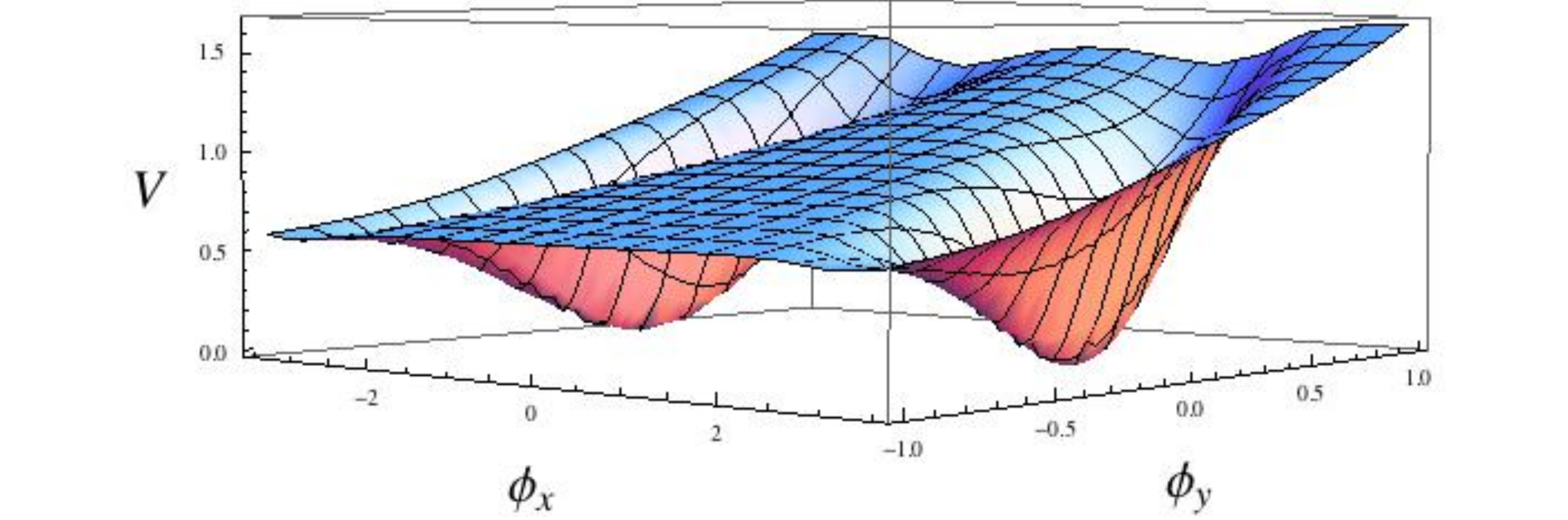} 
   \caption{The more complicated potential introduced in Eq~.(\protect\ref{ToyPotential}), with $q=0.5~, \ r_1 = -2.5~, \ r_2=2.5~.$  We will use two values of $S$, $1.1$ and $0.9$, but that makes no visual difference.}
   \label{WeirdPotential}
\end{figure}

In the end, we provide a much more complicated potential as in Fig.~(\ref{WeirdPotential}).  It has a general slope in the $\phi_x$ direction and two minima located at $(0.001,\pm2.498)$.  So, trivially, the interpolation path will always involve both fields.
\begin{equation}\label {ToyPotential}
	V(\phi_x,\phi_y) = e^{q \phi_x} \left\{1 - S \exp\left[ -4 \left(\phi_x - \sin (\frac{\phi_y-r_1}{r_2-r_1})\right)^2\right] \right\} \left[ \tanh^2\left( \frac{(\phi_y-r_1)(\phi_y-r_2)}{3}\right)\right]~.
\end{equation}

\begin{figure}[htbp] 
   \centering
   \includegraphics[width=3in]{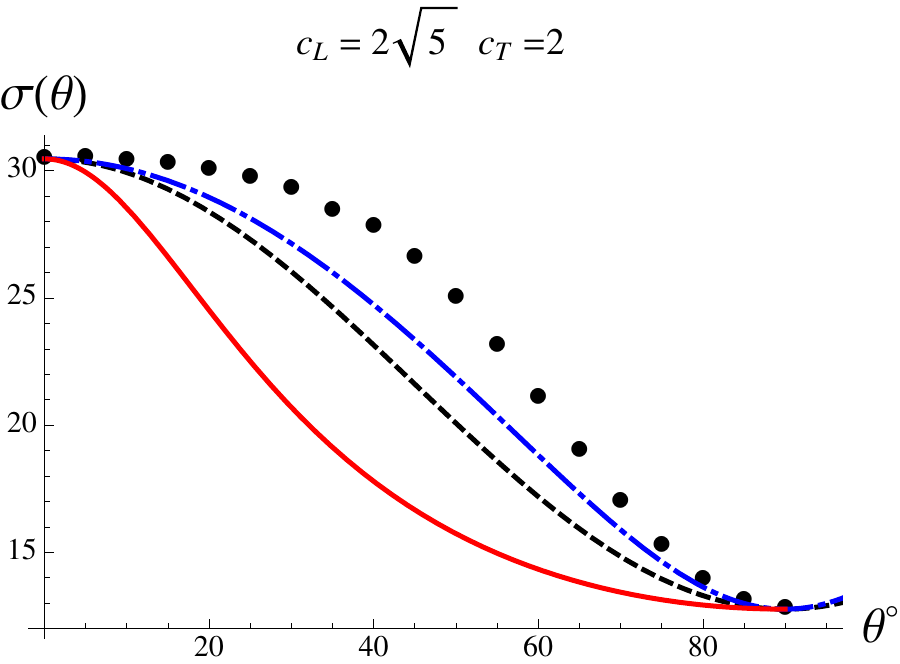} 
   \includegraphics[width=3in]{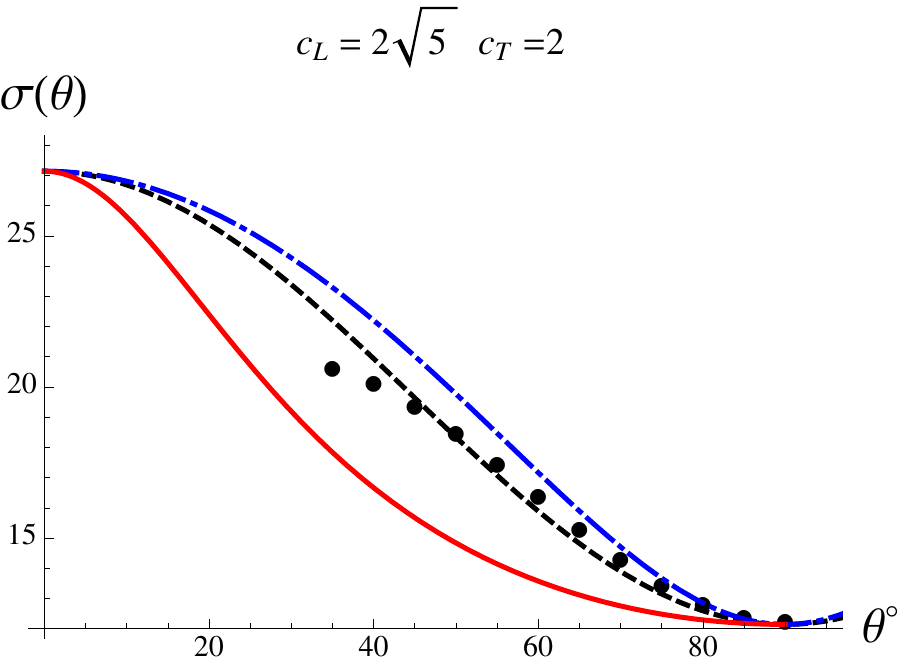}
   \caption{From the potential given by Eq~.(\protect\ref{ToyPotential}), we again compare the numerical $\sigma(\theta)$ with the three equations.  In the left figure we have $S=1.1$. In the right figure we have $S=1$. For some orientations the domain-wall does not exist because the path runs away toward the $-\phi_x$ direction.}
   \label{fig:example}
\end{figure}

We numerically evaluated the tension for various orientations and plotted it against the three analytical fits in Fig.~(\ref{fig:example}).   The overall shape can be quite different from any equation given in this paper.  In particular, note that in the right portion of Fig.~(\ref{fig:example}) the longitudinal domain-wall (actually an open set near $\theta=0$) does not exist.\footnote{This is for  the same reason as described in \cite{AguJoh09a}.  The interpolation path breaks into two parts, connecting each vacuum individually with the $-\phi_x$ region.  For vector fields, such runaway behavior also acquires an orientation dependence.} 

\chapter{Curvature of warped product of metrics.}
\label{Appendix-WarpedMetric}

In this appendix, we calculate the curvature tensor for a warped product of metrics. These results were used in Chapter \ref{EMLandscape} to calculate the effective potential of the lower dimensional spacetimes after compactification. We do these calculations in two different parameterizations of the metric. 
Consider the warped product of manifolds $\mathcal{M}\times \mathcal{N}^1\times \ldots \mathcal{N}^p$ which are respectively $m$, $n_1,n_2 \ldots$ and $n_p$ dimensional. The coordinates of  $\mathcal{M}$ are  $x_\mu$ and $y^{(a)}_i$ are the coordinates of $N^{(a)}$.

\section{Parametrization 1}
The metric in the first parameterization is 
\begin{equation}
	ds^2 = G_{MN} dx^M dx^N = e^{-2 A(x)} g_{\mu \nu}(x) dx^\mu dx^\nu + e^{2 B_a(x)} h^{(a)}_{ij}\left(y_{(a)}\right) dy_{(a)}^i dy_{(a)}^j ~,
\end{equation}
where the index $a$ runs from $1$ to $p$. Latin indices from beginning of the alphabet  specify which manifold $\mathcal{N}$ is used.
The nonzero components of the affine connection are

\begin{align}
	&\Gamma^{\mu}_{\nu\alpha}= \tilde{\Gamma}^\mu_{\nu \alpha}- 2 \delta^\mu\,_{(\alpha} \nabla_{\nu)}A + g_{\nu \alpha}\nabla^\mu A \\
	&\Gamma^\mu_{i_a j_a} = - e^{2(A+B_a)} h^{(a)}_{i_a j_a}\nabla^\mu B_a\\
	&\Gamma^{i_a}_{j_a l_a} = \tilde{\Gamma}^{i_a}_{j_al_a } \\
	&\Gamma^{i_a}_{\mu j_a} = \delta^{i_a}_{j_a} \nabla_\mu B_a 
\end{align}
All covariant derivatives with Greek indices are with respect to the affine connection obtained form $g_{\mu nu}$ . The nonzero components of the Riemann tensor are
\begin{align}
	&R^\mu_{\nu\alpha\beta}= \tilde{R}^\mu_{\nu\alpha\beta}+ 2 \delta^\mu\,_{[\alpha} \nabla_{\beta]} \nabla_\nu A+ 2 g_{\nu [\beta} \nabla_{\alpha]}\nabla^\mu A + 2 \delta^\mu \,_{[\alpha} \nabla_{\beta]} A \nabla_\nu A \nonumber \\
	& \qquad \qquad + 2 g_{\nu [\beta} \nabla_{\alpha ]} A \nabla^\mu A - 2 \delta^\mu \,_{[\alpha} g_{\beta]\nu} \left(\nabla A \right)^2~, \\
	&R^{i_a}_{j_ak_al_a} = \tilde{R}^{i_a}_{j_ak_al_a} -2 e^{2(A+ B_a)} \delta^{i_a} \,_{[k_a} h^{(a)}_{l_a]j_a}\left(\nabla B_a \right)^2 ~,\\
	&R^\mu_{i_a\nu j_a} = - h^{(a)}_{i_a j_a} e^{2(A+ B_a)} \left[ \nabla_\nu A \nabla^\mu B_a + \nabla_\nu B_a \nabla^\mu A + \nabla_\nu \nabla^\mu B_a + \nabla^\mu B_a \nabla_\nu B_a - \delta^\mu_\nu \nabla A \cdot \nabla B_a\right]~, \\
	&R^{i_a}_{\mu j_a \nu} = -\delta^{i_a}_{j_a} \left[ \nabla_\mu \nabla_\nu B_a + \nabla_\mu B_{a}\nabla_\nu A + \nabla_\nu B_a \nabla_\mu A + \nabla_\nu B_a \nabla_\mu B_a- g_{\mu \nu} \nabla_\rho B_a \nabla^\rho A \right]~, \\
	&R^{i_a}_{k_b j_a l_b} = - \delta^{i_a}_{j_a} h^{(b)}_{k_b l_b} e^{2(A + B_b)} \nabla B_a \cdot \nabla B_b~. \qquad \qquad (a\neq b)
 \end{align} 
The nonzero components of the Ricci tensor are
\begin{align}
	&R_{\mu \nu} = \tilde{R}_{\mu \nu} + (m-2)\left[  \nabla_\mu \nabla_\nu A + \nabla_\mu A \nabla_\nu A - g_{\mu \nu} \left(\nabla A\right)^2 \right] + g_{\mu \nu} \nabla^2 A \nonumber \\
	&\qquad \qquad - n_a \left[ \nabla_\mu \nabla_\nu B_a + \nabla_\mu B_a \nabla_\nu B_a + \nabla_\mu B_a  \nabla_\nu A + \nabla_\nu B_a \nabla_\mu A - g_{\mu \nu} \nabla A \cdot \nabla B_a\right]~, \\
	&R_{i_a j_a} = \tilde{R}_{i_a j_a} - e^{2(A+B_a)} h^{(a)}_{i_a j_a} \left[ - (m-2)\nabla A \cdot \nabla B_a + \nabla^2 B_a + n_b \nabla B_a \cdot \nabla B_b\right]~, 
\end{align}
and the scalar curvature is 
\begin{align}
	& R = e^{2 A} \tilde{R}^{(x)} + e^{-2 B_a} \tilde{R}^{(a)} + e^{2A} \left[ 2 (m-1) \nabla^2 A - (m-1)(m-2) \left(\nabla A\right)^2\right. \nonumber \\
	&\qquad \qquad\left.  - 2 n_a \nabla^2 B_a - n_a \left( \nabla B_a\right)^2 + 2 n_a (m-2)\nabla A \cdot \nabla B_a - n_a n_b \nabla B_a \cdot \nabla B_b\right]~.
\end{align}
The Laplacian operator on the scalar field $\phi(x, y_a)$ is 
\begin{equation}
	\nabla^2 \phi = e^{2 A} \nabla^2_{(x)} \phi + e^{-2 B_a} \nabla^2_{(a)} \phi + e^{2 B_a} \nabla_\mu \left[ n_a B_a - (m-2)A\right] \nabla^\mu \phi~.
\end{equation}
\section{Parametrization 2}
Another parametrization of the metric may be useful in other situations. Let's assume the warped metric has the  form

\begin{equation}
	ds^2 = G_{MN} dx^M dx^N = g_{\mu \nu}(x) dx^\mu dx^\nu +f_a(x) h^{(a)}_{ij}\left(y_{(a)}\right) dy_{(a)}^i dy_{(a)}^j ~.
\end{equation}
The non-zero components of the affine connection are
\begin{align}
	&\Gamma^{\mu}_{\nu\alpha}= \tilde{\Gamma}^\mu_{\nu \alpha}~, \\
	&\Gamma^\mu_{i_a j_a} = -\frac{1}{2} h^{(a)}_{i_a j_a} \nabla^\mu f_a~,\\
	&\Gamma^{i_a}_{j_a l_a} = \tilde{\Gamma}^{i_a}_{j_al_a } ~,\\
	&\Gamma^{i_a}_{\mu j_a} =\frac{1}{2} \delta^{i_a}_{j_a} f_a^{-1} \nabla_\mu f_a~.
\end{align}
The nonzero components of the Riemann  tensor are
\begin{align}
	&R^\mu_{\nu\alpha\beta}= \tilde{R}^\mu_{\nu\alpha\beta}~, \nonumber \\
	&R^{i_a}_{j_ak_al_a} = \tilde{R}^{i_a}_{j_ak_al_a}+\frac{1}{4}f_a^{-1} \left( \nabla f_a\right)^2 \left[ h_{j_a k_a} \delta^{i_a}_{l_a}-h_{j_a l_a} \delta^{i_a}_{k_a}  \right]~, \\
	&R^{\mu}_{i_a \nu j_a} = \frac{1}{2} h^{(a)}_{i_a j_a} \left[ \frac{1}{2}f_a^{-1} \nabla^\mu f_a \nabla_\nu f_a - \nabla^\mu \nabla_\nu f_a\right]~,\\
	&R^{i_a}_{\mu j_a \nu} =\frac{1}{2}\delta^{i_a}_{j_a} \left[ \frac{1}{2} f_a^{-2} \nabla_\mu f_a \nabla_\nu f_a - f_a^{-1} \nabla_\mu \nabla_\nu f_a \right] ~, \\
	&R^{i_a}_{k_b j_a l_b} = -\frac{1}{4} \delta^{i_a}_{j_a} f_a^{-1} h_{k_b l_b}^{(b)} \nabla f_a \cdot \nabla f_b~. \qquad \qquad (a\neq b)
 \end{align} 

The nonzero components of the Ricci tensor are 
\begin{align}
	&R_{\mu \nu} = \tilde{R}_{\mu \nu} + \frac{1}{2} n_a \left[ \frac{1}{2} f_a^{-2} \nabla_\mu f_a \cdot \nabla_\nu f_a - f_a^{-1} \nabla_\mu \nabla_\nu f_a\right] ~, \\
	&R_{i_a j_a} = \tilde{R}_{i_a j_a} - \frac{1}{2} h^{(a)}_{i_a j_a} \left[ \frac{1}{2} n_b f_b^{-1} \nabla f_a \cdot \nabla f_b - f_a^{-1} \left( \nabla f_a\right)^2 + \nabla^2 f_a\right]~,
\end{align}
and the scalar curvature is
\begin{align}
	& R = f_a^{-1} \tilde{R}^{(a)} + \tilde{R}^{(x)} - \frac{n_a n_b }{4} f_a^{-1} f_b^{-1} \nabla f_a \cdot \nabla f_b -n_a f_a^{-1} \nabla^2 f_a + \frac{3}{4} n_a f_a ^{-2} \left( \nabla f_a\right)^2~.
\end{align}
The Laplacian operator for a scalar field $\phi(x, y_a)$
\begin{align}
	&\nabla^2 \phi = \nabla^2_{(x)} \phi + f_a^{-1} \nabla_{(y_a)}^2 \phi + \frac{1}{2} n_a f_a^{-1} \nabla f_a \cdot \nabla \phi ~.
\end{align}

\chapter{Quartic supersymmetric potentials}
\label{SUSYAppendix}

In this appendix, we explain the supersymmetric potentials used in Chapter \ref{Tumbling} to model the dependence of tunneling rates on the number of fields. We use chiral superfields $\Phi_i$,
\begin{equation}
	\Phi(x,\theta) = A(x) + i \theta \sigma^m \bar{\theta} \partial_m A(x) + \frac{1}{4} \theta \theta \bar{\theta}\bar{\theta} \Box A(x) + \sqrt 2 \theta \psi(x) - \frac{i}{\sqrt 2} \theta \theta \partial_m \psi(x) \sigma^m \bar{\theta} + \theta \theta F(x)~.
\end{equation}

The general form of Lagrangian for a set of chiral fields $\Phi_1, \ldots, \Phi_N$ is \cite{wess1992supersymmetry}
\begin{eqnarray} 
	\mathcal{L} = \left.  \Phi_i \Phi_i^\dagger\right|_{\theta \theta \bar{\theta} \bar{\theta} {\rm }} + \left. \left[ \left( \frac{1}{2} m_{ij} \Phi_i \Phi_j + \frac{1}{3} g_{ijk} \Phi_i \Phi_j \Phi_k + \frac{1}{4} h_{ijmn} \Phi_i \Phi_j \Phi_m\Phi_n+  \lambda_i \Phi_i \right)\right|_{\theta \theta}+ {\rm h.c.}\right]~.
\end{eqnarray}
where $m$, $g$ and $h$ are symmetric tensors. In terms of the  component fields, this becomes 
\begin{align}\label{ChiralLagrangian}
	&\mathcal{L}= i \partial_m \bar{\psi}_i \bar{\sigma}^m \psi_i + A_i^* \Box A_i + F_i F_i^*  + \left[  m_{ij} \left(  A_i F_i - \frac{1}{2}\psi_i \psi_j \right) \right. \nonumber \\
	& \qquad  + g_{ijk} (A_i A_j F_k - \psi_i \psi_j A_k) + h_{ijmn} \left( F_i A_j A_m A_n - 3 A_i A_j \psi_m \psi_n\right)+  \lambda_i F_i+ { \rm h.c.} \bigg]~.
\end{align}
Varying with respect to the auxiliary fields $F_k$ and $F_k^*$, we find
\begin{eqnarray} \label{AuxiliaryFields}
	F_k &=& - \lambda_k^* - m_{ik}^* A_i^* - g^*_{ijk} A_i^* A_j^* - h^*_{ijmk} A^*_i A^*_j A^*_m ~, \nonumber \\
	F_k^* &=& - \lambda_k - m_{ik} A_i - g_{ijk} A_i A_j - h_{ijmk} A_i A_j A_m ~.
\end{eqnarray}
Plugging this back in Eq.\eqref{ChiralLagrangian} gives 

\begin{align}
	&\mathcal{L}= i \partial_m \bar{\psi}_i \bar{\sigma}^m \psi_i + A_i^* \Box A_i - \frac{1}{2} m_{ik} \psi_i \psi_k - \frac{1}{2} m^*_{ik} \psi_i \bar{\psi}_k - g_{ijk} \psi_i \psi_j A_k\nonumber \\
	& \qquad -g^*_{ijk} \bar{\psi_i} \bar{ \psi_j} A^*_k  - \mathcal{V} (A_i , A_j^*)~,
\end{align}
where
\begin{equation}
	\mathcal{V} = F_k^* F_k~.
\end{equation}
Using Eq.\eqref{AuxiliaryFields} gives
\begin{align} \label{AppendixSusyPotential1}
	&{\mathcal V} = \lambda_k \lambda^*_k + (\lambda^*_k m_{ik} A_i+ \lambda_k m^*_{ik} A^*_i)  \nonumber \\
	&\qquad + m_{ik} m^*_{jk}A_i A_j^* + \lambda^*_k g_{ijk} A_i A_j + \lambda_k g^*_{ijk} A^*_i A^*_j \nonumber \\
	&\qquad + m_{ik}g^*_{abk}A_i A^*_aA_b^*+ m^*_{ik}g_{abk}A_i^* A_aA_b + \lambda_k h^*_{ijmk} A_i^* A_j^*A^*_m+  \lambda^*_k h_{ijmk} A_i A_jA_m \nonumber \\
	&\qquad + g_{ijk} g^*_{abk} A_i A_j A^*_a A^*_b+ h^*_{ijmk} m_{nk} A^*_i A^*_j A^*_m A_n+ h_{ijmk} m^*_{nk} A_i A_j A_m A^*_n \nonumber \\
	&\qquad + g^*_{abk} h_{ijmk} A^*_a A^*_b A_i A_j A_m +  g_{abk} h^*_{ijmk} A_a A_b A^*_i A^*_j A^*_m \nonumber \\
	&\qquad + h_{ijmk} h^*_{abck} A^*_iA^*_jA^*_m A_a A_b A_c~. 
\end{align}

To make the linear part of the potential vanish, we need 
\begin{equation}
	\lambda^*_k=0~.
\end{equation}
Keeping only therm to quartic order in Eq.\eqref{AppendixSusyPotential1}
\begin{align} \label{Potential2}
	&{\mathcal V} =  m_{ik} m^*_{jk}A_i A_j^* +  m_{ik}g^*_{abk}A_i A^*_aA_b^*+ m^*_{ik}g_{abk}A_i^* A_aA_b \nonumber \\
	&\qquad  + g_{ijk} g^*_{abk} A_i A_j A^*_a A^*_b+ h^*_{ijmk} m_{nk} A^*_i A^*_j A^*_m A_n+ h_{ijmk} m^*_{nk} A_i A_j A_m A^*_n ~.
\end{align}
We can choose a basis $m_{ik} = \delta_{ik} M_k $. This further simplifies the potential
\begin{align} \label{Potential2}
	&{\mathcal V} = |M_k|^2 |A_k|^2 +  M_kg^*_{abk}A_k A^*_aA_b^*+ M^*_kg_{abk}A_k^* A_aA_b \nonumber \\
	&\qquad + g_{ijk} g^*_{abk} A_i A_j A^*_a A^*_b+ h^*_{ijmk} M_k A^*_i A^*_j A^*_m A_k+ h_{ijmk} M^*_k A_i A_j A_m A^*_k~.
\end{align}

Let's define $H_{ijmk} = h_{ijmk} M_k^*$. Please notice there is no summation here on $k$ and $H$ is still symmetric in $i,j,k$. Similarly let's define $G_{abk} = g_{abk} M_k^*$ where again there is no summation over $k$ and the tensor is still symmetric in $a,b$. Let's define  everything in terms of real quantities as follows 
\begin{eqnarray}
		A_{i} &=& \phi_i + i \pi_i~,  \cr  
		H_{ijmk}&=& F_{ijmk} + i L_{ijmk} ~,\cr
		G_{abc} &=& P_{abc} + i Q_{abc}~, \cr
		g_{abc} &=& R_{abc} +i S_{abc} ~.
\end{eqnarray}
The potential in Eq.\eqref{Potential2} becomes 
\begin{align} \label{SUSYPotentialExpansion}
	&{\mathcal V} = |M_k|^2 \phi_k^2 + |M_k|^2 \pi_k^2+2 P_{ijk} \left[ \phi_i \phi_j \phi_k - \pi_i \pi_j \phi_k +2 \phi_i \pi_j \pi_k\right] \nonumber \\
	&\qquad + 2 Q_{ijk} \left[ -\pi_i \pi_j \pi_k + \phi_i \phi_j \pi_k - 2 \pi_i \phi_j \phi_k\right] \nonumber \\
	&\qquad + \phi_i\phi_j\phi_m\phi_n \left[ R_{ijk} R_{mnk} + S_{ijk} S_{mnk} + 2 F_{ijmn}\right] + \pi_i \pi_j \pi_m \pi_n \left[ S_{ijk} S_{mnk}+ R_{ijk} R_{mnk} -2 F_{ijmn} \right]  \nonumber \\
	& \qquad +2\phi_i \phi_j \pi_m \pi_n \left[2 S_{imk} S_{njk} + 2 R_{imk}R_{njk} - R_{ijk}R_{mnk} - S_{ijk} S_{mnk}+3 F_{ijmn}- 3 F_{ijnm}\right] \nonumber \\
	& \qquad + 2\phi_i \pi_j \pi_m \pi_n \left[ 2R_{njk} S_{imk} -2 S_{njk}R_{imk}+ L_{ijnm}-3 L_{ijmn}\right] \nonumber \\
	& \qquad +2 \pi_i \phi_j \phi_m \phi_n  \left[ 2S_{njk} R_{imk} -2 R_{njk}S_{imk}+ L_{ijnm}-3L_{ijmn}\right]~.
\end{align}
This is the form we used for  the calculations in Chapter \ref{Tumbling}. We chose the real and imaginary parts of $g_{ijk}$ and $h_{ijkl}$ in the range $[-1,1]$.


%

\nocite{*}
\addcontentsline{toc}{chapter}{Bibliography}
\singlespace
\bibliography{refs}

\providecommand{\href}[2]{#2}\begingroup\raggedright\begin{thebibliography}{10}

\bibitem{Guth:1982pn}
A.~H. Guth and E.~J. Weinberg, {\em {Could the Universe Have Recovered from a
  Slow First Order Phase Transition?}\/},
\href{http://dx.doi.org/10.1016/0550-3213(83)90307-3}{Nucl.Phys. {\bf B212}
  (1983)  321}.

\bibitem{Winitzki:2006rn}
S.~Winitzki, {\em {Predictions in eternal inflation}\/},
  \href{http://dx.doi.org/10.1007/978-3-540-74353-8_5}{Lect.Notes Phys. {\bf
  738} (2008)  157--191},
\href{http://arxiv.org/abs/gr-qc/0612164}{{\tt arXiv:gr-qc/0612164 [gr-qc]}}.

\bibitem{Guth:2007ng}
A.~H. Guth, {\em {Eternal inflation and its implications}\/},
  \href{http://dx.doi.org/10.1088/1751-8113/40/25/S25}{J.Phys. {\bf A40} (2007)
   6811--6826},
\href{http://arxiv.org/abs/hep-th/0702178}{{\tt arXiv:hep-th/0702178
  [HEP-TH]}}.

\bibitem{Vilenkin:2006xv}
A.~Vilenkin, {\em {A Measure of the multiverse}\/},
  \href{http://dx.doi.org/10.1088/1751-8113/40/25/S22}{J.Phys. {\bf A40} (2007)
   6777},
\href{http://arxiv.org/abs/hep-th/0609193}{{\tt arXiv:hep-th/0609193
  [hep-th]}}.

\bibitem{Linde:2007fr}
A.~D. Linde, {\em {Inflationary Cosmology}\/},
  \href{http://dx.doi.org/10.1007/978-3-540-74353-8_1}{Lect.Notes Phys. {\bf
  738} (2008)  1--54},
\href{http://arxiv.org/abs/0705.0164}{{\tt arXiv:0705.0164 [hep-th]}}.

\bibitem{Coleman:1977th}
S.~R. Coleman, V.~Glaser, and A.~Martin, {\em {Action Minima Among Solutions to
  a Class of Euclidean Scalar Field Equations}\/},
\href{http://dx.doi.org/10.1007/BF01609421}{Commun.Math.Phys. {\bf 58} (1978)
  211}.

\bibitem{Loeffel:1970zz}
J.~Loeffel and A.~Martin,
{\em {Analytic properties of the anharmonic oscillator levels and convergence
  of the Pade approximations. }\/}, .

\bibitem{Loeffel:1970fe}
J.~Loeffel, A.~Martin, B.~Simon, and A.~Wightman, {\em {Pade approximants and
  the anharmonic oscillator}\/},
\href{http://dx.doi.org/10.1016/0370-2693(69)90087-2}{Phys.Lett. {\bf B30}
  (1969)  656--658}.

\bibitem{Bender:1969si}
C.~M. Bender and T.~T. Wu, {\em {Anharmonic oscillator}\/},
\href{http://dx.doi.org/10.1103/PhysRev.184.1231}{Phys.Rev. {\bf 184} (1969)
  1231--1260}.

\bibitem{Bender:1971gu}
C.~M. Bender and T.~T. WU, {\em {Large order behavior of Perturbation
  theory}\/},
\href{http://dx.doi.org/10.1103/PhysRevLett.27.461}{Phys.Rev.Lett. {\bf 27}
  (1971)  461}.

\bibitem{Bender:1973rz}
C.~M. Bender and T.~Wu, {\em {Anharmonic oscillator. ii. a study of
  perturbation theory in large order}\/},
\href{http://dx.doi.org/10.1103/PhysRevD.7.162}{Phys.Rev. {\bf D7} (1973)
  162O--1636}.

\bibitem{Banks:1973ps}
T.~Banks, C.~M. Bender, and T.~T. Wu, {\em {Coupled anharmonic oscillators. 1.
  Equal mass case}\/},
\href{http://dx.doi.org/10.1103/PhysRevD.8.3346}{Phys.Rev. {\bf D8} (1973)
  3346--3378}.

\bibitem{Banks:1974ij}
T.~Banks and C.~M. Bender, {\em {Coupled anharmonic oscillators. ii.
  unequal-mass case}\/},
\href{http://dx.doi.org/10.1103/PhysRevD.8.3366}{Phys.Rev. {\bf D8} (1973)
  3366--3378}.

\bibitem{Langer:1967ax}
J.~Langer, {\em {Theory of the condensation point}\/},
\href{http://dx.doi.org/10.1016/0003-4916(67)90200-X}{Annals Phys. {\bf 41}
  (1967)  108--157}.

\bibitem{Kobzarev:1974cp}
I.~Y. Kobzarev, L.~Okun, and M.~Voloshin, {\em {Bubbles in Metastable
  Vacuum}\/},
Sov.J.Nucl.Phys. {\bf 20} (1975)  644--646.

\bibitem{Coleman:1977py}
S.~R. Coleman, {\em {The Fate of the False Vacuum. 1. Semiclassical Theory}\/},
\href{http://dx.doi.org/10.1103/PhysRevD.15.2929,
  10.1103/PhysRevD.16.1248}{Phys.Rev. {\bf D15} (1977)  2929--2936}.

\bibitem{Callan:1977pt}
J.~Callan, Curtis~G. and S.~R. Coleman, {\em {The Fate of the False Vacuum. 2.
  First Quantum Corrections}\/},
\href{http://dx.doi.org/10.1103/PhysRevD.16.1762}{Phys.Rev. {\bf D16} (1977)
  1762--1768}.

\bibitem{coleman1988aspects}
S.~Coleman, {\em Aspects of Symmetry: Selected Erice Lectures}.
\newblock Cambridge University Press, 1988.
\newblock \url{http://books.google.com/books?id=PX2Al8LE9FkC}.

\bibitem{Affleck:1980ac}
I.~Affleck, {\em {Quantum Statistical Metastability}\/},
\href{http://dx.doi.org/10.1103/PhysRevLett.46.388}{Phys.Rev.Lett. {\bf 46}
  (1981)  388}.

\bibitem{Linde:1981zj}
A.~D. Linde, {\em {Decay of the False Vacuum at Finite Temperature}\/},
\href{http://dx.doi.org/10.1016/0550-3213(83)90293-6}{Nucl.Phys. {\bf B216}
  (1983)  421}.

\bibitem{Lee:1987qc}
K.-M. Lee and E.~J. Weinberg, {\em {DECAY OF THE TRUE VACUUM IN CURVED
  SPACE-TIME}\/},
\href{http://dx.doi.org/10.1103/PhysRevD.36.1088}{Phys.Rev. {\bf D36} (1987)
  1088}.

\bibitem{Hawking:1981fz}
S.~Hawking and I.~Moss, {\em {Supercooled Phase Transitions in the Very Early
  Universe}\/},
\href{http://dx.doi.org/10.1016/0370-2693(82)90946-7}{Phys.Lett. {\bf B110}
  (1982)  35}.

\bibitem{Coleman:1980aw}
S.~R. Coleman and F.~De~Luccia, {\em {Gravitational Effects on and of Vacuum
  Decay}\/},
\href{http://dx.doi.org/10.1103/PhysRevD.21.3305}{Phys.Rev. {\bf D21} (1980)
  3305}.

\bibitem{Jensen:1988zx}
L.~G. Jensen and P.~J. Steinhardt, {\em {BUBBLE NUCLEATION FOR FLAT POTENTIAL
  BARRIERS}\/},
\href{http://dx.doi.org/10.1016/0550-3213(89)90539-7}{Nucl.Phys. {\bf B317}
  (1989)  693--705}.

\bibitem{Hackworth:2004xb}
J.~C. Hackworth and E.~J. Weinberg, {\em {Oscillating bounce solutions and
  vacuum tunneling in de Sitter spacetime}\/},
  \href{http://dx.doi.org/10.1103/PhysRevD.71.044014}{Phys.Rev. {\bf D71}
  (2005)  044014},
\href{http://arxiv.org/abs/hep-th/0410142}{{\tt arXiv:hep-th/0410142
  [hep-th]}}.

\bibitem{Batra:2006rz}
P.~Batra and M.~Kleban, {\em {Transitions Between de Sitter Minima}\/},
  \href{http://dx.doi.org/10.1103/PhysRevD.76.103510}{Phys.Rev. {\bf D76}
  (2007)  103510},
\href{http://arxiv.org/abs/hep-th/0612083}{{\tt arXiv:hep-th/0612083
  [hep-th]}}.

\bibitem{Brown:2007sd}
A.~R. Brown and E.~J. Weinberg, {\em {Thermal derivation of the Coleman-De
  Luccia tunneling prescription}\/},
  \href{http://dx.doi.org/10.1103/PhysRevD.76.064003}{Phys.Rev. {\bf D76}
  (2007)  064003},
\href{http://arxiv.org/abs/0706.1573}{{\tt arXiv:0706.1573 [hep-th]}}.

\bibitem{WeinbergBook}
E.~J. Weinberg, {\em {Classical solutions in quantum field theory}}.
\newblock Cambridge University Press,
2012.
\newblock

\bibitem{Masoumi:2012yy}
A.~Masoumi and E.~J. Weinberg, {\em {Bounces with O(3) x O(2) symmetry}\/},
  \href{http://dx.doi.org/10.1103/PhysRevD.86.104029}{Phys.Rev. {\bf D86}
  (2012)  104029},
\href{http://arxiv.org/abs/1207.3717}{{\tt arXiv:1207.3717 [hep-th]}}.

\bibitem{Farhi:1989yr}
E.~Farhi, A.~H. Guth, and J.~Guven, {\em {IS IT POSSIBLE TO CREATE A UNIVERSE
  IN THE LABORATORY BY QUANTUM TUNNELING?}\/},
\href{http://dx.doi.org/10.1016/0550-3213(90)90357-J}{Nucl.Phys. {\bf B339}
  (1990)  417--490}.

\bibitem{Farhi:1986ty}
E.~Farhi and A.~H. Guth, {\em {AN OBSTACLE TO CREATING A UNIVERSE IN THE
  LABORATORY}\/},
\href{http://dx.doi.org/10.1016/0370-2693(87)90429-1}{Phys.Lett. {\bf B183}
  (1987)  149}.

\bibitem{Maeda:1981gw}
K.-i. Maeda, K.~Sato, M.~Sasaki, and H.~Kodama, {\em {CREATION OF DE
  SITTER-SCHWARZSCHILD WORMHOLES BY A COSMOLOGICAL FIRST ORDER PHASE
  TRANSITION}\/},
\href{http://dx.doi.org/10.1016/0370-2693(82)91151-0}{Phys.Lett. {\bf B108}
  (1982)  98}.

\bibitem{Samuel:1991dy}
D.~Samuel and W.~Hiscock, {\em {Effect of gravity on false vacuum decay rates
  for O(4) symmetric bubble nucleation}\/},
\href{http://dx.doi.org/10.1103/PhysRevD.44.3052}{Phys.Rev. {\bf D44} (1991)
  3052--3061}.

\bibitem{Berezin:1991qz}
V.~Berezin, V.~Kuzmin, and I.~Tkachev, {\em {O(3) invariant tunneling at false
  vacuum decay in general relativity}\/},
\href{http://dx.doi.org/10.1088/0031-8949/1991/T36/030}{Phys.Scripta {\bf T36}
  (1991)  269--275}.

\bibitem{Garriga:2004nm}
J.~Garriga and A.~Megevand, {\em {Decay of de Sitter vacua by thermal
  activation}\/},
  \href{http://dx.doi.org/10.1023/B:IJTP.0000048178.69097.fb}{Int.J.Theor.Phys.
  {\bf 43} (2004)  883--904},
\href{http://arxiv.org/abs/hep-th/0404097}{{\tt arXiv:hep-th/0404097
  [hep-th]}}.

\bibitem{Brown:1987dd}
J.~D. Brown and C.~Teitelboim, {\em {DYNAMICAL NEUTRALIZATION OF THE
  COSMOLOGICAL CONSTANT}\/},
\href{http://dx.doi.org/10.1016/0370-2693(87)91190-7}{Phys.Lett. {\bf B195}
  (1987)  177--182}.

\bibitem{Brown:1988kg}
J.~D. Brown and C.~Teitelboim, {\em {Neutralization of the Cosmological
  Constant by Membrane Creation}\/},
\href{http://dx.doi.org/10.1016/0550-3213(88)90559-7}{Nucl.Phys. {\bf B297}
  (1988)  787--836}.

\bibitem{Israel:1966rt}
W.~Israel, {\em {Singular hypersurfaces and thin shells in general
  relativity}\/},
\href{http://dx.doi.org/10.1007/BF02710419, 10.1007/BF02712210}{Nuovo Cim. {\bf
  B44S10} (1966)  1}.

\bibitem{Banados:1993qp}
M.~Banados, C.~Teitelboim, and J.~Zanelli, {\em {Black hole entropy and the
  dimensional continuation of the Gauss-Bonnet theorem}\/},
  \href{http://dx.doi.org/10.1103/PhysRevLett.72.957}{Phys.Rev.Lett. {\bf 72}
  (1994)  957--960},
\href{http://arxiv.org/abs/gr-qc/9309026}{{\tt arXiv:gr-qc/9309026 [gr-qc]}}.

\bibitem{Hawking:1995fd}
S.~Hawking and G.~T. Horowitz, {\em {The Gravitational Hamiltonian, action,
  entropy and surface terms}\/},
  \href{http://dx.doi.org/10.1088/0264-9381/13/6/017}{Class.Quant.Grav. {\bf
  13} (1996)  1487--1498},
\href{http://arxiv.org/abs/gr-qc/9501014}{{\tt arXiv:gr-qc/9501014 [gr-qc]}}.

\bibitem{Nariai1951}
H.~Nariai, {\em {}\/},  Science Reports of the Tohoku Univ. {\bf 35} (1951)
  62.

\bibitem{Ginsparg:1982rs}
P.~H. Ginsparg and M.~J. Perry, {\em {Semiclassical Perdurance of de Sitter
  Space}\/},
\href{http://dx.doi.org/10.1016/0550-3213(83)90636-3}{Nucl.Phys. {\bf B222}
  (1983)  245}.

\bibitem{Masoumi:2012aa}
A.~Masoumi, X.~Xiao, and I.-S. Yang, {\em {Bubble nucleation of spatial vector
  fields}\/},  \href{http://dx.doi.org/10.1103/PhysRevD.87.045008}{Phys.Rev.
  {\bf D87} (2013)  045008},
\href{http://arxiv.org/abs/1205.7052}{{\tt arXiv:1205.7052 [hep-th]}}.

\bibitem{Coleman:1987rm}
S.~R. Coleman, {\em {QUANTUM TUNNELING AND NEGATIVE EIGENVALUES}\/},
\href{http://dx.doi.org/10.1016/0550-3213(88)90308-2}{Nucl.Phys. {\bf B298}
  (1988)  178}.

\bibitem{GalFou95}
P.~Galatola and J.~B. Fournier, {\em New Features of Two-Dimensional Soft
  Matter Domains: Dips and Quasicusps\/},
  \href{http://dx.doi.org/10.1103/PhysRevLett.75.3297}{Phys. Rev. Lett. {\bf
  75} (1995)  3297--3300}.
  \url{http://link.aps.org/doi/10.1103/PhysRevLett.75.3297}.

\bibitem{Leg75}
A.~J. Leggett, {\em A theoretical description of the new phases of liquid
  $^{3}\mathrm{He}$\/},
  \href{http://dx.doi.org/10.1103/RevModPhys.47.331}{Rev. Mod. Phys. {\bf 47}
  (1975)  331--414}. \url{http://link.aps.org/doi/10.1103/RevModPhys.47.331}.

\bibitem{Whe75}
J.~C. Wheatley, {\em Experimental properties of superfluid
  $^{3}\mathrm{He}$\/},
  \href{http://dx.doi.org/10.1103/RevModPhys.47.415}{Rev. Mod. Phys. {\bf 47}
  (1975)  415--470}. \url{http://link.aps.org/doi/10.1103/RevModPhys.47.415}.

\bibitem{He3}
D.~Vollhardt and P.~Woelfle, {\em The Superfluid Phases Of Helium 3}.
\newblock Taylor \& Francis, 1990.
\newblock \url{http://books.google.com/books?id=t0Rw75gMuwIC}.

\bibitem{Fou95}
J.~B. Fournier, {\em Generalized Gibbs-Thomson Equation and Surface Stiffness
  for Materials with an Orientational Order Parameter\/},
  \href{http://dx.doi.org/10.1103/PhysRevLett.75.854}{Phys. Rev. Lett. {\bf 75}
  (1995)  854--857}. \url{http://link.aps.org/doi/10.1103/PhysRevLett.75.854}.

\bibitem{MacJia95}
J.~E. Maclennan, Q.~Jiang, and N.~A. Clark, {\em Computer simulation of domain
  growth in ferroelectric liquid crystals\/},
  \href{http://dx.doi.org/10.1103/PhysRevE.52.3904}{Phys. Rev. E {\bf 52}
  (1995)  3904--3914}. \url{http://link.aps.org/doi/10.1103/PhysRevE.52.3904}.

\bibitem{RudLoh99}
J.~Rudnick and K.-K. Loh, {\em Theory of monolayers with boundaries: Exact
  results and perturbative analysis\/},
  \href{http://dx.doi.org/10.1103/PhysRevE.60.3045}{Phys. Rev. E {\bf 60}
  (1999)  3045--3062}. \url{http://link.aps.org/doi/10.1103/PhysRevE.60.3045}.

\bibitem{SilPat06}
N.~M. Silvestre, P.~Patr\'icio, and M.~M. Telo~da Gama, {\em Elliptical soft
  colloids in smectic-$C$ films\/},
  \href{http://dx.doi.org/10.1103/PhysRevE.74.021706}{Phys. Rev. E {\bf 74}
  (2006)  021706}. \url{http://link.aps.org/doi/10.1103/PhysRevE.74.021706}.

\bibitem{RudBru95}
J.~Rudnick and R.~Bruinsma, {\em Shape of Domains in Two-Dimensional Systems:
  Virtual Singularities and a Generalized Wulff Construction\/},
  \href{http://dx.doi.org/10.1103/PhysRevLett.74.2491}{Phys. Rev. Lett. {\bf
  74} (1995)  2491--2494}.
  \url{http://link.aps.org/doi/10.1103/PhysRevLett.74.2491}.

\bibitem{AguJoh09a}
A.~Aguirre, M.~C. Johnson, and M.~Larfors, {\em {Runaway dilatonic domain
  walls}\/},  \href{http://dx.doi.org/10.1103/PhysRevD.81.043527}{Phys.Rev.
  {\bf D81} (2010)  043527},
\href{http://arxiv.org/abs/0911.4342}{{\tt arXiv:0911.4342 [hep-th]}}.

\bibitem{GibLam10}
J.~Giblin, John~T., L.~Hui, E.~A. Lim, and I.-S. Yang, {\em {How to Run Through
  Walls: Dynamics of Bubble and Soliton Collisions}\/},
  \href{http://dx.doi.org/10.1103/PhysRevD.82.045019}{Phys.Rev. {\bf D82}
  (2010)  045019},
\href{http://arxiv.org/abs/1005.3493}{{\tt arXiv:1005.3493 [hep-th]}}.

\bibitem{AhlGre10}
P.~Ahlqvist, B.~R. Greene, D.~Kagan, E.~A. Lim, S.~Sarangi, \textit{et al.},
  {\em {Conifolds and Tunneling in the String Landscape}\/},
  \href{http://dx.doi.org/10.1007/JHEP03(2011)119}{JHEP {\bf 1103} (2011)
  119},
\href{http://arxiv.org/abs/1011.6588}{{\tt arXiv:1011.6588 [hep-th]}}.

\bibitem{AleSur90}
I.~Aleiner and I.~Suris, {\em {The Shape and Activation Energy of Critical
  Two-dimensional Nuclei on the (001) Surface of a III-V Crystal during
  Epitaxial Growth,}\/},
Sov. Tech. Phys. Lett. {\bf 16} (1990)  547.

\bibitem{OshCro77}
D.~Osheroff and M.~Cross, {\em {Interfacial Surface Energy between the
  Superfluid Phases of He-3}\/},
\href{http://dx.doi.org/10.1103/PhysRevLett.38.905}{Phys.Rev.Lett. {\bf 38}
  (1977)  905--909}.

\bibitem{EasGib09}
R.~Easther, J.~Giblin, John~T., L.~Hui, and E.~A. Lim, {\em {A New Mechanism
  for Bubble Nucleation: Classical Transitions}\/},
  \href{http://dx.doi.org/10.1103/PhysRevD.80.123519}{Phys.Rev. {\bf D80}
  (2009)  123519},
\href{http://arxiv.org/abs/0907.3234}{{\tt arXiv:0907.3234 [hep-th]}}.

\bibitem{YanTye11}
I.-S. Yang, S.-H.~H. Tye, and B.~Shlaer, {\em {Classical Transitions in
  Superfluid $^3He$}\/},
\href{http://arxiv.org/abs/1110.2045}{{\tt arXiv:1110.2045 [cond-mat.other]}}.

\bibitem{Greene:2013ida}
B.~Greene, D.~Kagan, A.~Masoumi, E.~J. Weinberg, and X.~Xiao, {\em {Tumbling
  through a landscape: Evidence of instabilities in high-dimensional moduli
  spaces}\/},
\href{http://arxiv.org/abs/1303.4428}{{\tt arXiv:1303.4428 [hep-th]}}.

\bibitem{Strominger:1986uh}
A.~Strominger, {\em {Superstrings with Torsion}\/},
\href{http://dx.doi.org/10.1016/0550-3213(86)90286-5}{Nucl.Phys. {\bf B274}
  (1986)  253}.

\bibitem{Giddings:2001yu}
S.~B. Giddings, S.~Kachru, and J.~Polchinski, {\em {Hierarchies from fluxes in
  string compactifications}\/},
  \href{http://dx.doi.org/10.1103/PhysRevD.66.106006}{Phys.Rev. {\bf D66}
  (2002)  106006},
\href{http://arxiv.org/abs/hep-th/0105097}{{\tt arXiv:hep-th/0105097
  [hep-th]}}.

\bibitem{Polchinski:1995mt}
J.~Polchinski, {\em {Dirichlet Branes and Ramond-Ramond charges}\/},
  \href{http://dx.doi.org/10.1103/PhysRevLett.75.4724}{Phys.Rev.Lett. {\bf 75}
  (1995)  4724--4727},
\href{http://arxiv.org/abs/hep-th/9510017}{{\tt arXiv:hep-th/9510017
  [hep-th]}}.

\bibitem{Grana:2005jc}
M.~Grana, {\em {Flux compactifications in string theory: A Comprehensive
  review}\/},
  \href{http://dx.doi.org/10.1016/j.physrep.2005.10.008}{Phys.Rept. {\bf 423}
  (2006)  91--158},
\href{http://arxiv.org/abs/hep-th/0509003}{{\tt arXiv:hep-th/0509003
  [hep-th]}}.

\bibitem{Douglas:2006es}
M.~R. Douglas and S.~Kachru, {\em {Flux compactification}\/},
  \href{http://dx.doi.org/10.1103/RevModPhys.79.733}{Rev.Mod.Phys. {\bf 79}
  (2007)  733--796},
\href{http://arxiv.org/abs/hep-th/0610102}{{\tt arXiv:hep-th/0610102
  [hep-th]}}.

\bibitem{Kachru:2003sx}
S.~Kachru, R.~Kallosh, A.~D. Linde, J.~M. Maldacena, L.~P. McAllister,
  \textit{et al.}, {\em {Towards inflation in string theory}\/},
  \href{http://dx.doi.org/10.1088/1475-7516/2003/10/013}{JCAP {\bf 0310} (2003)
   013},
\href{http://arxiv.org/abs/hep-th/0308055}{{\tt arXiv:hep-th/0308055
  [hep-th]}}.

\bibitem{Ceresole:2006iq}
A.~Ceresole, G.~Dall'Agata, A.~Giryavets, R.~Kallosh, and A.~D. Linde, {\em
  {Domain walls, near-BPS bubbles, and probabilities in the landscape}\/},
  \href{http://dx.doi.org/10.1103/PhysRevD.74.086010}{Phys.Rev. {\bf D74}
  (2006)  086010},
\href{http://arxiv.org/abs/hep-th/0605266}{{\tt arXiv:hep-th/0605266
  [hep-th]}}.

\bibitem{Dine:2007er}
M.~Dine, G.~Festuccia, A.~Morisse, and K.~van~den Broek, {\em {Metastable
  Domains of the Landscape}\/},
  \href{http://dx.doi.org/10.1088/1126-6708/2008/06/014}{JHEP {\bf 0806} (2008)
   014},
\href{http://arxiv.org/abs/0712.1397}{{\tt arXiv:0712.1397 [hep-th]}}.

\bibitem{Sarangi:2007jb}
S.~Sarangi, G.~Shiu, and B.~Shlaer, {\em {Rapid Tunneling and Percolation in
  the Landscape}\/},
  \href{http://dx.doi.org/10.1142/S0217751X09042529}{Int.J.Mod.Phys. {\bf A24}
  (2009)  741--788},
\href{http://arxiv.org/abs/0708.4375}{{\tt arXiv:0708.4375 [hep-th]}}.

\bibitem{Tye:2007ja}
S.-H.~H. Tye, {\em {A Renormalization Group Approach to the Cosmological
  Constant Problem}\/},
\href{http://arxiv.org/abs/0708.4374}{{\tt arXiv:0708.4374 [hep-th]}}.

\bibitem{Podolsky:2008du}
D.~I. Podolsky, J.~Majumder, and N.~Jokela, {\em {Disorder on the
  landscape}\/},  \href{http://dx.doi.org/10.1088/1475-7516/2008/05/024}{JCAP
  {\bf 0805} (2008)  024},
\href{http://arxiv.org/abs/0804.2263}{{\tt arXiv:0804.2263 [hep-th]}}.

\bibitem{Brown:2010bc}
A.~R. Brown and A.~Dahlen, {\em {Small Steps and Giant Leaps in the
  Landscape}\/},  \href{http://dx.doi.org/10.1103/PhysRevD.82.083519}{Phys.Rev.
  {\bf D82} (2010)  083519},
\href{http://arxiv.org/abs/1004.3994}{{\tt arXiv:1004.3994 [hep-th]}}.

\bibitem{Brown:2007zzh}
A.~R. Brown, S.~Sarangi, B.~Shlaer, and A.~Weltman, {\em {A Wrinkle in
  Coleman-De Luccia}\/},
  \href{http://dx.doi.org/10.1103/PhysRevLett.99.161601}{Phys.Rev.Lett. {\bf
  99} (2007)  161601},
\href{http://arxiv.org/abs/0706.0485}{{\tt arXiv:0706.0485 [hep-th]}}.

\bibitem{Denef:2004ze}
F.~Denef and M.~R. Douglas, {\em {Distributions of flux vacua}\/},
  \href{http://dx.doi.org/10.1088/1126-6708/2004/05/072}{JHEP {\bf 0405} (2004)
   072},
\href{http://arxiv.org/abs/hep-th/0404116}{{\tt arXiv:hep-th/0404116
  [hep-th]}}.

\bibitem{Chen:2011ac}
X.~Chen, G.~Shiu, Y.~Sumitomo, and S.~H. Tye, {\em {A Global View on The Search
  for de-Sitter Vacua in (type IIA) String Theory}\/},
  \href{http://dx.doi.org/10.1007/JHEP04(2012)026}{JHEP {\bf 1204} (2012)
  026},
\href{http://arxiv.org/abs/1112.3338}{{\tt arXiv:1112.3338 [hep-th]}}.

\bibitem{Blum:1997:CRC:265020}
L.~Blum, F.~Cucker, M.~Shub, and S.~Smale, {\em Complexity and real
  computation}.
\newblock Springer-Verlag New York, Inc., Secaucus, NJ, USA, 1998.

\bibitem{Dedieu200889}
J.-P. Dedieu and G.~Malajovich, {\em On the number of minima of a random
  polynomial\/},  \href{http://dx.doi.org/10.1016/j.jco.2007.09.003}{Journal of
  Complexity {\bf 24} (2008) no.~2, 89 -- 108}.
  \url{http://www.sciencedirect.com/science/article/pii/S0885064X07001252}.

\bibitem{Ahlqvist:2010ki}
P.~Ahlqvist, B.~R. Greene, D.~Kagan, E.~A. Lim, S.~Sarangi, \textit{et al.},
  {\em {Conifolds and Tunneling in the String Landscape}\/},
  \href{http://dx.doi.org/10.1007/JHEP03(2011)119}{JHEP {\bf 1103} (2011)
  119},
\href{http://arxiv.org/abs/1011.6588}{{\tt arXiv:1011.6588 [hep-th]}}.

\bibitem{Danielsson:2006xw}
U.~H. Danielsson, N.~Johansson, and M.~Larfors, {\em {The World next door:
  Results in landscape topography}\/},
  \href{http://dx.doi.org/10.1088/1126-6708/2007/03/080}{JHEP {\bf 0703} (2007)
   080},
\href{http://arxiv.org/abs/hep-th/0612222}{{\tt arXiv:hep-th/0612222
  [hep-th]}}.

\bibitem{Johnson:2008kc}
M.~C. Johnson and M.~Larfors, {\em {Field dynamics and tunneling in a flux
  landscape}\/},  \href{http://dx.doi.org/10.1103/PhysRevD.78.083534}{Phys.Rev.
  {\bf D78} (2008)  083534},
\href{http://arxiv.org/abs/0805.3705}{{\tt arXiv:0805.3705 [hep-th]}}.

\bibitem{deAlwis:2006cb}
S.~de~Alwis, {\em {Transitions Between Flux Vacua}\/},
  \href{http://dx.doi.org/10.1103/PhysRevD.74.126010}{Phys.Rev. {\bf D74}
  (2006)  126010},
\href{http://arxiv.org/abs/hep-th/0605184}{{\tt arXiv:hep-th/0605184
  [hep-th]}}.

\bibitem{Freund:1980xh}
P.~G. Freund and M.~A. Rubin, {\em {Dynamics of Dimensional Reduction}\/},
\href{http://dx.doi.org/10.1016/0370-2693(80)90590-0}{Phys.Lett. {\bf B97}
  (1980)  233--235}.

\bibitem{RandjbarDaemi:1982hi}
S.~Randjbar-Daemi, A.~Salam, and J.~Strathdee, {\em {Spontaneous
  Compactification in Six-Dimensional Einstein-Maxwell Theory}\/},
\href{http://dx.doi.org/10.1016/0550-3213(83)90247-X}{Nucl.Phys. {\bf B214}
  (1983)  491--512}.

\bibitem{BrownA}
A.~Brown, A.~Dahlen, and A.~Masoumi, {\em {Trans-dimensional tunneling in
  Einstein-Maxwell theory}, Appears soon\/}, .

\bibitem{BlancoPillado:2009di}
J.~J. Blanco-Pillado, D.~Schwartz-Perlov, and A.~Vilenkin, {\em {Quantum
  Tunneling in Flux Compactifications}\/},
  \href{http://dx.doi.org/10.1088/1475-7516/2009/12/006}{JCAP {\bf 0912} (2009)
   006},
\href{http://arxiv.org/abs/0904.3106}{{\tt arXiv:0904.3106 [hep-th]}}.

\bibitem{Spradlin:2001pw}
M.~Spradlin, A.~Strominger, and A.~Volovich, {\em {Les Houches lectures on de
  Sitter space}\/},
\href{http://arxiv.org/abs/hep-th/0110007}{{\tt arXiv:hep-th/0110007
  [hep-th]}}.

\bibitem{Bekenstein:1973ur}
J.~D. Bekenstein, {\em {Black holes and entropy}\/},
\href{http://dx.doi.org/10.1103/PhysRevD.7.2333}{Phys.Rev. {\bf D7} (1973)
  2333--2346}.

\bibitem{Hawking:1974sw}
S.~Hawking, {\em {Particle Creation by Black Holes}\/},
\href{http://dx.doi.org/10.1007/BF02345020}{Commun.Math.Phys. {\bf 43} (1975)
  199--220}.

\bibitem{wess1992supersymmetry}
J.~Wess and J.~Bagger, {\em Supersymmetry and supergravity}.
\newblock Princeton series in physics. PRINCETON University Press, 1992.
\newblock \url{http://books.google.com/books?id=4QrQZ\_Rjq4UC}.

\bibitem{Bender:1968sa}
C.~M. Bender and T.~T. Wu, {\em {Analytic structure of energy levels in a field
  theory model}\/},
\href{http://dx.doi.org/10.1103/PhysRevLett.21.406}{Phys.Rev.Lett. {\bf 21}
  (1968)  406--409}.

\bibitem{Marsh:2011aa}
D.~Marsh, L.~McAllister, and T.~Wrase, {\em {The Wasteland of Random
  Supergravities}\/},  \href{http://dx.doi.org/10.1007/JHEP03(2012)102}{JHEP
  {\bf 1203} (2012)  102},
\href{http://arxiv.org/abs/1112.3034}{{\tt arXiv:1112.3034 [hep-th]}}.

\bibitem{Denef:2004dm}
F.~Denef, M.~R. Douglas, and B.~Florea, {\em {Building a better racetrack}\/},
  \href{http://dx.doi.org/10.1088/1126-6708/2004/06/034}{JHEP {\bf 0406} (2004)
   034},
\href{http://arxiv.org/abs/hep-th/0404257}{{\tt arXiv:hep-th/0404257
  [hep-th]}}.

\bibitem{Gukov:1999ya}
S.~Gukov, C.~Vafa, and E.~Witten, {\em {CFT's from Calabi-Yau four folds}\/},
  \href{http://dx.doi.org/10.1016/S0550-3213(00)00373-4}{Nucl.Phys. {\bf B584}
  (2000)  69--108},
\href{http://arxiv.org/abs/hep-th/9906070}{{\tt arXiv:hep-th/9906070
  [hep-th]}}.

\bibitem{deAlwis:2006am}
S.~de~Alwis, {\em {The Scales of brane nucleation processes}\/},
  \href{http://dx.doi.org/10.1016/j.physletb.2006.11.022}{Phys.Lett. {\bf B644}
  (2007)  77--82},
\href{http://arxiv.org/abs/hep-th/0605253}{{\tt arXiv:hep-th/0605253
  [hep-th]}}.

\end{thebibliography}\endgroup
\bibliographystyle{mybibstyle} 

\end{document}